\documentclass[12pt,a4paper]{article}
\pdfoutput=1
\usepackage{jheppub}
\usepackage{amsfonts}
\usepackage{bbm}
\usepackage{multirow}
\usepackage{enumitem}   
\usepackage{stmaryrd}
\usepackage{placeins}
\usepackage[none]{hyphenat}
\usepackage{makecell}
\setcellgapes{3pt}

\usepackage[dvipsnames]{xcolor}

\usepackage{tikz-cd}   
\usepackage{tikz}
\usepackage{pgfplots}
\usetikzlibrary{decorations.markings,calc,intersections,through,backgrounds,shapes,snakes,shapes.geometric,knots}
\usepgfplotslibrary{fillbetween}

\def \Z{\mathbb{Z}}
\def \R{\mathbb{R}}
\def \C{\mathbb{C}}
\def \Q{\mathbb{Q}}
\def \TT{\mathbb{T}}

\def \sch{{\mathfrak{ch}_2}}
\newcommand{\crep}[1]{{\llbracket #1 \rrbracket}}

\def \CL{\mathcal{L}}
\def \CA{\mathcal{A}}
\def \CB{\mathcal{B}}
\def \CC{\mathcal{C}}
\def \CH{\mathcal{H}}

\def \CF{\mathcal{F}}
\def \CQ{\mathcal{Q}}

\def \pt{{\text{pt}}}
\def \lk{{\ell k}}

\def \Spin{{\mathrm{Spin}}}
\def \Pin{{\mathrm{Pin}}}
\def \GB{{\Spin_G(X)}}
\def \MCG{{\mathrm{MCG}}}
\def \Tor{{\mathrm{Tor}\,}}
\def \Hom{{\mathrm{Hom}}}
\def \GL{{\mathrm{GL}}}
\def \ABK{{\mathrm{ABK}}}
\def \Arf{{\mathrm{Arf}}}
\def \PD{{\mathrm{PD}}}
\def \r{{\mathrm{R}}}
\def \ns{{\mathrm{NS}}}
\def \S{{\mathcal{S}}}
\def \T{{\mathcal{T}}}
\def \F{{(-1)^{\mathcal{F}}}}
\def \SL{{\mathrm{SL}}}
\def \Mp{{\widetilde{\mathrm{SL}}(2,\mathbb{Z})}}
\def \e{{\mathrm{e}}}
\def \c{{\mathrm{c}}}
\def \gap{{\mathrm{gap}}}
\def \scal{{\mathrm{scal}}}

\renewcommand{\tilde}{\widetilde}
\renewcommand{\hat}{\widehat}

\title{Spin-cobordisms, surgeries and fermionic modular bootstrap}

\author[1, 2]{Andrea Grigoletto,}
\author[3]{Pavel Putrov}

\affiliation[1]{SISSA, Via Bonomea 265, Trieste 34136, Italy }
\affiliation[2]{INFN, Sezione di Trieste, Via Valerio 2, 34127 Trieste, Italy}
\affiliation[3]{ICTP, Strada Costiera 11, Trieste 34151, Italy}

\abstract{We consider general fermionic quantum field theories with a global finite group symmetry $G$, focusing on the case of 2-dimensions and torus spacetime. The modular transformation properties of the family of partition functions with different backgrounds is determined by the 't Hooft anomaly of $G$ and fermion parity. For a general possibly non-abelian $G$ we provide a method to determine the modular transformations directly from the bulk 3d invertible topological quantum field theory (iTQFT) corresponding to the anomaly by inflow.  We also describe a method of evaluating the character map from the real representation ring of $G$ to the group which classifies anomalies. Physically the value of the map is given by the anomaly of free fermions in a given representation. We assume classification of the anomalies/iTQFTs by spin-cobordisms. As a byproduct, for all abelian symmetry groups $G$, we provide explicit combinatorial expressions for corresponding spin-bordism invariants in terms of surgery representation of arbitrary closed spin 3-manifolds. We work out the case of $G=\Z_2$ in detail, and, as an application, we consider the constraints that 't~Hooft anomaly puts on the spectrum of the infrared conformal field theory.}
\keywords{TQFT, SPT, Anomalies, Bootstrap}

\begin{document}
\tikzset{->-/.style={decoration={
  markings,
  mark=at position .5 with {\arrow{>}}},postaction={decorate}}}
\maketitle

\section{Introduction and summary}
\label{sec:intro}

Gravitational anomalies and anomalies of global symmetries provide one the few tools to analyze quantum field theories (QFTs) on non-perturbative level. In particular the anomalies contain certain robust information about the dynamics of a theory. They normally take values in a discrete set and are invariant under all possible continuous deformations, including renormalization group flow. Moreover, the set of possible values of anomalies has a natural abelian group structure with respect to stacking of theories. 

The anomalies\footnote{By default in this paper, by ``anomalies'' we mean 't Hooft anomalies of gravitational and global symmetries. In particular the considered quantum field theories with anomalies  are well defined, and the corresponding global symmetry is unbroken, when the background is trivial.} are usually understood as non-invariance under diffeomorphisms or gauge transformations of background gauge fields on the quantum level. The global anomalies exhibit themselves as non-invariance under ``large'' (meaning that they cannot be continuously deformed to identity) diffeomorphisms or gauge transformations.

On the other hand, it is well accepted that anomalies can be understood in terms of topological quantum field theories (TQFT) in one more dimensions, see e.g. \cite{Witten:2019bou}. If the anomalous quantum field theory is put on the boundary of the spacetime of the corresponding TQFT, the combined system becomes invariant under diffeomorphisms and gauge transformations. Consider a partition function of the theory on a closed manifold with some particular background metric and global symmetry gauge field. A combination of a diffeomorphism and a large gauge transformation that does not change the background (up to homotopy) then takes the partition function to itself (possibly with a deformed background), up to an anomalous phase. The extra phase equals to the partition function of the bulk TQFT on the corresponding mapping class torus \cite{Witten:1982fp,Witten:1985xe,Witten:2015aba}.

More generally, however, one can consider a diffeomorphism that changes the homotopy class of the background gauge fields of global symmetries (note that all metrics are homotopic to each other). Such a diffeomorphism then takes the partition function to a one with a different background, again, up to an anomalous phase. The phase is then determined by the linear evolution map that TQFT associates to the corresponding mapping cylinder.

The set of the phases that can appear also form an abelian group under stacking of decoupled theories together. The same can be said about the invertible bulk TQFTs, with a given global symmetry, that describe the anomalies by inflow. In this paper we will often call the abelian group, which classifies anomalies, the \textit{anomaly group} or the \textit{group of anomalies} for short, although it is not a very standard convention. Schematically we have the following homomorphism of abelian groups:
\begin{equation}
    \left\{
        \text{invertible TQFTs}
    \right\}
    \qquad \longrightarrow \qquad
    \left\{
        \text{anomalous phases}
    \right\}.
    \label{intro-map}
\end{equation}
One of the main goals of this work is to make this homomorphism and the group in the right hand side precise and explicit, using the classification of the invertible (unitary, or, equivalently, reflection-positive in the Euclidean setup) TQFTs by cobordisms \cite{Kapustin:2014tfa,Kapustin:2014dxa,Freed:2016rqq,Yonekura:2018ufj}. A similar construction in the case of bosonic TQFTs classified by group cohomology \cite{Dijkgraaf:1989pz,Freed:1991bn}, when the 2d spacetime is two-torus was considered in \cite{Freed:1991bn,ganter2009hecke,Yu:2021kzr}. Note that invertible unitary TQFTs are often also referred to as Symmetry Protected Topological states (SPTs), although in some definitions there are subtle differences (e.g. sometimes SPTs do not include invertible TQFTs corresponding to purely gravitational anomalies). In this paper we will use the term SPT as a synonym to invertible unitary TQFT. We will also  assume by default that all considered TQFTs are unitary.

One can use this map in two ways. First, it can be used to determine anomaly of a given theory,  if one knows how to calculate a partition and determine how it transforms under large diffemorphisms. The other way around, one can use this map to determine what constraints a given anomaly imposes on the dynamics of the theory, for example using modular bootstrap. 

Note that one can often determine the anomalous phases indirectly, by considering a particular theory that realizes a given anomaly. One of the most common ways to realize anomalies is to consider free fermions transforming in a non-trivial way under the considered symmetry. However it may happen that partition functions in certain background vanish (unless appropriate observables are inserted, as in the case of free fermions), which can lead to ambiguities in determining the phases. Calculating anomalies using directly invertible TQFTs leaves no ambiguity about which features are general and which are specific for a particular theory. 

Nevertheless, considering free fermion theories provides us with a natural homomorphism from the free abelian group generated by representations of the symmetry (also known as representation ring, if one also considers the tensor product operation) to the group of anomalies.  This map takes a representation to the anomaly of the theory of free fermions in this representation. So altogether we have a sequence of two homomorphisms, which schematically reads as follows: 
\begin{equation}
\left\{
        \text{representations}
    \right\}
      \longrightarrow
    \left\{
        \text{invertible TQFTs}
    \right\}
     \longrightarrow 
    \left\{
        \text{anomalous phases}
    \right\}.
    \label{intro-map-triple}
\end{equation}
The maps in the sequence are natural in the sense that, if one considers a homomorphism between two different symmetry groups (e.g. corresponding to breaking of a symmetry to a subgroup), the induced maps between the three groups in the sequence (\ref{intro-map-triple}) commute with the maps in the sequence. In mathematical terminology this means that we have a sequence of natural transformations between three (contravariant) functors from the category of groups to the category of abelian groups. As we will see this property is very constraining and can significantly help in doing explicit calculations. A version of this property has been used previously to determine global anomalies from local ones, with focus on the cases where at least one of the symmetry groups is continuous \cite{Witten:1983tw,Elitzur:1984kr,Ibanez:1991hv,Ibanez:1991pr,Davighi:2020bvi,Davighi:2020uab}. 

As a main example in this paper we consider the case of $\Z_2^f\times \Z_2$ global symmetry in two spacetime dimensions. Here $\Z_2^f$ is fermionic parity and $\Z_2$ a unitary global symmetry. Assuming absence of perturbative gravitational anomaly, the anomalies in this case are known to be classified by a $\Z_8$ group \cite{Ryu:2012he,Gu:2013azn}. In this case the map (\ref{intro-map}) turns out to be injective. Note that the anomalous phases due to perturbative gravitational anomalies were calculated from the corresponding bulk TQFT for example in  \cite{Golkar:2015oxw,Chowdhury:2016cmh}.
First we determine the anomalous phases directly from the values of the partition function of the bulk TQFT on certain closed manifolds. Then, as an application, following the techniques of \cite{Collier:2016cls,Lin:2019kpn,Benjamin:2020zbs,Lin:2021udi}, we perform the modular bootstrap of the spectrum of the conformal field theories with given anomaly. We also show how the anomalous action of the symmetry generators on the Hilbert space analyzed in  \cite{Delmastro:2021xox} can be geometrically interpreted in terms of the extension of the corresponding charge operators to the bulk TQFT.

Let us note that fermionic theories in principle can  always be studied in terms of corresponding bosonic theories via bosonization/fermionization map \cite{Gaiotto:2015zta,Novak:2015ela,Bhardwaj:2016clt,Thorngren:2018bhj,Karch:2019lnn,Ji_2020,Fukusumi:2021zme,Thorngren:2021yso, Gaiotto:2020iye}. It however becomes more subtle to distinguish fermionic theories with different anomalies of the global symmetries after the bosonization. In this work we will study the fermionic theories directly, without relating them to their bosonic counterparts. 

Finally, let us note that understanding the group of anomalous phases for a given symmetry group $G$ in two dimensions is relevant for description of the $G$-equivariant version of the generalized cohomology theory of topological modular forms (TMF) \cite{hopkins1995topological}. The latter physically classifies the minimally supersymmetric theories with global symmetry $G$ \cite{stolz2004elliptic,stolz2011supersymmetric,Gukov:2018iiq}.

The rest of the paper is organized as follows. In Section \ref{sec:mod-TQFT} we provide a broad overview about the description of modular transformations in terms of the bulk invertible TQFT that cancels the anomaly by inflow. In Section \ref{sec:Z2case} we consider in detail the case of $\Z_2^f\times \Z_2$ symmetry, where we consider three different (but closely related) approaches to determining anomalous phases on 2-torus $\TT^2$: 1) explicit calculation, using geometric definition of the corresponding spin-bordism invariant, 2) using operations on symmetry defects in 2d, 3) using surgery representation of closed 3-manifolds. In Section \ref{sec:gen-groups} we provide a method of explicit evaluation of the maps in the sequence (\ref{intro-map-triple}) for 2d theories on $\TT^2$ with general, possibly non-abelian, symmetry groups. In Section \ref{sec:bootstrap} we come back to the particular case of  $\Z_2^f\times \Z_2$ symmetry and use the modular transformations and the known modular bootstrap techniques to provide certain constrains on the spectrum of fermionic conformal theories with given global anomalies. 

\section{Modular transformations and cobordisms}
\label{sec:mod-TQFT}

In this section we collect relevant known facts that establish the relation between the modular transformation properties of partition functions of $d$-dimensional QFTs and $(d+1)$-dimensional cobordism invariants. 

\subsection{Anomalous phases and invertible TQFT}
\label{sec:anomalous-phases}

As it is well known, the global anomalies of a QFT are directly related to the non-invariance of a partition function $Z(X,g,a)$ under large diffeomorphisms and gauge transformations. Here $X$ denotes a closed spacetime manifold, $g$ is its metric, and $a$ is the background gauge field of the global symmetry. In practice the non-invariance means that in the process of defining $Z(X,g,a)$ (e.g. using a path integral) one has to make some auxiliary choices that are not respected by large diffeomorphisms or automorphisms of the gauge bundle.

For concreteness we assume that there is no time-reversal symmetry. However this is not crucial for our general discussion below, which can be easily generalized. We will work in a Euclidean spacetime. Let us denote the internal global symmetry of the theory as $G^f$, which we assume to be discrete. It includes the fermionic parity $\Z_2^f$, generated by $(-1)^F$, as a subgroup: $\Z_2^f\subset G^f$. Note that in our setup the symmetry is not required to act faithfully on the operators of the theory. In particular, the theory is allowed to be purely bosonic, so that $\Z_2^f$ acts trivially. In general we will denote the ``bosonic'' symmetry by $G:=G^f/
\Z_2^f$, so that $G^f$ is an extension of $G$ by $\Z_2^f$. The symmetry group that also includes local rotations is then $G_d:=G^f\times_{\Z_2^f} \Spin(d)$ and can be understood as an extension of $SO(d)$ by $G^f$:
\begin{equation}
    1\longrightarrow G^f\longrightarrow G_d\equiv G^f\times_{\Z_2^f} \Spin(d)\longrightarrow SO(d)\longrightarrow 1
\end{equation}
which is obtained from the standard extension
\begin{equation}
    1\longrightarrow \Z_2^f\longrightarrow \Spin(d)\longrightarrow SO(d)\longrightarrow 1
\end{equation}
by taking the fiber product of the first two groups with $G^f$ over $\Z_2^f$.

In the case of present time-reversal symmetry $\mathbf{T}$ the $\Spin(d)$ group appearing above and below should be replaced by $\text{Pin}^{\pm}(d)$ group, depending on whether $\mathbf{T}^2=1$ or $\mathbf{T}^2=(-1)^F$, the $SO(d)$ group replaced with $O(d)$, and $GL_+(d,\R)$ replaced with $GL(d,\R)$. 

The choice of background $a$ in $Z(X,g,a)$ can be understood as a choice of principal $G_d$ bundle, equipped with a bundle map to $O(d)$ principal bundle of oriented orthonormal frames in the tangent bundle. The bundle map should take the $G_d$ action to the $SO(d)$ action on the fibers according to the homomorphism $G_d\rightarrow SO(d)$.

By now it is widely accepted that the anomalies of general $d$-dimensional QFTs are governed by invertible reflection-positive (i.e. unitary) TQFTs, or equivalently SPTs, in $(d+1)$-dimensions that have the same global symmetries. In particular, if $X$, the spacetime of the QFT, is a boundary of a $(d+1)$-dimensional spacetime $Y$ of the TQFT, the total system should be invariant under diffeomorphisms and gauge transformations. That is anomaly cancels by ``inflow'' from the bulk.  

We consider a functorial definition of TQFT following \cite{Freed:2016rqq,Yonekura:2018ufj}. To make the bulk theory \textit{topological}, instead of the compact Lie group $G_d=G^f\times_{\Z_2^f} \Spin(d)$ one should consider the corresponding non-compact group
\begin{equation}
    H_d:=G^f\times_{\Z_2^f} \tilde{\GL}_+(d,\R)
\end{equation}
where $\tilde{\GL}_+(d,\R)$ is the double cover of $\GL_+(d,\R)$, the group of general linear transformations of $\R^d$ preserving orientation (i.e. given by matrices with positive determinant). Note that the group $\GL_+(d,\R)$ deformation retracts to $SO(d)$, while $\tilde{\GL}_+(d,\R)$ deformation retracts to $\Spin(d)$, so that the following diagramm commutes:
\begin{equation}
       \begin{tikzcd}
           & 
           &
          \Spin(d) \ar[r] & 
          SO(d) \ar[dr] &
           \\
          1\ar[r] & 
          \Z_2^f \ar[ur]\ar[dr] &
          &
          &
          1.
          \\
               & 
          &
          \tilde{\GL}_+(d) \ar[r]\ar[uu] & 
          \GL_+(d) \ar[ur]\ar[uu] &
    \end{tikzcd}
\end{equation}
Let
\begin{equation}
    \rho_d:\;H_d\equiv G^f\times_{\Z_2^f} \tilde{\GL}_+(d,\R) \longrightarrow \GL_+(d,\R)
\end{equation}
be the canonical projection map. In the functorial approach a $(d+1)$-dimensional TQFT with internal symmetry $G$ then associates a vector space to a smooth\footnote{There exists a version without requiring smooth structure.} $d$-manifold manifold with $H_d$ structure, or $H_d$-manifold for short. Namely, an $H_d$-manifold is a triple $(X,P,\theta)$, where: 1) $X$ is an oriented $d$-manifold, 2) $P\rightarrow X$ is a principal $H_d$ bundle, 3) $\theta$ is a bundle map to $F_+X$, the $\GL_+(d)$ principal bundle of oriented frames in the tangent bundle $TX$:
\begin{equation}
       \begin{tikzcd}
           H_d \ar[r] & 
           P \ar[r,"\theta"] \ar[d] & 
           F_+X \ar[dl] &
           \GL_+(d,\R) \ar[l]
                         \\
               & 
               X
          &
        &
    \end{tikzcd}.
\end{equation}
The bundle map is required to commute with the corresponding group actions on the fibers: $\theta(g\cdot p)=\rho_d(g)\cdot \theta(p)$ where $p\in P$, $g\in H_d$. 
An isomorphism between a pair of $H_d$-manifolds $(X,P,\theta)$ and $(X',P',\theta')$  is a smooth bundle map $\tilde{\psi}$ covering a diffeomorphism $\psi$ such that the following diagram is commutative:
\begin{equation}
    \begin{tikzcd}
        F_+X \ar[r,"\psi_*"]\ar[d] &  
        F_+X'\ar[d] \\
        X \ar[r,"\psi"]& 
        X' \\
        P \ar[r,"\tilde{\psi}"]\ar[u]\ar[uu,bend left=50,"\theta"] &
        P' \ar[u]\ar[uu,bend right=50,"\theta'",swap]
    \end{tikzcd}
\end{equation}
where $\psi_*$ is the pushforward (differential) of $\psi$. Sometimes, for the sake of brevity, we will denote an $H_d$-manifold simply as $X$ (instead of $(X,P,\theta)$), keeping in mind that it comes equipped with a particular choice of $H_d$-structure.

A TQFT then associates an isomorphism $\hat{\tilde{\psi}}$ between the pair of the corresponding vector spaces:
\begin{equation}
    \hat{\tilde{\psi}}:\;\CH(X,P,\theta) \longrightarrow    
    \CH(X',P',\theta')
    \label{TQFT-iso-map}
\end{equation}
which is functorial with respect to composition: $\widehat{\tilde{\psi}\circ \tilde{\phi}}=\hat{\tilde{\psi}}\circ \hat{\tilde{\phi}}$. More generally, a TQFT associates a linear map to a bordism between a pair of $H_d$ manifolds: an $H_{d+1}$-manifold equipped with isomorphism of its boundary (considered as an $H_d$-manifold) to the disjoint union of this pair, with orientation of one of the $H_d$-manifold flipped. The bordisms are considered modulo isomorphisms of $H_{d+1}$-manifolds identical at the boundary. The special case (\ref{TQFT-iso-map}) is then realized by taking the bordism to be $X\times [0,1]$ with the product $H_{d+1}$ structure and the boundary isomorphisms given by the identity and $\tilde{\psi}$.

In practice, for concrete calculations, it may be not convenient to work with all possible triples $(X,P,\theta)$. Instead, let us fix one particular oriented $d$-manifold $X$ in its orientation-preserving diffeomorphism class. For this fixed manifold consider equivalence relation between the triples given by the isomorphisms acting identically on $X$, i.e. the bundle isomorphism $\mu$ such that the following diagram is commutative:
\begin{equation}
    \begin{tikzcd}
    &
        F_+X \ar[d] &  
         \\
         &
        X & 
         \\
        P \ar[rr,"\mu"]\ar[ur]\ar[uur,bend left=20,"\theta"] &
        &
        P' \ar[lu]\ar[uul,bend right=20,"\theta'",swap]
    \end{tikzcd}.
\end{equation}
Denote the set of equivalence classes as $\GB$. For each class $a\in \GB$ we then fix a particular representative bundle $P_a\rightarrow X$ and a map $\theta_a\rightarrow F_+X$. A TQFT then provides us with a family of vector spaces
\begin{equation}
    \CH_a:=\CH(X,P_a,\theta_a),\qquad a\in \GB.
\end{equation}
The equivalence class $a$ can be understood as the choice of the background ${G^f\times_{\Z_2^f} \Spin(d)}$-structure for some fixed metric on $X$. When the group $G$ is finite, there is a finite number of equivalence classes. Moreover, if the internal symmetry is of the form $G^f=\Z_2^f\times G$, and $G$ is a discrete abelian group, one can understand it as a pair $a=(s,a')\in \GB$ where $a'\in H^1(X,G)$ is the background $G$ gauge field and $s\in \Spin(X)$ is a choice of spin structure on $X$. 

By $\psi^{-1*}a$ let us denote the pullback of the structure $a$ with respect to the inverse of the diffeomorphism $\psi:X\rightarrow X$. Namely, it is the equivalence class of the triple $(X,\psi^{-1*}P_a,\psi_*\theta_a \psi^*)$. We then restrict our attention to the isomorphisms between the finite set of triples $(X,P_a,\theta_a)$:
\begin{equation}
        \begin{tikzcd}
        F_+X \ar[r,"\psi_*"]\ar[d] &  
        F_+X\ar[d] \\
        X \ar[r,"\psi"]& 
        X \\
        P_a \ar[r,"\tilde{\psi}"]\ar[u]\ar[uu,bend left=50,"\theta_a"] &
        P_{\psi^{-1*}a} \ar[u]\ar[uu,bend right=50,"\theta_{\psi^{-1*}a}",swap]
    \end{tikzcd}.
    \label{covering-auto}
\end{equation}
A TQFT associates to them maps
\begin{equation}
    \hat{\tilde{\psi}}:\;\CH_a \longrightarrow        
    \CH_{\psi^{-1*}a}.
    \label{TQFT-iso-map-fixed}
\end{equation}
The maps only depend on the homotopy classes of the maps $\psi$ as the theory is topological. This is because one can relate the homotopic maps by an isomorphism inside the cylinder $X\times [0,1]$.

For a fixed diffeomorphism $\psi:X\rightarrow X$ the covering bundle map $\psi$ is fixed up to compositions automorphisms of $H_d$-structures:
\begin{equation}
    \begin{tikzcd}
    &
        F_+X \ar[d] &  
         \\
         &
        X & 
         \\
        P_a \ar[rr,"\mu"]\ar[ur]\ar[uur,bend left=20,"\theta_a"] &
        &
        P_a \ar[lu]\ar[uul,bend right=20,"\theta_a",swap]
    \end{tikzcd}
\end{equation}
which are finite number of, assuming again that $G$ is finite.  They similarly define the action
\begin{equation}
    \hat{\mu}:\;\CH_a \longrightarrow        
    \CH_{a}.
    \label{TQFT-auto-map-fixed}
\end{equation}
 All the isomorphisms (\ref{covering-auto}), considered up to homotopy, form a group, which we denote by $\widetilde\MCG(X)$. If $X$ is connected, the automorphisms (\ref{TQFT-auto-map-fixed}) form a subgroup group isomorphic to $G^f$. The total group $\widetilde\MCG(X)$ is then an extension of $\MCG(X)$, the orientation preserving mapping class group of $X$, by $G^f$:
\begin{equation}
    1\longrightarrow G^f\stackrel{i}{\longrightarrow}
    \widetilde{\MCG}(X) \stackrel{\pi}{\longrightarrow} \MCG(X) 
    \longrightarrow 1.
\end{equation}
When $G^f$ is abelian, the extension is central. Elements of $\MCG(X)$ act of the space isomorphism classes of structures $\GB$:
\begin{equation}
\begin{tikzcd}
   \MCG(X)\times \GB \ar[r]&
   \GB, 
   \\
    ([\psi],a) \ar[r] & {[\psi]\cdot a\equiv \psi^{-1*}a},
\end{tikzcd}
\end{equation}
where $\psi:X\rightarrow X$ is a diffeomorphism representing an element $[\psi]\in \MCG(X)$. The full $\widetilde\MCG(X)$ then also acts on $\Spin_G(X)$ through the projection $\pi$.

The partition function $Z(X,g,a)$ of the boundary $d$-dimensional QFT can be understood as an element of $\CH_a$, which is one-dimensional. To assign to it a particular number one still needs to choose a basis element $e_a$ in the vector space, so that 
\begin{equation}
    Z(X,g,a)e_a \in \CH_a.
\end{equation}
The linear maps (\ref{TQFT-iso-map-fixed}) then act on the basis elements as follows:
\begin{equation}
    \widehat{\tilde{\psi}}:\;\; e_a\;
    \longmapsto C_{[\tilde\psi]}(a)\,e_{[\psi]\cdot a}
\end{equation}
where $[\tilde\psi]\in \widetilde\MCG(X)$ and $[\psi]\equiv \pi([\tilde\psi])\in \MCG(X)$. We can assume that $C_{[\tilde\psi]}(a)\in U(1)\subset \C$ because of unitarity. The coefficients must satisfy the following cocycle condition:
\begin{equation}
    C_{[\tilde\psi][\tilde\phi]}(a)=
    C_{[\tilde\psi]}([\phi]\cdot a)\,
    C_{[\tilde\phi]}(a).
\label{eq:cocyclecondition}
\end{equation}
Moreover, the change of basis $e_a\rightsquigarrow \alpha(a)\,e_a$ results in the following redefinition of the coefficients $C$:
\begin{equation}
    C_{[\tilde\psi]}(a)\rightsquigarrow
    C_{[\tilde\psi]}(a)\,\frac{\alpha([\psi]\cdot a)}{\alpha(a)}.
    \label{C-redef}
\end{equation}
The coefficients $C$, regarded as a function\footnote{Due to unitarity of TQFT, without loss of generality we can assume that $C$ and $\alpha$ are valued in $U(1)\subset \C^*$.}
\begin{equation}
    C:\;\widetilde{\MCG}(X)\longrightarrow U(1)^{\GB}
\end{equation}
modulo redefinitions (\ref{C-redef}), define an element in the first group cohomology of $\widetilde\MCG(X)$:
\begin{equation}
    [C]\;\in\; H^1(\widetilde\MCG(X),U(1)^\GB).
    \label{MCG-cohomology}
\end{equation}
Here $U(1)^\GB$ is considered as a module of $\widetilde\MCG(X)$ with the action determined by the action on $\GB$ defined above. Equivalently, one can consider $C$ as a representation of the action groupoid $\widetilde\MCG(X) \ltimes \GB$. In its categorical description, the objects are elements $a\in \GB$ and the arrows (i.e. morphisms) between the pair of objects $(a_1,a_2)$ are elements $[\tilde\psi]\in \widetilde{\MCG}(X)$ such that $a_2=[\tilde\psi]\cdot a_1$. Recall that in the groupoid representation, to each object one associates a vector space and to each arrow a linear map. Therefore (\ref{TQFT-iso-map-fixed}) provides a representation of the groupoid where all vector spaces are one-dimensional. 

Representations of a grupoid are known to be described in terms of individual representations of its connected components. For each connected component, up to isomorphisms, there is a one-to-one correspondence between groupoid representations and representation of the group of closed loops in the groupoid starting and ending at some fixed ``base'' object $a_*$ (does not matter which). The correspondence is realized as follows. First, it is clear that a groupoid representation provides a representation of the group of loops. To construct a representation of a connected groupoid from the representation of the loop group one can proceed as follows. For each object $a$ in the groupoid choose an arrow $\ell_a$ to the base object $a_*$. Then, take $\CH_{a}:=\CH_{a_*}$ and to an arrow $g:a_1\rightarrow a_2$ assign a linear map which was assigned to the loop $\ell_{a_2}\circ g\circ \ell_{a_1}^{-1}$.

In our setup, the connected components of the groupoid correspond to the orbits of $\widetilde\MCG(X)$ action on $\Spin_G(X)$. Let $a_i\in\GB$ be some representatives of the orbits and $\widetilde\MCG_i:=\mathrm{Stab}_{a_i}(\widetilde\MCG(X))$ be their stabilizer subgroups in $\widetilde\MCG(X)$. We then have the following decomposition in terms of the product over the orbits:
\begin{multline}
    H^1(\widetilde\MCG(X),U(1)^\GB)\cong\\
    H^1(\widetilde\MCG(X) \ltimes \GB,U(1))\cong\\
    \prod_i H^1(\widetilde\MCG_i ,U(1)).
    \label{MCG-cohomogy-grupoid}
\end{multline}
 Note that although the right hand side of (\ref{MCG-cohomogy-grupoid}) is expressed in terms of group cohomology with trivial action on the coefficients (unlike the left hand side), in practice it is often easier to describe the group structure of full group $\widetilde\MCG(X)$ than of its stabilizer subgroups $\widetilde\MCG_i$. Note that representations of this groupoid can be also understood as line bundles on the corresponding stack $\GB//\MCG(X)$. 

The phases $C_{[\tilde\psi]}(a)$ determine the anomalous phases that appear in the partition function under the large gauge transformations and diffeomorphisms. If the automorphisms of the $H_d$ structure (i.e. the elements of $\widetilde\MCG(X)$ that project to a trivial element of $\MCG(X)$) act non-trivially on $\CH_a$, this means that  the partition function $Z(X,g,a)$ vanishes, unless one inserts some observables transforming non-trivially under the global symmetries, so that the anomalous phase is compensated. Otherwise, the partition function transforms as follows under a large diffeomorphism $\psi$:
\begin{equation}
    Z(X,\psi^{-1*}g,\psi^{-1*}a)=C_{[\tilde\psi]}(a)Z(X,g,a).
\end{equation}
The elements (\ref{MCG-cohomology}) in the group cohomology capture the robust information about the anomaly, invariant under redefinitions of the partition function.

\subsection{Modular transformations from cobordism invariants}

The invertible TQFTs with a given symmetry form a group under stacking (tensor product).  Under the assumption that the $d$-dimensional QFT is free of perturbative anomalies, one can restrict attention to the TQFTs responsible for non-perturbative anomalies. Their deformation classes form the torsion subgroup in the group of all $(d+1)$-dimensional TQFTs.  They are known to be classified by the Pontryagin dual of the torsion subgroup of the bordism group of $(d+1)$-manifolds with $G^f\times_{\Z_2^f}\Spin$ structure \cite{Kapustin:2014dxa,Freed:2016rqq,Yonekura:2018ufj}:
\begin{equation}
    \Tor \{\text{ref.-pos. $(d+1)$-dim iTQFTs}\}/\sim_\text{def}
    \;\cong\;
    \Hom( \Tor\Omega_{d+1}^{G^f\times_{\Z_2^f} \Spin},U(1)).
    \label{Tor-SPT-classification}
\end{equation}
In the case when $G^f=\Z_2^f\times G$, the relevant bordism group can be understood as the spin bordism group of the classifying space of $G$:
\begin{equation}
    \Omega_{d+1}^{G^f\times_{\Z_2^f} \Spin}
    \cong
    \Omega_{d+1}^{\Spin}(BG).
\end{equation}
The elements of the group in the right-hand side of (\ref{Tor-SPT-classification}) can be understood as bordism invariants. 

The classifying group on the right-hand side of (\ref{Tor-SPT-classification}) is canonically isomorphic to the group of connected components of the Pontryagin dual of the full bordism group\footnote{Note that
\begin{equation}
    \Tor\Omega_{d+1}^{G^f\times_{\Z_2^f} \Spin}\cong 
    \Omega_{d+1}^{G^f\times_{\Z_2^f} \Spin}
\end{equation}
when $G^f$ is discrete and $d\neq -1\mod 4$.
}:
\begin{equation}
    \Hom( \Tor\Omega_{d+1}^{G^f\times_{\Z_2^f} \Spin},U(1))\cong
     \pi_0\Hom( \Omega_{d+1}^{G^f\times_{\Z_2^f} \Spin},U(1)).
\end{equation}
As will be reviewed below (mostly following \cite{Yonekura:2018ufj}) to a fixed $U(1)$-valued bordism invariant
\begin{equation}
    \nu \in \Hom(\Omega_{d+1}^{G^f\times_{\Z_2^f} \Spin},U(1))
    \label{nu-nontorsion}
\end{equation}
one can construct an invertible $(d+1)$-dimensional TQFT with internal symmetry $G$ in the corresponding deformation class.

The bordism groups $\Omega_{d+1}^{G^f\times_{\Z_2^f} \Spin}$ (as well as their Pin versions), together with the corresponding invariants, have been calculated for various choices of the global symmetry group $G^f$ \cite{milnor1963spin,kirby_taylor_1991,Freed:2016rqq,Guo:2017xex,beaudry2018guide,Hsieh:2018ifc,Garcia-Etxebarria:2018ajm,Guo:2018vij,Wan:2018bns,Wan:2019oax,Wan:2019fxh}, in part motivated by the classification statement above. Considering a TQFT corresponding to an element of the group in the right hand side of (\ref{Tor-SPT-classification}) on a fixed $d$-manifold $X$ provides a homomorphism of abelian groups:
\begin{equation}
    \Hom(\Omega_{d+1}^{G^f\times_{\Z_2^f} \Spin},U(1))
    \longrightarrow
    H^1(\widetilde\MCG(X),U(1)^\GB).
    \label{anomaly-MCG-cohomology-map}
\end{equation}

An explicit functorial description of an invertible TQFT corresponding to a particular group element in the right hand side of (\ref{Tor-SPT-classification}) can be obtained by the method described in \cite{Yonekura:2018ufj}. This is a version of ``universal construction'' used in \cite{blanchet1995topological} to extend invariants of closed manifolds to a TQFT. Below we briefly review the main points of the construction, omitting some subtle details (in particular details related to ordering of manifolds in disjoint union and orientation reversal)  for the moment. The reader is welcome to follow the systematic discussion in \cite{Yonekura:2018ufj}. Such details will be however relevant for concrete calculations later on.

The first step of the construction is to fix a representative $H_d$-manifold $(X_\alpha,P_\alpha,\theta_\alpha)$ for each element $\alpha\in \Omega_d^{G^f\times_{\Z_2^f} \Spin}$ of the bordism group in one less dimension. Then a Hilbert space associated to an arbitrary $H_d$-manifold $(X,P,\theta)$ is defined as a quotient of an infinite dimensional vector space generated by all possible $(d+1)$-dimensional bordisms from it to the fixed representative $(X_\alpha,P_\alpha,\theta_\alpha)$ in its bordism class $\alpha=[(X,P,\theta)]\in\Omega_d^{G^f\times_{\Z_2^f} \Spin}$:
\begin{equation}
    \CH(X,P,\theta):=\frac{\bigoplus\limits_{Y\in \{\text{Bordism $(X_\alpha,P_\alpha,\theta_\alpha)\rightarrow (X,P,\theta)$\}}} |Y\rangle }{|Y_1\rangle \;\sim \;\nu([Y_1\cup \overline{Y_2}])\,|Y_2\rangle}
    \label{H-bordism-def}
\end{equation}
where $[Y_1\cup \bar{Y_2}]$ is the class a closed $H_{d+1}$-manifold obtained by gluing together $Y_1$ and $\overline{Y_2}$ along the common boundary $(X,P,\theta)\sqcup \overline{(X_\alpha,P,\theta)}$, where bar denotes reversing of orientation. The equivalence relation in used in the quotient identifies any two bordisms up to a phase given by the value of the invariant $\nu$ evaluated on the closed manifold obtained by such a gluing. It follows that the resulting vector space is one-dimensional. A basis vector can be fixed by choosing a particular bordism from $(X_\alpha,P_\alpha,\theta_\alpha)$ to $(X,P,\theta)$. Note that choosing a different representative $(X_\alpha',P'_\alpha,\theta_\alpha')$ in the same bordism class $\alpha$ would result in an isomorphic vector space. An explicit isomorphism can be constructed by choosing a particular bordism $(X_\alpha',P_\alpha',\theta_\alpha')\rightarrow (X_\alpha,P_\alpha,\theta_\alpha)$.

Using the definition (\ref{H-bordism-def}), the TQFT action of a bordism $W:(X,P,\theta)\rightarrow (X',P',\theta')$ on the Hilbert spaces is simply given by the composition of bordisms:
\begin{equation}
    \begin{array}{rrcl}
         \hat{W}:\; & 
         \CH(X,P,\theta) &
         \longrightarrow &
         \CH(X',P',\theta')
         \\
         &
         |Y\rangle
         & 
         \longmapsto
         &
         |W\circ Y\rangle
    \end{array}
\end{equation}
which is well defined, as the bordism classes of $(X,P,\theta)$ and $(X',P',\theta')$ are necessarily the same, so they have the same representative $(X_\alpha,P_\alpha,\theta_\alpha)$.

Consider then the setup described in Section (\ref{sec:anomalous-phases}). Let us realize the basis elements $e_a$ in terms of particular reference bordisms $Y_a$:
\begin{equation}
    e_a:=|Y_a\rangle\,\in \CH_a\equiv \CH(X,P_a,\theta_a),\qquad Y_a:(X_\alpha,P_\alpha,\theta_\alpha)\rightarrow (X,P_a,\theta_a)
\end{equation}
where $\alpha=[(X_a,P_a,\theta_a)]$. Note that in principle one can always choose $(X_\alpha,P_\alpha,\theta_\alpha)=(X,P_b,\theta_b)$ for some $b$ in the same orbit the mapping class group action as $a$. By considering an isomorphism $\tilde\psi$ in (\ref{covering-auto}) as a bordism
\begin{equation}
    \tilde{\psi}:\;(X,P_a,\theta_a)\longrightarrow 
    (X,P_{\psi^{-1*}a},\theta_{\psi^{-1*}a})
\end{equation}
it then follows that
\begin{equation}
    C_{[\tilde\psi]}(a)=
    \nu((\tilde\psi\circ Y_a)\cup \overline{Y_{\psi^{-1*}a}})
    \label{C-nu-formula}
\end{equation}
where the union means gluing of the bordisms along the common boundary $(X,P_{\psi^{-1}*a},\theta_{\psi^{-1*}a})\sqcup \overline{(X_\alpha,P_\alpha,\theta_\alpha)}$. Note that a change of various choices made above (such as choice of representative manifolds in the bordism classes and the choice of basis bordisms) will result in the redefinition of the coefficients $C$ of the form (\ref{C-redef}). This will not change the cohomology class $[C]$ in (\ref{MCG-cohomology}). Therefore the formula (\ref{C-nu-formula}) provides us with the map (\ref{anomaly-MCG-cohomology-map}). The bosonic version of such a map (in the case of $d=2$ and $X=\TT^2$, a 2-torus), when the left hand side is replaced by group cohomology $H^{3}(BG,U(1))$, was considered in \cite{Freed:1991bn,ganter2009hecke,Yu:2021kzr} (see also \cite{Dijkgraaf:1989pz,Gaberdiel:2012gf,Bantay:1990yr,Roche:1990hs}).
The generalization to the fermionic case that we consider also allows construction of representation $C$ using more general closed manifolds, not just mapping tori. As we will see, this can sometimes simplify calculations.

\section{The case of $\Z_2^f \times \Z_2$ symmetry}
\label{sec:Z2case}

In this section we discuss the case of $2$-dimensional fermionic QFTs with a $\Z_2$ global symmetry as a working example of the general framework presented above. This means that we will consider how theories behave under modular transformations depending on the anomaly that characterizes their symmetry group $G^f=\Z_2 \times \Z_2^f$ (with bosonic symmetry $G=\Z_2$). In order to do so we study explicitly their relation with $3$-dimensional invertible spin-TQFTs living on appropriate mapping tori and ultimately by $\Omega_3^{\Spin}(B\Z_2)\cong\Z_8$. In particular, we will reduce ourselves to study the case $X=\TT^2$, since, as we will see, this captures all the anomaly information present for a $2$-dimensional fermionic theory with $\Z_2$ symmetry.

Let us start by discussing what are the full mapping class group $\widetilde{\MCG}(\TT^2)$ and $H^1(\widetilde{\MCG}(\TT^2),U(1)^{\Spin_G(\TT^2)})$ in this case. The first is the fermionic central extension of $\MCG(\TT^2)=\SL(2,\Z)=\langle S,T|S^4=1,(ST)^3=S^2\rangle$ by $G^f=\Z_2\times \Z_2^f$. This means that the relations between $S$ and $T$ now close up to some elements of $G$. A redefinition of $S$ or $T$ always allow to fix the relation $(ST)^3=S^2$, while in principle the first equation can instead change. By the requirement of acting non-trivially with fermionic parity on $\SL(2,\Z)$, the relation $S^4=1$ will have to be properly modified. Therefore the full mapping class group will be the unique, up to an isomorphism, non-trivial extension $\widetilde{\MCG}(\TT^2)\cong \Mp\times \Z_2$, where
\begin{multline}
    \Mp = \\\langle S,T,\F|S^4=\F,(ST)^3=S^2, ((-1)^{\CF})^2=1,(-1)^{\CF}T=T(-1)^{\CF}\rangle.
\label{Mp}
\end{multline}
The group $\Mp$ is a non-trivial extension of the modular group $\SL(2,\Z)$ by $\Z_2$. It is often referred to as \textit{metaplectic} group and often also denoted as $\mathrm{Mp}_1(\Z)$. For a fermionic theory we expect the modular transformations to be a representation of \eqref{Mp} acting on the partition functions, with $\F$ playing the role of the non-trivial automorphism of the spin bundle over $\TT^2$, or, equivalently, the fermion parity operator of the 3d bulk TQFT considered on $\TT^2$ ``spatial'' slices. We will denote the generator of the $\Z_2$ factor in $\widetilde\MCG(\TT^2)$ which is not included in $\Mp$ as $(-1)^{\CQ}$. It has a meaning of the non-trivial automorphism of $G=\Z_2$ principal bundle over $\TT^2$, or, equivalently, the $\Z_2$ global symmetry charge operator of the 3d TQFT.

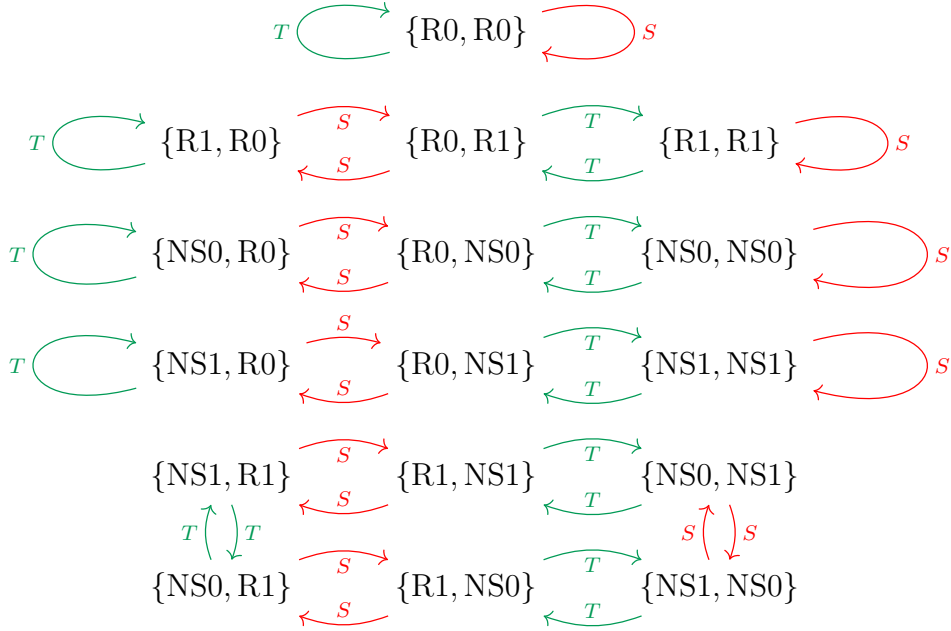
\begin{figure}
    \centering
\begin{equation}
\begin{tikzcd}
   &  \{\r0,\r0\} \ar[loop left,"T",ForestGreen]   \ar[loop right,"S",red] & 
   \\
    \{\r1,\r0\} \ar[loop left,"T",ForestGreen] \ar[r,"S",bend left=20,swap,red] & \{\r0,\r1\} \ar[r,"T",bend left=20,swap,ForestGreen] \ar[l,"S",bend left=20,swap,red] & \{\r1,\r1\} \ar[loop right,"S",red] \ar[l,"T",bend left=20,swap,ForestGreen]
    \\
    \{\ns0,\r0\} \ar[loop left,"T",ForestGreen] \ar[r,"S",bend left=20,swap,red] & \{\r0,\ns0\} \ar[r,"T",bend left=20,swap,ForestGreen] \ar[l,"S",bend left=20,swap,red] & \{\ns0,\ns0\} \ar[loop right,"S",red] \ar[l,"T",bend left=20,swap,ForestGreen]
    \\        
    \{\ns1,\r0\} \ar[loop left,"T",ForestGreen] \ar[r,"S",bend left=15,red] & 
    \{\r0,\ns1\} \ar[r,"T",bend left=20,swap,ForestGreen] \ar[l,"S",bend left=20,swap,red] & \{\ns1,\ns1\} \ar[loop right,"S",red] \ar[l,"T",bend left=20,swap,ForestGreen]
    \\
    \{\ns1,\r1\} \ar[r,"S",bend left=20,swap,red] \ar[d,"T",bend left=20,ForestGreen]  & \{\r1,\ns1\} \ar[r,"T",bend left=20,swap,ForestGreen] \ar[l,"S",bend left=20,swap,red] & \{\ns0,\ns1\}  \ar[l,"T",bend left=20,swap,ForestGreen] \ar[d,"S",bend left=20,red]
    \\
    \{\ns0,\r1\} \ar[r,"S",bend left=20,swap,red] \ar[u,"T",bend left=20,ForestGreen] & \{\r1,\ns0\} \ar[r,"T",bend left=20,swap,ForestGreen] \ar[l,"S",bend left=20,swap,red] & \{\ns1,\ns0\} \ar[l,"T",bend left=20,swap,ForestGreen] \ar[u,"S",bend left=20,red]
\end{tikzcd}
\nonumber
\end{equation}
    \caption{The structure of the action groupoid $\Mp\ltimes \Spin_{G}(\TT^2)$ for $G=\Z^2$. The $(-1)^\CF$ and $(-1)^\CQ$ generators always act from a groupoid object to itself and are not shown explicitly.}
    \label{fig:Z2-groupoid}
\end{figure}

For what regards the cohomology group $H^1(\widetilde{\MCG}(\TT^2),U(1)^{\Spin_G(\TT^2)})$, we can proceed with a direct computation. Note that the cocycle equation \eqref{eq:cocyclecondition} defines relations only between cocycles evaluated at elements $a\in \Spin_G(\TT^2)= \Spin(\TT^2)\times H^1(\TT^2,\Z_2)$ in the same orbits of $\SL(2,\Z)$. These are the set of couples of spin periodicity conditions $\ns,\r$ (Neveu-Schwarz and Ramond, or, equivalently, anti-periodic and periodic respectively) and $0,1\in\Z_2$ holonomies defined over a basis of $H^1(\TT^2,\Z_2)$. Therefore, in our case of interest there are three kinds of orbits, which are composed by one, three and six elements respectively. The total cohomology group is then the direct product of the cohomology groups generated by each orbit. The action groupoid is displayed in Figure \ref{fig:Z2-groupoid}.

The orbit of a single element is given by the cocycles evaluated at $\{\r0,\r0\}$ and its contribution to \eqref{MCG-cohomogy-grupoid} is $\Z_{24}\times\Z_2$.

The orbits composed by three elements are instead $\left\{\{\r0,\ns0\},\right.$ $\left.\{\ns0,\r0\},\{\ns0,\ns0\}\right\}$, $\left\{\{\r0,\r1\},\{\r1,\r0\},\{\r1,\r1\}\right\}$ and $\left\{\{\r0,\ns1\},\right.$ $\left.\{\ns1,\r0\},\{\ns1,\ns1\}\right\}$, with a contribution to the total group equal to $U(1)\times\Z_8\times \Z_2$ for each one.

Finally, the six elements left generates $U(1)^2\times\Z_4\times\Z_2$, so that the total group is
\begin{equation}
H^1(\widetilde{\MCG}(\TT^2),U(1)^{\Spin_G(X)})\cong U(1)^5\times \Z_8^4\times \Z_4\times \Z_3\times \Z_2^5.
\end{equation}
We anticipate here that the anomalous phases captured by the modular transformations lie in the diagonal subgroup $\Z_8$ for the two orbits of three elements with non-trivial $\Z_2$ holonomies and the $U(1)$ subgroup for the 6-dimensional orbit.

\subsection{Modular matrices from cobordism invariants}
\label{sec:matrices-from-invariants}

With these premises, we can now turn to study the modular transformations of the theories in discussion. In order to do so, we first review what are the cobordism invariants that one can construct for closed low dimensional manifolds equipped with a spin or pin$^-$ structure (see e.g. \cite{kirby_taylor_1991,benedetti1995spin,Guo:2018vij} for details). With this knowledge one is able to determine such transformations simply by knowing the anomaly of a theory and without further knowledge of its field content.

\subsubsection{Invariants of 1-manifolds}
The only connected and compact $1$-dimensional manifold is $S^1$, which we can equip with two different spin structures. They correspond to the two possible $\Z_2$ bundles covering the trivial $SO(1)$-bundle over $S^1$. More precisely, if such a bundle is non-trivial, then we denote the spin-manifold as $S^1_\ns$ and $S^1_\r$ otherwise. Only $S^1_\ns$ can be seen as the boundary of a disk, so that the spin structure on the former is the restriction of the unique spin structure admitted on the latter. The case $S^1_\r$, which is equivalent to having periodic boundary conditions on the circle, is instead the generator of $\Omega_1^{\Spin}(\pt)\cong \Z_2$. We will denote with $\eta$ the non-trivial $\Z_2$-valued bordism invariant defined by
\begin{equation}
    \eta(M)=\begin{cases}0& M=S^1_\ns,\\
    1,& M=S^1_\r,\\
    \sum_i \eta(N_i),& M=\sqcup_i N_i.
    \end{cases}
\end{equation}

\subsubsection{Invariants of 2-manifolds}
An oriented $2$-manifold $X$ with non-trivial genus can always be written as the connected sum of an arbitrary number of tori. The spin structures on the torus are canonically identified with $H^1(\TT^2,\Z_2)\cong \Z_2 \times \Z_2$, corresponding to the periodicity conditions on the two non-trivial generating $1$-cycles. From the argument above one can see that $\TT^2=\partial(D^2 \times S^1)$ as a spin-manifold if at least one of the two generators has an antiperiodic boundary condition, while the trivial element $(0,0)$ is the only one which does not bound. Extending the reasoning to surfaces of generic genus, it follows that $\Omega_2^{\Spin}(\pt)\cong \Z_2$ as well.

The bordism invariant which tells us in which element of $\Omega^{\Spin}_{2}(\pt)$ our manifold falls in is the $\Arf$ invariant. To build it we recall that spin structures on an oriented 2-manifold $X$ are in one-to-one correspondence with quadratic forms
\begin{equation}
    \tilde{q}:H^1(X,\Z_2)\rightarrow\Z_2
\end{equation}
such that 
\begin{equation}
    \tilde{q}(a+b)-\tilde{q}(a)-\tilde{q}(b)=\int_{X}a\cup b.
\end{equation}
Geometrically such form is equivalent to evaluating $\eta$ over the $1$-manifolds Poincar\'e duals to the $1$-cycles. 
From this we can define the $\Arf$ invariant as 
\begin{equation}
\begin{array}{@{}r@{}l@{}}
  \Arf\colon \Omega^{\Spin}_{2}(\pt) & {}\longrightarrow \Z_2 \\
          \left[X\right] & {}\longmapsto \Arf (X):=\sum_{i=1}^g \tilde{q}(\tilde{a}_i)\tilde{q}(\tilde{b}_i)\quad \mbox{mod 2},
 \end{array}
\end{equation}
where $\{\tilde{a}_i,\tilde{b}_i\}^g_{i=1}$ is a symplectic basis for $H^1(X,\Z_2)$ and $X$ the chosen representative of the bordism class.

Next we are interested in non-oriented closed surfaces. Although they can admit also a pin$^+$ structure, they always admit a pin$^-$ structure, which we will focus on. A generic non-oriented surface $X$ can be written as a direct sum of tori and projective spaces $\mathbb{RP}^2$. The cup product defines again a non-degenerate symmetric bilinear form on $H^1(X,\Z_2)$, allowing us to define, in similarity with the oriented case, the quadratic enhancement 
\begin{equation}
    q:H^1(X,\Z_2)\rightarrow\Z_4,
    \label{qnonoriented}
\end{equation}
that satisfies
\begin{equation}
    q(a,b)-q(a)-q(b)=2\int_{X}a\cup b.
\label{qenhancement}
\end{equation}
Such quadratic enhancements on $X$ are in one-to-one correspondence with the set of its pin$^-$ structures. One can also associate to any enhancement the Arf-Brown-Kervaire (ABK) invariant, defined by the Gaussian sum\footnote{Here $X$ represents a manifold together with a spin structure, which is usually clear by the context. Whenever this is not the case, we use the notation $\ABK(X,s)$ to specify the spin structure $s$ of the manifold $X$.}
\begin{equation}
e^{i\pi \ABK(X)/4}:=\frac{1}{\sqrt{|H^1(X,\Z_2)|}}\sum_{a\in H^1(X,\Z_2)}e^{i\pi q(a)/2},
\end{equation}
which in turns provides a $\Pin^-$-bordism invariant as well. Thus we can think of it as an isomorphism labelling the classes in $\Omega^{\Pin^-}_2(\pt)$, i.e. $\ABK:\;\Omega^{\Pin^-}_{2}(\pt) \rightarrow \Z_8$.

Note that in the case of orientable surfaces, the two enhancements and invariants match as one should expect, $q=2\tilde{q}\;\mbox{mod 4}$ and $\ABK(X)=4\Arf(X)\;\mbox{mod 8}$.

\subsubsection{Invariants of 3-manifolds}
Up to some differences, in the case of a 3-dimensional manifold $Y$ one can proceed by analogy and make use of the intersection form to build bordism invariants. When $d=3$, it defines a symmetric trilinear product which allows to define also the bilinear form \begin{equation}
\begin{array}{@{}r@{}l@{}}
  \lambda\colon H^1(Y,\Z_2)\times H^1(Y,\Z_2) & {}\longrightarrow \Z_2 \\
          (a,b) & {}\longmapsto \lambda (a,b):=\int_{Y} a\cup a\cup b
 \end{array}.
\end{equation}
One can think of $\lambda$ in the following way: given smooth surfaces representing\footnote{It is known that for a general $d$-manifold $X$ the elements of $H_{d-1}(X,\Z_2)$ can always be represented by \emph{smooth} codimension one submanifolds.} the Poincar\'e duals of $a$ and $b$, $\PD(a)$ and $\PD(b)$, their intersections will be a disjoint union of embedded circles. Restricting the normal bundle of $\PD(a)$ on these, $\lambda(a,b)$ counts the number of them mod $2$ on which such a bundle is non-trivial.

If the manifold $Y$ is actually equipped with a spin structure $s$, then it is possible to enhance  $\lambda$ to a function 
\begin{equation}
    \delta_s:H^1(Y,\Z_2) \times H^1(Y,\Z_2) \longrightarrow \Z_4.
    \label{delta3manifolds}
\end{equation}
Indeed, let $f:\Sigma\hookrightarrow Y$ be an embedded surface which describes the Poincar\'e dual of $a\in H^1(Y,\Z_2)$, i.e. $f(\Sigma)=\PD(a)$. Since $Y$ is orientable, it is possible to show that the pullback of the tangent bundle is of the form $f^* TY \cong T\Sigma \oplus \mathrm{det}T\Sigma$ and therefore, by means of $s$, in bijection with the $\Pin^-$ structures defined on $\Sigma$. 

At this point consider a generic smooth embedding $\iota:S^1\hookrightarrow Y$. Using the canonical spin structure $S^1_\ns$ and the spin structure on $Y$ it is possible to define a spin structure on the normal bundle $N_{Y/\iota(S^1)}$ as well. The framings of $N_{Y/\iota(S^1)}$ which are compatible with this choice are called \emph{even framings}, while the others are called \emph{odd}.\\
From this there is a unique quadratic enhancement $q_{\PD(a)}$ on a representative $\Sigma$ of $\PD(a)$ such that, for any $\tilde{b}\in H^1(\Sigma,\Z_2)$, $q_{\PD(a)}(\tilde{b})$ equals the number mod $4$ of left (positive) half turns which the restriction $N_{Y/\PD(a)}\rvert_{\iota(\tilde{b})}$ does with respect to any \emph{even} framing in $Y$\footnote{Note that this definition is independent on which direction we are moving along the circle.} \cite{benedetti1995spin}.
Then the definition of \eqref{delta3manifolds} follows simply as 
\begin{equation}
    \delta_s(a,b):=q_{\PD(a)}(\tilde{b}),
\end{equation}
with the obvious restriction $\tilde{b}=b\rvert_{\PD(a)}$. Geometrically it is clear why $\delta_s$ is the enhancement of $\lambda$ for $s$ fixed.\\
Since the pin$^-$ structures of $\PD(a)$ are classified by its $\ABK$ invariant, for any spin structure $s$ one can define without additional effort also the invariant
\begin{equation}
\begin{array}{@{}r@{}l@{}}
  \beta_s\colon H^1(Y,\Z_2) & {}\longrightarrow \Z_8 \\
          a & {}\longmapsto \beta_s(a):=\ABK(\PD(a),s\rvert_{\PD(a)})
\end{array}.
\label{beta-inv-def}
\end{equation}
From the definition it follows that it holds 
\begin{equation}
    \beta_s(a+b)=\beta_s(a)+\beta_s(b)+2\delta_s(a,b).
\end{equation}
Moreover, if one changes the spin structure acting on it with $c \in H^1(Y,\Z_2)$, then we have also the relation 
\begin{equation}
    \beta_{s+c}(a)=\beta_s(a)+\delta_s(a,c).
\end{equation}
Finally we remark that unlike the lower dimensional cases, the spin bordism group $\Omega^\Spin_3(\pt)$ is empty, since it can be shown that any spin $3$-manifold bounds a spin $4$-manifold \cite{kirby_1989}. Therefore the only invariants one can build arises from lower dimensional submanifolds, as we have just shown.

\subsubsection{Mapping tori and modular matrices}
As already mentioned, generic bordism groups $\Omega_n^\Spin(BG)$ can be computed by well known methods, for example via the Atiyah-Hirzebruch spectral sequence associated to the trivial fibration 
\begin{equation}
    \pt \longrightarrow BG \overset{p}{\longrightarrow} BG,
\end{equation}
or by other means \cite{milnor1963spin,kirby_taylor_1991,Freed:2016rqq,brumfiel2016pontrjagin,Guo:2017xex,beaudry2018guide,Hsieh:2018ifc,Garcia-Etxebarria:2018ajm,Guo:2018vij,brumfiel2018pontrjagin}. However, the case $G=\Z_2$ is simple enough that one can understand its structure geometrically by the basic tools we presented\footnote{By a simple generalization one can actually determine in this way the group for the generic case $G=\Z_2^k$ as well \cite{yu1995connective,bruner2010connective,Guo:2018vij}.}.

Let us focus on the $2$-dimensional case first. After choosing a proper spin structure $s$ on a surface $X$, the only thing one has left to fix is the map $g:X\rightarrow B\Z_2$, which can be equivalently seen as an element $a_g\in H^1(X,\Z_2)$. Therefore, for labelling elements in $\Omega_2^\Spin(B\Z_2)$ we can build only one additional invariant besides the ones of the previous section, i.e. $\eta_g:=\tilde{q}(a_g)$. This means that $\Omega^\Spin_2(B\Z_2)\cong \Z_2 \oplus \Z_2$, where the isomorphism is given by the couple $(\Arf,\eta_g)$.

By the same reasoning, it is easy to see that the only possible invariant which can label elements of $\Omega^\Spin_3(B\Z_2)$ is 
$\beta_s(a_g)$, where now $g:Y\rightarrow B\Z_2$. Thus $\Omega^\Spin_3(B\Z_2)$ is classified by the possible pin$^-$ structures that can be defined on $\PD(a_g)$ and therefore is isomorphic to $\Omega^{\Pin^-}_2(\pt)\cong \Z_8$. The partition function of the TQFT corresponding to the element $\nu\in \Z_8\cong \Hom(\Omega_3^\Spin(B\Z_2),U(1))$ on $Y$ is given by
\begin{equation}
    e^{\frac{\pi i\beta_s(a_g)}{4}}.
\end{equation}

In order to apply this general knowledge to the study of modular transformations, we now consider the $2$-manifold $X=\TT^2$. As previously mentioned, we denote $\{s_0 g_0,s_1 g_1\}$ the data that define $X$ as a manifold with extra structure, so that it represents an element in $\Omega_2^\Spin(B\Z_2)$. Here $s_0,\,s_1\in \{\r,\ns\}$ are the periodicity conditions along the time and space direction respectively (before acting with the $\Z_2$ gauge field), while $g_i\in\{0,1\}$ are the $\Z_2$ holonomies. With the notation $Z_{s_0 g_0}^{s_1 g_1}(\tau,\bar{\tau})$ we refer to the corresponding partition functions\footnote{In the case of an anomalous theory with $\nu=1\mod2$ and depending on the Hilbert space considered, it holds $[(-1)^F,(-1)^Q]\not = 0$, so one has to proceed with care defining $Z_{s_0 g_0}^{\ns 1}$. See Section \ref{subsec:Hilbertspace} for further discussion.} with appropriate insertions of $(-1)^{F,Q}$, the  2-dimensional fermion parity and $G=\Z_2$ global symmetry charge operators\footnote{Recall that we denote their 3d bulk counterparts as $(-1)^{\CF,\CQ}$.}. E.g. in the case of a CFT 
\begin{align}
    Z_{s_1 g_1}^{\ns g_0}(\tau,\bar{\tau}) &= \mathrm{Tr}_{\mathcal{H}_{s_1 g_1}}\left[(-1)^{g_0 Q} q^{L_0 - c/24}\bar{q}^{\bar{L}_0-\bar{c}/24}\right],\\
    Z_{s_1 g_1}^{\r g_0}(\tau,\bar{\tau}) &= \mathrm{Tr}_{\mathcal{H}_{s_1 g_1}}\left[(-1)^{F}(-1)^{g_0 Q} q^{L_0 - c/24}\bar{q}^{\bar{L}_0-\bar{c}/24}\right],
\end{align}
where $\mathcal{H}_{\ns /\r 0}$ and $\mathcal{H}_{\ns/\r 1}$ are respectively the untwisted and twisted Hilbert spaces with $\ns$ or $\r$ periodicity conditions and $q=\exp (2\pi i \tau)$. 

The modular transformations will mix only partition functions of tori which lie in the same classes of $\Omega^\Spin_2(B\Z_2)$. Therefore we put these together as entries of some vectors $(\mathbf{Z}_{(\Arf,\eta_g)})_i$, so that their transformations are described by the corresponding matrices $S_{(\Arf,\eta_g)},\, T_{(\Arf,\eta_g)}$:
\begin{equation}
\begin{aligned}
    \mathbf{Z}_{(0,0)}^T = &\;(Z^{\r 0}_{\ns 0},Z^{\ns 0}_{\r 0},Z^{\ns0}_{\ns0},Z^{\ns0}_{\ns1},Z^{\r1}_{\ns0},Z^{\ns0}_{\r1},Z^{\ns1}_{\ns0},Z^{\r1}_{\ns1},Z^{\ns1}_{\r1}),\\
    \mathbf{Z}_{(1,0)}^T = &\;(Z^{\r 0}_{\r 0}),\\
    \mathbf{Z}_{(0,1)}^T = &\;(Z^{\r 0}_{\ns 1},Z^{\ns 1}_{\r 0},Z^{\ns 1}_{\ns 1}),\\
    \mathbf{Z}_{(1,1)}^T = &\;(Z^{\r 0}_{\r 1},Z^{\r 1}_{\r 0},Z^{\r 1}_{\r 1}).
    \label{classesoftori} 
\end{aligned}
\end{equation}
In order to determine how the modular group acts we must first choose the reference frame $(X_\alpha,P_\alpha,\theta_\alpha)$ for each bordism class, call it $\tilde{X}_{(\Arf,\eta_g)}$, and the basis elements $e_a$ for the Hilbert space associated to each tori in \eqref{classesoftori}:
\begin{itemize}
\let\labelitemi\labelitemii
    \item For the trivial class the canonical choice is given by $\tilde{X}_{(0,0)} = \varnothing$, while the manifolds defining the basis elements $e_a$ are solid tori $Y_a\cong D^2\times S^1$ with $\partial Y_a =X_a$ and the radial direction given by the continuous contraction of one direction\footnote{There are multiple directions which satisfy this constraint. See Appendix \ref{app:matrices} for more details on how we set this choice.} $\ns 0$ of $X_a$ to a point;
    \item For $(1,0)$ there is no choice to do, $\tilde{X}_{(1,0)}=\{\r 0,\r 0\}$ and $Y_{\{\r 0,\r0\}}=\mathrm{id}_{\{\r 0,\r0\}}$;
    \item For $(0,1)$ we choose $\tilde{X}_{(0,1)}=\{\r0,\ns1\}$. The degrees of freedom we are left with allows us to choose\footnote{Actually for these last two classes one still has to choose a proper resolution of the Poincar\'e dual of $a_g\in H^1(X,\Z_2)$. See Appendix \ref{app:matrices} for our convention.}
    \begin{align}
    Y_{\{\ns1,\r0\}}&=\S_{\{\ns1,\r0\}}^{\{\r0,\ns1\}}:\{\r0,\ns1\}\overset{S}\rightarrow \{\ns1,\r0\},\\
    Y_{\{\ns1,\ns1\}}&=\T_{\{\ns1,\ns1\}}^{\{\r0,\ns1\}}:\{\r0,\ns1\}\overset{T}\rightarrow \{\ns1,\ns1\};
    \label{eq:NS1NS1-base}
    \end{align}
    \item Analogously to $(0,1)$ up to the change $\ns 1 \mapsto \r 1$, for the class $(1,1)$ we choose $\tilde{X}_{(1,1)}=\{\r0,\r1\}$ and
    \begin{align}
    Y_{\{\r1,\r0\}}&=\S_{\{\r1,\r0\}}^{\{\r0,\r1\}}:\{\r0,\r1\}\overset{S}\rightarrow \{\r1,\r0\},\\
    Y_{\{\r1,\r1\}}&=\T_{\{\r1,\r1\}}^{\{\r0,\r1\}}:\{\r0,\r1\}\overset{T}\rightarrow \{\r1,\r1\}.
    \label{eq:R1R1-base}
    \end{align}
\end{itemize}
Now the action of $\widetilde{\MCG}(X)$ is fixed, with the representation of its generators corresponding to the evaluation of the appropriate cobordism invariant $\nu \in \Z_8$ for the closed manifolds determined by \eqref{C-nu-formula}.

In order to understand how these closed manifolds are built, one needs to proceed with care since the theories in discussion have a dependence on the spin structure. 
Indeed, for a given isomorphism $\tilde{\psi}:X_a\rightarrow X_b$ and fixed basis on the respective Hilbert spaces by $Y_{a}:X_\alpha \rightarrow X_{a}$, $Y_{b}:X_\alpha \rightarrow X_{b}$,  the building blocks of the closed manifold are the bordisms $\tilde\psi\circ Y_a : X_\alpha \rightarrow X_b$ and $(\overline{Y_{\psi^{-1*}a}}) : \overline{X}_b \rightarrow \overline{X}_\alpha$. However, one still needs to connect these manifolds by two additional bordisms which properly reverse the structure of them. This operation is given by the \emph{evaluation} and \emph{coevaluation} bordisms, both defined for any $X$ as the manifold $I_X:=[0,1]\times X$, but with different directions: 
\begin{align}
    \mathrm{e}_X:= I_X: X \cup \overline{X} \rightarrow \varnothing,\\
    \mathrm{c}_X:= I_X:\varnothing \rightarrow \overline{X} \cup X.
\end{align}
With these tools the closed manifold corresponding to isomorphism $\tilde{\psi}$ can be defined more precisely by
\begin{equation}
    MT(\tilde{\psi}):= \e_{X_\alpha}\circ (\tilde{\psi}\circ Y_a \cup  (\overline{Y_{\psi^{-1*}a}}) ) \circ \c_{\overline{X}_b},
    \label{defmappingtori}
\end{equation}
where the presence of $e_{X_{\alpha}}$ and $c_{\overline{X}_b}$ are usually omitted. In our setup, for non-trivial classes $[X_\alpha]\in\Omega_2^\Spin(B\Z_2)$ these closed manifolds are mapping tori of $\TT^2$, that is fibrations over $S^1$ with fibers isomorphic to $\TT^2$. For a trivial classes these 3-manifolds are isomorphic to lens-spaces, i.e. manifolds of the form $(D^2\times S^1)\cup (D^2\times S^1)$ where the gluing along the common $\TT^2$ boundary is done using a non-trivial automorphism.

If the theory have a dependence on the spin structure the subtlety we mentioned arise in the identification $\overline{\overline{X}}_b\cong X_b$ which appears in the definition of $c_{\overline{X}_b}$. Indeed, by the requirement of unitarity it follows that such isomorphism is defined via action of $(-1)^\CF$ on one of the two manifolds (see \cite{Yonekura:2018ufj} for more details). Once this is properly taken into account for the description of the mapping tori, the evaluation of $\nu$ for \eqref{defmappingtori} defines the action of the modular transformations, which we are now ready to study. 
\begin{figure}
    \centering
\begin{tikzpicture}
[scale=1,baseline=(current  bounding  box.center),>=stealth]
    \draw[->,very thick] (0.2,2) -- (2.9,2) node[pos=0,anchor=south west]{$X_\alpha$} node[pos=0.5,anchor=north]{$Y_a$};
    \draw (3,2) node[anchor=south]{$X_a$};
    \draw[->,very thick] (3.1,2) -- (5.8,2) node[pos=0.5,anchor=north]{$\tilde{\psi}$} node[pos=1,anchor=south east]{$X_b$};
    \draw[->,very thick] (6,2) .. controls (7.5,2) and (7.5,0) .. (6,0) node[pos=0,anchor=south west]{$\overline{\overline{X}}_b$} node[pos=0.5,anchor=west]{$\c_{\overline{X}_b}$};
    \draw[->,very thick] (5.8,0) -- (0.2,0) node[pos=0,anchor=north]{$\overline{X}_b$} node[pos=0.5,anchor=south]{$\overline{Y}_b$} node[pos=1,anchor=north west]{$\overline{X}_\alpha$};
    \draw[->,very thick] (0,0) .. controls (-1.5,0) and (-1.5,2) .. (0,2) node[pos=0.5,anchor=east]{$\e_{X_\alpha}$};
\end{tikzpicture}
\caption{The structure of the closed manifolds \eqref{defmappingtori} (mapping tori for non-trivial classes in $\Omega_2^\Spin(B\Z_2)$), from which is clear one has to remember the identification $\overline{\overline{X}}\cong X$ via $\F$.}
\end{figure}
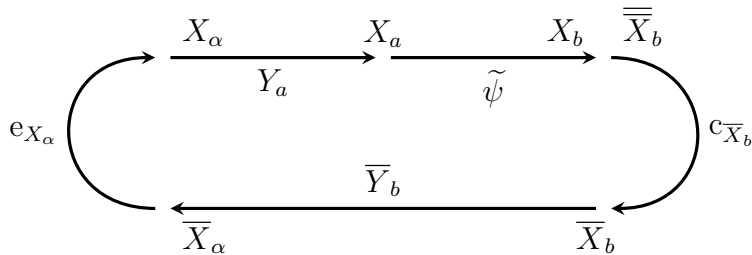

To illustrate how this framework determines them, here we present the simplest case out of the action of the three generators of $\Mp$, namely $\F$. Instead, we omit the detailed computations of the $S$ and $T$ matrices, leaving them to Appendix \ref{app:matrices}.
Fermionic parity maps each element $\{s_0g_0,s_1g_1\}$ to itself, which means that on the vectors of partition functions \eqref{classesoftori} it will act diagonally, with entries $(-1)^\CF_a$. Indeed, for a generic $d$-manifold $X_a$ one has  $MT((-1)^\CF_a)=X_a\times S^1_\r$, where $S^1$ is equipped with a periodic spin structure. This immediately tells something more, i.e. that $\F$ is a bordism invariant independently of the $H_d$-structure we might have on $d$-manifolds. In fact, if two elements $X_{a}$, $X_b$ lie in the same class of $\Omega_d^{H_d}$ then $X_a\times S^1_\r$ and $X_b\times S^1_\r$ lie in the same class of $\Omega_{d+1}^{H_{d+1}}$ as well, because any bordism $Y:X_a\rightarrow X_b$ can be extended to $Y\times S^1_\r:X_a\times S^1_\r \rightarrow X_b\times S^1_\r$. Since the action of $\F$ in given by the evaluation of a cobordism invariant $\nu \in \Omega^{H_{d+1}}_{d+1}$, then this must hold.

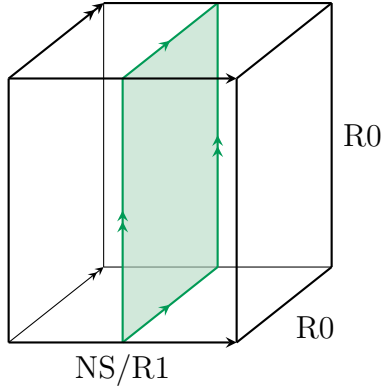
\begin{figure}
    \centering
\begin{tikzpicture}
[scale=0.5,baseline=(current  bounding  box.center),>=stealth]
    \coordinate (P1) at (-3,-3);
    \coordinate (P2) at (3,-3);
    \coordinate (P3) at (-0.5,-1);
    \coordinate (P4) at (5.5,-1);
    \coordinate (P5) at (-3,4);    
    \coordinate (P6) at (3,4);
    \coordinate (P7) at (-0.5,6);
    \coordinate (P8) at (5.5,6);
    \coordinate (C1) at ($(P1)!0.5!(P4)$);
    \coordinate (C2) at ($(P5)!0.5!(P8)$);
    \coordinate (Q1) at ($(P1)!0.5!(P2)$);
    \coordinate (Q2) at ($(P2)!0.5!(P4)$);
    \coordinate (Q3) at ($(P1)!0.5!(P3)$);
    \coordinate (Q4) at ($(P3)!0.5!(P4)$);
    \coordinate (Q5) at ($(P5)!0.5!(P6)$);
    \coordinate (Q6) at ($(P6)!0.5!(P8)$);
    \coordinate (Q7) at ($(P5)!0.5!(P7)$);
    \coordinate (Q8) at ($(P7)!0.5!(P8)$);
    \coordinate (X) at ($(Q5)+(0,10)$);
    \coordinate (Y) at ($(Q4)+(0,-10)$);
    
    \draw[->>,thin] (P1) -- (P3);
    \draw[thin] (P3) -- (P4);
    \draw[thin] (P3) -- (P7);

    \path[fill=ForestGreen!30,fill opacity=0.6] (Q1) -- (Q4) -- (Q8) -- (Q5) -- (Q1);
    \draw[ForestGreen,thick] (Q1) -- (Q4) node[pos=0.5,sloped,rotate=-90]{\tikz\draw[->,thick](0,0);};
    \draw[ForestGreen,thick] (Q1) -- (Q5) node[pos=0.5,sloped,rotate=-90]{\tikz\draw[->>,thick](0,0);};
    \draw[ForestGreen,thick] (Q4) -- (Q8) node[pos=0.5,sloped,rotate=-90]{\tikz\draw[->>,thick](0,0);};
    \draw[ForestGreen,thick] (Q5) -- (Q8) node[pos=0.5,sloped,rotate=-90]{\tikz\draw[->,thick](0,0);};
    
    \draw[->,thick] (P1) -- (P2) node[pos=0.5,anchor=north]{$\ns /\r 1$}; 
    \draw[thick] (P2) -- (P4) node[pos=0.5,anchor=north west]{$\r 0$};
    \draw[->,thick] (P5) -- (P6);
    \draw[thick] (P6) -- (P8);
    \draw[->>,thick] (P5) -- (P7);
    \draw[thick] (P7) -- (P8);
    \draw[thick] (P1) -- (P5);
    \draw[thick] (P2) -- (P6);
    \draw[thick] (P4) -- (P8) node[pos=0.5,anchor=west]{$\r 0$};
\end{tikzpicture}
\caption{The mapping tori describing the action of $\F$ on the bordism classes $(0/1,1)$ with the pin$^-$ surface $\Sigma^\F_{(0/1,1)}$ highlighted in green. The horizontal slices correspond to the tori $\{\r0,\ns/\r1\}$ while the vertical direction corresponds to $S^1_\r$, with its direction being from bottom to top.}
\label{fig:-Fmappingtori}
\end{figure}

Thus for our needs we simply evaluate the invariant $\nu\in \Omega^3_\Spin(B\Z_2)$ for the reference manifold in each bordism class $(\Arf,\eta_g)$. In the classes $(0,0)$ and $(0,1)$ it follows that $(-1)^\CF_{(0,0)}=(-1)^\CF_{(1,0)}=\mathrm{id}$. Indeed, while for $(0,0)$ the result is straightforward, the torus in $(1,0)$ has trivial $\Z_2$ holonomies and therefore there is no pin$^-$-surface associated with a non-trivial $\ABK$ invariant.
The two more interesting cases are $(0/1,1)$, of which the corresponding mapping tori are reported in Figure \ref{fig:-Fmappingtori}.

From the figure it is clear that the pin$^-$ surface $\Sigma^\F_{(0/1,1)}$ in question is simply a torus with a symplectic basis of $H^1(\Sigma^\F_{(0/1,1)},\Z_2)$ given by the two $\r 0$ directions depicted. Therefore we have 
\begin{equation}
    (-1)^\CF_{(0/1,1)}\equiv \nu (MT((-1)^\CF_{(0/1,1)})) = e^{\pi i\nu \,\Arf(\Sigma^\F_{(0/1,1)})}= (-1)^\nu \cdot \mathrm{id}.
\end{equation}
For the $S$ and $T$ matrices one can proceed in an analogous way, albeit the mapping tori become more elaborated and the computation of their bordism classes more involved. As already mentioned, we refer the interested reader to Appendix \ref{app:matrices} for their computation. Here we simply report in Figures \ref{fig:T00mappingtori}-\ref{fig:S101mappingtori} the tori associated to their non-trivial entries, together with the $\mathrm{pin}^-$ surfaces of which the $\ABK$ invariant determines the bordism classes.
\begin{figure}[tb]
    \centering
\begin{tikzpicture}
[scale=0.5,baseline=(current  bounding  box.center),>=stealth]
    \coordinate (P1) at (-3,-3);
    \coordinate (P2) at (3,-3);
    \coordinate (P3) at (-0.5,-1);
    \coordinate (P4) at (5.5,-1);
    \coordinate (P5) at (-3,4);    
    \coordinate (P6) at (3,4);
    \coordinate (P7) at (-0.5,6);
    \coordinate (P8) at (5.5,6);
    \coordinate (C1) at ($(P1)!0.5!(P4)$);
    \coordinate (C2) at ($(P5)!0.5!(P8)$);
    \coordinate (Q1) at ($(P1)!0.5!(P2)$);
    \coordinate (Q2) at ($(P2)!0.5!(P4)$);
    \coordinate (Q3) at ($(P1)!0.5!(P3)$);
    \coordinate (Q4) at ($(P3)!0.5!(P4)$);
    \coordinate (Q5) at ($(P5)!0.5!(P6)$);
    \coordinate (Q6) at ($(P6)!0.5!(P8)$);
    \coordinate (Q7) at ($(P5)!0.5!(P7)$);
    \coordinate (Q8) at ($(P7)!0.5!(P8)$);
    \coordinate (X) at ($(Q5)+(0,1)$);
    \coordinate (Y) at ($(Q4)+(0,-1)$);
    
    \path[name path=Q2Q3] (Q2) -- (Q3);
    \path[name path=Q6Q7] (Q6) -- (Q7);
    \path[name path=Q1Q5] (Q1) -- (Q5);
    \path[name path=Q4Q8] (Q4) -- (Q8);
    \path[name path=Q5X] (Q5) -- (X);
    \path[name path=Q4Y] (Q4) -- (Y);
    \path[name intersections={of=Q2Q3 and Q1Q5,by=A1}];
    \path[name intersections={of=Q2Q3 and Q4Y,by=A2}];
    \path[name intersections={of=Q6Q7 and Q5X,by=A3}];
    \path[name intersections={of=Q6Q7 and Q4Q8,by=A4}];   

    \path[name path=Q1A1] (Q1) .. controls (C1) .. (A1) coordinate[pos=0.5] (B1);
    \path[name path=Q2Q4] (A2) .. controls (C1) .. (Q4) coordinate[pos=0.5] (B2);
    \path[name path=Q5Q7] (A3) .. controls (C2) .. (Q5) coordinate[pos=0.5] (B3);
    \path[name path=Q6Q8] (A4) .. controls (C2) .. (Q8) coordinate[pos=0.5] (B4);

    \draw[thin] (P1) -- (P3) -- (P4);
    \draw[thin] (P3) -- (P7);
    \draw[Red,very thick,densely dashed] (P2) --(P3);

    \draw[ForestGreen,very thick,dashed] (B1) .. controls +(-0.1,0.1) .. (A1);
    \path[fill=ForestGreen!30,fill opacity=0.6] (B1) .. controls +(0,-.1) .. (Q1) -- (Q5) .. controls +(0,-.1) .. (B3) -- (B1);
    \path[fill=ForestGreen!30,fill opacity=0.1] (A1) -- (Q3) -- (Q7) -- (A3) .. controls +(-0.1,0.1) .. (B3) .. controls +(0,-.1) .. (Q5) -- (A1);
    \draw[ForestGreen,very thick] (B1) .. controls +(0,-.1) .. (Q1);
    \draw[ForestGreen,very thick] (B3) .. controls +(-0.1,0.1) .. (A3);
    \draw[ForestGreen,very thick] (B3) .. controls +(0,-.1) .. (Q5);
    
    \draw[ForestGreen,very thick,dashed] (B2) .. controls +(0,0.1) .. (Q4);
    \draw[ForestGreen,very thick, dashed] (A4) -- (Q4);        
    \path[fill=ForestGreen!30,fill opacity=0.6] (B2) .. controls +(0,-.1) .. (A2) -- (Q2) -- (Q6) -- (A4) .. controls +(0,-.1) .. (B4) -- (B2);
    \path[fill=ForestGreen!30,fill opacity=0.1] (A4) -- (Q8) .. controls +(0,0.1) .. (B4) .. controls +(0,-.1) .. (A4);
    \draw[ForestGreen,very thick] (B2) .. controls +(0,-.1) .. (A2);
    \draw[ForestGreen,very thick] (B4) .. controls +(0,0.1) .. (Q8);
    \draw[ForestGreen,very thick] (B4) .. controls +(0,-.1) .. (A4);
    
    \draw[ForestGreen,very thick] (Q3) -- (A1);
    \draw[ForestGreen,very thick] (Q2) -- (A2);
    \draw[ForestGreen,very thick] (Q7) -- (A3);
    \draw[ForestGreen,very thick] (Q6) -- (A4);
    \draw[ForestGreen,very thick] (Q1) -- (Q5);
    \draw[ForestGreen,very thick] (Q3) -- (Q7);
    \draw[ForestGreen,very thick] (Q2) -- (Q6);
    \draw[ForestGreen,very thick] (A4) -- (Q8);
    \draw[ForestGreen,very thick] (B1) -- (B3);
    \draw[ForestGreen,very thick] (B2) -- (B4);

    \draw[Red,very thick,densely dashed] (P5) --(P8);

    \draw[thick] (P1) -- (P2) node[pos=0.5,anchor=north west]{$\r 1$};
    \draw[thick] (P2) -- (P4) node[pos=0.5,anchor=north west]{$\ns 1$};
    \draw[thick] (P5) -- (P6) -- (P8) -- (P7) -- cycle;
    \draw[thick] (P1) -- (P5);
    \draw[thick] (P2) -- (P6);
    \draw[thick] (P4) -- (P8);
\end{tikzpicture}%
\qquad
\begin{tikzpicture}
[scale=0.5,>=stealth,baseline=(current  bounding  box.center)]
    \coordinate (P1) at (-3,-3);
    \coordinate (P2) at (3,-3);
    \coordinate (P3) at (-0.5,-1);
    \coordinate (P4) at (5.5,-1);
    \coordinate (P5) at (-3,4);    
    \coordinate (P6) at (3,4);
    \coordinate (P7) at (-0.5,6);
    \coordinate (P8) at (5.5,6);
    \coordinate (C1) at ($(P1)!0.5!(P4)$);
    \coordinate (C2) at ($(P5)!0.5!(P8)$);
    \coordinate (Q1) at ($(P1)!0.5!(P2)$);
    \coordinate (Q2) at ($(P2)!0.5!(P4)$);
    \coordinate (Q3) at ($(P1)!0.5!(P3)$);
    \coordinate (Q4) at ($(P3)!0.5!(P4)$);
    \coordinate (Q5) at ($(P5)!0.5!(P6)$);
    \coordinate (Q6) at ($(P6)!0.5!(P8)$);
    \coordinate (Q7) at ($(P5)!0.5!(P7)$);
    \coordinate (Q8) at ($(P7)!0.5!(P8)$);
    \coordinate (X) at ($(Q5)+(0,10)$);
    \coordinate (Y) at ($(Q4)+(0,-10)$);
    
    \draw[thin] (P1) -- (P3) -- (P4);
    \draw[thin] (P3) -- (P7);
    \draw[Red,very thick,densely dashed] (P2) -- (Q3);
    \draw[Red,very thick,densely dashed] (Q2) -- (P3);

    \path[fill=ForestGreen!30,fill opacity=0.6] (Q1) -- (Q4) -- (Q8) -- (Q5) -- (Q1);
    \draw[ForestGreen,very thick] (Q1) -- (Q4);
    \draw[ForestGreen,very thick] (Q1) -- (Q5);
    \draw[ForestGreen,very thick] (Q4) -- (Q8);
    \draw[ForestGreen,very thick] (Q5) -- (Q8);
    
    \draw[thick] (P1) -- (P2) node[pos=0.5,anchor=north]{$\ns 1$}; 
    \draw[thick] (P2) -- (P4) node[pos=0.5,anchor=north west]{$\ns 0$};
    \draw[thick] (P5) -- (P6) -- (P8) -- (P7) -- cycle;
    \draw[thick] (P1) -- (P5);
    \draw[thick] (P2) -- (P6);
    \draw[thick] (P4) -- (P8);
    
    \draw[Red,very thick,densely dashed] (P5) -- (P7);
\end{tikzpicture}%
\caption{The closed 3-manifolds (lens spaces) associated to non-trivial entries $(T_{(0,0)})^9_6$ and $(T_{(0,0)})^4_8$. The vertical direction represents the bordism from $\{s_0g_0,s_1 g_1\}$ on the bottom to $\{(s_0+s_1)(g_0+g_1),s_1 g_1\}$ on the top. The red dashed lines represent instead the directions which get contracted to single points.}
\label{fig:T00mappingtori}

\bigskip

\begin{tikzpicture}
[scale=0.5,baseline=(current  bounding  box.center),>=stealth]
    \coordinate (P1) at (-3,-3);
    \coordinate (P2) at (3,-3);
    \coordinate (P3) at (-0.5,-1);
    \coordinate (P4) at (5.5,-1);
    \coordinate (P5) at (-3,4);    
    \coordinate (P6) at (3,4);
    \coordinate (P7) at (-0.5,6);
    \coordinate (P8) at (5.5,6);
    \coordinate (C1) at ($(P1)!0.5!(P4)$);
    \coordinate (C2) at ($(P5)!0.5!(P8)$);
    \coordinate (Q1) at ($(P1)!0.5!(P2)$);
    \coordinate (Q2) at ($(P2)!0.5!(P4)$);
    \coordinate (Q3) at ($(P1)!0.5!(P3)$);
    \coordinate (Q4) at ($(P3)!0.5!(P4)$);
    \coordinate (Q5) at ($(P5)!0.5!(P6)$);
    \coordinate (Q6) at ($(P6)!0.5!(P8)$);
    \coordinate (Q7) at ($(P5)!0.5!(P7)$);
    \coordinate (Q8) at ($(P7)!0.5!(P8)$);
    \coordinate (X) at ($(Q5)+(0,10)$);
    \coordinate (Y) at ($(Q4)+(0,-10)$);
    
    \draw[thin] (P1) -- (P3) node[pos=0.5,sloped,rotate=-90]{\tikz\draw[->>,thin](0,0);};
    \draw[thin] (P3) -- (P4);
    \draw[thin] (P3) -- (P7);

    \path[fill=ForestGreen!30,fill opacity=0.6] (Q2) -- (Q3) -- (Q7) -- (Q6);
    \draw[ForestGreen,very thick] (Q2) -- (Q3);
    \draw[ForestGreen,very thick] (Q3) -- (Q7);
    \draw[ForestGreen,very thick] (Q6) -- (Q7);
    \draw[ForestGreen,very thick] (Q6) -- (Q2);

    \draw[thick] (P1) -- (P2) node[pos=0.5,sloped,rotate=-90]{\tikz\draw[->,thick](0,0);} node[pos=0.5,anchor=north]{$\r 0$}; 
    \draw[thick] (P2) -- (P4) node[pos=0.5,anchor=north west]{$\ns/\r 1$};
    \draw[thick] (P5) -- (P6) node[pos=0.5,sloped,rotate=-90]{\tikz\draw[->,thick](0,0);};
    \draw[thick] (P6) -- (P8) -- (P7);
    \draw[thick] (P5) -- (P7);
    \draw[thick] (P1) -- (P5);
    \draw[thick] (P2) -- (P6);
    \draw[thick] (P4) -- (P8) node[pos=0.5,anchor=west]{$\ns 0$};
    \draw[thick,dashed] (P6) -- (P7) node[pos=0.5,sloped,rotate=90]{\tikz\draw[->>,thick](0,0);};

\end{tikzpicture}
\qquad
\begin{tikzpicture}
[scale=0.5,>=stealth,baseline=(current  bounding  box.center)]
    \coordinate (P1) at (-3,-3);
    \coordinate (P2) at (3,-3);
    \coordinate (P3) at (3,4);
    \coordinate (P4) at (-3,4);
    \coordinate (P5) at (-0.5,-1);
    \coordinate (P6) at (5.5,-1);
    \coordinate (P7) at (5.5,6);
    \coordinate (P8) at (-0.5,6);
    \coordinate (R1) at (-3,1);
    \coordinate (R2) at (3,1);
    \coordinate (R3) at (-0.5,3);
    \coordinate (R4) at (5.5,3);
    \coordinate (R5) at (-2.375,1.5);
    \coordinate (R6) at (-1.125,2.5);
    \coordinate (R7) at (3.625,1.5);
    \coordinate (R8) at (4.875,2.5);
    \coordinate (R9) at (0,1);
    \coordinate (R10) at (2.5,3);
    \coordinate (A1) at (4.25,0);
    \coordinate (A2) at (-1.75,0);
    \coordinate (C1) at (0.625,1.5);
    \coordinate (C2) at (1.875,2.5);
    \coordinate (C3) at (-0.5,1.5);
    \coordinate (C4) at (0.5,2);
    \coordinate (C5) at (-2.375,-1);
    \coordinate (C6) at (-1.125,0);
    \coordinate (C7) at (3.625,-1);
    \coordinate (C8) at (4.875,0);
    \coordinate (C9) at (-0.1,3);
    \coordinate (C10) at (0.6,3);
    \coordinate (A3) at (-0.3,1.75);
    \coordinate (U1) at (-2.375,0.5);
    \coordinate (U2) at (-2.1,0);
    \coordinate (U3) at (3.625,0.5);
    \coordinate (U4) at (3.85,0);
    \coordinate (U5) at (4.55,0);
    \coordinate (U6) at (4.875,1.5);
    \draw[thick] (P1) -- (P2) coordinate[pos=0.5](Q1) node[pos=0.5,sloped,rotate=-90]{\tikz\draw[->,thick](0,0);} node[pos=0.5,anchor=north]{$\ns/\r 1$}; 
    \draw[thick] (P2) -- (P3);
    \draw[thick] (P3) -- (P4) coordinate[pos=0.5](Q2) node[pos=0.3,sloped,rotate=-90]{\tikz\draw[->,thick](0,0);};;
    \draw[thick] (P4) -- (P1);
    \draw[thick] (P4) -- (P8) coordinate[pos=0.25](Q5) coordinate[pos=0.75](Q6);
    \draw[thick] (P7) -- (P8) coordinate[pos=0.5](Q4);
    \draw[thick] (P7) -- (P6) node[pos=0.5,anchor=west]{$\ns 0$};
    \draw[thick] (P6) -- (P2) node[pos=0.5,anchor=north west]{$\r 0$};
    \draw[thick] (P3) -- (P7) coordinate[pos=0.25](Q7) coordinate[pos=0.75](Q8);
    \draw[thin] (P8) -- (P5); 
    \draw[thin] (P5) -- (P6) coordinate[pos=0.5](Q3);
    \draw[thin] (P1) -- (P5) node[pos=0.5,sloped,rotate=-90]{\tikz\draw[->>,thick](0,0);};;
    \draw[ForestGreen,very thick, name path=Q1Q2] (Q1) -- (Q2);
    \draw[ForestGreen,very thick, name path=Q2Q5] (Q5) -- (Q2) coordinate [pos=0.7](W1); 
    \draw[ForestGreen,very thick] (Q1) -- (Q3) coordinate[pos=0.25](W3) coordinate[pos=0.5](W4) coordinate[pos=0.75](W5) coordinate[pos=0.125](W6); 
    \draw[ForestGreen,thick,densely dashed, name path=Q3Q4] (Q3) -- (Q4); 
    \draw[ForestGreen,very thick] (Q4) -- (Q8); 
    \draw[ForestGreen,very thick] (Q8) -- (R8); 
    \draw[ForestGreen,very thick] (R7) -- (Q7);  
    \draw[ForestGreen,very thick, name path=Q6Q7] (Q7) -- (Q6) coordinate[pos=0.55](W2); 
    \draw[ForestGreen,thick,densely dashed, name path=Q6R6] (Q6) -- (R6); 
    \draw[ForestGreen,very thick] (R5) -- (Q5);
    \draw[ForestGreen,thick,dotted] (R9) .. controls (C1) .. (R7) coordinate[pos=0.35](W7);
    \draw[ForestGreen,thick,dotted] (R10) .. controls (C2) .. (R8);
    \draw[ForestGreen,very thick] (R8) .. controls (C8) and (C7) .. (R7) coordinate[pos=0.6](A1); 
    \draw[ForestGreen,thick, densely dashed] (R6) .. controls (C6) and (C5) .. (R5) coordinate[pos=0.6](A2);
    \draw[ForestGreen,thick, dotted] (R5) .. controls (C3) and (C4) .. (R6) coordinate[pos=0.6](A3);
    \draw[ForestGreen,very thick] (A2) to [bend right=30] (A3);
    \draw[ForestGreen,thick, dotted, name path=A3W7] (A3) .. controls (C9) and (C10) .. (W7) coordinate[pos=0.5](A4);
    \draw[ForestGreen,thick, dotted] (W6) to [bend right=2] (W7);
    \draw[ForestGreen,very thick, name path=A1W4] (A1) to [bend right=45] (W4);
    \draw[ForestGreen,thick, dotted] (W3) to [bend left=40] (R7);
    \draw[ForestGreen,thin, dotted] (W5) to [bend left=45] (R8);
    \draw[ForestGreen,thick, dotted] plot [smooth, tension=1] coordinates {(W1) (A4) (W2)};
    \path[name intersections={of=Q1Q2 and A3W7,by=A5}];
    \path[name intersections={of=Q3Q4 and Q6Q7,by=A6}];
    \path[name intersections={of=Q3Q4 and A1W4,by=A7}];
    \path[name intersections={of=Q2Q5 and Q6R6,by=A8}];
    \draw[ForestGreen, very thick] (A3) to [bend left=15] (A5);
    \draw[ForestGreen, very thin,fill=ForestGreen!30,fill opacity=0.1] (R5) .. controls (U1) and (U2) .. (A2) to [bend right=30] (A3) to [bend left=15] (A5) --(Q2) -- (Q5) -- (R5);
    \draw[ForestGreen, very thin,fill=ForestGreen!30,fill opacity=0.6] (Q1) -- (Q3) -- (A7) to [bend left=15] (A1) .. controls (U4) and (U3) .. (R7) -- (Q7) -- (Q6) -- (A8) -- (Q2) -- (Q1);
    \draw[ForestGreen, very thin,fill=ForestGreen!30,fill opacity=0.1] (A1) .. controls (U4) and (U3) .. (R7) -- (Q7) -- (A6) -- (Q4) -- (Q8) -- (R8) .. controls (U6) and (U5) .. (A1);
    \draw[ForestGreen,very thick] (Q3) -- (A7);
    \draw[ForestGreen,very thick] (Q4) -- (A6);
    \draw[ForestGreen,very thick] (R5) .. controls (U1) and (U2) .. (A2);
    \draw[ForestGreen,very thick] (Q6) -- (A8);
    
    \draw[thick,dashed] ($(P4)!0.5!(P8)$) -- (P3) node[pos=0.5,sloped,rotate=90]{\tikz\draw[->>,thick](0,0);};
    \draw[thick,dashed] ($(P3)!0.5!(P7)$) -- (P8) node[pos=0.5,sloped,rotate=90]{\tikz\draw[->>,thick](0,0);};
\end{tikzpicture} 
\caption{The mapping tori associated to 
$(T_{(0/1,1)})^3_1$ and $(T_{(0/1,1)})^2_2$. Here the torus on the bottom and the one on the top are identified. The dashed arrows represent how the generators of the bottom are transposed to the top.}
\label{fig:T101mappingtori}

\bigskip

\begin{tikzpicture}
[scale=0.5,>=stealth,baseline=(current  bounding  box.center)]
    \coordinate (P1) at (-3,-3);
    \coordinate (P2) at (3,-3);
    \coordinate (P3) at (-0.5,-1);
    \coordinate (P4) at (5.5,-1);
    \coordinate (P5) at (-3,4);    
    \coordinate (P6) at (3,4);
    \coordinate (P7) at (-0.5,6);
    \coordinate (P8) at (5.5,6);
    \coordinate (C1) at ($(P1)!0.5!(P4)$);
    \coordinate (C2) at ($(P5)!0.5!(P8)$);
    \coordinate (Q1) at ($(P1)!0.5!(P2)$);
    \coordinate (Q2) at ($(P2)!0.5!(P4)$);
    \coordinate (Q3) at ($(P1)!0.5!(P3)$);
    \coordinate (Q4) at ($(P3)!0.5!(P4)$);
    \coordinate (Q5) at ($(P5)!0.5!(P6)$);
    \coordinate (Q6) at ($(P6)!0.5!(P8)$);
    \coordinate (Q7) at ($(P5)!0.5!(P7)$);
    \coordinate (Q8) at ($(P7)!0.5!(P8)$);
    \coordinate (X) at ($(Q5)+(0,10)$);
    \coordinate (Y) at ($(Q4)+(0,-10)$);
    
    \draw[thin] (P1) -- (P3) node[pos=0.5,sloped,rotate=-90]{\tikz\draw[->>,thick](0,0);};
    \draw[thin] (P3) -- (P4);
    \draw[thin] (P3) -- (P7);

    \path[fill=ForestGreen!30,fill opacity=0.6] (Q1) -- (Q4) -- (Q8) -- (Q5) -- (Q1);
    \draw[ForestGreen,very thick] (Q1) -- (Q4);
    \draw[ForestGreen,very thick] (Q1) -- (Q5);
    \draw[ForestGreen,very thick] (Q4) -- (Q8);
    \draw[ForestGreen,very thick] (Q5) -- (Q8);
    
    \draw[thick] (P1) -- (P2) node[pos=0.5,anchor=north]{$\ns/\r 1$} node[pos=0.5,sloped,rotate=-90]{\tikz\draw[->,thick](0,0);}; 
    \draw[thick] (P2) -- (P4) node[pos=0.5,anchor=north west]{$\r 0$};
    \draw[thick] (P5) -- (P6) node[pos=0.5,sloped,rotate=90]{\tikz\draw[->,thick](0,0);};
    \draw[thick] (P6) -- (P8) -- (P7);
    \draw[thick] (P7) -- (P5) node[pos=0.5,sloped,rotate=90]{\tikz\draw[->>,thick](0,0);};
    \draw[thick] (P1) -- (P5);
    \draw[thick] (P2) -- (P6);
    \draw[thick] (P4) -- (P8) node[pos=0.5,anchor=west]{$\ns 0$};
    
\end{tikzpicture}%
\qquad
\begin{tikzpicture}
[scale=0.5,>=stealth,baseline=(current  bounding  box.center)]
   \coordinate (P1) at (-3,-3);
    \coordinate (P2) at (3,-3);
    \coordinate (P3) at (-0.5,-1);
    \coordinate (P4) at (5.5,-1);
    \coordinate (P5) at (-3,4);    
    \coordinate (P6) at (3,4);
    \coordinate (P7) at (-0.5,6);
    \coordinate (P8) at (5.5,6);
    \coordinate (C1) at ($(P1)!0.5!(P4)$);
    \coordinate (C2) at ($(P5)!0.5!(P8)$);
    \coordinate (C3) at ($(C1)!0.5!(C2)$);
    \coordinate (Q1) at ($(P1)!0.5!(P2)$);
    \coordinate (Q2) at ($(P2)!0.5!(P4)$);
    \coordinate (Q3) at ($(P1)!0.5!(P3)$);
    \coordinate (Q4) at ($(P3)!0.5!(P4)$);
    \coordinate (Q5) at ($(P5)!0.5!(P6)$);
    \coordinate (Q6) at ($(P6)!0.5!(P8)$);
    \coordinate (Q7) at ($(P5)!0.5!(P7)$);
    \coordinate (Q8) at ($(P7)!0.5!(P8)$);
    \coordinate (X) at ($(Q5)+(0,10)$);
    \coordinate (Y) at ($(Q4)+(0,-2)$);
    
    \draw[thin] (P1) -- (P3) node[pos=0.5,sloped,rotate=-90]{\tikz\draw[->>,thick](0,0);};
    \draw[thin] (P3) -- (P4);
    \draw[thin] (P3) -- (P7);
    
    \path[name path=Q1Q5] (Q1) -- (Q5);
    \path[name path=Q2Q6] (Q2) -- (Q6);  
    \path[name path=Q3Q7] (Q3) -- (Q7);
    \path[name path=Q4Q8] (Q4) -- (Q8);

    \path[name path=Q2Q3] (Q2) -- (Q3);
    \path[name path=Q4Y] (Q4) -- (Y);
    \path[name path=Q4Q8] (Q4) -- (Q8);     
    \path[name intersections={of=Q1Q5 and Q2Q3, by=A1}];
    \path[name intersections={of=Q2Q3 and Q4Y,by=A2}];

    \path (A1) .. controls (C1) .. (Q1) coordinate[pos=0.5](B1);
    \path[name path=Q5Q6] (Q5) .. controls (C2) .. (Q6) coordinate[pos=0.5](A3);
    \path (Q7) .. controls (C2) .. (Q8) coordinate[pos=0.5](A4);
    \path (Q4) .. controls (C1) .. (A2) coordinate[pos=0.5](B2);     
    
    \path[name intersections={of=Q4Q8 and Q5Q6,by=A5}];
    
    \path[fill=ForestGreen!30,fill opacity=0.1] (Q3) -- (A1) -- (Q5) .. controls (C2) .. (A5) -- (Q8) .. controls (C2) .. (Q7) -- (Q3);
    \path[fill=ForestGreen!30,fill opacity=0.6] (Q1) .. controls ($(B1)+(0,-0.15)$) .. (B1) .. controls ($(C3)+(-0.4,0.8)$) and ($(C3)+(0.4,0.8)$) .. (B2) .. controls ($(B2)+(0,-0.12)$) .. (A2) -- (Q2) -- (Q6) .. controls (C2) .. (Q5) -- (Q1);

    \draw[ForestGreen,very thick] (B1) .. controls ($(C3)+(-0.4,0.8)$) and ($(C3)+(0.4,0.8)$) .. (B2) coordinate[pos=0.5](C4);
    \draw[ForestGreen,very thick,dotted] plot [smooth, tension=1.7] coordinates {(A3) (C4) (A4)};
    
    \draw[ForestGreen,very thick] (Q1) -- (Q5);
    \draw[ForestGreen,very thick] (Q3) -- (Q7);
    \draw[ForestGreen,very thick] (Q3) -- (A1);    
    \draw[ForestGreen,very thick] (Q2) -- (A2);

    \draw[ForestGreen,very thick] (Q2) -- (Q6);
    \draw[ForestGreen,very thick] (Q5) .. controls (C2) .. (Q6);
    \draw[ForestGreen,very thick] (Q7) .. controls (C2) .. (Q8);    
    \draw[ForestGreen,very thick] (Q8) -- (A5);
    \draw[ForestGreen,very thick,dashed] (A5) -- (Q4);

    \draw[ForestGreen,very thick] (Q1) .. controls ($(B1)+(0,-0.15)$) .. (B1);
    \draw[ForestGreen,very thick,dashed] (B1) .. controls ($(B1)+(-0.1,0.1)$) .. (A1);
    
    \draw[ForestGreen,very thick] (A2) .. controls ($(B2)+(0,-0.12)$) .. (B2);
    \draw[ForestGreen,very thick,dashed] (B2) .. controls ($(B2)+(0,0.1)$) .. (Q4);
    
    \draw[thick] (P1) -- (P2) node[pos=0.5,anchor=north]{$\ns/\r 1$} node[pos=0.5,sloped,rotate=-90]{\tikz\draw[->,thick](0,0);};
    \draw[thick] (P2) -- (P4) node[pos=0.5,anchor=north west]{$\ns/\r 1$};
    \draw[thick] (P5) -- (P6) node[pos=0.5,sloped,rotate=-90]{\tikz\draw[->>,thick](0,0);};
    \draw[thick] (P6) -- (P8) -- (P7);
    \draw[thick] (P5) -- (P7) node[pos=0.5,sloped,rotate=90]{\tikz\draw[->,thick](0,0);};
    \draw[thick] (P1) -- (P5);
    \draw[thick] (P2) -- (P6);
    \draw[thick] (P4) -- (P8) node[pos=0.5,anchor=west]{$\ns 0$};
\end{tikzpicture}
\caption{Mapping tori associated to the non-trivial entries $(S_{(0/1,1)})^2_1$ and $(S_{(0/1,1)})^3_3$. The vertical direction has to be read from the bottom to the top as before.}
\label{fig:S101mappingtori}
\end{figure}

In the basis fixed by \eqref{classesoftori} the final result is
\begin{alignat}{4}    
    &S_{(0,0)}=
    \begin{pmatrix}
    0&1&0&0&0&0&0&0&0\\
    1&0&0&0&0&0&0&0&0\\
    0&0&1&0&0&0&0&0&0\\
    0&0&0&0&0&0&1&0&0\\
    0&0&0&0&0&1&0&0&0\\
    0&0&0&0&1&0&0&0&0\\
    0&0&0&1&0&0&0&0&0\\
    0&0&0&0&0&0&0&0&1\\
    0&0&0&0&0&0&0&1&0\\
    \end{pmatrix},
    &&\qquad T_{(0,0)}=
    \begin{pmatrix}
    0&0&1&0&0&0&0&0&0\\
    0&1&0&0&0&0&0&0&0\\
    1&0&0&0&0&0&0&0&0\\
    0&0&0&0&0&0&0&1&0\\
    0&0&0&0&0&0&1&0&0\\
    0&0&0&0&0&0&0&0&e^{-i \frac{\pi \nu}{4}}\\
    0&0&0&0&1&0&0&0&0\\
    0&0&0&e^{i \frac{\pi \nu}{4}}&0&0&0&0&0\\
    0&0&0&0&0&1&0&0&0\\
    \end{pmatrix},
    \label{ST00}
\end{alignat}
\begin{alignat}{4}        
    &S_{(0,1)}=
    \begin{pmatrix}
    0&(-i)^\nu&0\\
    1&0&0\\
    0&0&e^{-i\frac{\pi}{4}\nu}\\
    \end{pmatrix},
    &&\qquad T_{(0,1)}=
    \begin{pmatrix}
    0&0&e^{i\frac{\pi \nu}{4}}\\
    0&1&0\\
    1&0&0\\
    \end{pmatrix},\label{ST01}\\
    &S_{(1,1)}=
    \begin{pmatrix}
    0&i^\nu&0\\ 
    1&0&0\\
    0&0&e^{i\frac{\pi}{4}\nu}\\
    \end{pmatrix},
    &&\qquad T_{(1,1)}=
    \begin{pmatrix}
    0&0&e^{-i\frac{\pi \nu}{4}}\\
    0&1&0\\
    1&0&0\\
    \end{pmatrix}.\label{ST11}
\end{alignat}

For the class $(1,0)$ instead it follows immediately that the modular transformations act trivially on the single element $\{\r0,\r0\}$.
As one can immediately check, the matrices presented indeed satisfy the relations \eqref{Mp}.

Finally, to see the full $\widetilde{\MCG}(X)$ representation we need to compute the action of $(-1)^\CQ$ as well. With the previous results one can determine it for free. In fact, $(-1)^\CQ$ will act diagonally for a generic torus $\{s_0g_0,s_1g_1\}$ and the $3$-manifold describing it is again a $3$-torus, with the vertical direction now given by $\ns1$.
Moreover $(-1)^\CQ$ commutes with $\Mp$, so it is enough to compute its value for a single element in each orbit of it, as one can see by applying $S$ and $T$ to the result found.
All the orbits have at least one element with trivial $g_0=0$, so we can restrict to such cases. If we denote with $\{s_0g_0,s_1g_1,s'g'\}$ the spin periodicity and holonomies of a $
\TT^3$ associated to the bordisms $\mathrm{id}/\F:\{s_0g_0,s_1g_1\}\xrightarrow{\mathrm{id}/\F}\{s_0g_0,s_1g_1\}$, then we know that $g'=0$ and $s'=\ns/\r$ for $\mathrm{id}/\F$. Therefore a 3-torus $\{s_0 0,s_1g_1,\ns1\}$ that describes a $(-1)^\CQ$ action describes also the action of $\mathrm{id}$ or $\F$ for $\{s_1g_1,\ns1\}$, depending on the value of $s_0$. This means $(-1)^\CQ$ acts non-trivially only if the mapping torus describes $(-1)^\CF_{(0/1,1)}$ as well, which implies that necessarily $s_0=\r$ and $s_1g_1=\ns1,\,\r1$. From this follows that the action of $(-1)^\CQ$ is exactly the same of $\F$ and, not surprisingly, a bordism invariant, i.e.
\begin{equation}
    (-1)^\CQ_{(0,0)}=(-1)^\CQ_{(1,0)}=\mathrm{id},\qquad     (-1)^\CQ_{(0,1)}=(-1)^\CQ_{(1,1)}=(-1)^\nu \cdot \mathrm{id}.
\end{equation}

\subsection{Defect rules}
\label{sec:defect-rules}

As was used in the previous section, the $\Z_8$ valued cobordism invariant generating the group 3d iTQFTs, on a given closed spin 3-manifold $Y$ and $a\in H^1(Y,\Z_2)$, can be realized as
\begin{equation}
    \beta_s(a)=\ABK(\PD(a),s|_{\PD(a)}),
    \label{beta-ABK-invariant}
\end{equation}
where the right hand side is the Arf-Brown-Kervaire (ABK) invariant of a 2d smooth surface representing the Poincar\'e dual to $a$ in $Y$, with pin$^-$ structure induced from the ambient spin structure (see \cite{kirby_taylor_1991} for details). Physically such surface can be understood as the support of the $\Z_2$ global symmetry charge operator. On an arbitrary bordism the 3d spin$\times \Z_2$-TQFT can be described in terms of a 2d pin$^-$ iTQFT supported on such codimension one defects inside the bordism, possibly ending on codimension one defects in the 2d boundaries. The 2d pin$^-$ iTQFT has an action given by a multiple of the ABK invariant and physically corresponds to a stack of Kitaev spin chains. 

When the anomalous 2d QFT is considered on the boundary of the spacetime of the 3d TQFT, the boundaries of the $\Z_2$ bulk charge operator are identified with the $\Z_2$ charge operators in the 2d QFT.

A 3d cylinder $X\times [0,1]$ with a non-trivial surface defect inside can be understood as sequence of ``moves'' applied to the line defects in $X$ as one goes along the ``time'' direction $[0,1]$.
The calculation of the matrix elements of modular transformations that was reviewed in the previous section (with technical details in Appendix \ref{app:matrices}) then can be reformulated in terms of certain rules on the changes of the charge defects in 2d, similar to the rules used in \cite{chan2016topological,Lin:2019kpn,Lin:2021udi} in the bosonic case.  

Because of the fermionic nature of the 2d theory we decorate the topological $\Z_2$-charge line operators with additional information: a lift of the tangent vector to a $\Spin(2)$, a double cover of $SO(2)$ with respect to some fixed reference frame. That is, at each point of the line one has to specify a lift of the angle of the slope, defined modulo $2\pi$, to a value $\mod 4\pi$ in a continuous manner.  Below we will explicitly indicate such lifts in the diagrams near the relevant points.

Consider first configuration of such line operators on a torus with periodic boundary condition on fermions. This means that the transition functions for the $\Spin(2)$ bundle are trivial. For each connected component of the line defect it is then sufficient to fix a lift of the slope to $\Spin(2)$ at a single point. As we will see, in this case it is enough to consider the following single basic move:
\begin{equation}
\begin{tikzpicture}[scale=0.6,baseline=(current  bounding  box.center)]
\draw[ForestGreen,thick,->-] (0,2) to[out=0,in=90] node[pos=0,below] {$0$} node[pos=1,right] {$-\frac{\pi}{2}$} (2,0);
\draw[ForestGreen,thick,->-] (2,4) to[out=-90,in=180] node[pos=1,below] {$0$} node[pos=0,right] {$-\frac{\pi}{2}$} (4,2);
\node at (6,2) {$\sim \;\;\alpha\;\cdot$};
\begin{scope}[shift={(8,0)}]
\draw[ForestGreen,thick,->-] (0,2) to[out=0,in=-90] node[pos=0,below] {$0$} node[pos=1,right] {$\frac{\pi}{2}$} (2,4);
\draw[ForestGreen,thick,->-] (2,0) to[out=90,in=180] node[pos=1,below] {$0$} node[pos=0,right] {$\frac{\pi}{2}$} (4,2);
\end{scope}
\end{tikzpicture}
\end{equation}
where $\alpha$ is certain phase. The endpoints of the defects in the diagrams are pairwise identified. By applying this basic rule twice, we get the following rule for changing the orientation of the defect:
\begin{equation}
\begin{tikzpicture}[scale=0.6,baseline=(current  bounding  box.center)]
\draw[ForestGreen,thick,->-] (0,0) -- node[pos=0.7,above] {$0$} (4,0);
\node at (6,0) {$\sim \;\;\alpha^2\;\cdot$};
\begin{scope}[shift={(8,0)}]
\draw[ForestGreen,thick,->-] (4,0) -- node[pos=0.3,above] {$\pi$} (0,0);
\end{scope}
\end{tikzpicture}
\end{equation}
from which it follows that
\begin{equation}
\begin{tikzpicture}[scale=0.6,baseline=(current  bounding  box.center)]
\draw[ForestGreen,thick,->-] (0,0) -- node[pos=0.7,above] {$0$} (4,0);
\node at (6,0) {$\sim \;\;\alpha^4\;\cdot$};
\begin{scope}[shift={(8,0)}]
\draw[ForestGreen,thick,->-] (0,0) -- node[pos=0.7,above] {$2\pi$} (4,0);
\end{scope}
\end{tikzpicture}.
\label{move-lift-change}
\end{equation}
And, in particular
\begin{equation}
\begin{tikzpicture}[scale=0.6,baseline=(current  bounding  box.center)]
\draw[ForestGreen,thick,->-] (0,0) -- node[pos=0.7,above] {$0$} (4,0);
\node at (6,0) {$\sim \;\;\alpha^8\;\cdot$};
\begin{scope}[shift={(8,0)}]
\draw[ForestGreen,thick,->-] (0,0) -- node[pos=0.9,above] {$4\pi\sim 0$} (4,0);
\end{scope}
\end{tikzpicture}
\end{equation}
so that $\alpha^8=1$. Therefore, for self-consistency $\alpha$ is required to be an 8th root of unity. This is in agreement with classification of the anomalies by $\Z_8$.

The basis elements then correspond to particular configuration of the defects on $T^2$. As in the previous section, we can choose them as the results of the application of $\mathrm{id},S,T$ transforms to some particular configuration, as shown in Figure \ref{fig:defect-PP-basis}. The result of the application of the $S$ and $T$ transformation to the basis elements is shown in Figures \ref{fig:defect-S-11} and \ref{fig:defect-T-11}.

\begin{figure}
    \centering
\begin{equation}
\begin{array}{rcl}
\vspace{1ex}
\begin{tikzpicture}[scale=0.4,baseline=(current  bounding  box.center)]
\draw (0,0) rectangle (4,4);
\node[left] at (0,2) {$\r$};
\node[below] at (2,0) {$\r$};
\draw[ForestGreen,thick,->-] (2,0) -- (2,4) node[pos=.75,left] {$\frac{\pi}{2}$};
\end{tikzpicture} 
&
\qquad\stackrel{\mathrm{id}}{\longrightarrow}\qquad
&
\begin{tikzpicture}[scale=0.4,baseline=(current  bounding  box.center)]
\draw (0,0) rectangle (4,4);
\node[left] at (0,2) {$\r$};
\node[below] at (2,0) {$\r$};
\draw[ForestGreen,thick,->-] (2,0) -- (2,4) node[pos=.75,left] {$\frac{\pi}{2}$};
\end{tikzpicture}\;, 
\\
\vspace{1ex}
\begin{tikzpicture}[scale=0.4,baseline=(current  bounding  box.center)]
\draw (0,0) rectangle (4,4);
\node[left] at (0,2) {$\r$};
\node[below] at (2,0) {$\r$};
\draw[ForestGreen,thick,->-] (2,0) -- (2,4) node[pos=.75,left] {$\frac{\pi}{2}$};
\end{tikzpicture} 
&
\qquad\stackrel{S}{\longrightarrow}\qquad
&
\begin{tikzpicture}[scale=0.4,baseline=(current  bounding  box.center)]
\draw (0,0) rectangle (4,4);
\node[left] at (0,2) {$\r$};
\node[below] at (2,0) {$\r$};
\draw[ForestGreen,thick,->-] (0,2) -- (4,2) node[pos=.75,above] {$0$};
\end{tikzpicture}\;,
\\
\begin{tikzpicture}[scale=0.4,baseline=(current  bounding  box.center)]
\draw (0,0) rectangle (4,4);
\node[left] at (0,2) {$\r$};
\node[below] at (2,0) {$\r$};
\draw[ForestGreen,thick,->-] (2,0) -- (2,4) node[pos=.75,left] {$\frac{\pi}{2}$};
\end{tikzpicture} 
&
\qquad\stackrel{T}{\longrightarrow}\qquad
&
\begin{tikzpicture}[scale=0.4,baseline=(current  bounding  box.center)]
\draw (0,0) rectangle (4,4);
\node[left] at (0,2) {$\r$};
\node[below] at (2,0) {$\r$};
\draw[ForestGreen,thick,->-] (2,0) to[out=90,in=0] node[pos=0.2,right] {$\frac{\pi}{2}$} (0,2);
\draw[ForestGreen,thick,->-] (4,2) to[out=180,in=-90] node[pos=0.2,below] {${\pi}$} (2,4);
\end{tikzpicture}\;.
\end{array}
\nonumber
\end{equation}
    \caption{The choice of basis configurations of $\Z_2$ charge defects on $\TT^2$ with periodic-periodic spin structure.}
    \label{fig:defect-PP-basis}
\end{figure}
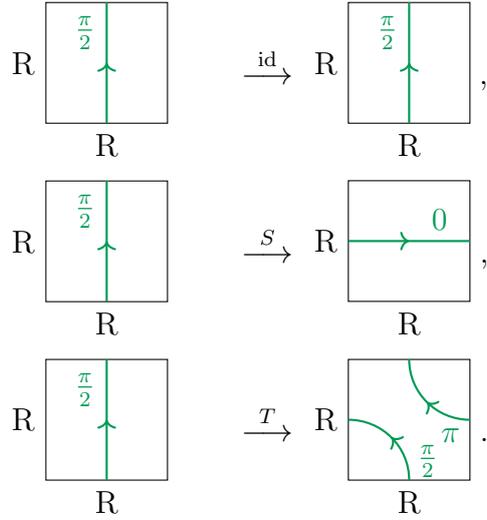

\begin{figure}
    \centering
\begin{equation}
\begin{array}{rcl}
\vspace{1ex}
\begin{tikzpicture}[scale=0.4,baseline=(current  bounding  box.center)]
\draw (0,0) rectangle (4,4);
\node[left] at (0,2) {$\r$};
\node[below] at (2,0) {$\r$};
\draw[ForestGreen,thick,->-] (2,0) -- (2,4) node[pos=.75,left] {$\frac{\pi}{2}$};
\end{tikzpicture}
&
\stackrel{S}{\longrightarrow}
&
\begin{tikzpicture}[scale=0.4,baseline=(current  bounding  box.center)]
\draw (0,0) rectangle (4,4);
\node[left] at (0,2) {$\r$};
\node[below] at (2,0) {$\r$};
\draw[ForestGreen,thick,->-] (0,2) -- (4,2) node[pos=.75,above] {$0$};
\end{tikzpicture}\;,
\\
\vspace{1ex}
\begin{tikzpicture}[scale=0.4,baseline=(current  bounding  box.center)]
\draw (0,0) rectangle (4,4);
\node[left] at (0,2) {$\r$};
\node[below] at (2,0) {$\r$};
\draw[ForestGreen,thick,->-] (0,2) -- (4,2) node[pos=.75,above] {$0$};
\end{tikzpicture}
&
\stackrel{S}{\longrightarrow}
&
\begin{tikzpicture}[scale=0.4,baseline=(current  bounding  box.center)]
\draw (0,0) rectangle (4,4);
\node[left] at (0,2) {$\r$};
\node[below] at (2,0) {$\r$};
\draw[ForestGreen,thick,->-] (2,4) -- (2,0) node[pos=.75,left] {$-\frac{\pi}{2}$};
\end{tikzpicture} 
\;\sim\; 
\alpha^2\cdot
\begin{tikzpicture}[scale=0.4,baseline=(current  bounding  box.center)]
\draw (0,0) rectangle (4,4);
\node[left] at (0,2) {$\r$};
\node[below] at (2,0) {$\r$};
\draw[ForestGreen,thick,->-] (2,0) -- (2,4) node[pos=.75,left] {$\frac{\pi}{2}$};
\end{tikzpicture}\;,
\\
\vspace{1ex}
\begin{tikzpicture}[scale=0.4,baseline=(current  bounding  box.center)]
\draw (0,0) rectangle (4,4);
\node[left] at (0,2) {$\r$};
\node[below] at (2,0) {$\r$};
\draw[ForestGreen,thick,->-] (2,0) to[out=90,in=0] node[pos=0.2,right] {$\frac{\pi}{2}$} (0,2);
\draw[ForestGreen,thick,->-] (4,2) to[out=180,in=-90] node[pos=0.2,below] {${\pi}$} (2,4);
\end{tikzpicture}
&
\stackrel{S}{\longrightarrow}
&
\begin{tikzpicture}[scale=0.4,baseline=(current  bounding  box.center)]
\draw (0,0) rectangle (4,4);
\node[left] at (0,2) {$\r$};
\node[below] at (2,0) {$\r$};
\draw[ForestGreen,thick,->-] (0,2) to[out=0,in=-90] node[pos=0.15,below] {$0$} node[pos=0.8,right] {$\frac{\pi}{2}$} (2,4);
\draw[ForestGreen,thick,->-] (2,0) to[out=90,in=180]  (4,2);
\end{tikzpicture}
\;\sim\;\alpha\cdot
\begin{tikzpicture}[scale=0.4,baseline=(current  bounding  box.center)]
\draw (0,0) rectangle (4,4);
\node[left] at (0,2) {$\r$};
\node[below] at (2,0) {$\r$};
\draw[ForestGreen,thick,->-] (2,0) to[out=90,in=0] node[pos=0.2,right] {$\frac{\pi}{2}$} (0,2);
\draw[ForestGreen,thick,->-] (4,2) to[out=180,in=-90] node[pos=0.2,below] {${\pi}$} (2,4);
\end{tikzpicture}\;.
\end{array}
\nonumber
\end{equation}
    \caption{The action of $S$ transformation on the basis elements chosen in Figure \ref{fig:defect-PP-basis}.}
    \label{fig:defect-S-11}
\end{figure}

\begin{figure}
    \centering
\begin{equation}
\begin{array}{rcl}
\vspace{1ex}
\begin{tikzpicture}[scale=0.4,baseline=(current  bounding  box.center)]
\draw (0,0) rectangle (4,4);
\node[left] at (0,2) {$\r$};
\node[below] at (2,0) {$\r$};
\draw[ForestGreen,thick,->-] (2,0) -- (2,4) node[pos=.75,left] {$\frac{\pi}{2}$};
\end{tikzpicture}
&
\stackrel{T}{\longrightarrow}
&
\begin{tikzpicture}[scale=0.4,baseline=(current  bounding  box.center)]
\draw (0,0) rectangle (4,4);
\node[left] at (0,2) {$\r$};
\node[below] at (2,0) {$\r$};
\draw[ForestGreen,thick,->-] (2,0) to[out=90,in=0] node[pos=0.2,right] {$\frac{\pi}{2}$} (0,2);
\draw[ForestGreen,thick,->-] (4,2) to[out=180,in=-90] node[pos=0.2,below] {${\pi}$} (2,4);
\end{tikzpicture}\;,
\\
\vspace{1ex}
\begin{tikzpicture}[scale=0.4,baseline=(current  bounding  box.center)]
\draw (0,0) rectangle (4,4);
\node[left] at (0,2) {$\r$};
\node[below] at (2,0) {$\r$};
\draw[ForestGreen,thick,->-] (0,2) -- (4,2) node[pos=.75,above] {$0$};
\end{tikzpicture}
&
\stackrel{T}{\longrightarrow}
&
\begin{tikzpicture}[scale=0.4,baseline=(current  bounding  box.center)]
\draw (0,0) rectangle (4,4);
\node[left] at (0,2) {$\r$};
\node[below] at (2,0) {$\r$};
\draw[ForestGreen,thick,->-] (0,2) -- (4,2) node[pos=.75,above] {$0$};
\end{tikzpicture}\;,
\\
\begin{tikzpicture}[scale=0.4,baseline=(current  bounding  box.center)]
\draw (0,0) rectangle (4,4);
\node[left] at (0,2) {$\r$};
\node[below] at (2,0) {$\r$};
\draw[ForestGreen,thick,->-] (2,0) to[out=90,in=0] node[pos=0.2,right] {$\frac{\pi}{2}$} (0,2);
\draw[ForestGreen,thick,->-] (4,2) to[out=180,in=-90] node[pos=0.2,below] {${\pi}$} (2,4);
\end{tikzpicture}
&
\stackrel{T}{\longrightarrow}
&
\begin{tikzpicture}[scale=0.4,baseline=(current  bounding  box.center)]
\draw (0,0) rectangle (4,4);
\node[left] at (0,2) {$\r$};
\node[below] at (2,0) {$\r$};
\draw[ForestGreen,thick,->-] (2,0) to[out=90,in=0] node[pos=0.2,right] {$\frac{\pi}{2}$} (0,1.3);
\draw[ForestGreen,thick,->-] (4,2.7) to[out=180,in=-90] node[pos=0.2,below] {${\pi}$} (2,4);
\draw[ForestGreen,thick,->-] (4,1.3) to[out=180,in=0] (0,2.7);
\end{tikzpicture}
\;\sim\;\alpha\cdot
\begin{tikzpicture}[scale=0.4,baseline=(current  bounding  box.center)]
\draw (0,0) rectangle (4,4);
\node[left] at (0,2) {$\r$};
\node[below] at (2,0) {$\r$};
\draw[ForestGreen,thick,->-] (0,1.3) to[out=0,in=90] node[pos=0.8,right] {$\frac{3\pi}{2}$} (2,0);
\draw[ForestGreen,thick,->-] (2,4) to[out=-90,in=0] (0,2.7);
\draw[ForestGreen,thick,->-] (4,2.7) to[out=180,in=180,looseness=3] node[pos=0.2,above] {${\pi}$} (4,1.3);
\end{tikzpicture}
\\
&
&
\;\sim\;\alpha\cdot
\begin{tikzpicture}[scale=0.4,baseline=(current  bounding  box.center)]
\draw (0,0) rectangle (4,4);
\node[left] at (0,2) {$\r$};
\node[below] at (2,0) {$\r$};
\draw[ForestGreen,thick,->-] (2,4) -- (2,0) node[pos=.75,left] {$\frac{3\pi}{2}$};
\end{tikzpicture} 
\;\sim\;\alpha^{-1}\cdot
\begin{tikzpicture}[scale=0.4,baseline=(current  bounding  box.center)]
\draw (0,0) rectangle (4,4);
\node[left] at (0,2) {$\r$};
\node[below] at (2,0) {$\r$};
\draw[ForestGreen,thick,->-] (2,0) -- (2,4) node[pos=.75,left] {$\frac{\pi}{2}$};
\end{tikzpicture}\;.
\end{array}
\nonumber
\end{equation}
    \caption{The action of $T$ transformation on the basis elements chosen in Figure \ref{fig:defect-PP-basis}.}
    \label{fig:defect-T-11}
\end{figure}
In terms of the basis elements in Figure \ref{fig:defect-PP-basis} the corresponding matrices read as follows:
\begin{equation}
    S=\left(
    \begin{array}{ccc}
        0 & \alpha^2 & 0 \\
        1 & 0 & 0 \\
        0 & 0 & \alpha 
    \end{array}
    \right),
    \qquad
        T=\left(
    \begin{array}{ccc}
        0 & 0 & \alpha^{-1} \\
        0 & 1 & 0 \\
        1 & 0 & 0
    \end{array}
    \right).
\label{eq:S-T-from-defects}
\end{equation}
They satisfy the conditions
\begin{equation}
    (ST)^3=S^2,\qquad S^4=\alpha^4\cdot \mathrm{id}.
\end{equation}
By comparing with the calculations in the previous section we fix $\alpha=e^{i\frac{\pi}{4}\nu}$. Note that so far we have considered the bordism class $(1,1)$, where all the boundary conditions are periodic.

Now turn to the case when at least one of the boundary conditions on $T^2$ is antiperiodic, that is spin-structure is different from the one considered before. Note that, in general, although the set of spin-structures on a manifold $X$ is not canonically isomorphic to $H^1(X,\Z_2)$, they form a torsor over this group. This means the differences between spin structures do canonically correspond to elements of $H^1(X,\Z_2)$. Physically this means that a change of spin structure corresponding to an element $b\in H^1(X,\Z_2)$ can be described by insertion of a $\Z_2^f$-charge operator (i.e. ``$(-1)^F$'') supported on a codimension-1 manifold representing Poincar\'e dual of $b$.

The change of the invariant (\ref{beta-ABK-invariant}) on a closed 3-manifold $Y$ with respect to the change of the spin structure by $b\in H^1(Y,\Z_2)$ has natural description in terms of such codimension-1 submanifold $\PD(b)$ (see \cite{kirby_taylor_1991} for details):
\begin{equation}
    \beta_{s+b}(a)-\beta_s(a)=
    2q_{\PD(a)}(b|_{\PD(a)})\mod 8
    \label{beta-spin-diff}
\end{equation}
where
\begin{equation}
    q_{\PD(a)}:\;H^1(\PD(a),\Z_2)\;\longrightarrow\;\Z_4    
\end{equation}
is the quadratic enhancement of the mod 2 intersection pairing on the pin$^-$-surface $\PD(a)$ (as before, the pin$^-$ structure is induced from the ambient spin structure $s$). This relation has already appeared in the previous section. The value of (\ref{beta-spin-diff}) then has a geometric meaning of counting the number of ``half-twists'' modulo 4 along the intersection $\PD(a)\cap \PD(b)$. With this interpretation it becomes explicitly symmetric under exchange of $a$ and $b$. Physically this means that the partition function of the 3d TQFT gets an extra contribution from the intersections of $(-1)^{\CF}$-defects and more usual $(-1)^{\CQ}$ charge defects. Namely, each closed orientation-preserving loop in the intersection contributes $(\pm 1)^\nu$, depending whether the induced spin-structure is even or odd. Each closed orientation-reversing loop contributes $(\pm i)^\nu$.

As in the case of $\Z_2$-charge defects, the bulk $(-1)^{\CF}$ surface defects end on $(-1)^F$-line defects of the boundary anomalous QFT.
Therefore the case of general spin structure on the torus can be understood in terms of moves on configurations of two types of defects: $\Z_2$-charge lines and $(-1)^F$ lines that are inserted on a torus with periodic-periodic spin structure. 

Let us depict $(-1)^F$-line operators by dashed blue lines. Note that the $\Spin(2)$ lift of the slope angle of $\Z_2$-charge lines is changed at the place of intersection. Thus the nontrivial $\Spin(2)$ transition function should be applied at the locus of the $(-1)^F$ operator. For example:
\begin{equation}
\begin{tikzpicture}[scale=0.6,baseline=(current  bounding  box.center)]
\draw[ForestGreen,thick,->-] (0,2) -- node[pos=0.2,above] {$0$} node[pos=0.85,above] {$2\pi$} (6,2);
\draw[Blue,thick,dashed] (4,0) -- (4,4);
\end{tikzpicture}\;.
\label{defect-crossing}
\end{equation}
Therefore, if a $\Z_2$-charge line has odd number of intersections with the $(-1)^F$ line, the lift is not globally well defined, and the corresponding decorations can be omitted. 

We will use the following three additional moves:
\begin{equation}
\begin{tikzpicture}[scale=0.6,baseline=(current  bounding  box.center)]
\draw[ForestGreen,thick,->-] (0,2) to[out=0,in=90] node[pos=0,above] {$0$} node[pos=1,right] {$-\frac{\pi}{2}$} (2,0);
\draw[Blue,thick,dashed] (0,1.7) to[out=0,in=90] (1.7,0);
\draw[ForestGreen,thick,->-] (2,4) to[out=-90,in=180] node[pos=1,above] {$0$} node[pos=0,right] {$-\frac{\pi}{2}$} (4,2);
\draw[Blue,thick,dashed] (1.7,4) to[out=-90,in=180] (4,1.7);
\end{tikzpicture}
\qquad \sim \qquad \beta\;\cdot\;
\begin{tikzpicture}[scale=0.6,baseline=(current  bounding  box.center)]
\draw[ForestGreen,thick,->-] (0,2) to[out=0,in=-90] node[pos=0,above] {$0$} node[pos=1,left] {$\frac{\pi}{2}$} (2,4);
\draw[Blue,thick,dashed] (0,1.7) to[out=0,in=-90] (2.3,4);
\draw[ForestGreen,thick,->-] (2,0) to[out=90,in=180] node[pos=1,above] {$0$} node[pos=0,left] {$\frac{\pi}{2}$} (4,2);
\draw[Blue,thick,dashed] (2.3,0) to[out=90,in=180] (4,1.7);
\end{tikzpicture}\;,
\label{move-beta}
\end{equation}
\begin{equation}
\begin{tikzpicture}[scale=0.6,baseline=(current  bounding  box.center)]
\draw[ForestGreen,thick,->-] (2,0) to[out=90,in=0]  (0,2);
\draw[ForestGreen,thick,->-] (4,2) to[out=180,in=-90] (2,4); \draw[Blue,thick,dashed] (3.5,0) -- (3.5,4);
\end{tikzpicture}
\qquad \sim \qquad \gamma\;\cdot\;
\begin{tikzpicture}[scale=0.6,baseline=(current  bounding  box.center)]
\draw[ForestGreen,thick,->-] (0,2) to[out=0,in=-90] (2,4);
\draw[ForestGreen,thick,->-] (2,0) to[out=90,in=180] (4,2);
\draw[Blue,thick,dashed] (3.5,0) -- (3.5,4);
\end{tikzpicture}\;,
\label{move-gamma}
\end{equation}
\begin{equation}
\begin{tikzpicture}[scale=0.6,baseline=(current  bounding  box.center)]
\draw[ForestGreen,thick,->-] (0,2) -- (4,2);
\draw[Blue,thick,dashed] (3,0) -- (3,4);
\end{tikzpicture}
\qquad \sim \qquad \delta\;\cdot\;
\begin{tikzpicture}[scale=0.6,baseline=(current  bounding  box.center)]
\draw[ForestGreen,thick,->-] (4,2) -- (0,2);
\draw[Blue,thick,dashed] (3,0) -- (3,4);
\end{tikzpicture}\;.
\label{move-delta}
\end{equation}
The self-consistency requires that $\beta^8=1$ and $\delta^2=1$. Note that unlike in the case without the $(-1)^F$ line, the move (\ref{move-delta}) cannot be obtained by applying the move (\ref{move-gamma}) twice. Also, since there is no pure fermion parity anomaly, there are no non-trivial rules for the moves involving only $(-1)^F$ lines. In particular:
\begin{equation}
\begin{tikzpicture}[scale=0.6,baseline=(current  bounding  box.center)]
\draw[Blue,thick,dashed] (2,0) to[out=90,in=0]  (0,2);
\draw[Blue,thick,dashed] (4,2) to[out=180,in=-90] (2,4); 
\end{tikzpicture}
\qquad \sim \qquad 
\begin{tikzpicture}[scale=0.6,baseline=(current  bounding  box.center)]
\draw[Blue,thick,dashed] (0,2) to[out=0,in=-90] (2,4);
\draw[Blue,thick,dashed] (2,0) to[out=90,in=180] (4,2);
\end{tikzpicture}\;.
\end{equation}

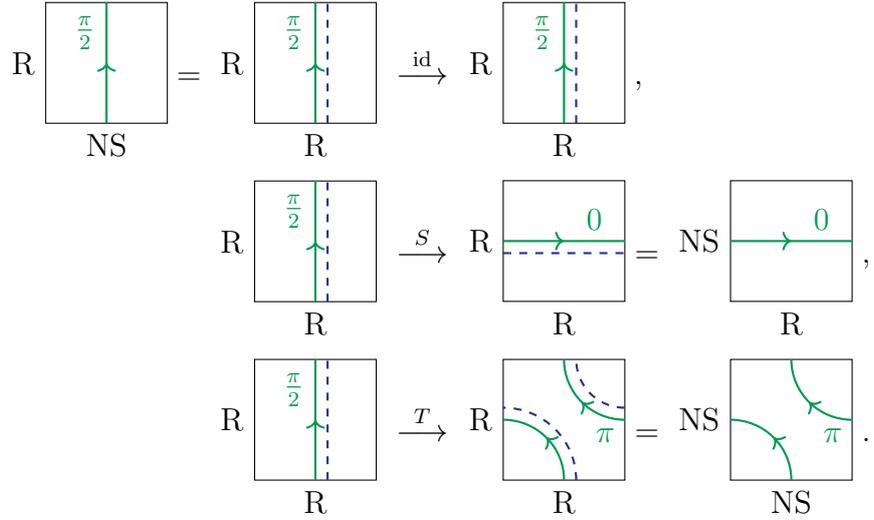
\begin{figure}
    \centering
\begin{equation}
\begin{array}{rcl}
\vspace{1ex}
\begin{tikzpicture}[scale=0.4,baseline=(current  bounding  box.center)]
\draw (0,0) rectangle (4,4);
\node[left] at (0,2) {$\r$};
\node[below] at (2,0) {$\ns$};
\draw[ForestGreen,thick,->-] (2,0) -- (2,4) node[pos=.75,left] {$\frac{\pi}{2}$};
\end{tikzpicture} 
=
\begin{tikzpicture}[scale=0.4,baseline=(current  bounding  box.center)]
\draw (0,0) rectangle (4,4);
\node[left] at (0,2) {$\r$};
\node[below] at (2,0) {$\r$};
\draw[ForestGreen,thick,->-] (2,0) -- (2,4) node[pos=.75,left] {$\frac{\pi}{2}$};
\draw[Blue,thick,dashed] (2.4,0) -- (2.4,4);
\end{tikzpicture} 
&
\;\stackrel{\mathrm{id}}{\longrightarrow}\;
&
\begin{tikzpicture}[scale=0.4,baseline=(current  bounding  box.center)]
\draw (0,0) rectangle (4,4);
\node[left] at (0,2) {$\r$};
\node[below] at (2,0) {$\r$};
\draw[ForestGreen,thick,->-] (2,0) -- (2,4) node[pos=.75,left] {$\frac{\pi}{2}$};
\draw[Blue,thick,dashed] (2.4,0) -- (2.4,4);
\end{tikzpicture}\;,
\\
\vspace{1ex}
\begin{tikzpicture}[scale=0.4,baseline=(current  bounding  box.center)]
\draw (0,0) rectangle (4,4);
\node[left] at (0,2) {$\r$};
\node[below] at (2,0) {$\r$};
\draw[ForestGreen,thick,->-] (2,0) -- (2,4) node[pos=.75,left] {$\frac{\pi}{2}$};
\draw[Blue,thick,dashed] (2.4,0) -- (2.4,4);
\end{tikzpicture} 
&
\;\stackrel{S}{\longrightarrow}\;
&
\begin{tikzpicture}[scale=0.4,baseline=(current  bounding  box.center)]
\draw (0,0) rectangle (4,4);
\node[left] at (0,2) {$\r$};
\node[below] at (2,0) {$\r$};
\draw[ForestGreen,thick,->-] (0,2) -- (4,2) node[pos=.75,above] {$0$};
\draw[Blue,thick,dashed] (0,1.6) -- (4,1.6);
\end{tikzpicture}
=
\begin{tikzpicture}[scale=0.4,baseline=(current  bounding  box.center)]
\draw (0,0) rectangle (4,4);
\node[left] at (0,2) {$\ns$};
\node[below] at (2,0) {$\r$};
\draw[ForestGreen,thick,->-] (0,2) -- (4,2) node[pos=.75,above] {$0$};
\end{tikzpicture}\;,
\\
\begin{tikzpicture}[scale=0.4,baseline=(current  bounding  box.center)]
\draw (0,0) rectangle (4,4);
\node[left] at (0,2) {$\r$};
\node[below] at (2,0) {$\r$};
\draw[ForestGreen,thick,->-] (2,0) -- (2,4) node[pos=.75,left] {$\frac{\pi}{2}$};
\draw[Blue,thick,dashed] (2.4,0) -- (2.4,4);
\end{tikzpicture} 
&
\;\stackrel{T}{\longrightarrow}\;
&
\begin{tikzpicture}[scale=0.4,baseline=(current  bounding  box.center)]
\draw (0,0) rectangle (4,4);
\node[left] at (0,2) {$\r$};
\node[below] at (2,0) {$\r$};
\draw[ForestGreen,thick,->-] (2,0) to[out=90,in=0] (0,2);
\draw[Blue,thick,dashed] (2.4,0) to[out=90,in=0] (0,2.4);
\draw[ForestGreen,thick,->-] (4,2) to[out=180,in=-90] node[pos=0.2,below] {${\pi}$} (2,4);
\draw[Blue,thick,dashed] (4,2.4) to[out=180,in=-90] (2.4,4);
\end{tikzpicture}
=
\begin{tikzpicture}[scale=0.4,baseline=(current  bounding  box.center)]
\draw (0,0) rectangle (4,4);
\node[left] at (0,2) {$\ns$};
\node[below] at (2,0) {$\ns$};
\draw[ForestGreen,thick,->-] (2,0) to[out=90,in=0] (0,2);
\draw[ForestGreen,thick,->-] (4,2) to[out=180,in=-90] node[pos=0.2,below] {${\pi}$} (2,4);
\end{tikzpicture}\;.
\end{array}
\nonumber
\end{equation}
    \caption{The choice of basis configurations of line defects for the case of $(0,1)\in \Omega_2^\Spin(B\Z_2)$ class.}
    \label{fig:defect-01-basis}
\end{figure}

Consider now the action of $S$ and $T$ transformations on the sector corresponding to $(0,1)\in \Omega^{\Spin}_2(B\Z_2)\cong \Z_2^2$ class in the 2-dimensional bordism group. That is the case when the torus has even spin structure, but the spin structure induced on the $\Z_2$-charge line is odd. Figure \ref{fig:defect-01-basis} displays the basis similar to the one in Figure \ref{fig:defect-PP-basis}.
By replacing the defect lines in Figures \ref{fig:defect-S-11} and \ref{fig:defect-T-11} with double lines we then get the following matrix elements for these three basis elements:
\begin{equation}
    S=\left(
    \begin{array}{ccc}
        0 & \beta^2 & 0 \\
        1 & 0 & 0 \\
        0 & 0 & \beta
    \end{array}
    \right),
    \qquad
        T=\left(
    \begin{array}{ccc}
        0 & 0 & \beta^{-1} \\
        0 & 1 & 0 \\
        1 & 0 & 0
    \end{array}
    \right).
\end{equation}
The matrices satisfy the similar conditions:
\begin{equation}
    (ST)^3=S^2,\qquad S^4=\beta^4\cdot \mathrm{id}.
\end{equation}
By comparing it with the bordism invariant calculation we have $\beta=e^{-i\frac{\pi}{4}\nu}$.

Finally we consider the trivial class in the 2-dimensional bordism group $\Omega^{\Spin}_2(B\Z_2)\cong \Z_2^2$. Excluding configurations without $\Z_2$-charge defects, we can choose the basis elements as displayed in Figure \ref{fig:defect-00-basis}. The $S$ and $T$ transformations then act as shown Figures \ref{fig:defect-00-T-action} and \ref{fig:defect-00-S-action}.

\begin{figure}
    \centering
\begin{equation}
\begin{array}{rcl}
\vspace{1ex}
\begin{tikzpicture}[scale=0.4,baseline=(current  bounding  box.center)]
\draw (0,0) rectangle (4,4);
\node[left] at (0,2) {$\ns$};
\node[below] at (2,0) {$\r$};
\draw[ForestGreen,thick,->-] (2,0) -- (2,4);
\end{tikzpicture} 
=
\begin{tikzpicture}[scale=0.4,baseline=(current  bounding  box.center)]
\draw (0,0) rectangle (4,4);
\node[left] at (0,2) {$\r$};
\node[below] at (2,0) {$\r$};
\draw[ForestGreen,thick,->-] (2,0) -- (2,4);
\draw[Blue,thick,dashed] (0,1) -- (4,1);
\end{tikzpicture} 
&
\;\stackrel{\mathrm{id}}{\longrightarrow}\;
&
\begin{tikzpicture}[scale=0.4,baseline=(current  bounding  box.center)]
\draw (0,0) rectangle (4,4);
\node[left] at (0,2) {$\r$};
\node[below] at (2,0) {$\r$};
\draw[ForestGreen,thick,->-] (2,0) -- (2,4);
\draw[Blue,thick,dashed] (0,1) -- (4,1);
\end{tikzpicture}\;,
\\
\vspace{1ex}
\begin{tikzpicture}[scale=0.4,baseline=(current  bounding  box.center)]
\draw (0,0) rectangle (4,4);
\node[left] at (0,2) {$\r$};
\node[below] at (2,0) {$\r$};
\draw[ForestGreen,thick,->-] (2,0) -- (2,4);
\draw[Blue,thick,dashed] (0,1) -- (4,1);
\end{tikzpicture} 
&
\;\stackrel{S}{\longrightarrow}\;
&
\begin{tikzpicture}[scale=0.4,baseline=(current  bounding  box.center)]
\draw (0,0) rectangle (4,4);
\node[left] at (0,2) {$\r$};
\node[below] at (2,0) {$\r$};
\draw[ForestGreen,thick,->-] (0,2) -- (4,2);
\draw[Blue,thick,dashed] (1,0) -- (1,4);
\end{tikzpicture}
=
\begin{tikzpicture}[scale=0.4,baseline=(current  bounding  box.center)]
\draw (0,0) rectangle (4,4);
\node[left] at (0,2) {$\r$};
\node[below] at (2,0) {$\ns$};
\draw[ForestGreen,thick,->-] (0,2) -- (4,2);
\end{tikzpicture}\;,
\\
\vspace{1ex}
\begin{tikzpicture}[scale=0.4,baseline=(current  bounding  box.center)]
\draw (0,0) rectangle (4,4);
\node[left] at (0,2) {$\r$};
\node[below] at (2,0) {$\r$};
\draw[ForestGreen,thick,->-] (2,0) -- (2,4);
\draw[Blue,thick,dashed] (0,1) -- (4,1);
\end{tikzpicture} 
&
\;\stackrel{T}{\longrightarrow}\;
&
\begin{tikzpicture}[scale=0.4,baseline=(current  bounding  box.center)]
\draw (0,0) rectangle (4,4);
\node[left] at (0,2) {$\r$};
\node[below] at (2,0) {$\r$};
\draw[ForestGreen,thick,->-] (2,0) to[out=90,in=0] (0,2);
\draw[ForestGreen,thick,->-] (4,2) to[out=180,in=-90] (2,4);
\draw[Blue,thick,dashed] (0,0.7) -- (4,0.7);
\end{tikzpicture}
=
\begin{tikzpicture}[scale=0.4,baseline=(current  bounding  box.center)]
\draw (0,0) rectangle (4,4);
\node[left] at (0,2) {$\ns$};
\node[below] at (2,0) {$\r$};
\draw[ForestGreen,thick,->-] (2,0) to[out=90,in=0] (0,2);
\draw[ForestGreen,thick,->-] (4,2) to[out=180,in=-90] (2,4);
\end{tikzpicture}\;,
\\
\vspace{1ex}
\begin{tikzpicture}[scale=0.4,baseline=(current  bounding  box.center)]
\draw (0,0) rectangle (4,4);
\node[left] at (0,2) {$\r$};
\node[below] at (2,0) {$\r$};
\draw[ForestGreen,thick,->-] (2,0) -- (2,4);
\draw[Blue,thick,dashed] (0,1) -- (4,1);
\end{tikzpicture} 
&
\;\stackrel{ST}{\longrightarrow}\;
&
\begin{tikzpicture}[scale=0.4,baseline=(current  bounding  box.center)]
\draw (0,0) rectangle (4,4);
\node[left] at (0,2) {$\r$};
\node[below] at (2,0) {$\r$};
\draw[ForestGreen,thick,->-] (0,2) to[out=0,in=-90] (2,4);
\draw[ForestGreen,thick,->-] (2,0) to[out=90,in=180] (4,2);
\draw[Blue,thick,dashed] (1,0) -- (1,4);
\end{tikzpicture}
=
\begin{tikzpicture}[scale=0.4,baseline=(current  bounding  box.center)]
\draw (0,0) rectangle (4,4);
\node[left] at (0,2) {$\r$};
\node[below] at (2,0) {$\ns$};
\draw[ForestGreen,thick,->-] (0,2) to[out=0,in=-90] (2,4);
\draw[ForestGreen,thick,->-] (2,0) to[out=90,in=180] (4,2);
\end{tikzpicture}\;,
\\
\vspace{1ex}
\begin{tikzpicture}[scale=0.4,baseline=(current  bounding  box.center)]
\draw (0,0) rectangle (4,4);
\node[left] at (0,2) {$\r$};
\node[below] at (2,0) {$\r$};
\draw[ForestGreen,thick,->-] (2,0) -- (2,4);
\draw[Blue,thick,dashed] (0,1) -- (4,1);
\end{tikzpicture} 
&
\;\stackrel{TST}{\longrightarrow}\;
&
\begin{tikzpicture}[scale=0.4,baseline=(current  bounding  box.center)]
\draw (0,0) rectangle (4,4);
\node[left] at (0,2) {$\r$};
\node[below] at (2,0) {$\r$};
\draw[ForestGreen,thick,->-] (2,0) -- (2,4);
\draw[Blue,thick,dashed] (3,0) to[out=90,in=0] (0,3);
\draw[Blue,thick,dashed] (4,3) to[out=190,in=-90] (3,4);
\end{tikzpicture}
=
\begin{tikzpicture}[scale=0.4,baseline=(current  bounding  box.center)]
\draw (0,0) rectangle (4,4);
\node[left] at (0,2) {$\ns$};
\node[below] at (2,0) {$\ns$};
\draw[ForestGreen,thick,->-] (2,0) -- (2,4);
\end{tikzpicture}\;,
\\
\begin{tikzpicture}[scale=0.4,baseline=(current  bounding  box.center)]
\draw (0,0) rectangle (4,4);
\node[left] at (0,2) {$\r$};
\node[below] at (2,0) {$\r$};
\draw[ForestGreen,thick,->-] (2,0) -- (2,4);
\draw[Blue,thick,dashed] (0,1) -- (4,1);
\end{tikzpicture} 
&
\;\stackrel{TS}{\longrightarrow}\;
&
\begin{tikzpicture}[scale=0.4,baseline=(current  bounding  box.center)]
\draw (0,0) rectangle (4,4);
\node[left] at (0,2) {$\r$};
\node[below] at (2,0) {$\r$};
\draw[Blue,thick,dashed] (1.5,0) to[out=90,in=0] (0,1.5);
\draw[Blue,thick,dashed] (4,1.5) to[out=180,in=-90] (1.5,4);
\draw[ForestGreen,thick,->-] (0,2) -- (4,2);
\end{tikzpicture}
=
\begin{tikzpicture}[scale=0.4,baseline=(current  bounding  box.center)]
\draw (0,0) rectangle (4,4);
\node[left] at (0,2) {$\ns$};
\node[below] at (2,0) {$\ns$};
\draw[ForestGreen,thick,->-] (0,2) -- (4,2);
\end{tikzpicture}\;.
\end{array}
\nonumber
\end{equation}
    \caption{The choice of basis for the trivial class in $\Omega^{\Spin}_2(B\Z_2)\cong \Z_2^2$, excluding configurations without $\Z_2$-charge defects.}
    \label{fig:defect-00-basis}
\end{figure}
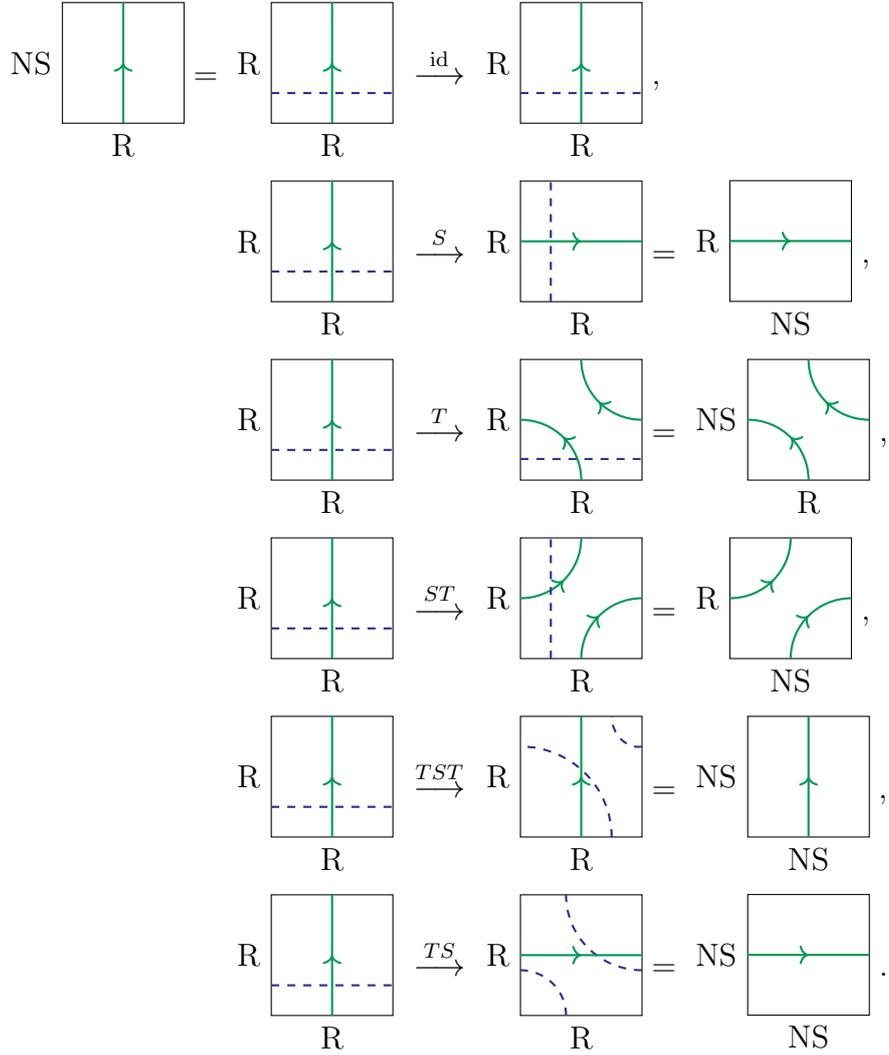

\begin{figure}
    \centering
\begin{equation}
\begin{array}{rcl}
\vspace{1ex}
\begin{tikzpicture}[scale=0.4,baseline=(current  bounding  box.center)]
\draw (0,0) rectangle (4,4);
\node[left] at (0,2) {$\r$};
\node[below] at (2,0) {$\r$};
\draw[ForestGreen,thick,->-] (2,0) -- (2,4);
\draw[Blue,thick,dashed] (0,1) -- (4,1);
\end{tikzpicture} 
&
\;\stackrel{T}{\longrightarrow}\;
&
\begin{tikzpicture}[scale=0.4,baseline=(current  bounding  box.center)]
\draw (0,0) rectangle (4,4);
\node[left] at (0,2) {$\r$};
\node[below] at (2,0) {$\r$};
\draw[ForestGreen,thick,->-] (2,0) to[out=90,in=0] (0,2);
\draw[ForestGreen,thick,->-] (4,2) to[out=180,in=-90] (2,4);
\draw[Blue,thick,dashed] (0,0.7) -- (4,0.7);
\end{tikzpicture}\;,
\\
\vspace{1ex}
\begin{tikzpicture}[scale=0.4,baseline=(current  bounding  box.center)]
\draw (0,0) rectangle (4,4);
\node[left] at (0,2) {$\r$};
\node[below] at (2,0) {$\r$};
\draw[ForestGreen,thick,->-] (0,2) -- (4,2);
\draw[Blue,thick,dashed] (1,0) -- (1,4);
\end{tikzpicture}
&
\;\stackrel{T}{\longrightarrow}\;
&
\begin{tikzpicture}[scale=0.4,baseline=(current  bounding  box.center)]
\draw (0,0) rectangle (4,4);
\node[left] at (0,2) {$\r$};
\node[below] at (2,0) {$\r$};
\draw[Blue,thick,dashed] (1.5,0) to[out=90,in=0] (0,1.5);
\draw[Blue,thick,dashed] (4,1.5) to[out=180,in=-90] (1.5,4);
\draw[ForestGreen,thick,->-] (0,2) -- (4,2);
\end{tikzpicture}\;,
\\
\vspace{1ex}
\begin{tikzpicture}[scale=0.4,baseline=(current  bounding  box.center)]
\draw (0,0) rectangle (4,4);
\node[left] at (0,2) {$\r$};
\node[below] at (2,0) {$\r$};
\draw[ForestGreen,thick,->-] (2,0) to[out=90,in=0] (0,2);
\draw[ForestGreen,thick,->-] (4,2) to[out=180,in=-90] (2,4);
\draw[Blue,thick,dashed] (0,0.7) -- (4,0.7);
\end{tikzpicture}
&
\;\stackrel{T}{\longrightarrow}\;
&
\begin{tikzpicture}[scale=0.4,baseline=(current  bounding  box.center)]
\draw (0,0) rectangle (4,4);
\node[left] at (0,2) {$\r$};
\node[below] at (2,0) {$\r$};
\draw[ForestGreen,thick,->-] (2,0) to[out=90,in=0] (0,1.3);
\draw[ForestGreen,thick,->-] (4,2.7) to[out=180,in=-90] (2,4);
\draw[ForestGreen,thick,->-] (4,1.3) to[out=180,in=0] (0,2.7);
\draw[Blue,thick,dashed] (0,0.7) -- (4,0.7);
\end{tikzpicture}
\;\sim\;\gamma^{-1}\delta\cdot\;
\begin{tikzpicture}[scale=0.4,baseline=(current  bounding  box.center)]
\draw (0,0) rectangle (4,4);
\node[left] at (0,2) {$\r$};
\node[below] at (2,0) {$\r$};
\draw[ForestGreen,thick,->-] (2,0) -- (2,4);
\draw[Blue,thick,dashed] (0,1) -- (4,1);
\end{tikzpicture} 
\\
\vspace{1ex}
\begin{tikzpicture}[scale=0.4,baseline=(current  bounding  box.center)]
\draw (0,0) rectangle (4,4);
\node[left] at (0,2) {$\r$};
\node[below] at (2,0) {$\r$};
\draw[ForestGreen,thick,->-] (0,2) to[out=0,in=-90] (2,4);
\draw[ForestGreen,thick,->-] (2,0) to[out=90,in=180] (4,2);
\draw[Blue,thick,dashed] (1,0) -- (1,4);
\end{tikzpicture}
&
\;\stackrel{T}{\longrightarrow}\;
&
\begin{tikzpicture}[scale=0.4,baseline=(current  bounding  box.center)]
\draw (0,0) rectangle (4,4);
\node[left] at (0,2) {$\r$};
\node[below] at (2,0) {$\r$};
\draw[ForestGreen,thick,->-] (2,0) -- (2,4);
\draw[Blue,thick,dashed] (3,0) to[out=90,in=0] (0,3);
\draw[Blue,thick,dashed] (4,3) to[out=190,in=-90] (3,4);
\end{tikzpicture}\;,
\\
\vspace{1ex}
\begin{tikzpicture}[scale=0.4,baseline=(current  bounding  box.center)]
\draw (0,0) rectangle (4,4);
\node[left] at (0,2) {$\r$};
\node[below] at (2,0) {$\r$};
\draw[ForestGreen,thick,->-] (2,0) -- (2,4);
\draw[Blue,thick,dashed] (3,0) to[out=90,in=0] (0,3);
\draw[Blue,thick,dashed] (4,3) to[out=190,in=-90] (3,4);
\end{tikzpicture}
&
\;\stackrel{T}{\longrightarrow}\;
&
\begin{tikzpicture}[scale=0.4,baseline=(current  bounding  box.center)]
\draw (0,0) rectangle (4,4);
\node[left] at (0,2) {$\r$};
\node[below] at (2,0) {$\r$};
\draw[ForestGreen,thick,->-] (2,0) to[out=90,in=0] (0,2);
\draw[ForestGreen,thick,->-] (4,2) to[out=180,in=-90] (2,4);
\draw[Blue,thick,dashed] (1,0) -- (1,4);
\end{tikzpicture}
\;\sim \;\gamma\cdot\;
\begin{tikzpicture}[scale=0.4,baseline=(current  bounding  box.center)]
\draw (0,0) rectangle (4,4);
\node[left] at (0,2) {$\r$};
\node[below] at (2,0) {$\r$};
\draw[ForestGreen,thick,->-] (0,2) to[out=0,in=-90] (2,4);
\draw[ForestGreen,thick,->-] (2,0) to[out=90,in=180] (4,2);
\draw[Blue,thick,dashed] (1,0) -- (1,4);
\end{tikzpicture}\;,
\\
\begin{tikzpicture}[scale=0.4,baseline=(current  bounding  box.center)]
\draw (0,0) rectangle (4,4);
\node[left] at (0,2) {$\r$};
\node[below] at (2,0) {$\r$};
\draw[Blue,thick,dashed] (1.5,0) to[out=90,in=0] (0,1.5);
\draw[Blue,thick,dashed] (4,1.5) to[out=180,in=-90] (1.5,4);
\draw[ForestGreen,thick,->-] (0,2) -- (4,2);
\end{tikzpicture}
&
\;\stackrel{T}{\longrightarrow}\;
&
\begin{tikzpicture}[scale=0.4,baseline=(current  bounding  box.center)]
\draw (0,0) rectangle (4,4);
\node[left] at (0,2) {$\r$};
\node[below] at (2,0) {$\r$};
\draw[ForestGreen,thick,->-] (0,2) -- (4,2);
\draw[Blue,thick,dashed] (1,0) -- (1,4);
\end{tikzpicture}\;.
\end{array}
\nonumber
\end{equation}
    \caption{The action of $T$ transformation on the basis elements chosen in Figure \ref{fig:defect-00-basis}.}
    \label{fig:defect-00-T-action}
\end{figure}


\begin{figure}
    \centering
\begin{equation}
\begin{array}{rcl}
\vspace{1ex}
\begin{tikzpicture}[scale=0.4,baseline=(current  bounding  box.center)]
\draw (0,0) rectangle (4,4);
\node[left] at (0,2) {$\r$};
\node[below] at (2,0) {$\r$};
\draw[ForestGreen,thick,->-] (2,0) -- (2,4);
\draw[Blue,thick,dashed] (0,1) -- (4,1);
\end{tikzpicture} 
&
\;\stackrel{S}{\longrightarrow}\;
&
\begin{tikzpicture}[scale=0.4,baseline=(current  bounding  box.center)]
\draw (0,0) rectangle (4,4);
\node[left] at (0,2) {$\r$};
\node[below] at (2,0) {$\r$};
\draw[ForestGreen,thick,->-] (0,2) -- (4,2);
\draw[Blue,thick,dashed] (1,0) -- (1,4);
\end{tikzpicture}\;,
\\
\vspace{1ex}
\begin{tikzpicture}[scale=0.4,baseline=(current  bounding  box.center)]
\draw (0,0) rectangle (4,4);
\node[left] at (0,2) {$\r$};
\node[below] at (2,0) {$\r$};
\draw[ForestGreen,thick,->-] (0,2) -- (4,2);
\draw[Blue,thick,dashed] (1,0) -- (1,4);
\end{tikzpicture}
&
\;\stackrel{S}{\longrightarrow}\;
&
\begin{tikzpicture}[scale=0.4,baseline=(current  bounding  box.center)]
\draw (0,0) rectangle (4,4);
\node[left] at (0,2) {$\r$};
\node[below] at (2,0) {$\r$};
\draw[ForestGreen,thick,->-] (2,4) -- (2,0);
\draw[Blue,thick,dashed] (0,1) -- (4,1);
\end{tikzpicture} 
\;\sim \;\delta\cdot\;
\begin{tikzpicture}[scale=0.4,baseline=(current  bounding  box.center)]
\draw (0,0) rectangle (4,4);
\node[left] at (0,2) {$\r$};
\node[below] at (2,0) {$\r$};
\draw[ForestGreen,thick,->-] (2,0) -- (2,4);
\draw[Blue,thick,dashed] (0,1) -- (4,1);
\end{tikzpicture} \;,
\\
\vspace{1ex}
\begin{tikzpicture}[scale=0.4,baseline=(current  bounding  box.center)]
\draw (0,0) rectangle (4,4);
\node[left] at (0,2) {$\r$};
\node[below] at (2,0) {$\r$};
\draw[ForestGreen,thick,->-] (2,0) to[out=90,in=0] (0,2);
\draw[ForestGreen,thick,->-] (4,2) to[out=180,in=-90] (2,4);
\draw[Blue,thick,dashed] (0,0.7) -- (4,0.7);
\end{tikzpicture}
&
\;\stackrel{S}{\longrightarrow}\;
&
\begin{tikzpicture}[scale=0.4,baseline=(current  bounding  box.center)]
\draw (0,0) rectangle (4,4);
\node[left] at (0,2) {$\r$};
\node[below] at (2,0) {$\r$};
\draw[ForestGreen,thick,->-] (0,2) to[out=0,in=-90] (2,4);
\draw[ForestGreen,thick,->-] (2,0) to[out=90,in=180] (4,2);
\draw[Blue,thick,dashed] (1,0) -- (1,4);
\end{tikzpicture}\;,
\\
\vspace{1ex}
\begin{tikzpicture}[scale=0.4,baseline=(current  bounding  box.center)]
\draw (0,0) rectangle (4,4);
\node[left] at (0,2) {$\r$};
\node[below] at (2,0) {$\r$};
\draw[ForestGreen,thick,->-] (0,2) to[out=0,in=-90] (2,4);
\draw[ForestGreen,thick,->-] (2,0) to[out=90,in=180] (4,2);
\draw[Blue,thick,dashed] (1,0) -- (1,4);
\end{tikzpicture}
&
\;\stackrel{S}{\longrightarrow}\;
&
\begin{tikzpicture}[scale=0.4,baseline=(current  bounding  box.center)]
\draw (0,0) rectangle (4,4);
\node[left] at (0,2) {$\r$};
\node[below] at (2,0) {$\r$};
\draw[ForestGreen,thick,->-] (0,2) to[out=0,in=90] (2,0);
\draw[ForestGreen,thick,->-] (2,4) to[out=-90,in=180] (4,2);
\draw[Blue,thick,dashed] (0,0.7) -- (4,0.7);
\end{tikzpicture}
\;\sim\;\delta\cdot
\begin{tikzpicture}[scale=0.4,baseline=(current  bounding  box.center)]
\draw (0,0) rectangle (4,4);
\node[left] at (0,2) {$\r$};
\node[below] at (2,0) {$\r$};
\draw[ForestGreen,thick,->-] (2,0) to[out=90,in=0] (0,2);
\draw[ForestGreen,thick,->-] (4,2) to[out=180,in=-90] (2,4);
\draw[Blue,thick,dashed] (0,0.7) -- (4,0.7);
\end{tikzpicture}\;,
\\
\vspace{1ex}
\begin{tikzpicture}[scale=0.4,baseline=(current  bounding  box.center)]
\draw (0,0) rectangle (4,4);
\node[left] at (0,2) {$\r$};
\node[below] at (2,0) {$\r$};
\draw[ForestGreen,thick,->-] (2,0) -- (2,4);
\draw[Blue,thick,dashed] (3,0) to[out=90,in=0] (0,3);
\draw[Blue,thick,dashed] (4,3) to[out=190,in=-90] (3,4);
\end{tikzpicture}
&
\;\stackrel{S}{\longrightarrow}\;
&
\begin{tikzpicture}[scale=0.4,baseline=(current  bounding  box.center)]
\draw (0,0) rectangle (4,4);
\node[left] at (0,2) {$\r$};
\node[below] at (2,0) {$\r$};
\draw[Blue,thick,dashed] (1.5,0) to[out=90,in=0] (0,1.5);
\draw[Blue,thick,dashed] (4,1.5) to[out=180,in=-90] (1.5,4);
\draw[ForestGreen,thick,->-] (0,2) -- (4,2);
\end{tikzpicture}\;,
\\
\begin{tikzpicture}[scale=0.4,baseline=(current  bounding  box.center)]
\draw (0,0) rectangle (4,4);
\node[left] at (0,2) {$\r$};
\node[below] at (2,0) {$\r$};
\draw[Blue,thick,dashed] (1.5,0) to[out=90,in=0] (0,1.5);
\draw[Blue,thick,dashed] (4,1.5) to[out=180,in=-90] (1.5,4);
\draw[ForestGreen,thick,->-] (0,2) -- (4,2);
\end{tikzpicture}
&
\;\stackrel{S}{\longrightarrow}\;
&
\begin{tikzpicture}[scale=0.4,baseline=(current  bounding  box.center)]
\draw (0,0) rectangle (4,4);
\node[left] at (0,2) {$\r$};
\node[below] at (2,0) {$\r$};
\draw[ForestGreen,thick,->-] (2,4) -- (2,0);
\draw[Blue,thick,dashed] (3,0) to[out=90,in=0] (0,3);
\draw[Blue,thick,dashed] (4,3) to[out=190,in=-90] (3,4);
\end{tikzpicture}
\;\sim\;\delta\cdot\;
\begin{tikzpicture}[scale=0.4,baseline=(current  bounding  box.center)]
\draw (0,0) rectangle (4,4);
\node[left] at (0,2) {$\r$};
\node[below] at (2,0) {$\r$};
\draw[ForestGreen,thick,->-] (2,0) -- (2,4);
\draw[Blue,thick,dashed] (3,0) to[out=90,in=0] (0,3);
\draw[Blue,thick,dashed] (4,3) to[out=190,in=-90] (3,4);
\end{tikzpicture}\;.
\end{array}
\nonumber
\end{equation}
    \caption{The action of $S$ transformation on the basis elements chosen in Figure \ref{fig:defect-00-basis}. }
    \label{fig:defect-00-S-action}
\end{figure}
The corresponding matrix elements are then as follows:
\begin{equation}
    S=\left(
    \begin{array}{cccccc}
       0 & \delta & 0 & 0 & 0 & 0 \\
       1 & 0 & 0 & 0 & 0 & 0 \\
       0 & 0 & 0 & \delta & 0 & 0 \\
       0 & 0 & 1 & 0 & 0 & 0 \\
       0 & 0 & 0 & 0 & 0 & \delta \\
       0 & 0 & 0 & 0 & 1 & 0 \\
    \end{array}
    \right),
    \qquad
        T=\left(
    \begin{array}{cccccc}
       0 & 0 & \delta/\gamma & 0 & 0 & 0 \\
       0 & 0 & 0 & 0 & 0 & 1 \\
       1 & 0 & 0 & 0 & 0 & 0 \\
       0 & 0 & 0 & 0 & \gamma & 0 \\
       0 & 0 & 0 & 1 & 0 & 0 \\
       0 & 1 & 0 & 0 & 0 & 0 \\
    \end{array}
    \right).
\end{equation}
For any $\gamma$ and $\delta$ (such that $\delta^2=1$) the matrices satisfy:
\begin{equation}
    (ST)^3=S^2,\qquad S^4=\mathrm{id}.
\end{equation}
By comparing it with the calculation of cobordism invariants in the previous section we have $\delta=1$, $\gamma=e^{\frac{\pi i\nu}{4}}$.

\subsection{Interpretation in terms of the Hilbert space on a circle}
\label{subsec:Hilbertspace}

In \cite{Delmastro:2021xox} it was analyzed how the anomaly of $\Z_2\times \Z_2^f$ symmetry exhibits itself on the Hilbert space of a 2-dimensional theory. The authors made analysis by considering a basic theory that realizes such anomaly, namely a theory of $\nu$ Majorana fermions, where only the left-moving ones are non-trivially charged with respect to $Q$. In this section we demonstrate how it is in agreement with description of the anomalies using the bulk iTQFT or defect rules described above. As those descriptions are universal, there is no ambiguity about which features are generic and which could be specific for a particular two-dimensional theory. 

In the geometric description the action of the operators $(-1)^Q$ and $(-1)^F$ on the Hilbert space is realized by the corresponding defect lines going along the spatial circle. Consider first the Hilbert space of the 2d theory in Ramond untwisted sector. A continuous process exchanging the two operators can be realized by the configuration of the surface defects in the bulk TQFT as depicted on the left side of Figure \ref{fig:commutation-F-Q}. The surfaces intersect along a circle with induced periodic spin structure, which contributes $(-1)^\nu$. This implies that for $\nu=1\mod 2$ the operators anticommute:
\begin{equation}
    (-1)^F(-1)^Q=(-1)^\nu\cdot(-1)^Q(-1)^F\qquad \text{on $\CH_{\text{R}0}$}.
\end{equation}
This indeed agrees with the previous analysis in the literature \cite{Delmastro:2021xox,Thorngren:2018bhj,Karch:2019lnn}.

\begin{figure}
    \centering
\begin{tikzpicture}[scale=0.5,baseline=(current  bounding  box.center)]
    \draw[thick] (-1,5) -- (-1,-3) -- (5,-3);
    \draw[thick] (-1,-3) -- (-3,-4);
    \draw[ForestGreen,very thick,fill=ForestGreen!30,fill opacity=0.7] 
        (0,0) to[out=120,in=-90] (-0.7,4) -- (1.3,5) to[out=-90,in=120] (2,1) -- cycle;
    \draw[Blue,very thick,fill=Blue!30,fill opacity=0.7] 
        (0,0) to[out=60,in=-90] (0.7,4) -- (2.7,5) to[out=-90,in=60] (2,1) -- cycle;
    \draw[Blue,very thick,fill=Blue!30,fill opacity=0.7] 
        (0,0) to[out=-120,in=90] (-0.7,-4) -- (1.3,-3) to[out=90,in=-120] (2,1) -- cycle;
    \draw[ForestGreen,very thick,fill=ForestGreen!30,fill opacity=0.7]
        (0,0) to[out=-60,in=90] (0.7,-4) -- (2.7,-3) to[out=90,in=-60] (2,1) -- cycle;
    \draw[thick] (-1,5) -- (5,5) -- (3,4) -- (-3,4) -- cycle;
    \draw[thick] (-3,4) -- (-3,-4) -- (3,-4) -- (3,4);
    \draw[thick] (5,5) -- (5,-3) -- (3,-4);
    \node[below] at (4,4.5) {$\r$};
\end{tikzpicture}
\qquad
\begin{tikzpicture}[scale=0.5,baseline=(current  bounding  box.center)]
    \draw[thick] (-1,5) -- (-1,-3) -- (5,-3);
    \draw[thick] (-1,-3) -- (-3,-4);
    \draw[Blue,very thick,fill=Blue!30,fill opacity=0.7] 
        (-2,0.5) -- (-1,1) -- (5,1) -- (4,0.5) -- cycle;
    \draw[ForestGreen,very thick,fill=ForestGreen!30,fill opacity=0.7]
        (-2,4.5) -- (4,4.5) -- (4,-3.5) -- (-2,-3.5) -- cycle;
    \draw[Blue,very thick,fill=Blue!30,fill opacity=0.7] 
        (-3,0) -- (-2,0.5) -- (4,0.5) -- (3,0) -- cycle;    
    \draw[thick] (-1,5) -- (5,5) -- (3,4) -- (-3,4) -- cycle;
    \draw[thick] (-3,4) -- (-3,-4) -- (3,-4) -- (3,4);
    \draw[thick] (5,5) -- (5,-3) -- (3,-4);
    \node[below] at (0,4) {$\r$};
\end{tikzpicture}
    \caption{Left: illustration of anti-commutativity of $(-1)^F$ and $(-1)^Q$ operators in the Ramond sector of the 2d boundary QFT for $\nu =1 \mod 2$ in terms of the bulk TQFT. Right: illustration of sign ambiguity of the trace over the Ramond or NS twisted Hilbert spaces with $(-1)^F$ inserted. In both cases the spacetime of the boundary theory lies in the horizontal place, with the time going from left to right. The intersection of $(-1)^{\CF}$ and $(-1)^{\CQ}$ surface defects in the bulk contributes is an orientation-preserving loop with odd induced spin structure. Therefore it contributes  $(-1)^\nu$ (see discussion below (\ref{beta-spin-diff})). }
    \label{fig:commutation-F-Q}
\end{figure}

This phenomenon can be also seen from the rules on the defect moves considered in the Section \ref{sec:defect-rules}. When the $(-1)^Q$ defect is passed through the $(-1)^F$ defect, the lift of the slope angle to $\Spin(2)$ shifts by $2\pi$ (cf. (\ref{defect-crossing}):
\begin{equation}
    \begin{tikzpicture}[scale=0.6,baseline=1ex]
\draw[ForestGreen,thick,->-] (0,0) -- (4,0) node[pos=.75,above] {$0$};
\draw[Blue,thick,dashed] (0,1) -- (4,1);
\end{tikzpicture}
\;\;\sim \;\;
    \begin{tikzpicture}[scale=0.6,baseline=1ex]
\draw[ForestGreen,thick,->-] (0,1) -- (4,1) node[pos=.75,above] {$2\pi$};
\draw[Blue,thick,dashed] (0,0) -- (4,0);
\end{tikzpicture}
\;\;\sim\;\; (-1)^\nu\cdot\,
    \begin{tikzpicture}[scale=0.6,baseline=1ex]
\draw[ForestGreen,thick,->-] (0,1) -- (4,1) node[pos=.75,above] {$0$};
\draw[Blue,thick,dashed] (0,0) -- (4,0);
\end{tikzpicture}
\end{equation}
where in the last relation we applied the rule (\ref{move-lift-change}) and used that $\alpha^4=(-1)^\nu$.

In the Neveu-Schwarz untwisted sector $\CH_{\text{NS}0}$ the operators commute for any $\nu$. The corresponding configuration of the surface defects in the bulk TQFT will be as the left part of Figure \ref{fig:commutation-F-Q}, but the induced spin-structure on the intersection will be anti-periodic instead, so it will contribute $1^\nu=1$. In terms of the defect rules, the NS sector has a $(-1)^F$ line inserted along the time direction, so the $(-1)^Q$ line will intersect once with it and will have no globally defined lift of the slope angle to $\Spin(2)$. Therefore moving it through the $(-1)^F$ defect does not change it:
\begin{equation}
    \begin{tikzpicture}[scale=0.6,baseline=(current  bounding  box.center)]
\draw[ForestGreen,thick,->-] (0,1) -- (4,1);
\draw[Blue,thick,dashed] (0,2) -- (4,2);
\draw[Blue,thick,dashed] (1,0) -- (1,4);
\end{tikzpicture}
\;\;\sim \;\;
    \begin{tikzpicture}[scale=0.6,baseline=(current  bounding  box.center)]
\draw[ForestGreen,thick,->-] (0,3) -- (4,3);
\draw[Blue,thick,dashed] (0,2) -- (4,2);
\draw[Blue,thick,dashed] (1,0) -- (1,4);
\end{tikzpicture}\;.
\end{equation}

On the other hand, the twisted Ramond or Neveu-Schwarz sector has a $(-1)^Q$ line inserted along the time direction, for which, at a given time slice, one has to choose the slope angle to $\Spin(2)$. If the $(-1)^Q$ line does not intersect a $(-1)^F$ line, the change of the lift results in the extra phase $\alpha^4=(-1)^\nu$. This in particular results in the ambiguity of the partition function with periodic boundary condition trace over the Hilbert space with $(-1)^F$ inserted for $\nu=1\mod 2$. Another way to see the sign ambiguity is through the configuration of the surface defects in the bulk TQFT depicted on the right part of Figure \ref{fig:commutation-F-Q}. The $(-1)^{\CF}$ defect corresponds to an automorphism of spin structure of the spacetime of the 2d QFT. It intersects with $(-1)^{\CQ}$ defect along a circle with periodic induced spin structure and thus gives an extra $(-1)^\nu$ phase.

Suppose one chooses to fix a particular lift to $\Spin(2)$ for the $(-1)^Q$ defect inserted along the time direction in 2d and a Hilbert space to it. Then the $(-1)^F$ operator will not act within this Hilbert space, but will transform it to a different one:
\begin{equation}
\begin{tikzpicture}[scale=0.6,baseline=(current  bounding  box.center)]
\draw[ForestGreen,thick,->-] (2,0) -- (2,2)  node[pos=.25,right] {$\frac{\pi}{2}$};
\draw[ForestGreen,thick,->-] (2,2) -- (2,4) node[pos=.25,right] {$\frac{5\pi}{2}$};
\draw[Blue,thick,dashed] (0,2) -- (4,2);
\end{tikzpicture}\;.
\end{equation}
This is again in the agreement with the analysis of \cite{Delmastro:2021xox} for free fermions.

\subsection{Surgery approach}
\label{sec:Z2-surgery}
In this section we will present yet another approach to calculating the elements of the modular $S$ and $T$ matrices. This approach is probably the least geometrical and physical, but allows a rather straightforward and algorithmic calculation, due to its combinatorial nature. Also, because of its simplicity, this is the approach that we will follow when we consider generalization to other groups in Section \ref{sec:gen-groups}.

As we have already seen, the calculation of matrix elements always boils down to calculation of the values of the  invariant $\beta_s(a)$ on closed 3-dimensional manifolds. This invariant is known to be directly related to the mod 16 valued Rokhlin invariant (also known as $\mu$-invariant) of spin manifolds \cite{kirby_taylor_1991} (see also \cite{Guo:2018vij} for physics interpretation of this relation):
\begin{equation}
    \beta_s(a)=\frac{\mu_s-\mu_{s+a}}{2}\mod 8.
    \label{beta-via-Rokhlin}
\end{equation}
Note that the difference of the values of the Rokhilin invariant for different choices of spin structure is always even.

The Rokhlin invariant $\mu_s$ of the 3-manifold $Y$ with spin structure $s\in \Spin(Y)$ is defined as the signature mod 16 of a spin 4-manifold $U$ which has $Y$ as its boundary, with $s$ induced from the spin structure of the 4-manifold:
\begin{equation}
    \mu_s:=\sigma(U)\mod 16\;.
\end{equation}
There is however also a formula for the value of the Rokhlin invariant in terms of Dehn surgery representation of 3-manifold \cite{kirby19913} which we will review below. 

First we recall that any closed oriented manifold can be obtained by a Dehn surgery on a framed link $\CL$ in a 3-sphere $S^3$. The framing can be geometrically understood as a choice of a non-vanishing normal vector field on each link component, considered up to isotopy (continuous deformation that respects the non-vanishing condition). Combinatorially, the framing can be described by assigning an integer number $p_I\in \Z$ to each link component $\CL_I\subset \CL$ ($I=1,\ldots,V$) -- the self-linking number. The self-linking number is the linking number between the link component and its push-off towards the framing vector field. The surgery operation amounts to removing tubular neighborhoods for each link component and then gluing back solid tori so that their meridians (contractible cycles) are mapped to the curves traced by the framing vectors on the boundaries of tubular neighborhoods. The results of such operation is usually denoted by $S^3(\CL)$.

Any 3-manifold is spin and the choice of spin structure has a natural description in terms of the surgery representation of 3-manifold. Namely, the spin-structures are in canonical one-to-one correspondence with \textit{characteristic sublinks} of the framed link representing a 3-manifold. A characteristic sublink $\CC$ is any sublink $\CC\subset \CL$ that satisfies:
\begin{equation}
    \lk(\CC,\CL_I)=\lk(\CL_I,\CL_I)\equiv p_I\mod 2,\qquad \forall I
    \label{char-sublink-property}
\end{equation}
where $\lk$ denotes the linking number\footnote{Recall that $\lk(\CA,\CB)$ for a pair of oriented links $\CA,\CB$ can be defined as the algebraic number of intersection points in $S\cap \CB$, where $S$ is any surface such that $\partial S=\CB$.}. To make this condition even more explicit, consider the $V\times V$ linking matrix $B$ with components
\begin{equation}
    B_{IJ}:=\lk(\CL_I,\CL_J).
\end{equation}
Then characteristic sublinks (and, therefore, spin structures on the corresponding 3-manifold $S^3(\CL)$) are in one-to-one correspondence with mod 2 vectors $s\in \Z_2^V$ such that
\begin{equation}
    \sum_{J}B_{IJ}s_J=B_{II}\mod 2,\qquad\forall I.
    \label{spin-vector-condition}
\end{equation}
The characteristic sublink corresponding to given $s$ is the union of components of $\CL$ for which $s_I=1\mod 2$:
\begin{equation}
    \CC=\bigsqcup\limits_{I:\;s_I=1\mod 2} \CL_{I}.
\end{equation}
The Rokhlin invariant of the 3-manifold $Y=S^3(\CL)$ is then given by the following formula in terms of the surgery data  \cite{kirby19913}:
\begin{equation}
    \mu_s(Y)=\sigma(B)-\lk(\CC_s,\CC_s)+8\,\Arf(\CC_s)\mod 16
    \label{Rokhlin-surgery-formula}
\end{equation}
where $\CC_s$ is the characteristic sublink corresponding to spin-structure $s\in \Spin(Y)$, $\sigma(B)$ is the signature of the linking matrix $B$, and $\Arf(\CC_s)$ is the mod 2 valued Arf invariant of link $\CC_s$. Note that the Arf invariant is only defined for \textit{proper} links, that is links such that the sum of the linking numbers
of any component with all the other components is even. For any characteristic sublink $\CC_s$ this property is satisfied automatically due to (\ref{char-sublink-property}). 

Similarly to spin structures, the elements $a\in H^1(Y,\Z_2)$ can be described combinatorially in terms of surgery data as mod 2 vectors $a\in \Z_2^V$ satisfying 
\begin{equation}
    \sum_{J}B_{IJ}a_J=0\mod 2,\qquad\forall I.
\end{equation}
This realization is in agreement with the action of elements of $H^1(Y,\Z_2)$ on $\Spin(Y)$ by $s\mapsto s+a$. Combining (\ref{beta-via-Rokhlin}) with (\ref{Rokhlin-surgery-formula}) we have the following explicit formula for the $\beta$-invariant of $Y=S^3(\CL)$ in terms of surgery data:
\begin{equation}
    \beta_s(a)=\frac{(s+a)^TB(s+a)-s^TBs}{2}+4(\Arf(\CC_{s+a})-\Arf(\CC_{s}))\mod 8,
    \label{beta-surgery-formula}
\end{equation}
Note that here $s+a$ is to be intended as an element of $\Z_2$. The Arf invariant of any link $\CL$ can be computed combinatorially using its relation to the Jones polynomial $V_L(q)$ (unframed version of, with normalization $V_{\text{unknot}}(q)=1$) at $q=i$ (with $q^{1/2}=e^{\frac{\pi i}{4}}$) \cite{jones1997polynomial,murakami1986recursive}:
\begin{equation}
    V_\CL(i)=\left\{
    \begin{array}{ll}
        (-\sqrt{2})^{\#(\CL)-1}\,(-1)^{\Arf(\CL)}, & \CL\text{ is proper},  \\
        0, & \CL\text{ is not proper},
    \end{array}
    \right.
\end{equation}
where $\#(\CL)$ is the number of components of $\CL$. The Jones polynomial $V_{\CL}(q)$ can be calculated by applying skein relations to the link. Moreover, its values for common links are readily available in the literature. Altogether this provides us with an explicit way to calculate the values of $\beta_s$ for any closed spin 3-manifold.

As we have already seen in order to determine the modular matrices it is enough to determined the values of $\beta_s$ for mapping tori $MT(\phi)$ of $\TT^2$ for certain elements  $\phi\in\SL(2,\Z)$ and also manifolds of the form\footnote{In principle one could use only mapping tori by choosing the representative of the trivial class in $\Omega_2^\Spin(B\Z_2)$ to be also a 2-torus instead of an empty space.} $(D^2\times S^1)\cup_{\phi} (D^2\times S^1)$. These 3-manifolds have canonical surgery realization given by a word presentation of an element $\phi\in \SL(2,\Z)$ in terms of generators $T$ and $S$.

Consider first the case $Y=(D^2\times S^1)\cup_{\phi} (D^2\times S^1)$. Without loss of generality one can assume that $\phi$ is of the form $\phi=ST^{p_V}\ldots ST^{p_2}ST^{p_1}S$. The manifold is realized by surgery on the following link: 
\begin{equation}
\CL\;\;=\;\;
\begin{tikzpicture}[scale=0.25,baseline=0]
\begin{knot}[flip crossing=2,flip crossing=4,flip crossing=6]
\strand[ultra thick] (2, 0) 
  .. controls ++(90:1) and ++(0:-1) .. (5,1.5) node[pos=1,above] {$p_1$}
  .. controls ++(0:1) and ++(90:1) .. (8,0)
  .. controls ++(90:-1) and ++(0:1) .. (5,-1.5)
  .. controls ++(0:-1) and ++(90:-1) .. (2, 0);
  \strand[ultra thick] (6, 0) 
  .. controls ++(90:1) and ++(0:-1) .. (9,1.5) node[pos=1,above] {$p_2$}
  .. controls ++(0:1) and ++(90:1) .. (12,0)
  .. controls ++(90:-1) and ++(0:1) .. (9,-1.5)
  .. controls ++(0:-1) and ++(90:-1) .. (6, 0);
 \strand[ultra thick]  (13,-1.5)
  .. controls ++(0:-1) and ++(90:-1) .. (10,0)
  .. controls ++(90:1) and ++(0:-1) .. (13,1.5);
 \strand[ultra thick]   (17,1.5)
  .. controls ++(0:1) and ++(90:1) .. (20,0)
  .. controls ++(90:-1) and ++(0:1) .. (17,-1.5);
 \strand[ultra thick] (18, 0) 
  .. controls ++(90:1) and ++(0:-1) .. (21,1.5) node[pos=1,above] {$p_V$}
  .. controls ++(0:1) and ++(90:1) .. (24,0)
  .. controls ++(90:-1) and ++(0:1) .. (21,-1.5)
  .. controls ++(0:-1) and ++(90:-1) .. (18, 0);
  \node at (15,0) {$\ldots$};
\end{knot}
\end{tikzpicture}
\label{linear-plumbing-link}
\end{equation}
where $p_I$ are framings. The linking matrix is the following:
\begin{equation}
    B=\left(
    \begin{array}{ccccc}
        p_1 & -1 & 0 & \ldots & 0 \\
        -1 & p_2 & -1 & 0 & \ldots \\
        0 & -1 & p_3 & -1 & \ldots \\
        \vdots & & \ddots & & \vdots \\
        0 & \ldots & 0 & -1 & p_V 
    \end{array}
    \right).
\end{equation}
Due to its form it is clear that a solution of (\ref{spin-vector-condition}) is completely fixed by the value of $s_1\in \Z_2$, which can be arbitrary. The value $s_1=1$ ($s_1=0$) corresponds to periodic (antiperiodic) spin structure along the non-contractible circle on $\TT^2=\partial(D^2\times S^1)$ on which $\phi\in \SL(2,\Z)$ acts. Note that in this case the corresponding characteristic sublink $\CC_s$ is always a disjoint union of unlinked unknots. Its Arf invariant is always trivial: $\Arf(\CC_s)=0\mod 2$.

Consider now the case $Y=MT(\phi)$. Again, without a loss of generality we can assume that $\phi$ is of the form $\phi=ST^{p_{V-1}}\dots ST^{p_2} ST^{p_1}$. This manifold can be similarly realized by surgery on a framed link $\CL$ which can be described as follows. Instead of the linear ``chain'' of unknots (framed according to the powers of $T$) we now have a circular chain passing through one extra unknot with framing zero:
\begin{equation}
\CL\;\;=\;\;
\begin{tikzpicture}[scale=0.25,baseline=0]
\begin{knot}
\strand[ultra thick] 
    (0.5,0) .. controls ++(90:1) and ++(0:-1) .. 
    (2,3) node[pos=0,left]{$p_1$} .. controls ++(0:1) and ++(90:1) .. 
    (3.5,0) .. controls ++(90:-1) and ++(0:1) .. 
    (2,-3) .. controls ++(0:-1) and ++(90:-1) .. 
    (0.5,0);
\strand[ultra thick,shift={(2,4)},rotate around={-45:(2,0)}] 
    (0.5,0) .. controls ++(90:1) and ++(0:-1) .. 
    (2,3) node[pos=0,left]{$p_2$} .. controls ++(0:1) and ++(90:1) .. 
    (3.5,0) .. controls ++(90:-1) and ++(0:1) .. 
    (2,-3) .. controls ++(0:-1) and ++(90:-1) .. 
    (0.5,0);
\strand[ultra thick,shift={(2,-4)},rotate around={45:(2,0)}] 
    (0.5,0) .. controls ++(90:1) and ++(0:-1) .. 
    (2,3) node[pos=0,left]{$p_{V-1}$}.. controls ++(0:1) and ++(90:1) .. 
    (3.5,0) .. controls ++(90:-1) and ++(0:1) .. 
    (2,-3) .. controls ++(0:-1) and ++(90:-1) .. 
    (0.5,0);
\strand[ultra thick,shift={(6.5,6)}] 
    (0.5,0) .. controls ++(90:1) and ++(0:-1) .. 
    (2,3) node[pos=1,above]{$0$} .. controls ++(0:1) and ++(90:1) .. (3.5,0);
\strand[white,line width=1.6mm] (6,7.2) -- (8,7.6);    
\strand[ultra thick,shift={(6,6)},rotate around={90:(2,0)}] 
    (0.5,-1) to[bend left=5] (0.5,0) .. controls ++(90:1) and ++(0:-1) .. (2,3) .. controls ++(0:1) and ++(90:1) .. 
    (3.5,0) to[bend left=5] (3.5,-1);
\strand[ultra thick,shift={(6,-6)},rotate around={90:(2,0)}] 
    (0.5,-1) to[bend left=5] (0.5,0) .. controls ++(90:1) and ++(0:-1) .. 
    (2,3) node[pos=0,below]{$p_{V-2}$} .. controls ++(0:1) and ++(90:1) .. 
    (3.5,0) to[bend left=5] (3.5,-1);
\strand[ultra thick,shift={(6.5,6)}]    
    (3.5,0) .. controls ++(90:-1) and ++(0:1) .. 
    (2,-3) .. controls ++(0:-1) and ++(90:-1) .. 
    (0.5,0);
\node at (6,8.2) {$p_3$};    
\node at (12,6) {$\cdots$};
\node at (12,-6) {$\cdots$};
\flipcrossings{2,4,8,6}
\end{knot}
\end{tikzpicture}
\label{circular-plumbing-link}
\end{equation}
The linking matrix now has the following form:
\begin{equation}
    B=\left(
    \begin{array}{cccccc}
        p_1 & -1 & 0 & \ldots & -1 & 0\\
        -1 & p_2 & -1 & 0 & \ldots & 0\\
        0 & -1 & p_3 & -1 & \ldots & 0\\
        \vdots & & \ddots & & & \vdots \\
        -1 & \ldots & 0 & -1 & p_{V-1} & 0 \\
        0 & \ldots & 0 & 0 & 0 & 0
    \end{array}
    \right).
\end{equation}
The solution to the characteristic sublink condition (\ref{spin-vector-condition}) is completely fixed by the values of $s_1,s_2$ and $s_V$. The values of $s_{1,2}$ correspond to the spin structures on the meridian and longitude of the 2-torus on which $\phi\in \SL(2,\Z)$ acts. As before, the value $1$ ($0$) corresponds to periodic (anti-periodic) condition on spinors. The value of $s_V$ is always arbitrary and independent of other values $s_I$. It corresponds to the choice of spin structure on the base circle of the mapping torus, if the latter is considered as a fibration over $S^1$ with torus in the fibers.

Note that in both cases one can consider more general words that also contain $S^{-1}$ (without expressing it through $S$ and $T$, e.g. $S^{-1}=ST^0ST^0S$) by flipping the links as follows:
\begin{equation}
\begin{array}{rcl}
S & \rightsquigarrow & S^{-1}\;, \\
\begin{tikzpicture}[scale=0.3,baseline=0]
\begin{knot}[flip crossing = 2]
  \strand[ultra thick] (9,1.5) .. controls ++(0:1) and ++(90:1) .. (12,0)
  .. controls ++(90:-1) and ++(0:1) .. (9,-1.5);
 \strand[ultra thick]  (13,-1.5)
  .. controls ++(0:-1) and ++(90:-1) .. (10,0)
  .. controls ++(90:1) and ++(0:-1) .. (13,1.5);
\end{knot}
\end{tikzpicture}
&
\rightsquigarrow
&
\begin{tikzpicture}[scale=0.3,baseline=0]
\begin{knot}[flip crossing = 1]
  \strand[ultra thick] (9,1.5) .. controls ++(0:1) and ++(90:1) .. (12,0)
  .. controls ++(90:-1) and ++(0:1) .. (9,-1.5);
 \strand[ultra thick]  (13,-1.5)
  .. controls ++(0:-1) and ++(90:-1) .. (10,0)
  .. controls ++(90:1) and ++(0:-1) .. (13,1.5);
\end{knot}
\end{tikzpicture}
\;,
\\
B_{I,I+1}=-1
&
\rightsquigarrow
&
B_{I,I+1}=1\,.
\end{array}
\end{equation}
Of course, making this replacement at any place in the linear chain of unknots (\ref{linear-plumbing-link}) will always result in an equivalent link. This, however, is not the case for a circular chain (\ref{circular-plumbing-link}).

As an example of using this method, consider the mapping torus of $\phi=
T^2\in \SL(2,\Z)$, which is needed to determine some of the matrix elements of the $T$ modular matrix. To represent $Y=MT(\phi)$ by surgery we write it in the form $\phi=ST^2S^{-1}T^0$, so
\begin{equation}
    \CL\;\;=\;\;
\begin{tikzpicture}[scale=0.25,baseline=0]
\begin{knot}
\strand[ultra thick] 
    (-3,0) .. controls ++(90:1) and ++(180:1) ..
    (0,1.5) node[pos=0,left]{$0$} .. controls ++(0:1) and ++(90:1) ..
    (3,0) .. controls ++(-90:1) and ++(0:1) ..
    (0,-1.5) .. controls ++(180:1) and ++(-90:1) ..
    (-3,0);
\strand[ultra thick] 
    (5.5,0) .. controls ++(90:1) and ++(180:1) ..
    (7,3) .. controls ++(0:1) and ++(90:1) ..
    (8.5,0) node[pos=1,right]{$2$} .. controls ++(-90:1) and ++(0:1) ..
    (7,-3) .. controls ++(180:1) and ++(-90:1) ..
    (5.5,0);
\strand[ultra thick]
    (4,5) .. controls ++(0:2) and ++(90:1) ..
    (9,3) .. controls ++(-90:1) and ++(0:1) ..
    (7,2) .. controls ++(180:1) and ++(0:1) ..
    (4,3) .. controls ++(180:1) and ++(90:1) ..
    (1.5,0) .. controls ++(-90:1) and ++(180:1) ..
    (4,-3) .. controls ++(0:1) and ++(180:1) ..
    (7,-2) .. controls ++(0:1) and ++(90:1) ..
    (9,-3) .. controls ++(-90:1) and ++(0:2) ..
    (4,-5) .. controls ++(180:2) and ++(-90:2) ..
    (-1.5,0) .. controls ++(90:2) and ++(180:2) ..
    (4,5) node[pos=0.5,above]{$0$};
\flipcrossings{1,2,6,7}
\end{knot}
\end{tikzpicture}\;.
\end{equation}
This 3-component link is commonly known as \textit{Borromean rings} and has a property that if either of the components is removed it becomes an unlink. The linking matrix is the following:
\begin{equation}
    B=\left(
        \begin{array}{ccc}
            0 & 0 & 0 \\
            0 & 2 & 0 \\
            0 & 0 & 0
        \end{array}
    \right).
\end{equation}
Therefore any sublink of $\CL$ is a characteristic sublink, which is consistent with the fact that there are $|H^1(Y,\Z_2)|=8$ different spin structures on $Y=S^3(\CL)$. Table \ref{table:surgery-T2-mapping-torus} shows different choices of characteristic sublinks $\CC_s$ corresponding to different spin structures $s\in \Spin(Y)$ and the corresponding values of the Rokhlin invariant $\mu_s$. Note that the Arf invariant is non-trivial only for the complete Borromean rings. 

\begin{table}[h!]
\begin{center}
\makegapedcells
\begin{tabular}{|c||p{0.4\textwidth}|p{0.4\textwidth}|}
\hline
$(s_1,s_2)$ & $s_3=0$ & $s_3=1$ \\
\hline 
\hline 
$(0,0)$ &
$\begin{array}{c}
\CC_s=\varnothing
\\
\lk(\CC_s,\CC_s)=0, \;\Arf(\CC_s)=0,
\\
\mu_s=1\mod 16
\end{array}$
&
$\begin{array}{c}
\CC_s= \begin{tikzpicture}[scale=0.25,baseline=0]
\begin{knot}
\strand[ultra thick] 
    (-3,0) .. controls ++(90:1) and ++(180:1) ..
    (0,1.5) node[pos=0,left]{$0$} .. controls ++(0:1) and ++(90:1) ..
    (3,0) .. controls ++(-90:1) and ++(0:1) ..
    (0,-1.5) .. controls ++(180:1) and ++(-90:1) ..
    (-3,0);
\end{knot}
\end{tikzpicture}
\\
\lk(\CC_s,\CC_s)=0,\; \Arf(\CC_s)=0,
\\
\mu_s=1\mod 16
\end{array}$
\\ 
\hline
$(1,0)$ 
& 
$\begin{array}{c}
\CC_s=\begin{tikzpicture}[scale=0.25,baseline=0]
\begin{knot}
\strand[ultra thick]
    (4,5) .. controls ++(0:2) and ++(90:1) ..
    (9,3) .. controls ++(-90:1) and ++(0:1) ..
    (7,2) .. controls ++(180:1) and ++(0:1) ..
    (4,3) .. controls ++(180:1) and ++(90:1) ..
    (1.5,0) .. controls ++(-90:1) and ++(180:1) ..
    (4,-3) .. controls ++(0:1) and ++(180:1) ..
    (7,-2) .. controls ++(0:1) and ++(90:1) ..
    (9,-3) .. controls ++(-90:1) and ++(0:2) ..
    (4,-5) .. controls ++(180:2) and ++(-90:2) ..
    (-1.5,0) .. controls ++(90:2) and ++(180:2) ..
    (4,5) node[pos=0.5,above]{$0$};
\end{knot}
\end{tikzpicture}
\\
\lk(\CC_s,\CC_s)=0,\; \Arf(\CC_s)=0,
\\
\mu_s=1\mod 16
\end{array}$
& 
$\begin{array}{c}
\CC_s=\begin{tikzpicture}[scale=0.25,baseline=0]
\begin{knot}
\strand[ultra thick] 
    (-3,0) .. controls ++(90:1) and ++(180:1) ..
    (0,1.5) node[pos=0,left]{$0$} .. controls ++(0:1) and ++(90:1) ..
    (3,0) .. controls ++(-90:1) and ++(0:1) ..
    (0,-1.5) .. controls ++(180:1) and ++(-90:1) ..
    (-3,0);
\strand[ultra thick]
    (4,5) .. controls ++(0:2) and ++(90:1) ..
    (9,3) .. controls ++(-90:1) and ++(0:1) ..
    (7,2) .. controls ++(180:1) and ++(0:1) ..
    (4,3) .. controls ++(180:1) and ++(90:1) ..
    (1.5,0) .. controls ++(-90:1) and ++(180:1) ..
    (4,-3) .. controls ++(0:1) and ++(180:1) ..
    (7,-2) .. controls ++(0:1) and ++(90:1) ..
    (9,-3) .. controls ++(-90:1) and ++(0:2) ..
    (4,-5) .. controls ++(180:2) and ++(-90:2) ..
    (-1.5,0) .. controls ++(90:2) and ++(180:2) ..
    (4,5) node[pos=0.5,above]{$0$};
\flipcrossings{1,2}
\end{knot}
\end{tikzpicture}
\\
\lk(\CC_s,\CC_s)=0,\; \Arf(\CC_s)=0,
\\
\mu_s=1\mod 16
\end{array}
$
\\
\hline
$(0,1)$ 
& 
$\begin{array}{c}
\CC_s=\begin{tikzpicture}[scale=0.25,baseline=0]
\begin{knot}
\strand[ultra thick] 
    (5.5,0) .. controls ++(90:1) and ++(180:1) ..
    (7,3) .. controls ++(0:1) and ++(90:1) ..
    (8.5,0) node[pos=1,right]{$2$} .. controls ++(-90:1) and ++(0:1) ..
    (7,-3) .. controls ++(180:1) and ++(-90:1) ..
    (5.5,0);
\end{knot}
\end{tikzpicture}
\\
\lk(\CC_s,\CC_s)=2,\; \Arf(\CC_s)=0,
\\
\mu_s=-1\mod 16
\end{array}
$
&
$\begin{array}{c}
\CC_s=\begin{tikzpicture}[scale=0.25,baseline=0]
\begin{knot}
\strand[ultra thick] 
    (-3,0) .. controls ++(90:1) and ++(180:1) ..
    (0,1.5) node[pos=0,left]{$0$} .. controls ++(0:1) and ++(90:1) ..
    (3,0) .. controls ++(-90:1) and ++(0:1) ..
    (0,-1.5) .. controls ++(180:1) and ++(-90:1) ..
    (-3,0);
\strand[ultra thick] 
    (5.5,0) .. controls ++(90:1) and ++(180:1) ..
    (7,3) .. controls ++(0:1) and ++(90:1) ..
    (8.5,0) node[pos=1,right]{$2$} .. controls ++(-90:1) and ++(0:1) ..
    (7,-3) .. controls ++(180:1) and ++(-90:1) ..
    (5.5,0);
\end{knot}
\end{tikzpicture}
\\
\lk(\CC_s,\CC_s)=2,\; \Arf(\CC_s)=0,
\\
\mu_s=-1\mod 16
\end{array}
$
\\
\hline
$(1,1)$ 
& 
$\begin{array}{c}
\CC_s=\begin{tikzpicture}[scale=0.25,baseline=0]
\begin{knot}[flip crossing =2, flip crossing =3]
\strand[ultra thick] 
    (5.5,0) .. controls ++(90:1) and ++(180:1) ..
    (7,3) .. controls ++(0:1) and ++(90:1) ..
    (8.5,0) node[pos=1,right]{$2$} .. controls ++(-90:1) and ++(0:1) ..
    (7,-3) .. controls ++(180:1) and ++(-90:1) ..
    (5.5,0);
\strand[ultra thick]
    (4,5) .. controls ++(0:2) and ++(90:1) ..
    (9,3) .. controls ++(-90:1) and ++(0:1) ..
    (7,2) .. controls ++(180:1) and ++(0:1) ..
    (4,3) .. controls ++(180:1) and ++(90:1) ..
    (1.5,0) .. controls ++(-90:1) and ++(180:1) ..
    (4,-3) .. controls ++(0:1) and ++(180:1) ..
    (7,-2) .. controls ++(0:1) and ++(90:1) ..
    (9,-3) .. controls ++(-90:1) and ++(0:2) ..
    (4,-5) .. controls ++(180:2) and ++(-90:2) ..
    (-1.5,0) .. controls ++(90:2) and ++(180:2) ..
    (4,5) node[pos=0.5,above]{$0$};
\end{knot}
\end{tikzpicture}
\\
\lk(\CC_s,\CC_s)=2,\; \Arf(\CC_s)=0,
\\
\mu_s=-1\mod 16
\end{array}
$
& 
$
\begin{array}{c}
\CC_s=\begin{tikzpicture}[scale=0.25,baseline=0]
\begin{knot}
\strand[ultra thick] 
    (-3,0) .. controls ++(90:1) and ++(180:1) ..
    (0,1.5) node[pos=0,left]{$0$} .. controls ++(0:1) and ++(90:1) ..
    (3,0) .. controls ++(-90:1) and ++(0:1) ..
    (0,-1.5) .. controls ++(180:1) and ++(-90:1) ..
    (-3,0);
\strand[ultra thick] 
    (5.5,0) .. controls ++(90:1) and ++(180:1) ..
    (7,3) .. controls ++(0:1) and ++(90:1) ..
    (8.5,0) node[pos=1,right]{$2$} .. controls ++(-90:1) and ++(0:1) ..
    (7,-3) .. controls ++(180:1) and ++(-90:1) ..
    (5.5,0);
\strand[ultra thick]
    (4,5) .. controls ++(0:2) and ++(90:1) ..
    (9,3) .. controls ++(-90:1) and ++(0:1) ..
    (7,2) .. controls ++(180:1) and ++(0:1) ..
    (4,3) .. controls ++(180:1) and ++(90:1) ..
    (1.5,0) .. controls ++(-90:1) and ++(180:1) ..
    (4,-3) .. controls ++(0:1) and ++(180:1) ..
    (7,-2) .. controls ++(0:1) and ++(90:1) ..
    (9,-3) .. controls ++(-90:1) and ++(0:2) ..
    (4,-5) .. controls ++(180:2) and ++(-90:2) ..
    (-1.5,0) .. controls ++(90:2) and ++(180:2) ..
    (4,5) node[pos=0.5,above]{$0$};
\flipcrossings{1,2,6,7}
\end{knot}
\end{tikzpicture}
\\
\lk(\CC_s,\CC_s)=2,\; \Arf(\CC_s)=1,
\\
\mu_s=7\mod 16
\end{array}
$
\\
\hline
\end{tabular}
\end{center}
\caption{Different characteristic sublinks of the link realizing the mapping torus of $T^2\in\SL(2,\Z)$ and the corresponding values of the Rokhlin invariant. We used the fact that $\sigma(B)=1$ which gives a spin-structure independent contribution to $\mu_s$.}
\label{table:surgery-T2-mapping-torus}
\end{table}

This table can be used to determine all the matrix elements of the diagonal matrix $T^2$ (which, in turn, determines off-diagonal elements of the matrix $T$ after fixing the basis). For example:
\begin{equation}
    T^2|_{\{R0,R1\}}=e^{\frac{\pi i\nu}{4}\beta_{(1,1,0)}((0,1,0))}=
    e^{\frac{\pi i\nu}{8}(\mu_{(1,1,0)}-\mu_{(1,0,0)})}=
    e^{-\frac{\pi i \nu}{4}}
\end{equation}
which is in agreement with the geometric calculation in Section \ref{sec:matrices-from-invariants}. One can easily determine other matrix elements of $S$ and $T$ in a similar way.

\section{Generalization to other symmetry groups in two dimensions}
\label{sec:gen-groups}

In this section we will outline a method to explicitly determine the maps in the sequence (\ref{intro-map-triple}) for the case of two-dimensional theories on a 2-torus with arbitrary finite group symmetry. To make the formulas more transparent we will assume that the symmetry is always of the form $G^f=G\times \Z_2^f$, although the more general case can be analyzed in a similar way.

\subsection{Naturality property}

In such a setup the sequence (\ref{intro-map-triple}) reads explicitly as follows:
\begin{equation}
    RO(G)\longrightarrow
    \Hom(\Omega_3^\Spin(BG),U(1))
    \longrightarrow 
    H^1((\Mp\times G)\ltimes \Spin_G(\TT^2),U(1))
    \label{three-term-seq-torus}
\end{equation}
where $RO(G)$ is the real representation ring of $G$ (considered just as an abelian group under addition). Note that in the second place in the sequence we only consider the subgroup of the total anomaly group corresponding to non-perturbative anomalies. That is we assume that perturbative anomalies (which can be only purely gravitational when $G$ is discrete and are formally classified by $\Hom(\Omega_4^\Spin(\text{pt}),U(1))\cong \Z$) are absent.  The first map is physically realized as follows. The elements of $RO(G)$, by definition, are formal differences $\rho_1-\rho_2$ of (isomorphism classes of) representations. We then first consider the theory of free right-moving (or anti-holomorphic) Majorana-Weyl fermions in representation $\rho_1$ and left-moving (or holomorphic) Majorana-Weyl fermions in representation $\rho_2$. We then add to it an appropriate number of left- or right-moving free fermions in the trivial representation to ensure that the chiral central charge of the theory is zero. That is we always add or substract copies of a trivial representation to $\rho_1-\rho_2$ to make the total virtual dimension zero. The anomaly of the resulting theory is then the value of the first map of the sequence. 

Mathematically the first map can be understood as a spin-cobordism version of a characteristic class map $RU(G)\rightarrow H^4(BG,\Z)\cong H^3(BG,U(1))$ \cite{atiyah1961characters,evens1965chern,kroll1987algebraic}  (here $RU(G)$ is the complex representation ring). A real representation $\rho$ of dimension $n$ can be understood as a homomorphism $\rho:G\rightarrow O(n)$. The value of its image in $\Hom(\Omega_3^\Spin(BG),U(1))$ then can be understood as a pullback under $\rho$ of a certain ``universal characteristic class'' in\footnote{The splitting of this group (Anderson dual of spin-bordism of $BO$) into direct sum is non-canonical.} $\Hom(\Omega_3^\Spin(BO),U(1))\oplus \Hom(\Omega_4^\Spin(BO),\Z)$ that vanishes under the forgetful map to $\Hom(\Omega_4^\Spin(\pt),\Z)$. This latter description is however hard to use in practice to evaluate the map for a given representation and we will use a different approach.

 The second map is just the special case of a more general map (\ref{anomaly-MCG-cohomology-map}) described in Section \ref{sec:anomalous-phases}. The evaluation of this map for a given element of the anomaly group boils down to calculation of the value of the corresponding spin-bordism invariant on certain closed spin 3-manifolds.

As was already mentioned in the introduction all three groups in the sequence (\ref{three-term-seq-torus}) can be understood as values of contravariant functors from the category of groups to the category of abelian groups. The sequence (\ref{three-term-seq-torus}) can then be understood as a sequence of natural transformation between the corresponding functors, evaluated on object $G$. Namely, for any group homomorphism
\begin{equation}
    G\stackrel{f}{\longrightarrow} G'
\end{equation}
 we have induced maps (pullbacks) for the corresponding abelian groups in (\ref{three-term-seq-torus}), so that they fit into the following commutative diagram:
\begin{equation}
\begin{tikzcd}
    RO(G) \ar[d] & RO(G') \ar[l,"f^*",swap] \ar[d] 
    \\
    \Hom(\Omega_3^\Spin(BG),U(1)) \ar[d] & \Hom(\Omega_3^\Spin(BG'),U(1)) \ar[l,"f^*",swap] \ar[d] \\ 
  H^1((\Mp\times G)\ltimes\Spin_G(\TT^2),U(1))  &
   H^1((\Mp\times G')\ltimes\Spin_{G'}(\TT^2),U(1)).  \ar[l,"f^*",swap] 
\end{tikzcd}
\label{torus-naturality-diagram}
\end{equation}
Let us clarify how the horizontal maps in this diagram are defined. The first horizontal map in the diagram is the standard pullback of representations. 

For the second map, a homomorphism $f:G\rightarrow G'$ first induces a map $BG\rightarrow BG'$ between the corresponding classifying space. This map then induces the pushforward of the corresponding spin-bordism group $\Omega_3^\Spin(BG)\rightarrow \Omega_3^\Spin(BG')$. Finally, taking the Pontryagin dual $\Hom(\cdot,U(1))$ to both of this group produces the dual map $f^*$ in the second line of the diagram (\ref{torus-naturality-diagram}). More explicitly, one can describe this map in terms of supercohomology representation of Pontryagin duals to the bordism groups \cite{cheng2018classification,Gaiotto:2015zta,brumfiel2016pontrjagin,Wang:2017moj,Wang:2018pdc}. In this approach the elements of the anomaly group $\Omega_3^\Spin(BG)$ are given by equivalence classes of a collection of functions on the products of copies of $G$ satisfying certain generalized cocycle conditions. These functions are then pulled back under $f:G\longrightarrow G'$.

Physically the second horizontal map in (\ref{torus-naturality-diagram}) can be understood as follows. Consider a theory with symmetry $G'$ that has an anomaly corresponding to a certain element of $\Hom(\Omega_3^\Spin(BG'),U(1))$. By using the homomorphism $f:G\rightarrow G'$ one can consider this as a theory with symmetry $G$ instead\footnote{Here we have no requirement of action being faithful.} (i.e. $g\in G$ acts on the operators of the theory as $f(g)\in G'$). Its anomaly then gives an element in $\Hom(\Omega_3^\Spin(BG),U(1))$.

To define the third horizontal map in (\ref{torus-naturality-diagram}) first note that explicitly we have
\begin{equation}
    \Spin_G(\TT^2)=\Spin(\TT^2)\times \Hom(\pi_1(\TT^2),G)\cong
    \Spin(\TT^2)\times \Hom(\Z^2,G)
\end{equation}
where the last isomorphism is achieved by choosing a basis in $\pi_1(\TT^2)\cong H_1(\TT^2,\Z)$. The elements of $\Hom(\Z^2,G)$ can be understood as pairs of commuting elements in $G$. We have a natural induced pushforward map
\begin{equation}
    f_*:\;\Spin_{G}(\TT^2)\longrightarrow \Spin_{G'}(\TT^2)   
\end{equation}
that acts on elements of $\Hom(\Z^2,G)$ by composition with $f$. Recall that an element of $H^1((\Mp\times G')\ltimes\Spin_{G'}(\TT^2),U(1))$ can be represented by a representation $\rho$ of the action groupoid $(\Mp\times G')\ltimes\Spin_{G'}(\TT^2)$, that is a collection of one-dimensional complex vector spaces $\CH_a$ for each $a\in \Spin_{G'}(\TT^2)$ and linear maps 
\begin{equation}
    \rho(g'):\;\CH_{a'}\longrightarrow \CH_{g'\cdot a'}
\end{equation}
for each $g'\in G'$ which satisfy the 1-cocycle condition $\rho(g_1g_2)=\rho(g_1)\circ \rho(g_2)$. To construct a pullback of this representation under $f$, one takes $f^*\CH_a:=\CH_{f_*(a)}$ for each $a\in \Spin_G(\TT^2)$ and the maps
\begin{equation}
    (f^*\rho)(g):=\rho(f(g)):\;f^*\CH_{a}\equiv \CH_{f_*(a)}\longrightarrow \CH_{f(g)\cdot f_*(a)}=
    \CH_{f_*(g\cdot a)}\equiv
    f^*\CH_{g\cdot a}.
\end{equation}
As we will see, the commutativity of the diagram (\ref{torus-naturality-diagram}) for any group homomorphism $f:G\rightarrow G'$ is a very constraining property that can be used to essentially fix the vertical maps for any $G$ from the knowledge of these maps for some basic cases.

\subsection{The map from the representation ring to the anomaly group}

As was already mentioned, the ``Chern character'' map
\begin{equation}
    \sch:RO(G)\longrightarrow \Hom(\Omega_3^{\Spin}(BG),U(1))
    \label{spin-ch2-map}
\end{equation}
can be understood as a spin-bordism\footnote{To be precise, Anderson dual to spin-bordism generalized cohomology theory.} analog of the maps $c_i:RU(G)\rightarrow H^{2i}(BG,\Z)$ and $w_i:RO(G)\rightarrow H^i(BG,\Z_2)$ given respectively by Chern and Stiefel-Whitney classes of linear representations of $G$ \cite{atiyah1961characters,evens1965chern,kroll1987algebraic} . This in part motivates the choice of the notation and terminology, as the map $\sch$ reduces to the map $2\mathrm{ch}_2=c_1^2-2c_2:RU(G)\rightarrow  H^4(BG,\Z) \cong H^3(BG,U(1))\subset \Hom(\Omega_3^\Spin(BG),U(1))$ when the corresponding anomaly is bosonic and realized by taking a \textit{double}\footnote{Note that $\mathrm{ch}_2=c_1^2/2-c_2$ in general does not give a well-defined element in integral cohomology because of $1/2$ factor.} of \textit{complex} Weyl fermions in the same representation. I.e. the following diagram is commutative:
\begin{equation}
\begin{tikzcd}
RU(G) \ar[r,"2\text{ch}_2=c_1^2-2c_2"] 
\ar[d,"(\,\cdot\,)_\R^{\oplus 2}"]
& H^3(BG,U(1)) \ar[d,hook] \\
RO(G) \ar[r,"\sch"] & \Hom(\Omega_3^{\Spin}(BG),U(1))
\end{tikzcd}
\end{equation}
where the left vertical map is realized by taking two copies of a complex representation, considered as a real representation (e.g. one-dimensional complex representation is mapped to a four-dimensional real one). The right vertical map physically corresponds to embedding the group of bosonic anomalies into the group of fermionic anomalies\footnote{Note that in 3 dimensions this map is injective, so it can be understood as an embedding, but this is not true in general}. Mathematically it is realized as the dual to the forgetful map $\Omega_3^\Spin(BG)\rightarrow H_3(BG,\Z)$. 

In this cases it is known that one can define these classes axiomatically, by specifying their properties which determine them uniquely. Although we do not prove this, we would like to claim that one can axiomatically define $\sch$ in a similar way. Namely we consider the following three axioms (which are in parallel to the axioms for Chern and Stiefel-Whitney characteristic classes):

\begin{enumerate}
    \item Linearity:
\begin{equation}
    \sch(\rho_1\oplus \rho_2)=\sch(\rho_1)+\sch(\rho_2).
\end{equation}
    \item Naturality: for any group homomorphism $f:G\rightarrow G'$
\begin{equation}
    f^*\sch(\rho) =\sch(f^*\rho)
\end{equation}
i.e. commutativity of the upper square in the diagram (\ref{torus-naturality-diagram}).
\item Normalization: for an abelian $G$ the value of $\sch$ is given according to Tables \ref{table:rep-map-A}, \ref{table:rep-map-B}, and \ref{table:rep-map-C} (see Section \ref{sec:abelian-rep-map} for details).  
\end{enumerate}
We conjecture that these properties uniquely fix $\sch$. In the Section \ref{sec:ch2-S3-example} we give an example on how one can use this axioms to evaluate this map in the case of a non-abelian $G$. 
Let us note that in \cite{Davighi:2020uab} the similar naturality property was used for the homomorphism $\Z_2\rightarrow U(1)$ (in this case the target group $U(1)$ is continuous and has non-trivial perturbative anomaly group).

\subsubsection{Abelian groups}
\label{sec:abelian-rep-map}
Consider a general abelian group $G$, written in the form \begin{equation}
    G=\Z_{p_1^{r_1}}\times \Z_{p_2^{r_2}}\times \ldots \times \Z_{p_N^{r_N}}
\end{equation}
where $p_i$ are prime numbers and $r_i\geq 1$. From the results of \cite{cheng2018classification,Wang:2017moj,Guo:2018vij,Wang:2018pdc} one can conclude that the anomaly group has the following form:
\begin{equation}
    \Hom(\Omega_3^\Spin(BG),U(1)) 
    \cong 
    \prod_{i} A_{p_i;r_i}\,\times\,
    \prod_{\substack{i<j\\ p_i=p_j}} B_{p_i;r_i,r_j}\,\times\,
    \prod_{\substack{i<j<k\\ p_i=p_j=p_k}} C_{p_i;r_i,r_j,r_k}
    \label{anomaly-group-ABC-decomposition}
\end{equation}
where 
\begin{equation}
    A_{p;r}=\left\{
    \begin{array}{cl}
     \Z_8, & p=2,r=1, \\
     \Z_2\times \Z_{2^{r+1}}, & p=2,r>1, \\
     \Z_{p^r}, & p>2,
     \end{array}
    \right.
\end{equation}
\begin{equation}
    B_{p;r_1,r_2}=\left\{
    \begin{array}{cl}
     \Z_{4}, & p=2,\;r_1=r_2=1, \\
     \Z_{2}\times \Z_{2^{\min(r_1,r_2)}}, & p=2,\; \max(r_1,r_2)>1, \\
     \Z_{p^{\min(r_1,r_2)}}, & p>2,
     \end{array}
    \right.
\end{equation}
\begin{equation}
    C_{p;r_1,r_2,r_3}=
     \Z_{p^{\min(r_1,r_2,r_3)}}\;.
\end{equation}
The decomposition into factors (\ref{anomaly-group-ABC-decomposition}) is determined by considering the pullbacks of the projections of $G$ on the factors of the form $\Z_{p_i^{r_i}}$,  $\Z_{p_i^{r_i}}\times \Z_{p_j^{r_j}}$, $\Z_{p_i^{r_i}}\times \Z_{p_j^{r_j}}\times \Z_{p_k^{r_k}}$. In particular the subgroup $A_{p_i;r_i}$ in the decomposition is the image of the anomaly group $\Hom(\Omega_3^\Spin(B\Z_{p_i^{r_i}}))$ under the projection $G\rightarrow \Z_{p_i^{r_i}})$. For comparison, the group of bosonic anomalies is given by
\begin{equation}
    H^3(BG,U(1))
    \cong 
    \prod_{i} \Z_{p_i^{r_i}}\,\times\,
    \prod_{\substack{i<j\\ p_i=p_j}} \Z_{p_i^{\min(r_i,r_j)}}\,\times\,
    \prod_{\substack{i<j<k\\ p_i=p_j=p_k}} \Z_{p_i^{\min(r_i,r_j,r_k)}}
\end{equation}
To describe the map (\ref{spin-ch2-map}) for a general abelian $G$ it is then enough to describe the maps
\begin{equation}
\begin{array}{rcl}
RO(\Z_{p^{r}}) & \longrightarrow & A_{p;r} \\
RO(\Z_{p^{r_1}}\times \Z_{p^{r_2}}) & \longrightarrow & B_{p;r_1,r_2} \\
RO(\Z_{p^{r1}}\times \Z_{p^{r_2}}\times \Z_{p^{r_3}})  & \longrightarrow & C_{p;r_1,r_2,r_3} 
\end{array}
\label{basic-rep-maps}
\end{equation}
for arbitrary $p$ and $r_i$. Without loss of generality we can assume that $r_1\leq r_2\leq r_3$.

The representation theory of abelian groups is simple. Consider first the case of a single cyclic group $\Z_n$. Consider first complex 1-dimensional representations $\crep{q}$ of a given ``mod $n$ charge'' $q$ which is the pullback of the standard charge $q$ representation of $U(1)$ under the embedding $\Z_n\subset U(1)$. That is the generator of $\Z_n$ acts as multiplication by $e^\frac{2\pi iq}{n}$. This representation can be considered as a 2-dimensional real representation of charge $\min \{q,n-q\}$. As a real representation, it is irreducible unless $q=0$ or $q=n/2$. The second case is only possible if $n$ is even. In both of these cases $\crep{q}$ decomposes into direct sum of two copies of an irreducible 1-dimensional representation, which we will denote using single brackets instead:
\begin{eqnarray}
    \crep{0} & = & [0]\oplus [0], \label{irrep0}\\
    \crep{n/2} & = & [n/2]\oplus [n/2]. \label{irrepn2}
\end{eqnarray}
The representations $[0]$ and $[n/2]$ are also known as \textit{trivial} and \textit{sign} representation respectively. Whenever we are going to consider a 2-dimensional irrep different than \eqref{irrep0} and \eqref{irrepn2}, we are going to denote it with the correspondent complex representation $\crep{q}$, where $0 < q \le n/2$. There are no other irreducible representations over $\R$. 

Similarly, for a product $\Z_{n_1}\times \Z_{n_2}\times\ldots $ we have the complex 1-dimensional representations
\begin{equation}
    \crep{q_1,q_2,\ldots }=\crep{q_1}\otimes_{\C}\crep{q_2}\otimes_{\C}\ldots,\;\; \text{where }q_i\in \Z_{n_i}\,,
\end{equation}
which are irreducible over $\R$ for generic charges $q_i$, and reduce to two copies of a real 1-dimensional representation
\begin{equation}
    [q_1,q_2,\ldots ]=[q_1]\otimes_{\R} [q_2]\otimes_{\R}\ldots
\end{equation}
when $q_i=0$ or $n_i/2$ for each $i$.

The values of the first two maps in (\ref{basic-rep-maps}) then can be determined, \textit{up to an automorphism} in the target groups, using the following facts: 
\begin{enumerate}
    \item The representation $\crep{q}$ of $\Z_{p^r}$ is the pullback of the representation with charge $q$ of $U(1)$ under the standard inclusion. The group of fermionic anomalies for $U(1)$ is classified by $\Z$ (the generator corresponds to the anomaly polynomial $c_1^2/2$) and the corresponding value is well known to be $q^2$. The image under the pullback of the anomaly group is the subgroup $\Z_{2^{r+1}}$ for $p=2$ and $\Z_{p^r}$ for odd $p$ in $A_{p;r}$.
    \item Similarly, the representation $\crep{q_1,q_2}$ of $\Z_{p^{r_1}}\times \Z_{p^{r_2}}$ is the pullback of representation with charges $(q_1,q_2)$ of $U(1)^{(1)}\times U(1)^{(2)}$ under the standard inclusion. We introduced the extra superscripts $(1)$ and $(2)$ to distinguish the $U(1)$ subgroups in the discussion below. The group of mixed fermionic anomalies between two copies of $U(1)$ is classified by $\Z$ (the generator corresponds to the anomaly polynomial $c_1^{(1)}c_1^{(2)}$) and the corresponding value is well known to be $q_1q_2$. The image under the pullback of the mixed anomaly group is the subgroup $\Z_{p^{\min(r_1,r_2)}}$ in $B_{p;r_2,r_2}$.
    \item From the supercohomology description of the anomaly group for a finite $G$ we have
    \begin{equation}
        \Hom(\Omega_3^\Spin(BG),U(1)\stackrel{\text{set}}{\cong} H^3(BG,U(1))\times SH^2(BG,\Z_2)\times H^1(BG,\Z_2),
    \end{equation}
    where $SH^2(BG,\Z_2)$ is a certain subgroup of $H^2(BG,\Z_2)$ (kernel of $(\cdot\cup \cdot)$ operation composed with the canonical map  $H^4(BG,\Z_2)\rightarrow H^4(BG,U(1))$)
    and, moreover there is a canonical inclusion homomorphism 
    \begin{equation}
        H^3(BG,U(1))\longrightarrow \Hom(\Omega_3^\Spin(BG),U(1))
    \end{equation}
    and a projection homomorphism
    \begin{equation}
        \pi:\Hom(\Omega_3^\Spin(BG),U(1))\longrightarrow H^1(BG,\Z_2)
    \end{equation}
    such that $\pi\sch(\rho)=w_1(\rho)$ where $w_1$ is the first Stiefel-Whitney class of the representation $\rho$ \cite{chen2019free}. The latter is realized by the 1-cocycle $\det \rho$, where $\rho$ is considered as a map $G\rightarrow O(n)$ for some $n$ and $\det:O(n)\rightarrow \Z_2$. Moreover, the component of $\sch(\rho)$ in $SH^2(BG,\Z_2)$ is given by the second Stiefel-Whitney class $w_2(\rho)$, possibly up to $w_1(\rho)^2$.
    
\end{enumerate}

 The result is presented in the Tables \ref{table:rep-map-A} and \ref{table:rep-map-B}. The freedom related to automorphisms can be fixed by assigning particular bordism invariants to the generators of the anomaly groups. This will be done in the Section \ref{sec:abelian-surgeries}. However, let us note that for the questions like anomaly cancellations for the theory of free fermions it is sufficient to consider the map to the anomaly group up to automorphisms (since zero element is invariant).

\begin{table}[h!]
\begin{center}
\begin{tabular}{c|c|c}
$(p;r)$ & $RO(\Z_{p^r})\longrightarrow A_{p;r}$ & $A_{p;r}$ \\
\hline
\hline
$(p>2;r)$ & $\crep{q}\longmapsto q^2$ & $\Z_{p^r}$ \\
\hline 
$(2;r>1)$ & 
$\begin{array}{rcl}
    \crep{q} & \longmapsto & (0,q^2)  \\
    {[2^{r-1}]} & \longmapsto & (1,2^{2r-3}) \\
\end{array}$ 
& $\Z_2\times\Z_{2^{r+1}}$ \\
\hline 
$(2;1)$ & 
$[1] \longmapsto  1$
& $\Z_8$ \\
\end{tabular}
\end{center}
\caption{The values of the map $RO(\Z_{p^r})\rightarrow A_{p;r}$ in terms of the generators. The trivial representation $[0]$ is always mapped to zero, so it is not written.}
\label{table:rep-map-A}
\end{table}

\begin{table}[h!]
\begin{center}
\begin{tabular}{c|c|c}
$(p;r_1,r_2)$, $r_1\leq r_2$ & $RO(\Z_{p^{r_1}}\times \Z_{p^{r_2}})\longrightarrow B_{p;r_1,r_2}$ & $B_{p;r_1,r_2}$ \\
\hline
\hline
$(p>2;r_1,r_2)$  & $\crep{q_1,q_2}\longmapsto q_1q_2$ & $\Z_{p^{r_1}}$ \\
\hline
$(2;1,1)$  & $\begin{array}{rcl}
    {[1,1]} & \longmapsto & 1 \\
\end{array}$  & $\Z_{4}$ \\
\hline
$(2;r_1,r_2>1)$  & $\begin{array}{rcl}
    \crep{q_1,q_2} & \longmapsto & (0,q_1q_2) \\
    {[2^{r_1-1},2^{r_2-1}]} & \longmapsto & (1,2^{r_1+r_2-3}) \\\\
\end{array}$ & $\Z_2\times\Z_{2^{r_1}}$ \\
\end{tabular}
\end{center}
\caption{The values of the map $RO(\Z_{p^{r_1}}\times \Z_{p^{r_2}})\rightarrow B_{p;r_1,r_2}$ in terms of the generators. The representations that contain a zero charge are always mapped to zero in $B_{p;r_1,r_2}$, so they are not written.}
\label{table:rep-map-B}
\end{table}

\begin{table}[h!]
\begin{center}
\begin{tabular}{c|c|c}
$(p;r_1,r_2,r_3)$, $r_1\leq r_2\leq r_3$ & $RO(\Z_{p^{r_1}}\times \Z_{p^{r_2}}\times \Z_{p^{r_3}})\longrightarrow C_{p;r_1,r_2,r_3}$ & $C_{p;r_1,r_2,r_3}$ \\
\hline
\hline
$(p>2,r_1,r_2,r_3)$  & $\crep{q_1,q_2,q_3}\longmapsto 0$ & $\Z_{p^{r_1}}$ \\
\hline
$(p=2,r_1,r_2,r_3)$  & $\begin{array}{rcl}
    \crep{q_1,q_2,q_3} & \longmapsto & 0 \\
    {[2^{r_1-1},2^{r_2-1},2^{r_3-1}]} & \longmapsto & 2^{r_1-1} \\\\
\end{array}$ & $\Z_{2^{r_1}}$ \\
\end{tabular}
\end{center}
\caption{The values of the map $RO(\Z_{p^{r_1}}\times \Z_{p^{r_2}}\times \Z_{p^{r_3}})\rightarrow C_{p;r_1,r_2,r_3}$ in terms of the generators. The representations that contain a zero charge are always mapped to zero in $C_{p;r_1,r_2,r_3}$, so they are not written.}
\label{table:rep-map-C}
\end{table}

 As an illustration of the use of the properties listed above to fixing the maps, we explain in detail a couple of cases. 
 
 As a first example, consider the second line in Table \ref{table:rep-map-A} for $r=2$. The symmetry group is $\Z_{4}$ and the anomaly group is $\Z_{2}\times \Z_8$. The image of the $U(1)$ anomaly group is the $\Z_8$ subgroup. The projection on $\Z_2$ factor can be identified with the projection on $H^1(B\Z_4,\Z_2)\cong \Z_2$ in the supercohomology description. Consider the representation $\crep{q}$. It can be lifted to a representation of $U(1)$ and has $w_1(\crep{q})=0\in H^1(B\Z_4,\Z_2)$. Therefore its image is $(0,q^2)$. Next consider the real 1-dimensional representation $[2]$. Its image is necessarily of the form $(1,a)$ where $2a=4\mod 8$. This is because $w_1([2])=1\in H^1(B\Z_4,\Z_2)$ and $[2]\oplus[2]=\crep{2}$. There are two choices satisfying this condition: $a=2\mod 8$ or $a=-2\mod 8$. However they are related by an automorphism $\Z_2\times \Z_8\rightarrow \Z_2\times \Z_8$ where $(c,d)\mapsto (c,d+4c)$.
 
 As a second example, consider the third line in Table \ref{table:rep-map-B} for $r_2=2$ and $r_1=1$. The symmetry group is $G=\Z_2\times \Z_4$ and the anomaly group is $A_{2;1}\times A_{2;2}\times B_{2;1,2}$, where $A_{2;1}=\Z_8$, $A_{2;2}=\Z_4\times\Z_2$ are the anomaly group of the individual factors $\Z_2$ and $\Z_4$ in $G$, and $B_{2;1,2}=\Z_2^{(1)}\times \Z_2^{(2)}$ are the group of ``mixed'' anomalies. We introduced the extra subscripts $(1)$ and $(2)$ to distinguish the subgroups in the discussion below. In particular we take $\Z_2^{(2)}$ to be the image of the pullback of the mixed anomaly group for $U(1)\times U(1)$ under the inclusion $\Z_2\times \Z_4\subset U(1)\times U(1)$. This in particular implies that $\sch(\crep{q_1,q_2})=(0,q_1q_2)\in B_{2;1,2}$, which belongs to the subgroup of bosonic anomalies. Next, note that $H^1(BG,\Z_2)\cong \Z_2^2$ is generated by $w_1([1,0])$ and $w_1([0,2])$, while $SH^2(BG,\Z_2)\cong H^2(BG,\Z_2)\cong \Z_2^3$ is generated by $w_2(\crep{1,0})=w_1([1,0])^2$, $w_2(\crep{0,1})$ and $w_1([1,0])w_1([0,2])$. Note that $\sch[1,0]$, $\sch[0,2]$ and $\sch\crep{0,1}$ generate the subgroup $A_{2;1}\times A_{2;2}$ under addition, which corresponds to taking direct sum of representations. It is left to determine the value of $\sch[1,2]$. We have $w_1([1,2])=w_1([1,0])+w_1([0,2])$ and $w_2([1,2])=0$. We can see that these values of $(w_1,w_2)$ are non-trivial and never realized by any direct sum of representations $[1,0]$, $[0,2]$ and $\crep{0,1}$. This implies that $\sch[1,2]$ is a non-trivial element not belonging to the $A_{2;1}\times A_{2;2}$ subgroup. We also know that it is an order $2$ element, since $2\sch[1,2]=\sch\crep{1,2}=(0,2)\equiv(0,0)\in B_{2;1}$. It also does not lie in $\Z_2^{(2)}$ subgroup since $[1,2]$ cannot be lifted to a complex representation. Therefore, up an automorphism of $A_{2;1}\times A_{2;2}\times B_{2;2,1}$ we can assume that $\sch[1,2]=(1,1)\in B_{2;1}$.
 
 The arguments for the other cases in Tables \ref{table:rep-map-A} and \ref{table:rep-map-B} are similar.
  The last map in (\ref{basic-rep-maps}) will be determined in Section \ref{sec:abelian-surgeries} using different arguments. The result is presented in Table \ref{table:rep-map-C}.

\subsubsection{Non-abelian example: $G=S_3$}
\label{sec:ch2-S3-example}

In this section we will show how one can use the axiomatic properties of the $\sch$ map to determine its values on the example of the smallest non-abealian group $G=S_3$, the group of permutation of 3 elements, or, equivalently the dihedral group of order 6. It can also be realized as a semiproduct $G=\Z_2\ltimes \Z_3$, where the generator of $\Z_2$ acts on the elements of $\Z_3$ by inversion. It has three real irreducible representations, which we will denote as $\rho_{0,1,2}$. First, $\rho_0$ is the 1-dimensional \textit{trivial} representation. Second, $\rho_1$ is the 1-dimensional \text{sign} representation, such that $\rho_2(a,b)=(-1)^a$, where $(a,b)\in \Z_2\ltimes \Z_3$. Third one, $\rho_2$ is the \textit{standard} 2-dimensional representation realized by considering $S_3$ as the subgroup of $O(2)$ preserving an equilateral triangle. 

Consider the following three homomorphisms to and from $S_3$:
\begin{equation}
\begin{tikzcd}
    \Z_2\ar[r,"i"]\ar[rr, bend left =30,dashed,"\text{id}"]
    & S_3=\Z_2\ltimes \Z_3 \ar[r,"\pi"] & 
    \Z_2 \\
    \Z_3\ar[ru,"j"]
\end{tikzcd}
\end{equation}
The maps $i$ and $j$ are the standard inclusions in the semi-direct product and the map $\pi$ is the projection on the quotient over $\Z_3$ normal subgroup. Note that we have $\pi\circ i=\text{id}$. The maps have the following pullbacks between the corresponding anomaly groups:
\begin{equation}
\begin{tikzcd}
    \Z_8
    & \Hom(\Omega_3^\Spin(BS_3),U(1)) \ar[l,"i^*",swap]\ar[ld,"j^*",swap] & 
    \Z_8 \ar[ll, bend right =30,dashed,"\text{id}",swap]\ar[l,"\pi^*",swap] \\
    \Z_3
\end{tikzcd}
\label{S3-pullbacks-diagram}
\end{equation}
The cohomology groups relevant for supercohomology description of $\Hom(\Omega_3^\Spin(BS_3),U(1))$ are $H^3(BS_3,U(1))\cong H^4(BS_3,\Z)\cong \Z_2\times \Z_3$, $H^2(BS_3,\Z_2)\cong \Z_2$, $H^1(BS_3,\Z_2)\cong \Z_2$. Therefore, without determining specifics of the supercohomology structure, we know that the anomaly group must be of order $2^3\cdot 3$ or $2^2\cdot 3$. The commutativity of the diagram (\ref{S3-pullbacks-diagram}) immediately implies that $\pi^*$ is injective, therefore we necessarily have 
\begin{equation}
    \Hom(\Omega_3^\Spin(BS_3),U(1))\cong \Z_8\times \Z_3
\end{equation}
The corresponding pullbacks of the representations are the following:
\begin{equation}
\begin{array}{rcl}
RO(\Z_2)& \stackrel{i^*}{\longleftarrow} & RO(S_3), \\
{[0]} & \longmapsfrom & \rho_0, \\
{[1]} & \longmapsfrom & \rho_1, \\
{[0]\oplus [1]} & \longmapsfrom & \rho_2, \\
\end{array}
\end{equation}
\begin{equation}
\begin{array}{rcl}
RO(\Z_3)& \stackrel{j^*}{\longleftarrow} & RO(S_3), \\
{[0]} & \longmapsfrom & \rho_0, \\
{[0]} & \longmapsfrom & \rho_1, \\
\crep{1} & \longmapsfrom & \rho_2, \\
\end{array}
\end{equation}
\begin{equation}
\begin{array}{rcl}
RO(S_3)& \stackrel{\pi^*}{\longleftarrow} & RO(\Z_2), \\
\rho_0 & \longmapsfrom & {[0]}, \\
\rho_1 & \longmapsfrom & {[1]}. \\
\end{array}
\end{equation}
From this and the Tables \ref{table:rep-map-A} and \ref{table:rep-map-B}   it is easy to conclude that:
\begin{equation}
\begin{array}{rcl}
RO(S_3)& \stackrel{\sch}{\longrightarrow} & \Hom(\Omega_3^\Spin(BS_3),U(1))\cong \Z_8\times \Z_3, \\
\rho_0 & \longmapsto & (0,0), \\
\rho_1 & \longmapsto & (1,0), \\
\rho_2 & \longmapsto & (1,1). \\
\end{array}
\end{equation}

\subsection{The map from the anomaly group to the group of anomalous phases}
Consider now again an arbitrary finite group $G$ ask the question of evaluation of the second map in the sequence (\ref{three-term-seq-torus}). 
\subsubsection{Reducing to abelian groups}
Suppose first we are interested to determine only the anomalous phases for modular transformations, but not the large gauge transformations. This means we consider only the following subgroupoid:
\begin{equation}
    \Mp\ltimes \Spin_{G}(\TT^2) 
    \qquad \hookrightarrow \qquad
     (\Mp\times G)\ltimes \Spin_{G}(\TT^2)
\end{equation}
This is the action of the subgroup $\Mp$ acting on the same set $\Spin_G(\TT^2)$. This inclusion induces the forgetful map on the groups classifying 1-dimensional  representations of the groupoids:
\begin{equation}
    H^1( (\Mp\times G)\ltimes \Spin_{G}(\TT^2),U(1))\;
    \longrightarrow \;
    H^1( \Mp\ltimes \Spin_{G}(\TT^2),U(1)).
\end{equation}
As was already mentioned before, to describe a representation of the groupoid one can consider its connected components independently. A connected component of $\Mp\ltimes \Spin_{G}(\TT^2)$ contains elements of $\Spin_G(\TT^2)=\Spin(\TT^2)\times \Hom(\Z^2,G)$ of the form $(s,(g_1^ag_2^b,g_1^cg_2^d))$ where $s\in \Spin(\TT^2)$, $ab-cd=1$, and $g_1,g_2$ are a pair of fixed commuting elements of $G$. Let $G'\subset G$ be the abelian subgroup of $G'$ generated by $g_1$ and $g_2$. It follows that $G'$ is an abelian group with at most two \textit{independent} generators. The considered connected component then can be identified with a connected component of the groupoid $\Mp\ltimes \Spin_{G'}(\TT^2)$. The restriction to this connected component of the representation corresponding to a given anomaly of the symmetry $G$ then can be determined by considering the corresponding anomaly for the subgroup $G'$ and determining the representation in this case. Namely, one can use the commutativity of the following diagram:
\begin{equation}
\begin{tikzcd}
    \Hom(\Omega_3^\Spin(BG),U(1))\ar[r]\ar[d] & \Hom(\Omega_3^\Spin(BG'),U(1)) \ar[d]\\
    H^1(\Mp\ltimes \Spin_G(\TT^2),U(1))\ar[r] &
    H^1(\Mp\ltimes \Spin_{G'}(\TT^2),U(1))
\end{tikzcd}
\end{equation}
combined with the fact the bottom horizontal map becomes an isomorphism when restricted on the considered connected components, so this restriction can be inverted. Therefore the problem of finding anomalous modular transformations can be reduced to the case of abelian symmetry group. Note that however one in general will need to consider different possibly non-isomorphic subgroups $G'$ for a given $G$.

To determine in addition the action of $G$ on the elements of $\Spin_G(\TT^2)$ it is enough to evaluate the cobordism invariants on a 3-torus $\TT^3$ in a non-trivial background $G$ gauge field. Since the space $\Hom(\pi_1(\TT^3),G)\cong \Hom(\Z^3,G)$ consists of commuting triples of elements of $G$, for this purpose it is enough to consider the abelian subgroups with at most three independent generators.

\subsubsection{Abelian SPTs on closed 3-manifolds via surgery}
\label{sec:abelian-surgeries}
Having reduced the (simplified version of) problem of determining the anomalous phases to the case of abelian groups with at most two independent generators, in this section we describe the method of computing the corresponding spin-bordism invariants on mapping tori for this class of groups. In fact, we will provide a method of calculating the invariants on arbitrary closed 3-manifold $Y$  using its surgery representation. The mapping tori can be realized by surgeries of particular type, already described in Section \ref{sec:Z2-surgery}. 

First we assign particular spin-bordism invariants to each generator of the groups $A_{p;r}$ and $B_{p;r_1,r_2}$ in the factorization (\ref{anomaly-group-ABC-decomposition}) of the anomaly group for an abelian $G$. The assignment is presented in Tables \ref{table:invariants-A} and \ref{table:invariants-B}. In principle we do not need to consider $C_{p;r_1,r_2,r_3}$ for the purpose of determining modular transformation, since, as explained in the previous section we can assume the symmetry group has at most two generators. We however still consider it for completeness and for the purpose of fixing the maps from the representation ring in Table \ref{table:rep-map-C}. In this case the corresponding invariants are purely bosonic (i.e. do not depend on spin structure and can be described in terms of Dijkgraaf-Witten action for certain subgroup of $H^3(BG,U(1))$) and well known. They are summarized in Table \ref{table:invariants-C}.

\begin{table}[h!]
\begin{center}
\begin{tabular}{c|c|c}
$(p;r)$ & bordism invariants & $A_{p;r}$ \\
\hline
\hline
$(p>2;r)$ &  $\begin{array}{rcl}
    1 & : & \lk(a,a) \\
\end{array}$ 
 & $\Z_{p^r}$ \\
\hline 
$(2;r>1)$ & 
$\begin{array}{rcl}
    (0,1) & : & \gamma_s(a) \\
    (1,0) & : & \omega_s(a) \\
\end{array}$ 
& $\Z_2\times\Z_{2^{r+1}}$ \\
\hline 
$(2;1)$ & 
$\begin{array}{rcl}
    1 & : & \beta_s(a) \\
\end{array}$ 
& $\Z_8$ \\
\end{tabular}
\end{center}
\caption{The correspondence between the generators of $A_{p;r}$ and particular bordism invariants, considered on a closed manifold $Y$ with spin structure $s\in \Spin(Y)$, and depending on the background gauge field $a\in H^1(Y,\Z_{p^r})$.}
\label{table:invariants-A}
\end{table}

\begin{table}[h!]
\begin{center}
\begin{tabular}{c|c|c}
$(p;r_1,r_2)$, $r_1\leq r_2$ & bordism invariants & $B_{p;r_1,r_2}$ \\
\hline
\hline
$(p>2;r_1,r_2)$  & $\begin{array}{rcl}
    1 & : & \lk(a,b\mod p^{r_1}) \\
\end{array}$ 
 & $\Z_{p^{r_1}}$ \\
\hline
$(2;1,1)$  & $\begin{array}{rcl}
    1 & : & \delta_s(a,b) \\
\end{array}$ 
  & $\Z_{4}$ \\
\hline
$(2;r_1,r_2>1)$  & $\begin{array}{rcl}
    (1,0) & : & \varepsilon_s(a,b) \\
    (0,1) & : & \lk(a,b\mod 2^{r_1}) \\
\end{array}$ & $\Z_2\times \Z_{2^{r_1}}$
 \\
\end{tabular}
\end{center}
\caption{The correspondence between the generators of $B_{p;r_1,r_2}$ and particular bordism invariants, considered on a closed manifold $Y$ with spin structure $s\in \Spin(Y)$, and depending on the background gauge fields $a\in H^1(Y,\Z_{p^{r_1}})$ and $b\in H^1(Y,\Z_{p^{r_2}})$.}
\label{table:invariants-B}
\end{table}

\begin{table}[h!]
\begin{center}
\begin{tabular}{c|c|c}
$(p;r_1,r_2,r_3)$, $r_1\leq r_2\leq r_3$ & bordism invariants & $C_{p;r_1,r_2,r_3}$ \\
\hline
\hline
$(p;r_1,r_2,r_3)$  & $1\;:\;\int_{Y}a\cup (b\mod p^{r_1})\cup (c\mod p^{r_1})$ & $\Z_{p^{r_1}}$
 \\
\end{tabular}
\end{center}
\caption{The correspondence between the generators of $C_{p;r_1,r_2,r_3}$ and particular bordism invariants, considered on a closed manifold $Y$ with spin structure $s\in \Spin(Y)$, and depending on the background gauge fields $a\in H^1(Y,\Z_{p^{r_1}})$, $b\in H^1(Y,\Z_{p^{r_2}})$, and $c\in H^1(Y,\Z_{p^{r_3}})$.}
\label{table:invariants-C}
\end{table}

As we will see shortly, the are relations between different invariants in the Tables \ref{table:invariants-A} and \ref{table:invariants-B}. In particular, all invariants can expressed using only $\lk$, $\gamma_s$ and $\beta_s$. We will also insure that the assignment of the invariants to the particular generators is consistent with the relations and the maps from the representations rings considered in Section \ref{sec:abelian-rep-map}.

The invariant $\beta_s$, has been already described and used earlier in detail. In particular a formula for it in terms of surgery realization of $Y$ was given in (\ref{beta-surgery-formula}).

The invariants
\begin{equation}
    \lk: H^1(Y,\Z_n)\otimes H^1(Y,\Z_n) \;
    \longrightarrow \Z_n
\end{equation}
and its quadratic spin-refinement
\begin{equation}
    \gamma_s: H^1(Y,\Z_n) \;
    \longrightarrow \Z_{2n}
\end{equation}
can be expressed using the standard linking paring and its spin-refinement \cite{kirby_taylor_1991} which are defined on $\mathrm{Tor}\,H_1(Y,\Z)$ and valued in $\Q/\Z$. The relation is obtained using the isomorphisms $H^1(Y,\Z_n)\cong \Hom(H_1(Y,\Z),\Z_n)\subset \Hom(H_1(Y,\Z),\Q/\Z)$. These invariants can also be understood as the restriction of the abelian Chern-Simons action and spin-version under the inclusion $\Z_n \subset U(1)$ \cite{Guo:2018vij}. Note that we also have $\lk(a,b)=\int_{Y} a\cup \mathcal{B}b$ where $\mathcal{B}:H^1(Y,\Z_p)\rightarrow H^2(Y,\Z_p)$ is the Bockstein map. The refinement $\lk$ from $\gamma_s$ by using the relation
\begin{equation}
\gamma_s(a+b)=\gamma_s(a)+\gamma_s(b)+2\lk(a,b).
\label{gamma-lk-relation}
\end{equation}

In terms of the surgery representation (see Section \ref{sec:Z2-surgery} for conventions) of the spin 3-manifold $M$ we have the following explicit formulas:
\begin{equation}
    \lk(a,b)=\frac{a^TBb}{n}\mod n,
    \label{lk-surgery-formula}
\end{equation}
\begin{equation}
    \gamma_s(a)=\frac{a^TBa}{n}+s^TBa \mod 2n,
    \label{gamma-surgery-formula}
\end{equation}
Here, as before, an element $a\in H^1(Y,\Z_n)$ is represented by a mod $n$ vector $a\in \Z_n^V$ satisfying $Ba=0\mod n$. Note that we have
\begin{equation}
    \beta_s(a)=\gamma_s(a)\mod 4
\end{equation}
for $a\in H^1(Y,\Z_2)$.

To obtain our first relations between the invariants consider the following sequence of maps between the symmetry groups for $r>1$:
\begin{equation}
    \begin{array}{ccccc}
        \Z_2 & \stackrel{2^{r-1}\cdot}\longrightarrow & 
        \Z_{2^r} & \stackrel{\mod 2}\longrightarrow &
        \Z_2
    \end{array}
\end{equation}
The pullbacks on the representation rings read:
\begin{equation}
    \begin{array}{ccccc}
        RO(\Z_2) & \longleftarrow & 
        RO(\Z_{2^r}) & \longleftarrow &
        RO(\Z_2) \\
        {[0]} & \longmapsfrom & {[2^{r-1}]} & \longmapsfrom & {[1]} \\
        2{[q\mod 2]} & \longmapsfrom & \crep{q} &  &  \\
    \end{array}
\end{equation}
Using the Table \ref{table:invariants-A} we then obtain the following maps between the corresponding anomaly groups:
\begin{equation}
    \begin{array}{ccccc}
        \Z_8 & \longleftarrow & 
        \Z_2\times \Z_{2^{r+1}} & \longleftarrow &
        \Z_8 \\
         &  & (1,2^{2r-3}) & \longmapsfrom & 1 \\
        2 & \longmapsfrom & (0,1) &  &  \\
        -2^{2r-2} & \longmapsfrom & (1,0) & &
    \end{array}
\end{equation}
From this we obtain the following relations involving the invariant
\begin{equation}
    \omega_s:\;H^1(Y,\Z_{2^r})\longrightarrow \Z_2
\end{equation}
and the invariants $\beta_s$, $\gamma_s$:
\begin{equation}
    \beta_s(a) = \frac{1}{2^{r-1}}\cdot\gamma_s(2^{r-1}\,a) \mod 4,\qquad a\in H^1(Y,\Z_{2}),
    \label{beta-gamma-relation}
\end{equation}
and
\begin{equation}
    \beta_s(a\mod 2) = 4\omega_s(a)+2^{r-1}\gamma_s(a)\mod 8,\qquad a\in H^1(Y,\Z_{2^r}).
    \label{omega-beta-gamma-relation}
\end{equation}
The relation (\ref{beta-gamma-relation}) is indeed consistent with the surgery formula (\ref{beta-surgery-formula}) and (\ref{gamma-surgery-formula}) for $\beta_s$ and $\gamma_s$ respectively. The relation (\ref{beta-gamma-relation}) can be used to actually \textit{define} $\omega_s$ via $\beta_s$ and $\gamma_s$:
\begin{equation}
    \omega_s(a)=\frac{\beta_s(a\mod 2)-2^{r-1}\gamma_s(a)}{4}\mod 2,\qquad a\in H^1(Y,\Z_{2^r}).
\end{equation}
Similarly, considering the pullback of the sum map
\begin{equation}
    \Z_2\times \Z_2\; \stackrel{+}\longrightarrow \;
    \Z_2
\end{equation}
between the symmetry groups we obtain another relation:
\begin{equation}
    \beta_s(a+b)=\beta_s(a)+\beta_s(b)+2\delta_s(a,b),\qquad a,b\in H^1(Y,\Z_2).
\end{equation}
It can be used to express $\delta_s$ via $\beta_s$:
\begin{equation}
    \delta_s(a,b)=\frac{\beta_s(a+b)-\beta_s(a)-\beta_s(b)}{2},\qquad a,b\in H^1(Y,\Z_2).
\end{equation}
One can conclude that this $\delta_s$ is the same as $\delta_s$ invariant already reviewed in Section \ref{sec:Z2case} from a more geometric point of view. This is in agreement with the description of the spin-bordism invariants in \cite{Guo:2018vij}.

More generally, one can consider the pullback of the map
\begin{equation}
    \Z_{2^{r_1}}\times \Z_{2^{r_2}}\; \stackrel{(\,\cdot\,)+(\,\cdot\mod 2^{r_1})}\longrightarrow \;
    \Z_{2^{r_1}}
\end{equation}
for $1\leq r_1\leq r_2>1$ to obtain the expressions for the invariant
\begin{equation}
    \varepsilon_s:\;H^1(Y,\Z_{2^{r_1}})\times 
    H^1(Y,\Z_{2^{r_2}})\;\longrightarrow \Z_2
\end{equation}
via already known invariants. When  $1<r_1\leq r_2$ we have
\begin{equation}
    \varepsilon_s(a,b)=\omega_s(a+(b\mod 2^{r_1}))
    +\omega_s(a)+\omega_s(b)\mod 2,
\end{equation}
And for $1=r_1<r_2=r$:
\begin{multline}
    \varepsilon_s(a,b)= \omega_s(b)
    +\\
    \frac{\beta_s(a+(b\mod 2))-\beta_s(a)-2^{r-1}\gamma_s(b)-2^{r}\lk(a,b\mod 2)}{4}
    \mod 2.
\end{multline}
This provides the expression for all the spin-bordism invariants via the invariants $\beta_s$, $\gamma_s$ and $\lk$ (the latter can be also expressed via $\gamma_s$), for which the explicit expressions in terms of surgery representation are given in (\ref{beta-surgery-formula}), (\ref{gamma-surgery-formula}) and (\ref{lk-surgery-formula}).

We have found relations between different invariants corresponding to the subgroups $A_{p;r}$, $B_{p;r_1,r_2}$ using the homomorphism between different symmetry groups and the maps from the representation rings listed in Tables \ref{table:rep-map-A} and \ref{table:rep-map-B}. For the subgroups $C_{p;r_1,r_2,r_3}$ we already know the invariants (listed in Table \ref{table:rep-map-C}) so we can reverse the logic and use the homomorphisms between different symmetry groups to determine the map from the representation ring. Namely, one can consider the homomorphisms of the form
\begin{equation}
        \Z_{2^{r_1}}\times \Z_{2^{r_2}}\times \Z_{2^{r_3}}\; \stackrel{q_1(\,\cdot\,)+q_2(\,\cdot\mod 2^{r_1})+q_3(\,\cdot\mod 2^{r_1})}\longrightarrow \;
    \Z_{2^{r_1}}
\end{equation}
for $r_1\leq r_2\leq r_3$ and some integer $q_i$, and use the relation (\ref{gamma-lk-relation}) together with the known relation
\begin{multline}
    \beta_s(a+b+c)=\beta_s(a)+\beta_s(b)+\beta_s(c)
    +2\delta_s(a,b)+2\delta_s(b,c)+2\delta_s(a,c)\\
    +4\int_Y{a\cup b\cup c}
\end{multline}
for $a,b,c\in H^1(Y,\Z_2)$ to arrive at the result listed in Table \ref{table:rep-map-C}.

Finally let us note that the invariants $\omega_s(a)$ and $\varepsilon_s(a,b)$ can be given rather simple a geometrical interpretations (similar to the ones of $\beta_s(a)$ and $\delta_s(a,b)$) when $a,b$ can be lifted to classes in integral cohomology $H^1(Y,\Z)$ \cite{Guo:2018vij}.

\section{Modular bootstrap}
\label{sec:bootstrap}

The results of the previous sections describe how a generic theory with fermions transform under modular transformations, showing that it is possible to fully capture such behaviour by simply knowing the value of its anomaly.

Since 't Hooft anomalies are robust under Renormalization Group (RG) flow, one can make use of this in order to constrain the spectrum of the infrared CFTs.

To reach this goal, we apply techniques of the conformal bootstrap, which allow us to see how the presence of a $\Z_2$ global symmetry affects the internal consistency of a fermionic CFT. This work extends the original results of \cite{Collier:2016cls,Benjamin:2020zbs}, where bounds on the spectrum of CFTs are found without any hypothesis on their characterizing global symmetry, and \cite{Lin:2019kpn,Lin:2021udi}, which determine instead how a $\Z_N$ global symmetry constrains the spectrum of a bosonic CFT.

Our case aims then to complete such analysis by looking at how the presence of a global symmetry for a CFT is affected by a dependence of the theory on the spin structure. In other terms, this also offers a first example of how not only the group cohomological classification of anomalies, which usually have a more transparent action on the Hilbert space, is able to put significant consistency conditions and constraints on the allowed spectra of theories, but also the full cobordism one.

In the following we start by presenting the general setup needed in order to apply techniques of modular bootstrap and then proceed to discuss the bounds implied by modular crossing equations.

\subsection{General setup}

For our working case we will consider unitary theories with no gravitational anomalies, which means we are going to assume central charges $c=c_R=c_L\ge 1$. 
This is reflected on the possible terms which compose the partition functions to work with. Indeed, the (anti)holomorphic sector is then characterized by one degenerate Virasoro representation with $h=0$, i.e. the vacuum, together with a continuous family of non-degenerate representations with $h>1$. We recall here that their Virasoro characters are given by
\begin{equation}
    \chi_0(\tau)= (1-q)\frac{q^{-\frac{c-1}{24}}}{\eta(\tau)}, \qquad  \chi_{h>1}(\tau)= \frac{q^{h-\frac{c-1}{24}}}{\eta(\tau)},
\end{equation}
where, as usual, $q=\exp(2\pi i \tau)$ and $\eta(\tau)=q^{1/24}\prod_{n\geq 1}(1-q^n)$. These will be the building blocks of the partition functions: in the case of no  space-like defects, they are simply defined as a positive sum over combinations of Virasoro characters, i.e.
\begin{align}
    Z^{\ns0}_{s_0 g_0}(\tau,\bar{\tau})&= \sum_{(h,\overline{h}) \in \mathcal{H}_{s_0g_0}} n_{h,\overline{h}}^{s_0g_0} \chi_h(\tau) \bar{\chi}_{\bar{h}} (\bar{\tau}).
\end{align}
Note that this holds for all the possible Hilbert spaces, be them twisted or not. The integers $n_{h,\overline{h}}^{s_0g_0} \in \Z_{\ge 0}$ are the degeneracies of the representations.\\ 
However, partition functions with insertion of (Euclidean) space-like defects will not admit a positive sum decomposition, since the effect of their insertion will generally change the sign (and value) of the degeneracies, depending on the representation under which the Virasoro characters transform. Hence the \emph{twist basis} \eqref{classesoftori} used so far is of no use.

To apply the modular bootstrap technique we need instead a basis of partition functions where the positivity property holds. This is granted by using the natural grading of the Hilbert spaces given by the charges of the operators with respect to $(-1)^F$ and $(-1)^Q$, i.e.
\begin{equation}
    \mathcal{H}_{s_1g_1} =\bigoplus_{F,Q} \mathcal{H}_{s_1 g_1}^{(F,Q)}.
    \label{Hgrading}
\end{equation}
We can thus divide the basis of partition functions into two parts. For the first, i.e. those defined on Hilbert spaces of which we are interested to analyze the various sectors, we consider the respective partition functions with the insertion of the projectors for the proper charges of the operators. For the second instead we can simply consider the full partition functions with no space-like defects insertion. We will call this a \emph{charge basis}. Note that by construction we are reducing the dimensionality of the problem. However, as we will later explain, this poses no problem and the bounds produced for our case of interest will still be as strict as possible.

We will work on finding the bounds for the sectors of $\mathcal{H}_{\ns0}$. Thus if we group the elements of the corresponding charge basis as entries $\tilde{{Z}}_i$ of a vector, we have
\begin{equation}
    \tilde{\mathbf{Z}}^T=(Z_{\ns0}^{+F+Q},Z_{\ns0}^{+F-Q},Z_{\ns0}^{-F+Q},Z_{\ns0}^{-F-Q},Z_{\r0}^{\ns0},Z_{\ns1}^{\ns0},Z_{\r1}^{\ns0}),
    \label{chargebase}
\end{equation}
where we defined
\begin{align}
\begin{aligned}
    Z_{\ns 0}^{+F \pm Q}(\tau,\bar{\tau}) &= \mathrm{Tr}_{\mathcal{H}_{\ns 0}}\left[\left(\frac{1+(-1)^F}{2}\right)\left(\frac{1\pm(-1)^Q}{2}\right)q^{L_0 - c/24}\bar{q}^{\bar{L}_0-\bar{c}/24}\right]\\
    &=\frac{1}{4}\left(Z_{\ns0}^{\ns0}+Z_{\ns0}^{\r0}\pm Z_{\ns0}^{\ns1}\pm Z_{\ns0}^{\r1}\right),\\
\end{aligned}
\label{eq:Z+F+-Q}\\
\begin{aligned}
Z_{\ns 0}^{-F \pm Q}(\tau,\bar{\tau}) &= \mathrm{Tr}_{\mathcal{H}_{\ns0}}\left[\left(\frac{1-(-1)^F}{2}\right)\left(\frac{1\pm(-1)^Q}{2}\right)q^{L_0 - c/24}\bar{q}^{\bar{L}_0-\bar{c}/24}\right]\\
    &=\frac{1}{4}\left(Z_{\ns0}^{\ns0}-Z_{\ns0}^{\r0}\pm Z_{\ns0}^{\ns1}\mp Z_{\ns0}^{\r1}\right).
\end{aligned}
\end{align}
It will be useful in the following to denote with $\mathcal{H}_i$ the Hilbert spaces corresponding to the entries $\tilde{{Z}}_i$.

Note that in principle one can make use of the same split of $\mathcal{H}_{\ns0}$ for the other spaces as well, particularly for $\mathcal{H}_{\r0}$, remembering that for a non-trivial anomaly the projectors in the twisted cases must be appropriately modified.

\subsubsection{Spin selection rule and modular crossing equation}
The most efficient way one can signal the presence of an anomaly for a theory on a torus is by analyzing the spins allowed for the operators in the defect Hilbert spaces.\\
Given $n=2$ the maximum order of elements of $G^f=\Z_2^f\times\Z_2$, then $T^n$ must act diagonally by adding a phase to the partition functions proportional to the spins $s=h-\bar{h}$ of the states over which the trace is taken:
\begin{equation}
    T^n[Z^{\ns0}_{s_1g_1}(\tau,\bar{\tau})]=Z^{\ns0}_{s_1g_1}(\tau+n,\bar{\tau}+n)= \mathrm{Tr}_{\mathcal{H}^{\ns0}_{s_1g_1}}\left[e^{2\pi i n s}q^{L_0 - c/24}\bar{q}^{\bar{L}_0-\bar{c}/24}\right].
\label{Tn-action}
\end{equation}
One can find the same result with the use of crossing relations of symmetry defects \cite{Chang:2018iay}.\\
By confronting \eqref{Tn-action} with $T^2$ from \eqref{ST01}-\eqref{ST00} we can infer the spins allowed in the various Hilbert spaces are\footnote{For $\mathcal{H}_{\r0}$ the action of $T$ is already diagonal.}
\begin{equation}
\begin{aligned}
    \mathcal{H}_{\ns0}:& \quad s \in \Z/2,\\
    \mathcal{H}_{\r0}:&\quad s \in \Z,\\
    \mathcal{H}_{\ns1}:&\quad s \in \nu/16 +\Z/2,\\
    \mathcal{H}_{\r1}:&\quad s \in -\nu/16 +\Z/2.   
\end{aligned}
\end{equation}
In particular $T$ is already diagonal on each sector \eqref{Hgrading} of $\mathcal{H}_{\ns0}$, so, as one might expect\footnote{Although it will not be relevant for later, note that such grading of $\mathcal{H}_{\r 0}$ holds only for $\nu = 0\mod2$.},
\begin{equation}
\begin{aligned}
    \mathcal{H}^{(+,+)}_{\ns0}:& \quad s \in \Z,\\
    \mathcal{H}^{(-,+)}_{\ns0}:&\quad s \in 1/2+\Z,\\
    \mathcal{H}^{(+,-)}_{\ns0}:&\quad s \in \Z,\\
    \mathcal{H}^{(-,-)}_{\ns0}:&\quad s \in 1/2+\Z.   
\end{aligned}
\end{equation}

At this point we have a set of observables $\tilde{{Z}}_i$ that satisfy a positive expansion in terms of Virasoro characters and their spin selection rule is determined, so the next step is to find their modular crossing equation.

The full action of $S$ splits into four pieces, each mixing together partition functions with different topological defects, of which the corresponding tori fall in the same bordism class of $\Omega_2^\Spin(B\Z_2)$. Therefore the complete crossing equation is defined by the $16$-dimensional matrix $S=S_{(0,0)}\oplus S_{(0,1)}\oplus S_{(1,0)} \oplus S_{(1,1)}$, i.e.
\begin{equation}
    \mathbf{Z}(-1/\tau,-1/\bar{\tau})=S\mathbf{Z}(\tau,\bar{\tau}), 
\end{equation}
where $\mathbf{Z}(\tau,\bar{\tau})=\mathbf{Z}_{(0,0)}\oplus \mathbf{Z}_{(0,1)}\oplus \mathbf{Z}_{(1,0)}\oplus \mathbf{Z}_{(1,1)}$. The reduction to the $7$-dimensional basis \eqref{chargebase} is then given by the modular crossing matrix
\begin{equation}
    \tilde{S}=\begin{pmatrix}
    \frac{1}{4}&\frac{1}{4}&\frac{1}{4}&\frac{1}{4}&\frac{1}{4}&\frac{1}{4}&\frac{1}{4}\\
    \frac{1}{4}&\frac{1}{4}&\frac{1}{4}&\frac{1}{4}&-\frac{1}{4}&\frac{1}{4}&-\frac{1}{4}\\       \frac{1}{4}&\frac{1}{4}&\frac{1}{4}&\frac{1}{4}&\frac{1}{4}&-\frac{1}{4}&-\frac{1}{4}\\
    \frac{1}{4}&\frac{1}{4}&\frac{1}{4}&\frac{1}{4}&-\frac{1}{4}&-\frac{1}{4}&\frac{1}{4}\\
    1&-1&1&-1&0&0&0\\
    1&1&-1&-1&0&0&0\\
    1&-1&-1&1&0&0&0\\
    \end{pmatrix}.
    \label{S-chargebase}
\end{equation}
Note that the dependence on the anomaly vanished, so its presence will play a role only on the spin selection rule for each $\tilde{{Z}}_i$.

Since we have not considered the split into the charge sectors of other Hilbert spaces, for example  $\mathcal{H}_{\r0}$, this raises the question whether it is actually possible that the bounds found by using \eqref{S-chargebase} are as strict as possible. It turns out that, considering $R$, the reduction matrix that satisfies $\tilde{\mathbf{Z}}(\tau,\bar{\tau})=R\mathbf{Z}(\tau,\bar{\tau})$, then one can always lift a solution of the modular crossing equation for $\tilde{\mathbf{Z}}$ to the full system via the lift matrix $L=R^t(RR^t)^{-1}$ \cite{Lin:2021udi}.\\

\subsection{The linear functional method}
In order to find bounds on the spectrum, we will apply the linear functional method to the modular crossing equation defined on the basis $\tilde{\mathbf{Z}}$.

From now on it will be useful to remember that we can divide any partition function as a sum of three kinds of combinations of Virasoro characters. These are 
\begin{enumerate}
    \item The vacuuum, the unique character we are going to assume to exist with $h=\bar{h}=0$, living in $\mathcal{H}_{\ns0}$;
    \item The conserved currents, namely combinations of characters with either $h=0$ or $\bar{h}=0$. These we assume to be a priori present in all sectors;
    \item The non-degenerate (ND) primaries, the characters with both $h,\,\bar{h}>0$, present in all sectors as well.
\end{enumerate}
With this premise, we quickly recall the main idea of this technique and leave the details for Appendix \ref{app:bootstrap}.
The modular crossing equation related to \eqref{S-chargebase} is 
\begin{equation}
    \delta_i^j \tilde{Z}_j(-1/\tau,-1/\bar{\tau})- \tilde{S}^j_i \tilde{Z}_j(\tau,\bar{\tau})=0.
    \label{eq:crossingequation}
\end{equation}
After applying a linear functional to it the identity will still hold. If, by taking a putative spectrum of Virasoro characters that may compose each $\tilde{Z}_i$, there exists instead a functional $\alpha$ that returns
\begin{equation}
     \alpha[\delta_i^j \tilde{Z}_j(-1/\tau,-1/\bar{\tau})]-  \alpha[\tilde{S}^j_i \tilde{Z}_j(\tau,\bar{\tau})]> 0, \qquad\forall\,i,
\end{equation}
then such a spectrum is not a consistent one and thus is ruled out.
In practice, instead of choosing some particular spectra, it is much more effective to start from a generic sum of Virasoro characters and then look for bounds from above on the scaling dimension $\Delta=h+\bar{h}$ of the lightest operators in the sectors we decide to analyze. In particular, in our analysis we will focus on two possible bounds:
\begin{enumerate}
    \item The maximal gap $\Delta^j_{\gap}$ in the scaling dimension of the lightest non-degenerate primary in a given $\mathcal{H}_j$.
    \item The maximal gap $\Delta^j_{\scal}$ in the scaling dimension of the lightest scalar primary in a given  $\mathcal{H}_j$. Note that this bound make sense only in the sectors where $s=0$ is an allowed value of the spin.
\end{enumerate}

Since the search for a functional $\alpha$ is numerical, it is convenient to work with the reduced partition functions
\begin{equation}
    \hat{Z}_i(\tau,\bar{\tau}):=|\tau|^{1/2} |\eta(\tau)|^2 \tilde{Z}_i(\tau,\bar{\tau}),
\end{equation}
so that the Virasoro characters one has to deal with are the reduced versions
\begin{equation}
    \hat{\chi}_0(\tau)= \tau^{1/4}(1-q)q^{-\frac{c-1}{24}}, \qquad  \hat{\chi}_h(\tau)= \tau^{1/4}q^{h-\frac{c-1}{24}}.
\end{equation}
Indeed under $S$ the combination $|\tau|^{1/2} |\eta(\tau)|^2$ is invariant and will not change the results.

\subsection{Bounds on $\mathcal{H}_{\ns 0}$ for layers of anomaly}
\label{subsec:results}

We now investigate what are the generic bounds present on the spectrum of $\mathcal{H}_{\ns0}$. As anticipated, in order to make full use of the constraints that the modular crossing equation and the linear functional method offer to us, a numerical study is needed. In order to do so, the analysis has been carried out with the use of the SDPB package \cite{simmonsduffin2015semidefinite,landry2019scaling} for CFTs with value of central charge in the range $1\le c\le10$.

We stress again that the cobordism groups $\Hom(\Omega_3^\Spin(BG),U(1))$ can be described by a set of three group cohomology layers, each giving a different contribution to the anomaly and with a precise physical meaning. In our case this means the value of anomaly can be recast as \begin{equation}
    \nu = 4w+2p+a \:\mathrm{mod}\,8, 
\end{equation}
where $w,\,p$ and $a$ are elements of $H^3(B\Z_2,U(1)),\,SH^2(B\Z_2,\Z_2)$ and $H^1(B\Z_2,\Z_2)$. Here $H^3(B\Z_2,U(1))\cong H^2(B\Z_2,\Z_2)\cong SH^2(B\Z_2,\Z_2)\cong H^1(B\Z_2,\Z_2)\cong \Z_2$.\\ 
Therefore it is interesting to see how spectra of fermionic theories are affected by the presence of each layer. We are going to focus precisely on this, which means we will focus our attention on theories where a single layer of anomaly is present, namely when $\nu=0,1,2,4\mod 8$. Note that since we are looking at quantities that are not sensitive to the swap $s\mapsto -s$, a theory with anomaly $\nu$ and its complex conjugate with anomaly $\bar{\nu}=-\nu \mod 8$ will produce the same bounds $\Delta^j_{\mathrm{gap}}$ and $\Delta^j_{\mathrm{scal}}$. Therefore our results will hold also for $\nu=6,7\mod 8$.

One last comment is in order. Usually for the $\nu=1 \mod 2$ case is argued that no proper Hilbert space $\mathcal{H}_{\ns 0}$ exists, due to the partition functions being proportional to the formal $\mathrm{dim}\,Cl(1) =\sqrt{2}$ \cite{Delmastro:2021xox}. However, one can still assume the case when the partition function of the CFT is defined as $\sqrt{2}$ times the trace over a well-defined Hilbert space. In our analysis for $\nu=1\mod 8$ we will precisely explore the space of fermionic CFTs that satisfy such hypothesis, which is still a valid and sensible thing to do.

\subsubsection{The starting point: $c=1$}
\label{subsec:c1}

The first thing that is instructive to do is to ask for which sectors we actually expect the presence of a bound. In this regard it is illuminating to look at CFTs with $c=1$, for which we have two main interesting examples at our disposal, namely the free compact boson and the Dirac fermion.

For bosonic theories with $\Z_2$ non-anomalous symmetry, it is known that no bound is expected in either the charged sector $\mathcal{H}^{-Q}_0$ of the untwisted Hilbert space and in the twisted Hilbert space, $\mathcal{H}_{1}$. We briefly recall here the main argument of \cite{Lin:2019kpn}. By stacking a free compact boson of radius $R$ with a generic non-anomalous CFT uncharged with respect to one of the two non-anomalous $\Z_2$ subgroups of $U(1)_n$ and $U(1)_w$ (symmetries corresponding to the winding and momentum quantum numbers), one can always produce a non-anomalous CFT with $c>1$ and arbitrarily high lightest non-degenerate state on $\mathcal{H}_0^{-Q}$ (and $\mathcal{H}_{1}$) by taking the radius to be arbitrarily large.
By the same argument, if the uncharged CFT is fermionic, then we see that for $\nu =0\mod8$ there is no bound for both $\mathcal{H}_{\ns 0}^{\pm F -Q}$. 

The next interesting example is given by a free massless Dirac fermion. Consider two Majorana fermions with holomorphic sectors ${\psi}_a,\;a=1,2$,  described by
\begin{equation}
    \mathcal{L} = \sum_{a=1}^2 \psi_a \bar{\partial} \psi_a +  \bar{\psi}_a \partial \bar{\psi}_a 
\end{equation}
and arrange them as the single complex fermion\footnote{Here $\overline{\Psi}$ will denote the antiholomorphic counterpart of $\Psi$ that composes the complex fermion. Instead, we will denoted with $\Psi^*=\psi_1-i\psi_2$ the conjugate field.} $\Psi=\psi_1+i \psi_2$. The theory presents the chiral and non-chiral fermionic parities acting as 
\begin{align}
    (-1)^{F}:&\quad\Psi \rightarrow -\Psi,\quad \overline{\Psi} \rightarrow -\overline{\Psi},\\
    (-1)^{F_L}:&\quad\Psi \rightarrow -\Psi,\quad \overline{\Psi} \rightarrow \overline{\Psi}.
\end{align}
By looking at the spin selection rule of the twisted sectors of this theory, one can find that this presents an anomaly $\nu=2$ (see Appendix \ref{app:bootstrap}).

Unlike before, a priori now there is no natural parameter that can drive the bound for some sectors to infinity. However, one has still a sensible procedure to take into account. Indeed, note that the system has the more generic global symmetry $U(1)\times \overline{U(1)}$, where each component acts only on one of the sectors, and gauging some diagonal subgroup $\Z_p$ with $p$ odd still preserves the global symmetry $G^f=\Z_2^f\times \Z_2$. In fact, there is no mixed anomaly between fermionic parity and $\Z_p$, since $\Hom(\Omega_3^\Spin(B\Z_p),U(1)) \cong H^3(B\Z_p,U(1))$. Moreover, by construction $\Psi$ and $\Psi^*$ have opposite charges, so it follows that $\Z_p$ is anomaly-free and can be gauged.

A generic state on the holomorphic sector for the $\ns$ untwisted Hilbert space of the gauged theory will be of the form
\begin{equation}
    \prod_{n=1}^N \Psi_{-k^+_n} \prod_{m=1}^M \Psi^*_{-k^-_m} \prod_{\bar{n}=1}^{\overline{N}} \overline{\Psi}_{-\overline{k}^+_{\overline{n}}} \prod_{\overline{m}=1}^{\overline{M}} \overline{\Psi}^*_{-\overline{k}^-_{\overline{m}}} |0\rangle,
    \label{eq:generic-state-Zp-gauging}
\end{equation}
where 
\begin{equation}
N-M+\overline{N}-\overline{M}=0\mod p,\qquad k^\pm_i,\overline{k}^\pm_{\overline{i}} = \frac{1}{2}\pm\frac{1}{p} \; \mathrm{mod}\,1.
\label{eq:condition-gauging}
\end{equation}
By requiring for a state to be charged under fermionic parity one needs to refine the first constraint
\begin{equation}
    N-M+\overline{N}-\overline{M}=s p,\qquad s \in 2\Z+1.
\end{equation}
Without loss of generality, we can assume $s>0$. 
Let us focus on the contribution from the set $\Psi_{-k^+_n}$: the states with the lowest possible scaling dimension are defined by towers of operators with increasing dimension
\begin{equation}
    k^+_n= \frac{1}{2}+\frac{1}{p}+n-1,\qquad 1\le n \le N.
\end{equation}
Applying the same logic to the other sets, one find that the lowest non-degenerate states are characterized by 
\begin{align}
    \Delta &= \frac{N^2+M^2+\overline{M}^2+\overline{N}^2}{2}+\frac{N-M+\overline{N}-\overline{M}}{p}\\
    &\ge \frac{N+M+\overline{N}+\overline{M}}{2}+s\\
    &\ge \frac{sp}{2}.
\end{align}
Therefore, by gauging $\Z_p$ with $p$ arbitrarily large, we can build a fermionic CFT with $c=1,\,\nu=2$ and no bound on\footnote{Note that the conditions $N+M=0,1\mod2$, necessary for considering operators specifically in the uncharged and charged sectors with respect to $(-1)^Q$, do not change the argument. Therefore one can apply it to reach the same conclusion for both $\mathcal{H}_{\ns 0}^{-F\pm Q}$.} $\mathcal{H}_{\ns 0}^{-F\pm Q}$. 
By using the same logic already explained in the bosonic case, the stacking of such theory with a bosonic theory produce generic theories with $c>1$ and anomaly $\nu=2\mod4$ with no bounds on the sectors $\mathcal{H}_{\ns 0}^{-F\pm Q}$.

We are now ready to discuss the numerical results and see what are the bounds for the sectors we are left with to analyze.

\subsubsection{Bounds on $\Z_2$ even operators}

\begin{figure}
    \centering
    \includegraphics[scale=0.55]{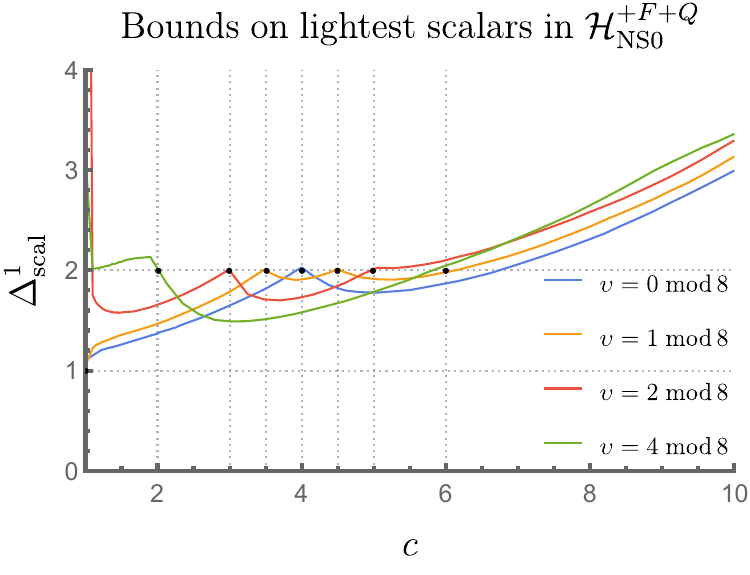}
    \qquad
    \includegraphics[scale=0.55]{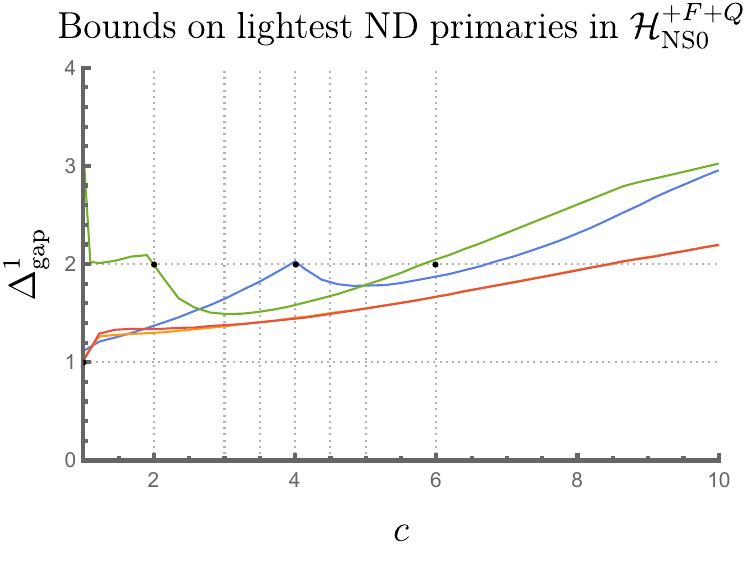}
    \caption{Upper bounds in the sector  $\mathcal{H}_{\ns0}^{+F+Q}$ for values of anomaly $\nu=0,1,2,4\mod8$. Bounds on the lightest non-degenerate primaries are represented by the same colored lines used to denote the bounds for the lightest scalars. The black dots mark the free fermion theories which almost saturate the bounds at the kinks we found.}
    \label{fig:J1}

\bigskip

    \includegraphics[scale=0.55]{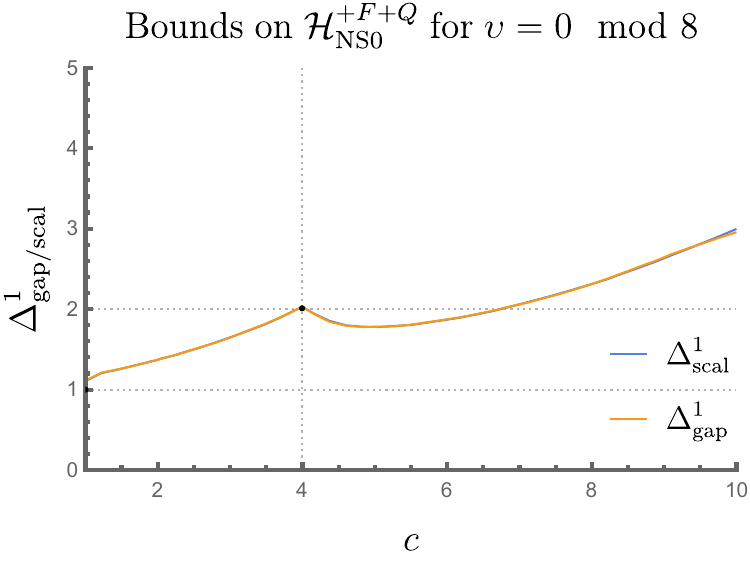}
    \qquad
    \includegraphics[scale=0.55]{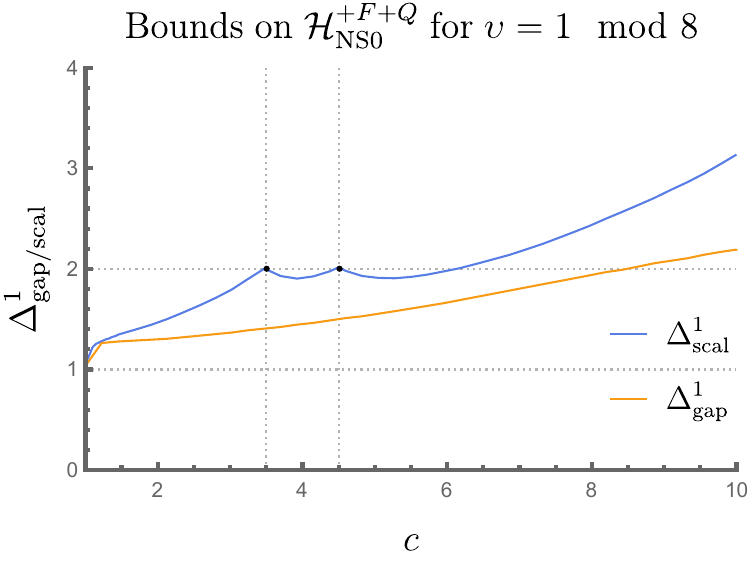}
    \caption{Left: upper bounds for lightest scalars and non-degenerate primaries confronted for theories with $\nu=0\mod8$. Right: upper bounds for lightest scalars and non-degenerate primaries confronted for theories with $\nu=1\mod8$.}
    \label{fig:V01J1}

\bigskip

    \includegraphics[scale=0.55]{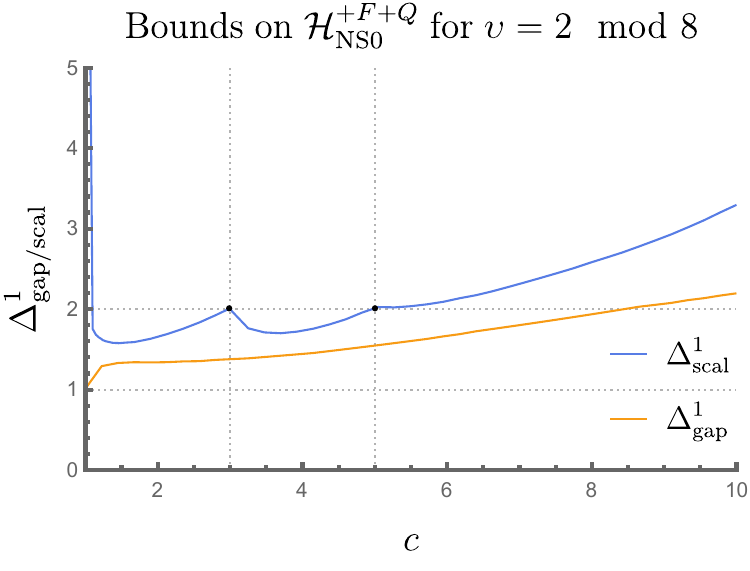}
    \qquad
    \includegraphics[scale=0.55]{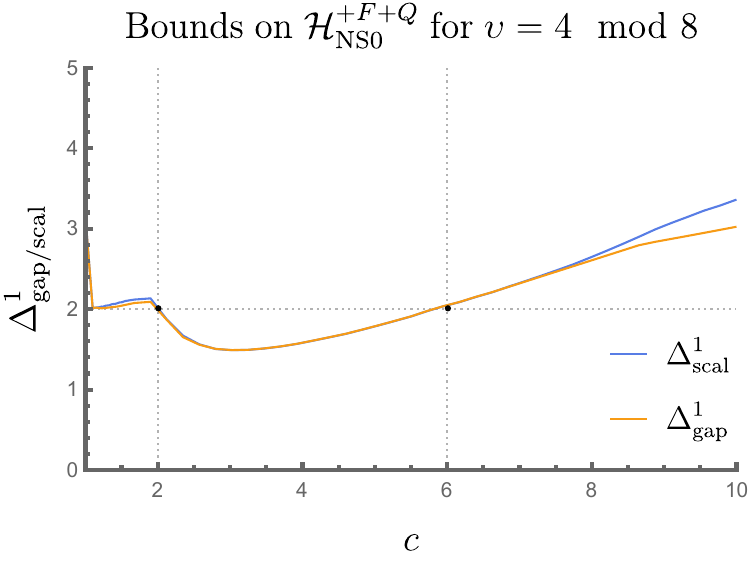}
    \caption{Left: upper bounds for lightest scalars and non-degenerate primaries confronted for theories with $\nu=2\mod8$. Right: upper bounds for lightest scalars and non-degenerate primaries confronted for theories with $\nu=4\mod8$.}
    \label{fig:V24J1}
\end{figure}

The results obtained from the analysis of the various sectors are reported in Figures \ref{fig:J1}-\ref{fig:V24J1}. At certain values of the central charge the bounds are almost saturated by known theories, most of which are different stacking combinations of free Majorana fermions. We marked these points with black dots in the figures. We summarize in the following all our results, while we refer the reader to Appendix \ref{app:bootstrap} for details on the limit cases.

\begin{itemize}
\let\labelitemi\labelitemii
    \item As a consistency check of our numerical method, one can see that all the bounds for fermionic theories with $\Z_2$ symmetries are higher than the bounds of bosonic ones found by \cite{Lin:2019kpn};
    \item Additionally to our previous discussion, our findings suggest that there is no actual bound for the sectors $\mathcal{H}_{\ns 0}^{-F\pm Q}$ for any value of the anomaly. This hints at the possibility of a generalization of the argument of section \ref{subsec:c1} which is not sensible to the anomaly of the theory;
    \item At $c=1$ all (finite) bounds are almost saturated by known theories. In the case $\nu=0,1,2\mod 8$ these are stacks of free fermions, while for $\nu=4\mod 8$ the ND bound has as limiting case the bosonic WZW model $su(2)_1$, which already saturated the bound for bosonic CFTs with $\Z_2$ anomaly \cite{Lin:2019kpn};
    \item For $c>1$ the bounds on scalars on theories with $\nu=0,4\mod8$ are almost as strict as bounds for generic non-degenerate primary states, suggesting that indeed the two coincide. In fact, this is the case for the free fermion theories with $\nu=0\mod 8,\,c=4$ and $\nu=4\mod 8,\,c=2,6$;
    \item  The bounds on non-degenerate primaries coincide for anomalies $\nu=1,2\mod 8$, so that they can not be used as a way to tell the two cases apart;
    \item Most interestingly, relevant operators always allowed in the following intervals of the central charge
    \begin{table}[h]
        \centering
        \begin{tabular}{cc}
            $\nu$ & $c$ interval \\
            \hline
            \hline
            $0\mod 8$ & $1\le c\lesssim 6.63$\\ 
            $1\mod 8$ & $1\le c\lesssim 5.95$\\ 
            $2\mod 8$ & $1.09\lesssim c\lesssim 4.93$\\ 
            $4\mod 8$ & $2.05\lesssim c\lesssim 5.73$\\ 
            \hline
        \end{tabular}
        \label{tab:marginal-operators}
    \end{table}
    
    Such limits can possibly be extended considering functionals with higher order of derivatives.\\
    The limits of marginal operators are instead almost saturated by free fermions with central charge $c=4 \pm \nu/2$.
\end{itemize} 

\subsubsection{Bounds on $\Z_2$ odd operators}
\begin{figure}
    \centering
    \includegraphics[scale=0.55]{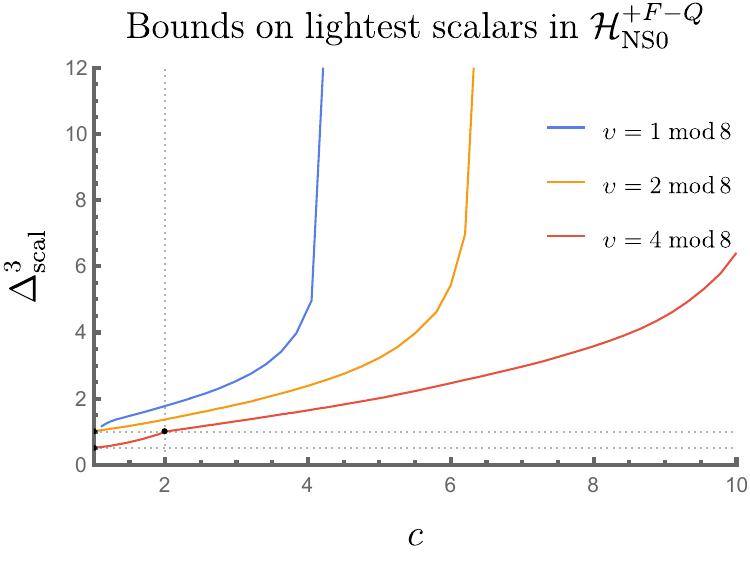}
    \qquad
    \includegraphics[scale=0.55]{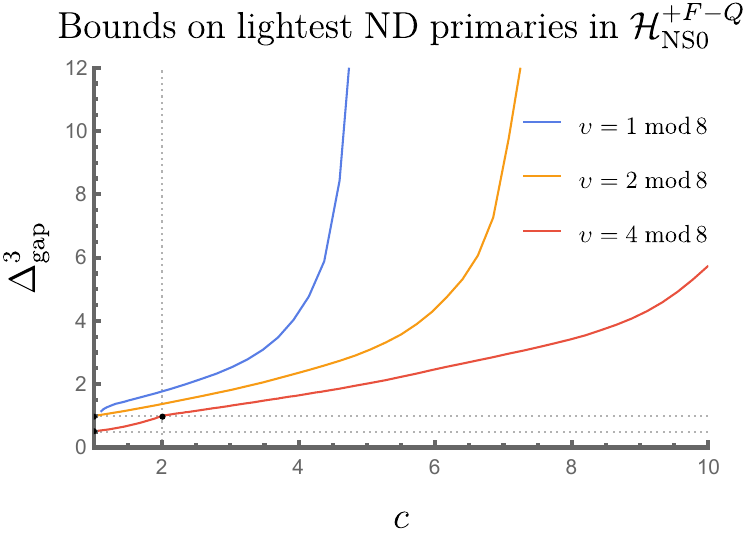}
    \caption{Upper bounds in the sector  $\mathcal{H}_{\ns0}^{+F-Q}$ for values of anomaly $\nu=1,2,4\mod8$. Bounds on the lightest non-degenerate primaries are represented by the same colored lines used to denote the bounds for the lightest scalars. The black dots mark the free fermion theories which almost saturate the bounds at the kinks we found.}
    \label{fig:J3}

\bigskip

    \includegraphics[scale=0.55]{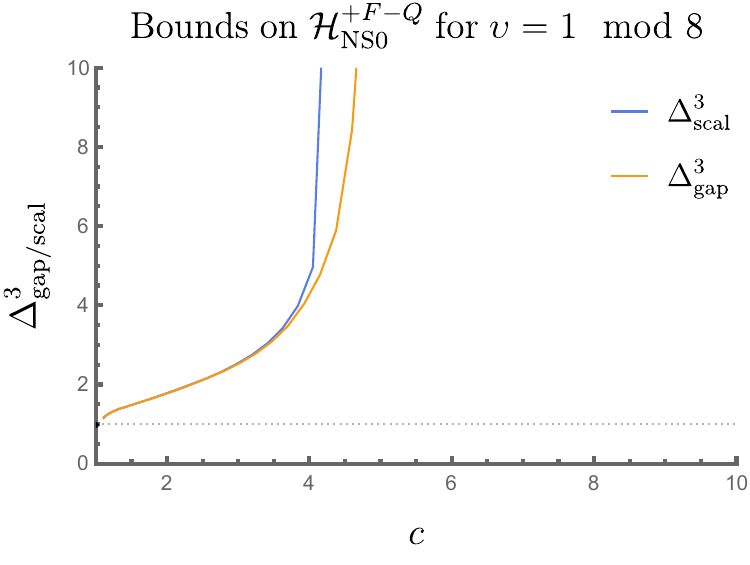}
    \qquad
    \includegraphics[scale=0.55]{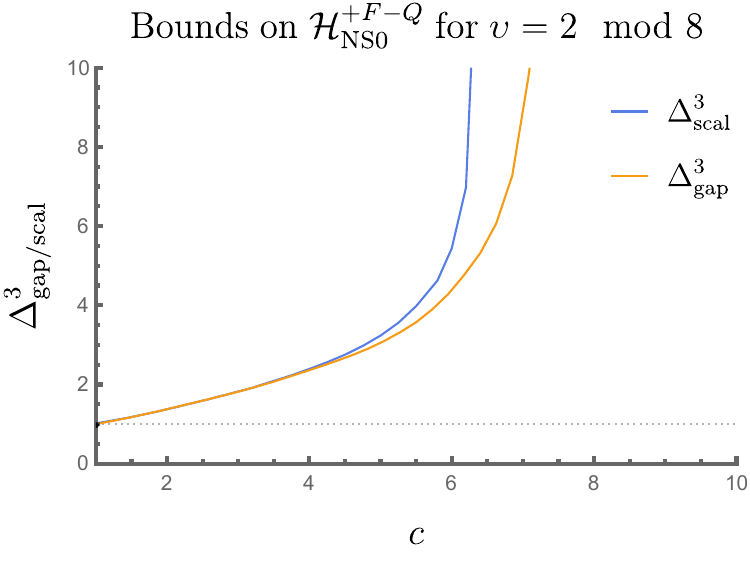}
    \caption{Left: upper bounds for lightest scalars and non-degenerate primaries confronted for theories with $\nu=1\mod8$. Right: upper bounds for lightest scalars and non-degenerate primaries confronted for theories with $\nu=2\mod8$.}
    \label{fig:V12J3}

\bigskip

    \includegraphics[scale=0.55]{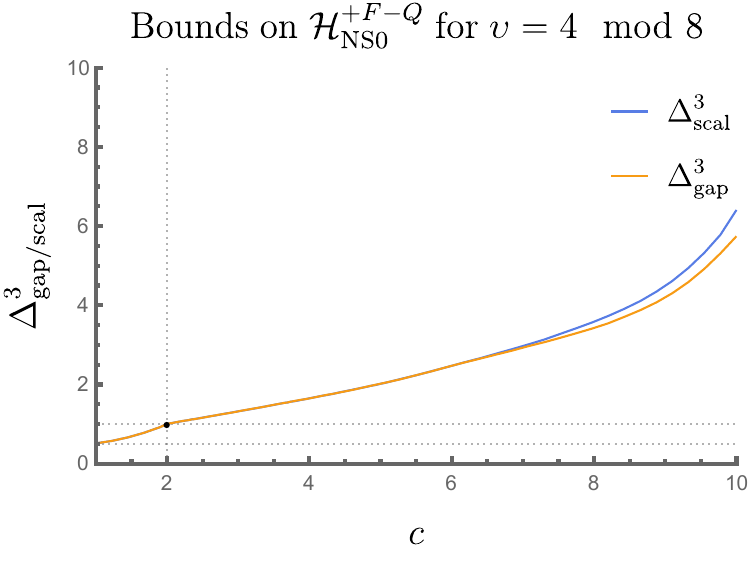}
    \caption{Upper bounds for lightest scalars and non-degenerate primaries confronted for theories with anomalies $\nu=4\mod8$.}
    \label{fig:V4J3}
\end{figure}

The behaviour of the bounds for charged operators is milder than for the even ones: we briefly summarize below the notable things that the numerical analysis highlighted. We report the full results in Figures \ref{fig:J3}-\ref{fig:V4J3}.

\begin{itemize}
\let\labelitemi\labelitemii
    \item Our findings are again coherent with the results founds from the analysis of the bosonic CFTs, i.e. they are always higher;
    \item Interestingly enough, the same behaviour of before is found for the operators charged under $\F$ and $(-1)^Q$ as well. This indeed seems to strength the hypothesis of before;
    \item The bounds present a single kink for $c>1$, namely a stack of free fermions with $c=2$ and anomaly $\nu=4\mod 8$;
    \item The other limit cases are again at $c=1$, which reproduces like before known theories;
    \item The overlap of the bounds for generic  non-degenerate states and scalars is present for all values of the anomalies analyzed, with a divergence between the two only for higher values of $c$.
\end{itemize}

\section*{Acknowledgements}

We would like to thank Lorenzo di Pietro, Shu-Heng Shao, Juven Wang for useful discussions. The work of A.G. is partially supported by the INFN Iniziativa Specifica ST\&FI.

\appendix
\section{Details on the derivation of $S$ and $T$ matrix elements}
\label{app:matrices}
Here we report the computation for the $S$ and $T$ matrices.

Let us start from the class $(0,0)$ and explain what are the $\ns 0$ directions we chose to contract in order to define the basis elements of $Z^{(0,0)}_i$.
Under $S$ the only element of this vector mapped to itself is $Z^{(0,0)}_3=Z^{\ns0}_{\ns0}$, while the others are mapped to themselves under $S^2$. This means that we can choose deliberately the direction of contraction for $4$ out of these $8$ tori in such a way that they are mapped to themselves under $S^2$ with no additional phase. The basis of the $4$ remaining ones will follow by using as reference bordism the $S$ transformation itself. As a consequence, 
we are able to fix all the non-zero entries of $S_{(0,0)}$ to $1$ with the exception of $\{\ns0,\ns0\}$. However this particular case has no $\Z_2$ holonomies and no pin$^-$ surface. Therefore $(S_{(0,0)})^3_3=1$ as well, which completes $S_{(0,0)}$.
The basis we chose that satisfy this property is
\begin{gather}
    \begin{tikzpicture}[scale=0.3,baseline=(current  bounding  box.center)]
    \draw (0,0) rectangle (4,4);
    \node[left] at (0,2) {$\ns0$};
    \node[below] at (2,0) {$\ns0$};
    \draw[Red,thick,dashed] (0,2) -- (4,2);
    \end{tikzpicture}\;,
    \label{fig:00base1}\\
    \begin{tikzpicture}[scale=0.3,baseline=(current  bounding  box.center)]
    \draw (0,0) rectangle (4,4);
    \node[left] at (0,2) {$\r0$};
    \node[below] at (2,0) {$\ns0$};
    \draw[Red,thick,dashed] (0,2) -- (4,2);
    \end{tikzpicture} 
    \quad
    \begin{tikzpicture}[scale=0.3,baseline=(current  bounding  box.center)]
    \draw (0,0) rectangle (4,4);
    \node[left] at (0,2) {$\ns0$};
    \node[below] at (2,0) {$\ns1$};
    \draw[Red,thick,dashed] (2,0) -- (2,4);
    \end{tikzpicture} 
    \quad
    \begin{tikzpicture}[scale=0.3,baseline=(current  bounding  box.center)]
    \draw (0,0) rectangle (4,4);
    \node[left] at (0,2) {$\ns0$};
    \node[below] at (2,0) {$\r1$};
        \draw[Red,thick,dashed] (2,0) -- (2,4);
    \end{tikzpicture}
    \quad
    \begin{tikzpicture}[scale=0.3,baseline=(current  bounding  box.center)]
    \draw (0,0) rectangle (4,4);
    \node[left] at (0,2) {$\r1$};
    \node[below] at (2,0) {$\ns1$};
    \draw[Red,thick,dashed] (4,0) -- (0,4);
    \end{tikzpicture}\;,
    \label{fig:00base2}\\
    \begin{tikzpicture}[scale=0.3,baseline=(current  bounding  box.center)]
    \draw (0,0) rectangle (4,4);
    \node[left] at (0,2) {$\ns0$};
    \node[below] at (2,0) {$\r0$};
    \draw[Red,thick,dashed] (2,0) -- (2,4);
    \end{tikzpicture} 
    \quad
    \begin{tikzpicture}[scale=0.3,baseline=(current  bounding  box.center)]
    \draw (0,0) rectangle (4,4);
    \node[left] at (0,2) {$\ns1$};
    \node[below] at (2,0) {$\ns0$};
    \draw[Red,thick,dashed] (0,2) -- (4,2);
    \end{tikzpicture} 
    \quad
    \begin{tikzpicture}[scale=0.3,baseline=(current  bounding  box.center)]
    \draw (0,0) rectangle (4,4);
    \node[left] at (0,2) {$\r1$};
    \node[below] at (2,0) {$\ns0$};
    \draw[Red,thick,dashed] (0,2) -- (4,2);
    \end{tikzpicture}
    \quad
    \begin{tikzpicture}[scale=0.3,baseline=(current  bounding  box.center)]
    \draw (0,0) rectangle (4,4);
    \node[left] at (0,2) {$\ns1$};
    \node[below] at (2,0) {$\r1$};
    \draw[Red,thick,dashed] (0,0) -- (4,4);
    \end{tikzpicture}\;,
    \label{fig:00base3}
\end{gather}
where the red dashed lines represent the direction of each torus contracted to a point in the bounding solid $3$-torus.

Next we turn to the computation of $T_{(0,0)}$. 
One can see that for the set \eqref{fig:00base1}-\eqref{fig:00base3} each element of basis $e_{a}$ ends up to the proper basis element $e_{T\cdot a}$ with the exception of two cases, namely for $(T_{(0,0)})^9_6$ and $(T_{(0,0)})^4_8$.

\begin{figure}
    \centering
\begin{tikzpicture}
[scale=0.5,baseline=(current  bounding  box.center),>=stealth]
    \coordinate (P1) at (-3,-3);
    \coordinate (P2) at (3,-3);
    \coordinate (P3) at (-0.5,-1);
    \coordinate (P4) at (5.5,-1);
    \coordinate (P5) at (-3,4);    
    \coordinate (P6) at (3,4);
    \coordinate (P7) at (-0.5,6);
    \coordinate (P8) at (5.5,6);
    \coordinate (C1) at ($(P1)!0.5!(P4)$);
    \coordinate (C2) at ($(P5)!0.5!(P8)$);
    \coordinate (Q1) at ($(P1)!0.5!(P2)$);
    \coordinate (Q2) at ($(P2)!0.5!(P4)$);
    \coordinate (Q3) at ($(P1)!0.5!(P3)$);
    \coordinate (Q4) at ($(P3)!0.5!(P4)$);
    \coordinate (Q5) at ($(P5)!0.5!(P6)$);
    \coordinate (Q6) at ($(P6)!0.5!(P8)$);
    \coordinate (Q7) at ($(P5)!0.5!(P7)$);
    \coordinate (Q8) at ($(P7)!0.5!(P8)$);
    \coordinate (X) at ($(Q5)+(0,1)$);
    \coordinate (Y) at ($(Q4)+(0,-1)$);
    
    \path[name path=Q2Q3] (Q2) -- (Q3);
    \path[name path=Q6Q7] (Q6) -- (Q7);
    \path[name path=Q1Q5] (Q1) -- (Q5);
    \path[name path=Q4Q8] (Q4) -- (Q8);
    \path[name path=Q5X] (Q5) -- (X);
    \path[name path=Q4Y] (Q4) -- (Y);
    \path[name intersections={of=Q2Q3 and Q1Q5,by=A1}];
    \path[name intersections={of=Q2Q3 and Q4Y,by=A2}];
    \path[name intersections={of=Q6Q7 and Q5X,by=A3}];
    \path[name intersections={of=Q6Q7 and Q4Q8,by=A4}];   

    \path[name path=Q1A1] (Q1) .. controls (C1) .. (A1) coordinate[pos=0.5] (B1);
    \path[name path=Q2Q4] (A2) .. controls (C1) .. (Q4) coordinate[pos=0.5] (B2);
    \path[name path=Q5Q7] (A3) .. controls (C2) .. (Q5) coordinate[pos=0.5] (B3);
    \path[name path=Q6Q8] (A4) .. controls (C2) .. (Q8) coordinate[pos=0.5] (B4);

    \draw[thin] (P1) -- (P3) -- (P4);
    \draw[thin] (P3) -- (P7);
    \draw[Red,very thick,densely dashed] (P2) --(P3);

    \draw[ForestGreen,very thick,dashed] (B1) .. controls +(-0.1,0.1) .. (A1);
    \path[fill=ForestGreen!30,fill opacity=0.6] (B1) .. controls +(0,-.1) .. (Q1) -- (Q5) .. controls +(0,-.1) .. (B3) -- (B1);
    \path[fill=ForestGreen!30,fill opacity=0.1] (A1) -- (Q3) -- (Q7) -- (A3) .. controls +(-0.1,0.1) .. (B3) .. controls +(0,-.1) .. (Q5) -- (A1);
    \draw[ForestGreen,very thick] (B1) .. controls +(0,-.1) .. (Q1);
    \draw[ForestGreen,very thick] (B3) .. controls +(-0.1,0.1) .. (A3);
    \draw[ForestGreen,very thick] (B3) .. controls +(0,-.1) .. (Q5);
    
    \draw[ForestGreen,very thick,dashed] (B2) .. controls +(0,0.1) .. (Q4);
    \draw[ForestGreen,very thick, dashed] (A4) -- (Q4) node[pos=0.5,sloped,rotate=90]{\tikz\draw[->,thick](0,0);};        
    \path[fill=ForestGreen!30,fill opacity=0.6] (B2) .. controls +(0,-.1) .. (A2) -- (Q2) -- (Q6) -- (A4) .. controls +(0,-.1) .. (B4) -- (B2);
    \path[fill=ForestGreen!30,fill opacity=0.1] (A4) -- (Q8) .. controls +(0,0.1) .. (B4) .. controls +(0,-.1) .. (A4);
    \draw[ForestGreen,very thick] (B2) .. controls +(0,-.1) .. (A2);
    \draw[ForestGreen,very thick] (B4) .. controls +(0,0.1) .. (Q8);
    \draw[ForestGreen,very thick] (B4) .. controls +(0,-.1) .. (A4);
    
    \draw[ForestGreen,very thick] (Q3) -- (A1);
    \draw[ForestGreen,very thick] (Q2) -- (A2);
    \draw[ForestGreen,very thick] (Q7) -- (A3);
    \draw[ForestGreen,very thick] (Q6) -- (A4);
    \draw[ForestGreen,very thick,->-] (Q1) -- (Q5);
    \draw[ForestGreen,very thick] (Q3) -- (Q7) node[pos=0.5,sloped,rotate=-90]{\tikz\draw[->,thick](0,0);};
    \draw[ForestGreen,very thick] (Q2) -- (Q6) node[pos=0.5,sloped,rotate=-90]{\tikz\draw[->,thick](0,0);};
    \draw[ForestGreen,very thick] (A4) -- (Q8);
    \draw[ForestGreen,very thick] (B1) -- (B3);
    \draw[ForestGreen,very thick] (B2) -- (B4);

    \draw[Red,very thick,densely dashed] (P5) --(P8);

    \draw[thick] (P1) -- (P2) node[pos=0.5,anchor=north west]{$\r 1$};
    \draw[thick] (P2) -- (P4) node[pos=0.5,anchor=north west]{$\ns 1$};
    \draw[thick] (P5) -- (P6) -- (P8) -- (P7) -- cycle;
    \draw[thick] (P1) -- (P5);
    \draw[thick] (P2) -- (P6);
    \draw[thick] (P4) -- (P8);
    \node[anchor=south east] at (Q1) {$x$};
    \node[anchor=north] at (Q4) {$x$};
    \node[anchor=north east] at (Q5) {$x'$};
    \node[anchor=south west] at (Q8) {$x'$};  
    \node[anchor=west] at (Q6) {$x''$};
    \node[anchor=south east] at (Q7) {$x''$};
\end{tikzpicture}\qquad
\begin{tikzpicture}
[scale=0.5,>=stealth,baseline=(current  bounding  box.center)]
    \coordinate (P1) at (-3,-3);
    \coordinate (P2) at (3,-3);
    \coordinate (P3) at (-0.5,-1);
    \coordinate (P4) at (5.5,-1);
    \coordinate (P5) at (-3,4);    
    \coordinate (P6) at (3,4);
    \coordinate (P7) at (-0.5,6);
    \coordinate (P8) at (5.5,6);
    \coordinate (C1) at ($(P1)!0.5!(P4)$);
    \coordinate (C2) at ($(P5)!0.5!(P8)$);
    \coordinate (Q1) at ($(P1)!0.5!(P2)$);
    \coordinate (Q2) at ($(P2)!0.5!(P4)$);
    \coordinate (Q3) at ($(P1)!0.5!(P3)$);
    \coordinate (Q4) at ($(P3)!0.5!(P4)$);
    \coordinate (Q5) at ($(P5)!0.5!(P6)$);
    \coordinate (Q6) at ($(P6)!0.5!(P8)$);
    \coordinate (Q7) at ($(P5)!0.5!(P7)$);
    \coordinate (Q8) at ($(P7)!0.5!(P8)$);
    \coordinate (X) at ($(Q5)+(0,10)$);
    \coordinate (Y) at ($(Q4)+(0,-10)$);
    
    \draw[thin] (P1) -- (P3) -- (P4);
    \draw[thin] (P3) -- (P7);
    \draw[Red,very thick,densely dashed] (P2) -- (Q3);
    \draw[Red,very thick,densely dashed] (Q2) -- (P3);

    \path[fill=ForestGreen!30,fill opacity=0.6] (Q1) -- (Q4) -- (Q8) -- (Q5) -- (Q1);
    \draw[ForestGreen,very thick] (Q1) -- (Q4);
    \draw[ForestGreen,very thick,->-] (Q1) -- (Q5);
    \draw[ForestGreen,very thick,->-] (Q4) -- (Q8);
    \draw[ForestGreen,very thick] (Q5) -- (Q8);
    
    \draw[thick] (P1) -- (P2) node[pos=0.5,anchor=north]{$\ns 1$}; 
    \draw[thick] (P2) -- (P4) node[pos=0.5,anchor=north west]{$\ns 0$};
    \draw[thick] (P5) -- (P6) -- (P8) -- (P7) -- cycle;
    \draw[thick] (P1) -- (P5);
    \draw[thick] (P2) -- (P6);
    \draw[thick] (P4) -- (P8);
    \node[anchor=south east] at (Q1) {$x$};
    \node[anchor=south east] at (Q4) {$x$};
    \node[anchor=north west] at (Q5) {$x'$};
    \node[anchor=south west] at (Q8) {$x'$};    
    
    \draw[Red,very thick,densely dashed] (P5) -- (P7);
\end{tikzpicture}%
    \caption{Left: a mapping torus for which evaluation of $\nu$ gives the element $(T_{(0,0)})^9_6$.  Right: a mapping torus for which evaluation of $\nu$ gives the element $(T_{(0,0)})^4_8$.}
    \label{fig:T0096and48torus} 
\end{figure}
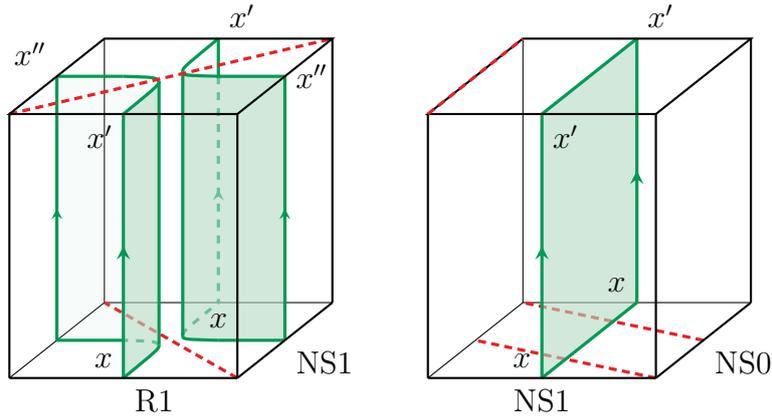

Let us start from the first. By applying equation \eqref{defmappingtori} it follows that the bounding $3$-manifold $Y=MT((T_{(0,0)})^9_6)$ is given by joining the two solid tori:
\begin{equation}
    \begin{tikzpicture}[scale=0.3,baseline=(current  bounding  box.center)]
    \draw (0,0) rectangle (4,4);
    \node[left] at (0,2) {$\ns1$};
    \node[below] at (2,0) {$\r1$};
    \draw[Red,thick,dashed] (0,4) -- (4,0);
    \end{tikzpicture} \quad\cup\quad
    \overline{\left(
    \begin{tikzpicture}[scale=0.3,baseline=(current  bounding  box.center)]
    \draw (0,0) rectangle (4,4);
    \node[right] at (4,2) {$\ns1$};
    \node[below] at (2,0) {$\r1$};
    \draw[Red,thick,dashed] (0,0) -- (4,4);
    \end{tikzpicture}\right)}.
\label{eq:T0096}
\end{equation}
At this point one needs to find a surface $\PD(a_g)$ associated to the $1$-cocycle $a_g\in H^1(Y,\Z_2)$, which is determined by appropriately extending over the $3^{\mathrm{rd}}$ direction the curve Poincar\'e dual of the restriction  $a_g|_{\{\ns1,\r1\}}$. This curve $\PD(a_g|_{\{\ns1,\r1\}})$ is found by a $T$ transformation of $\PD(a_g|_{\{\ns0,\r1\}})$, i.e.
\begin{equation}
    \begin{tikzpicture}[scale=0.4,baseline=(current  bounding  box.center)]
    \draw (0,0) rectangle (4,4);
    \node[left] at (0,2) {$\ns0$};
    \node[below] at (2,0) {$\r1$};
    \draw[ForestGreen,thick] (2,0) -- (2,4);
    \end{tikzpicture}    \quad \xrightarrow{T} \quad
    \begin{tikzpicture}[scale=0.4,baseline=(current  bounding  box.center)]
    \draw (0,0) rectangle (4,4);
    \node[left] at (0,2) {$\ns1$};
    \node[below] at (2,0) {$\r1$};
    \draw[ForestGreen,thick] (2,0) to[out=90,in=0] (0,2);
    \draw[ForestGreen,thick] (4,2) to[out=180,in=-90] (2,4);
    \end{tikzpicture}\;.
\label{eq:T009648torus}
\end{equation}
Therefore the $3$-manifold is the one on the left in Figure \ref{fig:T0096and48torus}. One can reach anologous conclusions for $(T_{(0,0)})^4_8$ and find the surface on the right of Figure \ref{fig:T0096and48torus}.\\
The closed 3-manifold in discussion is (in both cases) $\mathbb{RP}^3$.  Since $\mathbb{RP}^2$ is not orientable, there is no possibility to draw the full $\mathrm{pin}^-$ surface, but only some open subsets of it. To understand which kind of surface we are talking about, we need to inspect how it twists once we approach the radial directions of the tori that compose the full $3$-manifold.
\begin{figure}[tb]
    \centering
\begin{tikzpicture}
[scale=0.5,baseline=(current  bounding  box.center),>=stealth]
    \coordinate (P1) at (-3,-3);
    \coordinate (P2) at (3,-3);
    \coordinate (P3) at (-0.5,-1);
    \coordinate (P4) at (5.5,-1);
    \coordinate (X1) at ($(P1)!0.5!(P3)$);
    \coordinate (X2) at ($(P2)!0.5!(P4)$);
    \coordinate (Q1) at ($(P1)!0.5!(P2)$);
    \coordinate (Q2) at ($(P3)!0.5!(P4)$);
    \coordinate (P5) at ($(X1)+(0,-4)$);
    \coordinate (P6) at ($(X2)+(0,-4)$);    
    \coordinate (Q3) at ($(P5)!0.5!(P6)$);

    \draw[thin] (P3) -- (P5);

    \path[fill=ForestGreen!30,fill opacity=0.6] (Q1) -- (Q2) -- (Q3);
    \draw[ForestGreen,very thick,dashed,->-] (Q2) -- (Q3);
    \draw[ForestGreen,very thick] (Q1) -- (Q2);
    \draw[ForestGreen,very thick,->-] (Q1) -- (Q3);

    \draw[thick] (P1) -- (P2) node[pos=0.5,anchor=north east]{$x$};
    \draw[thick] (P3) -- (P4) node[pos=0.5,anchor=south east]{$\r 1$} node[pos=0.5,anchor=north west]{$x$}; \draw[thick] (P1) -- (P3) node[pos=0.5,anchor=south east]{$\ns 0$};
    \draw[thick] (P2) -- (P4);    
    \draw[thick] (P1) -- (P5);
    \draw[thick] (P2) -- (P6);
    \draw[thick] (P4) -- (P6); 
    \draw[thick] (P5) -- (P6);
\end{tikzpicture}
\qquad
\begin{tikzpicture}
[scale=0.5,baseline=(current  bounding  box.center),>=stealth]
    \coordinate (P1) at (-3,-3);
    \coordinate (P2) at (3,-3);
    \coordinate (P3) at (-0.5,-1);
    \coordinate (P4) at (5.5,-1);
    \coordinate (X1) at ($(P1)!0.5!(P3)$);
    \coordinate (X2) at ($(P2)!0.5!(P4)$);
    \coordinate (X3) at ($(X1)!0.5!(X2)$);
    \coordinate (Q1) at ($(P1)!0.5!(P2)$);
    \coordinate (Q2) at ($(P3)!0.5!(P4)$);
    \coordinate (P5) at ($(X1)+(0,-4)$);
    \coordinate (P6) at ($(X2)+(0,-4)$);    
    \coordinate (Q3) at ($(P5)!0.5!(P6)$);
    \coordinate (Q4) at ($(P1)!0.5!(X1)$);
    \coordinate (Q5) at ($(P3)!0.5!(X1)$);
    \coordinate (Q6) at ($(P2)!0.5!(X2)$);
    \coordinate (Q7) at ($(P4)!0.5!(X2)$);
    \coordinate (X4) at ($(Q1)!0.5!(X3)$);
    \coordinate (X5) at ($(Q2)!0.5!(X3)$);

    \draw[thin] (P3) -- (P5);
    
    \path[name path=P5Q5] (P5) -- (Q5);
    \path[name path=Q2Q3] (Q2) -- (Q3);

    \path[fill=ForestGreen!30,fill opacity=0.3] (Q1) .. controls (X4) .. (Q6) -- (P6) -- (Q3) -- cycle;
    \path[fill=ForestGreen!30,fill opacity=0.2] (Q4) .. controls (X4) .. (X3) .. controls (X5) .. (Q7) -- (P6) -- (P5) -- cycle;
    \path[fill=ForestGreen!30,fill opacity=0.1] (Q5) .. controls (X5) .. (Q2) -- (Q3) -- (P5) -- cycle;
    
    \draw[ForestGreen,very thick] (Q1) .. controls (X4) .. (Q6);
    \draw[ForestGreen,very thick,name path=Q4X3] (Q4) .. controls (X4) .. (X3);   
    \draw[ForestGreen,very thick,name path=Q7X3] (X3) .. controls (X5) .. (Q7);
    \draw[ForestGreen,very thick] (Q5) .. controls (X5) .. (Q2);
    
    \path[name intersections={of=P5Q5 and Q4X3,by=A1}];
    \path[name intersections={of=Q2Q3 and Q7X3,by=A2}];    

    \draw[ForestGreen,very thick] (Q1) -- (Q3);
    \draw[ForestGreen,very thick] (Q2) -- (A2);
    \draw[ForestGreen,very thick,dashed] (A2) -- (Q3);
    \draw[ForestGreen,very thick] (Q4) -- (P5);
    \draw[ForestGreen,very thick] (Q5) -- (A1);
    \draw[ForestGreen,very thick,dashed] (A1) -- (P5);
    \draw[ForestGreen,very thick] (Q6) -- (P6);
    \draw[ForestGreen,very thick] (Q7) -- (P6);
    \draw[ForestGreen,very thick,->-] (P5) -- (Q3);
    \draw[ForestGreen,very thick,->-] (Q3) -- (P6);

    \draw[thick] (P1) -- (P2) node[pos=0.5,anchor=north east]{$x'$};
    \draw[thick] (P3) -- (P4) node[pos=0.5,anchor=south east]{$\r 1$} node[pos=0.5,anchor=south west]{$x'$}; \draw[thick] (P1) -- (P3) node[pos=0.5,anchor=south east]{$\ns 0$};
    \draw[thick] (P2) -- (P4);    
    \draw[thick] (P1) -- (P5);
    \draw[thick] (P2) -- (P6);
    \draw[thick] (P4) -- (P6); 
    \draw[thick] (P5) -- (P6) node[pos=0.75,anchor=south]{$b$};
    
    \filldraw[ForestGreen] (X3) circle (2pt);
    \filldraw[ForestGreen] (Q3) circle (2pt);
    \node[anchor=east] at (X3) {$x''$};
    
\end{tikzpicture}
\caption{Representation of the two solid tori describing the manifold \eqref{eq:T0096}. On the torus on the right is represented the generator $b$ of $H^1(\mathbb{RP}^2,\Z_2)$, given by the contraction of the path going from $x'$ to $x''$ to the core.}
\label{fig:solidtorus1}
\end{figure}
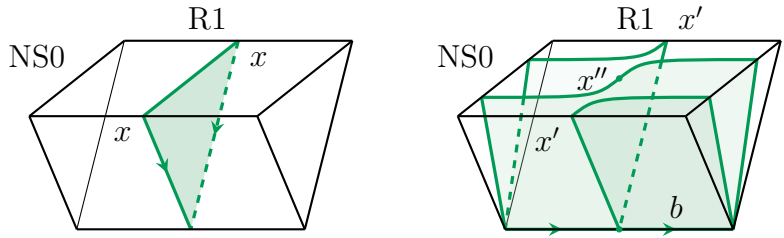
We start from the first of \eqref{eq:T009648torus}. We can draw the solid torus as on the left of Figure \ref{fig:solidtorus1}, where the symplectic basis depicted of the $2$-torus sections changed, so that one of the generators is the $\ns0$ direction which is being contracted to a point. In the figure the radial direction is represented by the two near-vertical lines, which are identified.
It is clear that in this case there is no twist and thus this particular subset of the surface can be represented as a disk.
By repeating the same reasoning with the second solid torus one arrives to the different situation represented on the right of Figure \ref{fig:solidtorus1}. Here there is a clear twist of the surface so this part of its $2$-skeleton is represented by a M\"obius band with core circle $\mathbb{RP}^1$. This means that the surface we are working with is $\mathbb{RP}^2$, for which $\ABK$ is determined by the value $q(b)$ of its generator $b$. To compute it, we look more carefully at the contraction of $b$ to the core of this solid torus. We can consider as even framing along $b$ a framing which does a full $2 \pi$ rotation while going around the loop. As one can see from Figure \ref{fig:3torus}, the normal bundle on the surface does a single negative half twist with respect to such framing, so that 
\begin{equation}
    q(b)=-1 \quad\Rightarrow\quad \ABK = 7\quad\mathrm{mod}\,8.
\end{equation}

By repeating the same reasoning we see that $\mathbb{RP}^2$ for $(T_{(0,0)})^4_8$ has instead $\ABK=1\mod8$. Indeed the procedure is almost the same, with the main difference being that the core of the torus now has $\ns$ periodicity, which is equivalent to changing the quadratic enhancement of the generator as $q(b)\rightarrow q(b) +2\mod4$.

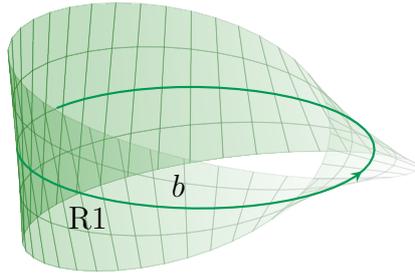
\begin{figure}
    \centering
\begin{tikzpicture}[>=stealth,baseline=0pt]
\begin{axis}[
    hide axis,
    view={20}{30}
]
\addplot3 [
    surf, shader=faceted interp,
    point meta=x,
    opacity=0.3,
    colormap={greenwhite}{rgb255=(34,139,34) color=(white)},
    samples=40,
    samples y=5,
    z buffer=sort,
    domain=0:360,
    y domain=-0.5:0.5
] (
    {(1+0.5*y*cos(x/2)))*cos(x)},
    {(1+0.5*y*cos(x/2)))*sin(x)},
    {-0.5*y*sin(x/2)});

\addplot3 [ForestGreen,
    samples=50,
    domain=-155:160, 
    samples y=0,
    thick,->-
] (
    {cos(x)},
    {sin(x)},
    {0}) node[pos=0.15,below,black] {$\mathrm{R}1$} node[pos=0.25,anchor=south,black]{$b$};
    
\end{axis}
\end{tikzpicture}
    \caption{The M\"obius strip embedded into the solid torus depicted on the right side of Figure \ref{fig:solidtorus1}.}
    \label{fig:3torus}
\end{figure}
Next we continue with the computation of $S_{(0/1,1)}$. Since these classes behave almost identically with the exception of changing the periodicity $\ns1 \leftrightarrow \r1$, we will use from now on a special notation to identify the two simultaneously. Depending on the class we are working with, the surfaces $\{X_1,\,X_2,\,X_3\}$ will stand for $\{\{\r0,\ns1\},\{\ns1,\r0\},\{\ns1,\ns1\}\}$ or $\{\{\r0,\r1\},\{\r1,\r0\},\{\r1,\r1\}\}$ for $(0,1)$ and $(1,1)$ respectively. The pin$^-$-surfaces for which $\ABK$ invariant determines the value of the matrix entries $(S_{(0/1,1)})^i_j$ and $(T_{(0/1,1)})^i_j$ will instead be denoted as $\Sigma^{S/T}_{ij}$.

\begin{figure}[tb]
    \centering
\begin{tikzpicture}
[scale=0.5,>=stealth,baseline=(current  bounding  box.center)]
    \coordinate (P1) at (-3,-3);
    \coordinate (P2) at (3,-3);
    \coordinate (P3) at (-0.5,-1);
    \coordinate (P4) at (5.5,-1);
    \coordinate (P5) at (-3,4);    
    \coordinate (P6) at (3,4);
    \coordinate (P7) at (-0.5,6);
    \coordinate (P8) at (5.5,6);
    \coordinate (C1) at ($(P1)!0.5!(P4)$);
    \coordinate (C2) at ($(P5)!0.5!(P8)$);
    \coordinate (Q1) at ($(P1)!0.5!(P2)$);
    \coordinate (Q2) at ($(P2)!0.5!(P4)$);
    \coordinate (Q3) at ($(P1)!0.5!(P3)$);
    \coordinate (Q4) at ($(P3)!0.5!(P4)$);
    \coordinate (Q5) at ($(P5)!0.5!(P6)$);
    \coordinate (Q6) at ($(P6)!0.5!(P8)$);
    \coordinate (Q7) at ($(P5)!0.5!(P7)$);
    \coordinate (Q8) at ($(P7)!0.5!(P8)$);
    \coordinate (X) at ($(Q5)+(0,10)$);
    \coordinate (Y) at ($(Q4)+(0,-10)$);
    
    \coordinate (AxO) at (-6,-3);
    \coordinate (Ax1) at (-4.5,-3);
    \coordinate (Ax2) at ($(AxO)+(1.171,0.937)$);
    \coordinate (Ax3) at (-6,-1.5);
    
    \coordinate (AxO') at ($(AxO)+(0,7)$);
    \coordinate (Ax1') at ($(Ax1)+(0,7)$);
    \coordinate (Ax2') at ($(Ax2)+(0,7)$);
    \coordinate (Ax3') at ($(Ax3)+(0,7)$);    
    
    \draw[RawSienna,thick,->] (AxO) -- (Ax1) node[pos=0.5,anchor=north]{$\partial_\theta$};
    \draw[RawSienna,thick,->] (AxO) -- (Ax2) node[pos=0.9,anchor=south]{$\partial_\chi$};
    \draw[RawSienna,thick,->] (AxO) -- (Ax3) node[pos=0.5,anchor=east]{$\partial_z$};    
    
    \draw[RawSienna,thick,->] (AxO') -- (Ax1') node[pos=0.5,anchor=north]{$\partial'_\theta$};
    \draw[RawSienna,thick,->] (AxO') -- (Ax2') node[pos=0.9,anchor=south]{$\partial'_\chi$};
    \draw[RawSienna,thick,->] (AxO') -- (Ax3') node[pos=0.5,anchor=east]{$\partial'_z$};        
    
    \draw[thin] (P1) -- (P3) node[pos=0.5,sloped,rotate=-90]{\tikz\draw[->>,thick](0,0);};
    \draw[thin] (P3) -- (P4);
    \draw[thin] (P3) -- (P7);

    \path[fill=ForestGreen!30,fill opacity=0.6] (Q1) -- (Q4) -- (Q8) -- (Q5) -- (Q1);
    \draw[ForestGreen,very thick,->-] (Q1) -- (Q4) node[black,pos=0.5,anchor=north west]{$a$};
    \draw[ForestGreen,very thick] (Q1) -- (Q5) node[pos=0.5,sloped,rotate=-90]{\tikz\draw[->,thick](0,0);} node[black,pos=0.5,anchor=west]{$b$};
    \draw[ForestGreen,very thick] (Q4) -- (Q8) node[pos=0.5,sloped,rotate=-90]{\tikz\draw[->,thick](0,0);} node[black,pos=0.5,anchor=east]{$b$};
    \draw[ForestGreen,very thick,->-] (Q8) -- (Q5) node[black,pos=0.5,anchor=south east]{$a$};
    
    \draw[thick] (P1) -- (P2) node[pos=0.5,anchor=north]{$\ns/\r 1$} node[pos=0.5,sloped,rotate=-90]{\tikz\draw[->,thick](0,0);}; 
    \draw[thick] (P2) -- (P4) node[pos=0.5,anchor=north west]{$\r 0$};
    \draw[thick] (P5) -- (P6) node[pos=0.5,sloped,rotate=90]{\tikz\draw[->,thick](0,0);};
    \draw[thick] (P6) -- (P8) -- (P7);
    \draw[thick] (P7) -- (P5) node[pos=0.5,sloped,rotate=90]{\tikz\draw[->>,thick](0,0);};
    \draw[thick] (P1) -- (P5);
    \draw[thick] (P2) -- (P6);
    \draw[thick] (P4) -- (P8) node[pos=0.5,anchor=west]{$\ns 0$};
    
\end{tikzpicture}
\caption{Mapping torus $MT((S_{(0/1,1)})^1_2)$: the bottom slice is $X_1\times\{0\}$, while the top one is the image of $S^2$, i.e. $X_1\times\{1\}$. The figure also displays cycles $a$ and $b$ along which one moves to determine the $\ABK$ invariant of the Klein bottle.}
\label{fig:MTS12}
\end{figure}
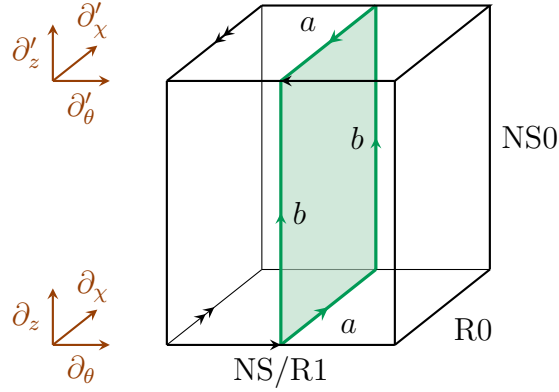

The first entry one has to compute is $(S_{(0/1,1)})^1_2$. The mapping torus $MT((S_{(0/1,1)})^1_2)$ associated to it is given by applying $S$ twice to $X_1$, which is our reference 2-manifold, and then identifying the boundaries $X_1 \times\{0\} \sim X_1 \times \{1\}$ (remembering of the additional $(-1)^{\mathcal{F}}$ action); see Figure \ref{fig:MTS12}. By the same argument of before one can see that the $\mathrm{pin}^-$ surface we are interested in is just a curve parallel to the time direction of $X_1$ spanned along the vertical direction. Therefore the surface $\Sigma^S_{12}$ in question is a Klein bottle $K=\mathbb{RP}^2_b\#\mathbb{RP}^2_{c=a+b}$. Here $b$ and $c$ denote the generators of $H_1(K,\mathbb{Z}_2)$ related to each component of its connected sum.

After identifying the surface, we have to fix the transition function from the top slice $X_1\times \{1\}$ to the bottom one $X_1\times \{0\}$. In this case the two canonical basis are related by
\begin{equation}
    \begin{pmatrix}
    \partial'_\chi & \partial'_\theta
    \end{pmatrix} = \begin{pmatrix}
    \partial_\chi & \partial_\theta
    \end{pmatrix} \begin{pmatrix}
    0 & -1 \\ 1&0 \\
    \end{pmatrix}^T
    ,\quad    
    \begin{pmatrix}
    \theta'\\ \chi'
    \end{pmatrix} = \begin{pmatrix}
    -1 & 0 \\ 0&-1 \\
    \end{pmatrix} \begin{pmatrix}
    \theta\\ \chi
    \end{pmatrix}.
\end{equation}
Here we denoted $(\theta,\chi)$ and $(\theta',\chi')$ the canonical coordinates on the tori $X_1\times\{0\}$ and $X_1\times\{1\}$ respectively. They are determined by identifying $(\partial_\theta,\partial_\chi)=(\omega_1,\omega_2)$ and $(\partial'_\theta,\partial'_\chi)=(\omega'_1,\omega'_2)$, where $\omega_1$ and $\omega_2$ are the lattice generators that define the modular parameter $\tau=\omega_2/\omega_1$.

Therefore if we start from a vector $v \in T_p MT((S_{(1,0/1)})^1_2)$ for any point ${p \in X_1\times\{0\}}\cong X_1 \times\{1\}$ and write it in the canonical basis of $X_1\times\{1\}$, then its entries in the canonical basis used for $X_1\times\{0\}$ are given by applying the transition function $\mathsf{T}_{S^2} = \mathsf{T}_S \mathsf{T}_S =R_3(\pi/2) R_3(\pi/2)$,
\begin{equation}
\begin{pmatrix}
v_\theta\\ v_\chi\\ v_z
\end{pmatrix}= R_3(\pi/2)R_3(\pi/2)\begin{pmatrix}
v'_\theta\\ v'_\chi\\ v'_z
\end{pmatrix}.
\end{equation}
Here $z$ denotes the third direction (i.e. the vertical $\partial_z$, with positive sign going from bottom to top in Figure \ref{fig:MTS12}) and $R_i(\alpha)$ denotes the rotation around $i$-th axis by angle $\alpha$.

The next step is choosing a lift for the transition function to $\Spin(3)\cong SU(2)$. Obviously we have
\begin{equation}
\mathsf{T}_S= R_3(\pi/2) \xrightarrow{\mathrm{lift}} \widetilde{ \mathsf{T}}^\pm_S =\pm e^{-i\frac{\pi}{4}\sigma_3}.
\label{strans}
\end{equation}
Since it will be useful also for the next computations\footnote{Note: the choice of the lift do not actually play any significant role as long as one is consistent in using it. A different choice of the lift for $S$ transformation is equivalent to the redefinition $S\rightsquigarrow S(-1)^\CF$ of the generator $S$ in the metaplectic group (\ref{Mp}), and similarly for $T$ transformation. Such redefinitions in general change the group relations. We will check a posteriori that our choice of the lifts is consistent with the relations presented in (\ref{Mp}).}, we choose from now on to use the lift $\widetilde{\mathsf{T}}_S:=\widetilde{\mathsf{T}}_S^+$.\\
Lastly, for $(0,1)$ the transition functions in $\Spin(2)$ for the identifications of the points $(\theta +1,\chi)\sim (\theta,\chi)$ and $(\theta,\chi+1)\sim (\theta,\chi)$ are respectively $-\mathrm{id}$ and $+\mathrm{id}$, which are properly lifted to $\mp\mathrm{id}\in \Spin(3)\cong SU(2)$ for the mapping torus. For $(1,1)$ we do not have to keep track of this since it is always $+\mathrm{id}$.

At this point we can compute the value of the enhancement $q$ for the $1$-cycles $a$ and $b$.
\begin{itemize}
\let\labelitemi\labelitemii
    \item  For the cycle $a$ a framing is given by doing a $2\pi n$ rotation while going from the point $x\in X_1\times \{0\}$ with coordinates $(\theta,\chi)=(1/2,0)$ to itself along the direction depicted. This means that a vector $v\in T_x MT((S_{(0/1,1)})^1_2)$ goes back to itself via
\begin{equation}
    v'=R_2(2\pi n)v=v.
\end{equation}
The lift in $SU(2)$ is $(-\mathrm{id})^n$, which implies that $n=1$ is needed for the framing to be even. In this way the normal bundle is doing two negative half twists around the framing chosen and $q(a)=2\mod4$.
    \item For the cycle $b$ we start again from the point $x$. Going along $b$ the generic vector does the transformation 
\begin{equation}
    v'=\underbrace{R_3(\pi/2)^2}_{\mathsf{T}_{S^2}}R_3((2n+1)\pi)v.
\end{equation}
The ending point is $x'$, with coordinates $(\theta',\chi')=(1/2,0)$, equivalent to $(\theta,\chi)=(-1/2,0)$. Thus to be back at $x$ one has an additional $-\mathrm{id}$ transformation to apply in the case of $\{\r 0,\ns1\}$. We will keep track of it by writing it inside square brackets.\\
In $SU(2)$ this means that the transformation is 
\begin{equation}
    [-\mathrm{id}]\underbrace{(-\mathrm{id})}_{(-1)^\CF}\underbrace{e^{-i \frac{\pi}{2}\sigma_3}}_{\widetilde{\mathsf{T}}_{S^2}}e^{-i(2n+1)\frac{\pi}{2}\sigma_3}= (-\mathrm{id})^{n[+1]}.
\end{equation}
Consider now the two different classes. For $(1,1)$ the even framing is found for $n=1$. In this case the framing is then defined by doing three 
\textit{positive} $\pi$ rotations, which means the normal bundle is doing three \textit{negative} half rotations and $q(b)=-3\mod4=1\mod4$. For $(0,1)$ instead we have $n=0$ and accordingly $q(b)=3\mod4$.
\end{itemize}
Now we can use \eqref{qenhancement} and find that $q(a)=q(b)$. Therefore we arrive to the values
\begin{equation}
    (S_{(1,1)})^1_2= i^\nu, \qquad
    (S_{(0,1)})^1_2=(-i)^\nu.
\end{equation}

The next matrix element we are going to compute is $(S_{(0/1,1)})^3_3$. In this case the manifold describes a single $S$ transformation of $X_3$. Proceeding as usual, the $\mathrm{pin}^-$ surface is a smooth curve Poincar\'e dual to the $\Z_2$ holonomies spanned along the vertical direction. In this case one has to pay attention when choosing such curve. Indeed, this amounts to choosing a resolution of its singular representative, which in general will not be invariant under $S$. From consistency with \eqref{eq:NS1NS1-base}, \eqref{eq:R1R1-base} we have

\begin{equation}
    \begin{tikzpicture}[scale=0.4,baseline=(current  bounding  box.center)]
    \draw (0,0) rectangle (4,4);
    \node[below] at (2,0) {Singular cycle};
    \draw[ForestGreen,thick] (2,0) -- (2,4);
    \draw[ForestGreen,thick] (0,2) -- (4,2);    
    \end{tikzpicture}    \quad \rightarrow \quad\begin{tikzpicture}[scale=0.4,baseline=(current  bounding  box.center)]
    \draw (0,0) rectangle (4,4);
    \node[below] at (2,0) {Smooth resolution};
    \draw[ForestGreen,thick] (2,0) to[out=90,in=0] (0,2);
    \draw[ForestGreen,thick] (4,2) to[out=180,in=-90] (2,4);
    \end{tikzpicture} \quad \xrightarrow{S} \quad
    \begin{tikzpicture}[scale=0.4,baseline=(current  bounding  box.center)]
    \draw (0,0) rectangle (4,4);
    \node[below] at (2,0) {$S$ transformation};
    \draw[ForestGreen,thick] (2,0) to[out=90,in=180] (4,2);
    \draw[ForestGreen,thick] (0,2) to[out=0,in=-90] (2,4);
    \end{tikzpicture}. 
\end{equation}

With this fixed, it follows that $\PD(a_g)$ is the saddle surface in Figure \ref{fig:MTS33}, represented by the string of $1$-chains
\begin{equation}
    acb^{-1}db^{-1}cad\sim eeffhh,
\end{equation}
where $e=cb^{-1}db^{-1}$, $f=bd^{-1}$ and $h=da$.
Thus $\Sigma^S_{33}=\mathbb{RP}^2_e\#\mathbb{RP}^2_f\#\mathbb{RP}^2_h$, where again the indices represent the generators of the first $\Z_2$ homology group of each $\mathbb{RP}^2$ component. By knowing the value of $q$ for three independent elements in $H^1(\Sigma_{33}^S,\Z_2)$ we can determine the value of its $\ABK$ invariant. 

\begin{figure}
    \centering
\begin{tikzpicture}
[scale=0.5,>=stealth,baseline=(current  bounding  box.center)]
   \coordinate (P1) at (-3,-3);
    \coordinate (P2) at (3,-3);
    \coordinate (P3) at (-0.5,-1);
    \coordinate (P4) at (5.5,-1);
    \coordinate (P5) at (-3,4);    
    \coordinate (P6) at (3,4);
    \coordinate (P7) at (-0.5,6);
    \coordinate (P8) at (5.5,6);
    \coordinate (C1) at ($(P1)!0.5!(P4)$);
    \coordinate (C2) at ($(P5)!0.5!(P8)$);
    \coordinate (C3) at ($(C1)!0.5!(C2)$);
    \coordinate (Q1) at ($(P1)!0.5!(P2)$);
    \coordinate (Q2) at ($(P2)!0.5!(P4)$);
    \coordinate (Q3) at ($(P1)!0.5!(P3)$);
    \coordinate (Q4) at ($(P3)!0.5!(P4)$);
    \coordinate (Q5) at ($(P5)!0.5!(P6)$);
    \coordinate (Q6) at ($(P6)!0.5!(P8)$);
    \coordinate (Q7) at ($(P5)!0.5!(P7)$);
    \coordinate (Q8) at ($(P7)!0.5!(P8)$);
    \coordinate (X) at ($(Q5)+(0,10)$);
    \coordinate (Y) at ($(Q4)+(0,-2)$);
    
    \coordinate (AxO) at (-6,-3);
    \coordinate (Ax1) at (-4.5,-3);
    \coordinate (Ax2) at ($(AxO)+(1.171,0.937)$);
    \coordinate (Ax3) at (-6,-1.5);
    
    \coordinate (AxO') at ($(AxO)+(0,7)$);
    \coordinate (Ax1') at ($(Ax1)+(0,7)$);
    \coordinate (Ax2') at ($(Ax2)+(0,7)$);
    \coordinate (Ax3') at ($(Ax3)+(0,7)$);    
    
    \draw[RawSienna,thick,->] (AxO) -- (Ax1) node[pos=0.5,anchor=north]{$\partial_\theta$};
    \draw[RawSienna,thick,->] (AxO) -- (Ax2) node[pos=0.9,anchor=south]{$\partial_\chi$};
    \draw[RawSienna,thick,->] (AxO) -- (Ax3) node[pos=0.5,anchor=east]{$\partial_z$};    
    
    \draw[RawSienna,thick,->] (AxO') -- (Ax1') node[pos=0.5,anchor=north]{$\partial'_\theta$};
    \draw[RawSienna,thick,->] (AxO') -- (Ax2') node[pos=0.9,anchor=south]{$\partial'_\chi$};
    \draw[RawSienna,thick,->] (AxO') -- (Ax3') node[pos=0.5,anchor=east]{$\partial'_z$};      
    
    \draw[thin] (P1) -- (P3) node[pos=0.5,sloped,rotate=-90]{\tikz\draw[->>,thick](0,0);};
    \draw[thin] (P3) -- (P4);
    \draw[thin] (P3) -- (P7);
    
    \path[name path=Q1Q5] (Q1) -- (Q5);
    \path[name path=Q2Q6] (Q2) -- (Q6);  
    \path[name path=Q3Q7] (Q3) -- (Q7);
    \path[name path=Q4Q8] (Q4) -- (Q8);

    \path[name path=Q2Q3] (Q2) -- (Q3);
    \path[name path=Q4Y] (Q4) -- (Y);
    \path[name path=Q4Q8] (Q4) -- (Q8);     
    \path[name intersections={of=Q1Q5 and Q2Q3, by=A1}];
    \path[name intersections={of=Q2Q3 and Q4Y,by=A2}];

    \path (A1) .. controls (C1) .. (Q1) coordinate[pos=0.5](B1);
    \path[name path=Q5Q6] (Q5) .. controls (C2) .. (Q6) coordinate[pos=0.5](A3);
    \path (Q7) .. controls (C2) .. (Q8) coordinate[pos=0.5](A4);
    \path (Q4) .. controls (C1) .. (A2) coordinate[pos=0.5](B2);     
    
    \path[name intersections={of=Q4Q8 and Q5Q6,by=A5}];
    
    \path[fill=ForestGreen!30,fill opacity=0.1] (Q3) -- (A1) -- (Q5) .. controls (C2) .. (A5) -- (Q8) .. controls (C2) .. (Q7) -- (Q3);
    \path[fill=ForestGreen!30,fill opacity=0.6] (Q1) .. controls ($(B1)+(0,-0.15)$) .. (B1) .. controls ($(C3)+(-0.4,0.8)$) and ($(C3)+(0.4,0.8)$) .. (B2) .. controls ($(B2)+(0,-0.12)$) .. (A2) -- (Q2) -- (Q6) .. controls (C2) .. (Q5) -- (Q1);

    \draw[ForestGreen,very thick] (B1) .. controls ($(C3)+(-0.4,0.8)$) and ($(C3)+(0.4,0.8)$) .. (B2) coordinate[pos=0.5](C4);
    \draw[ForestGreen,very thick,dotted] plot [smooth, tension=1.7] coordinates {(A3) (C4) (A4)};
    
    \draw[ForestGreen,very thick,->-] (Q1) -- (Q5) node[black,pos=0.5,anchor=west]{$c$};
    \draw[ForestGreen,very thick,->-] (Q7) -- (Q3) node[black,pos=0.5,anchor=east]{$d$};
    \draw[ForestGreen,very thick,->-] (Q3) -- (A1) node[black,pos=0.5,anchor=north]{$a$};    
    \draw[ForestGreen,very thick,->-] (A2) -- (Q2)  node[black,pos=0.5,anchor=south]{$b$};

    \draw[ForestGreen,very thick,->-] (Q6) -- (Q2) node[black,pos=0.5,anchor=west]{$d$};
    \draw[ForestGreen,very thick,->-] (Q6) .. controls (C2) .. (Q5) node[black,pos=0.2,anchor=south west]{$b$};
    \draw[ForestGreen,very thick,->-] (Q8) .. controls (C2) .. (Q7) node[black,pos=0.5,anchor=south east]{$a$};    
    \draw[ForestGreen,very thick] (Q8) -- (A5);
    \draw[ForestGreen,very thick,dashed,->-] (Q4) -- (A5) node[black,pos=0.5,anchor=east]{$c$};

    \draw[ForestGreen,very thick] (Q1) .. controls ($(B1)+(0,-0.15)$) .. (B1);
    \draw[ForestGreen,very thick,dashed] (B1) .. controls ($(B1)+(-0.1,0.1)$) .. (A1);
    
    \draw[ForestGreen,very thick] (B2) .. controls ($(B2)+(0,-0.12)$) .. (A2);
    \draw[ForestGreen,very thick,dashed] (B2) .. controls ($(B2)+(0,0.1)$) .. (Q4);
    
    \draw[thick] (P1) -- (P2) node[pos=0.5,anchor=north]{$\ns/\r 1$} node[pos=0.5,sloped,rotate=-90]{\tikz\draw[->,thick](0,0);};
    \draw[thick] (P2) -- (P4) node[pos=0.5,anchor=north west]{$\ns/\r 1$};
    \draw[thick] (P5) -- (P6) node[pos=0.5,sloped,rotate=-90]{\tikz\draw[->>,thick](0,0);};
    \draw[thick] (P6) -- (P8) -- (P7);
    \draw[thick] (P5) -- (P7) node[pos=0.5,sloped,rotate=90]{\tikz\draw[->,thick](0,0);};
    \draw[thick] (P1) -- (P5);
    \draw[thick] (P2) -- (P6);
    \draw[thick] (P4) -- (P8) node[pos=0.5,anchor=west]{$\ns 0$};
\end{tikzpicture}\qquad
\begin{tikzpicture}[scale=0.5,>=stealth,baseline=(current  bounding  box.center)]

\node[draw=none,minimum size=3.5cm,regular polygon,regular polygon sides=8,fill=ForestGreen!30,fill opacity=0.4] (SIGMA) {};

\draw[ForestGreen,very thick,->-] (SIGMA.corner 1) -- (SIGMA.corner 2) node[black,pos=0.5,anchor=south]{$b$};
\draw[ForestGreen,very thick,->-] (SIGMA.corner 3) -- (SIGMA.corner 2) node[black,pos=0.5,anchor=south east]{$c$};
\draw[ForestGreen,very thick,->-] (SIGMA.corner 4) -- (SIGMA.corner 3) node[black,pos=0.5,anchor=east]{$a$};
\draw[ForestGreen,very thick,->-] (SIGMA.corner 5) -- (SIGMA.corner 4) node[black,pos=0.5,anchor=north east]{$d$};
\draw[ForestGreen,very thick,->-] (SIGMA.corner 6) -- (SIGMA.corner 5) node[black,pos=0.5,anchor=north]{$a$};
\draw[ForestGreen,very thick,->-] (SIGMA.corner 7) -- (SIGMA.corner 6) node[black,pos=0.5,anchor=north west]{$c$};
\draw[ForestGreen,very thick,->-] (SIGMA.corner 7) -- (SIGMA.corner 8) node[black,pos=0.5,anchor=west]{$b$};
\draw[ForestGreen,very thick,->-] (SIGMA.corner 1) -- (SIGMA.corner 8) node[black,pos=0.5,anchor=south west]{$d$};

\draw[RawSienna,very thick,->-] (SIGMA.corner 3) -- (SIGMA.corner 1) node[black,pos=0.5,anchor=north]{$e+f$};
\draw[RawSienna,very thick,->-] (SIGMA.corner 7) -- (SIGMA.corner 1) node[black,pos=0.5,anchor=east]{$f$};
\draw[RawSienna,very thick,->-] (SIGMA.corner 3) -- (SIGMA.corner 7) node[black,pos=0.5,anchor=north]{$e$};
\draw[RawSienna,very thick,->-] (SIGMA.corner 5) -- (SIGMA.corner 3) node[black,pos=0.5,anchor=west]{$h$};

\end{tikzpicture}
\caption{Left: the mapping torus $MT((S_{(0/1,1)})^3_3)$ with the embedded surface $\Sigma_{33}^S$. It is understood we are using the same convention of the other figures. Right: the polygon representation of $\Sigma_{33}^S$ and its generators. }
\label{fig:MTS33}
\end{figure}
In this case the change of coordinates in $\TT^2\times\{0\}\sim \TT^2\times \{1\}$ is 
\begin{equation}
    \begin{pmatrix}
    \partial'_\chi & \partial'_\theta
    \end{pmatrix} = \begin{pmatrix}
    \partial_\chi & \partial_\theta
    \end{pmatrix} \begin{pmatrix}
    0 & -1 \\ 1&0 \\
    \end{pmatrix}^T,
    \quad
    \begin{pmatrix}
    \theta'\\ \chi'
    \end{pmatrix} = \begin{pmatrix}
    0 & 1 \\ -1&0 \\
    \end{pmatrix} \begin{pmatrix}
    \theta\\ \chi
    \end{pmatrix},
\end{equation}
while the transition functions in $SU(2)$ for the identifications $(\theta+1,\chi) \sim (\theta,\chi)$ and $(\theta,\chi+1)\sim(\theta,\chi)$ are both $-\mathrm{id}$ for $(0,1)$ and $+\mathrm{id}$ for $(1,1)$.\\
We now compute the value of $q$ for the cycles $h,\, f$ and $e+f$.
\begin{itemize}
\let\labelitemi\labelitemii
    \item We can move along the cycle $h$ in $4$ steps. First we start at $x$ with coordinates $(\theta,\chi)=(1/2,0)$ and arrive to the point $y$ with $(\theta,\chi)=(0,1/2)$ by moving along $a^{-1}$ and doing a $\pi/2$ rotation around the $3^{\mathrm{rd}}$ axis $\partial_z$. Then at $y$ the cycle can be smoothed so that by following it we do a rotation of $\pi/2$ around $\partial_\chi$. We then go to $x'$ with coordinates $(\theta',\chi')=(0,1/2)$ along $d^{-1}$ by doing a $(2n+1)\pi$ rotation around $\partial_z$. Finally, we do a $\pi/2$ rotation around $\partial'_\chi$ before applying the transition function. Since $x'=(\theta=-1/2,\chi=0)$, in order to go completely back to the starting basis of the spin bundle on $x$ there is an additional $[-\mathrm{id}]\in SU(2)$ for the manifold $\{\ns1,\ns1\}$.\\
    The total transformation for a vector $v$ is
\begin{equation}
    v' = \underbrace{R_3(\pi/2)}_{\mathsf{T}_S}R_2(\pi/2)R_3((2n+1)\pi )R_2(\pi/2)R_3(\pi/2)v = v,
\end{equation}
so that the requirement of an even framing is equivalent to the condition 
\begin{equation}
    [-\mathrm{id}]\underbrace{(-\mathrm{id})}_{(-1)^\CF}\underbrace{ e^{-i \frac{\pi}{4}\sigma_3}}_{\tilde{ \mathsf{T}}_S}e^{-i\frac{\pi}{4}\sigma_2}e^{-i\frac{2n+1}{2}\pi \sigma_3}e^{-i\frac{\pi}{4}\sigma_2}e^{-i\frac{\pi}{4}\sigma_3} =(-\mathrm{id})^{n[+1]} \overset{!}{=} -\mathrm{id}.
\end{equation}
\item Dividing in similar steps the framing of cycle $f$, a generic $v\in T_x MT((S_{0/1,1})^2_2)$ transforms as 
\begin{equation}
    v' = \underbrace{R_3(\pi/2)}_{\mathsf{T}_S}R_2(-\pi/2)R_3((2n+1)\pi )R_2(-\pi/2)R_3(\pi/2)v = v.
    \label{fs22}
\end{equation}
With the aid of Figure \ref{fig:MTS33} it is clear that by starting at $x$ considered with coordinates\footnote{Here and in the following we will sometimes rewrite the same points with different coordinates, making use of the identification $(\theta,\chi)\sim(\theta+n,\chi+m)$ for $n,m\in\Z$. As long as one is consistent, this has no effect other than making easier the visualization of the path by looking at the mapping tori.} $(\theta=1/2, \chi=1)$, then we arrive at $x'$ with coordinates $(\theta',\chi')=(1,1/2)\sim(\theta,\chi)=(-1/2,1)$. Thus the lift of the full rotation in $\Spin(3)$ gets again an additional $[-\mathrm{id}]$ in the case $\{\ns1,\ns1\}$. Therefore the even framing condition for $f$ is
\begin{equation}
    [-\mathrm{id}]\underbrace{(-\mathrm{id})}_{(-1)^\CF}\underbrace{ e^{-i \frac{\pi}{4}\sigma_3}}_{\tilde{ \mathsf{T}}_S}e^{i\frac{\pi}{4}\sigma_2}e^{-i\frac{2n+1}{2}\pi \sigma_3}e^{i\frac{\pi}{4}\sigma_2}.e^{-i\frac{\pi}{4}\sigma_3} =(-\mathrm{id})^{n[+1]} \overset{!}{=} -\mathrm{id}.
\end{equation}
\item Finally we look at the cycle $e+f\sim cb^{-1}$. Starting at $x$ with coordinates $(\theta,\chi)=(1/2,0)$ a vector $v$ transforms as
\begin{equation}
    v'=\underbrace{R_3(\pi/2)}_{\mathsf{T}_S}R_2(-\pi/2)R_3(-\pi/2)R_1(-\pi/2)R_3(2 \pi n) = v.
\end{equation}
The arrival point of this cycle now is $x'$ with coordinates $(\theta',\chi')=(1,1/2)\sim(\theta,\chi)=(-1/2,1)$, so there is no additional sign to keep track of. Therefore in $\Spin(3)$ we have
\begin{equation}
    \underbrace{(-\mathrm{id})}_{(-1)^\CF}\underbrace{e^{-i \frac{\pi}{4}\sigma_3}}_{\tilde{ \mathsf{T}}_S}e^{i\frac{\pi}{4}\sigma_2}e^{i\frac{\pi}{4} \sigma_3}e^{-i\frac{\pi}{4}\sigma_1}e^{i\pi n \sigma_3} = (-\mathrm{id})^{n+1} \overset{!}{=} (-\mathrm{id}).
\end{equation}
\end{itemize}
From these results we see that for the class $(1,1)$ one must set $n=1$ for the loops $h$ and $f$. Thus, in both cases the normal bundle does three negative half twists with respect to the framing chosen, or, in other words, $q(f)=q(h)=-3\mod4=1\mod4$. Instead, $n=0$ is necessary for $e+f$, so that on this cycle the normal bundle does not do any twist and $q(e)=q(f)+2\mod4$.

By the same logic, for the class $(0,1)$ one must impose $n=0$ for all the loops $h,\,f$ and $e+f$. This means that in this case the enhancement has the values $q(h)=q(f)=3\mod4$ and $q(e)=q(f)+2\mod4$.\\
Therefore the final result is 
\begin{equation}
    (S_{(0,1)})^2_2 = e^{-i\frac{\pi}{4}\nu}, \qquad    (S_{(1,1)})^2_2 = e^{i \frac{\pi}{4}\nu}.
\end{equation}

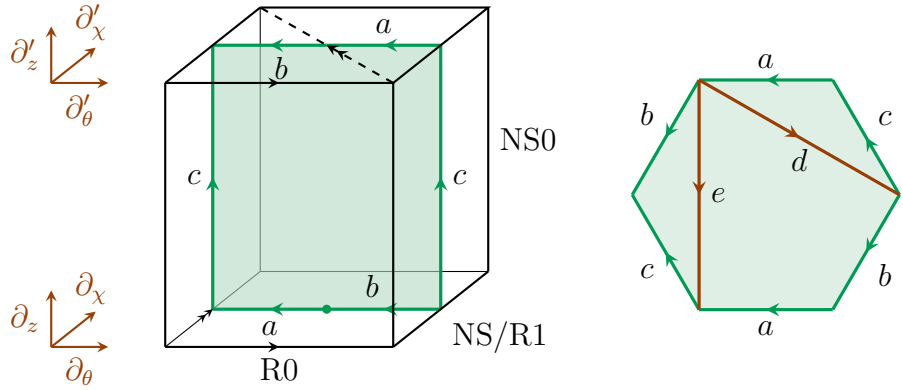
\begin{figure}[tb]
\centering
\begin{tikzpicture}
[scale=0.5,baseline=(current  bounding  box.center),>=stealth]
    \coordinate (P1) at (-3,-3);
    \coordinate (P2) at (3,-3);
    \coordinate (P3) at (-0.5,-1);
    \coordinate (P4) at (5.5,-1);
    \coordinate (P5) at (-3,4);    
    \coordinate (P6) at (3,4);
    \coordinate (P7) at (-0.5,6);
    \coordinate (P8) at (5.5,6);
    \coordinate (C1) at ($(P1)!0.5!(P4)$);
    \coordinate (C2) at ($(P5)!0.5!(P8)$);
    \coordinate (Q1) at ($(P1)!0.5!(P2)$);
    \coordinate (Q2) at ($(P2)!0.5!(P4)$);
    \coordinate (Q3) at ($(P1)!0.5!(P3)$);
    \coordinate (Q4) at ($(P3)!0.5!(P4)$);
    \coordinate (Q5) at ($(P5)!0.5!(P6)$);
    \coordinate (Q6) at ($(P6)!0.5!(P8)$);
    \coordinate (Q7) at ($(P5)!0.5!(P7)$);
    \coordinate (Q8) at ($(P7)!0.5!(P8)$);
    \coordinate (X) at ($(Q5)+(0,10)$);
    \coordinate (Y) at ($(Q4)+(0,-10)$);
    \coordinate (ypoint) at ($(Q2)!0.5!(Q3)$);
    \coordinate (xpoint) at ($(Q6)!0.5!(Q7)$);

    \coordinate (AxO) at (-6,-3);
    \coordinate (Ax1) at (-4.5,-3);
    \coordinate (Ax2) at ($(AxO)+(1.171,0.937)$);
    \coordinate (Ax3) at (-6,-1.5);
    
    \coordinate (AxO') at ($(AxO)+(0,7)$);
    \coordinate (Ax1') at ($(Ax1)+(0,7)$);
    \coordinate (Ax2') at ($(Ax2)+(0,7)$);
    \coordinate (Ax3') at ($(Ax3)+(0,7)$);    
    
    \draw[RawSienna,thick,->] (AxO) -- (Ax1) node[pos=0.5,anchor=north]{$\partial_\theta$};
    \draw[RawSienna,thick,->] (AxO) -- (Ax2) node[pos=0.9,anchor=south]{$\partial_\chi$};
    \draw[RawSienna,thick,->] (AxO) -- (Ax3) node[pos=0.5,anchor=east]{$\partial_z$};    
    
    \draw[RawSienna,thick,->] (AxO') -- (Ax1') node[pos=0.5,anchor=north]{$\partial'_\theta$};
    \draw[RawSienna,thick,->] (AxO') -- (Ax2') node[pos=0.9,anchor=south]{$\partial'_\chi$};
    \draw[RawSienna,thick,->] (AxO') -- (Ax3') node[pos=0.5,anchor=east]{$\partial'_z$};           
    
    \draw[thin] (P1) -- (P3) node[pos=0.5,sloped,rotate=-90]{\tikz\draw[->>,thin](0,0);};
    \draw[thin] (P3) -- (P4);
    \draw[thin] (P3) -- (P7);

    \path[fill=ForestGreen!30,fill opacity=0.6] (Q2) -- (Q3) -- (Q7) -- (Q6);
    \draw[ForestGreen,very thick,->-] (Q2) -- (ypoint) node[black,pos=0.6,anchor=south]{$b$};
    \draw[ForestGreen,very thick,->-] (ypoint) -- (Q3) node[black,pos=0.5,anchor=north]{$a$};
    \draw[ForestGreen,fill] (ypoint) circle (3pt);
    \draw[ForestGreen,very thick,->-] (Q3) -- (Q7) node[black,pos=0.5,anchor=east]{$c$};
    \draw[ForestGreen,very thick,->-] (Q6) -- (xpoint) node[black,pos=0.5,anchor=south]{$a$};
    \draw[ForestGreen,very thick,->-] (xpoint) -- (Q7) node[black,pos=0.4,anchor=north]{$b$};
    \draw[ForestGreen,very thick,->-] (Q2) -- (Q6) node[black,pos=0.5,anchor=west]{$c$};

    \draw[thick] (P1) -- (P2) node[pos=0.5,sloped,rotate=-90]{\tikz\draw[->,thick](0,0);} node[pos=0.5,anchor=north]{$\r 0$}; 
    \draw[thick] (P2) -- (P4) node[pos=0.5,anchor=north west]{$\ns/\r 1$};
    \draw[thick] (P5) -- (P6) node[pos=0.5,sloped,rotate=-90]{\tikz\draw[->,thick](0,0);};
    \draw[thick] (P6) -- (P8) -- (P7);
    \draw[thick] (P5) -- (P7);
    \draw[thick] (P1) -- (P5);
    \draw[thick] (P2) -- (P6);
    \draw[thick] (P4) -- (P8) node[pos=0.5,anchor=west]{$\ns 0$};
    \draw[thick,dashed] (P6) -- (P7) node[pos=0.5,sloped,rotate=90]{\tikz\draw[->>,thick](0,0);};
    
\end{tikzpicture}\qquad
\begin{tikzpicture}[scale=0.5,>=stealth,baseline=(current  bounding  box.center)]

\node[draw=none,minimum size=3.5cm,regular polygon,regular polygon sides=6,fill=ForestGreen!30,fill opacity=0.4] (SIGMA) {};

\draw[ForestGreen,very thick,->-] (SIGMA.corner 1) -- (SIGMA.corner 2) node[black,pos=0.5,anchor=south]{$a$};
\draw[ForestGreen,very thick,->-] (SIGMA.corner 2) -- (SIGMA.corner 3) node[black,pos=0.5,anchor=south east]{$b$};
\draw[ForestGreen,very thick,->-] (SIGMA.corner 4) -- (SIGMA.corner 3) node[black,pos=0.5,anchor=north east]{$c$};
\draw[ForestGreen,very thick,->-] (SIGMA.corner 5) -- (SIGMA.corner 4) node[black,pos=0.5,anchor=north]{$a$};
\draw[ForestGreen,very thick,->-] (SIGMA.corner 6) -- (SIGMA.corner 5) node[black,pos=0.5,anchor=north west]{$b$};
\draw[ForestGreen,very thick,->-] (SIGMA.corner 6) -- (SIGMA.corner 1) node[black,pos=0.5,anchor=south west]{$c$};

\draw[RawSienna,very thick,->-] (SIGMA.corner 2) -- (SIGMA.corner 4) node[black,pos=0.5,anchor=west]{$e$};
\draw[RawSienna,very thick,->-] (SIGMA.corner 2) -- (SIGMA.corner 6) node[black,pos=0.5,anchor=north]{$d$};

\end{tikzpicture}
\caption{Left: the mapping torus $MT((T_{(0/1,1)})^2_2)$ with the embedded surface $\Sigma_{22}^T$.\\ Right: the plain representation of $\Sigma_{22}^T$ and its generators.}
\label{fig:MTT22}
\end{figure}

The last two mapping tori we need to look at are the ones related to the entries $(T_{(0/1,1)})^2_2$ and $(T_{(0/1,1)})^3_1$. By applying \eqref{defmappingtori} it follows that the manifold describing the first of these two entries is given by the $\mathcal{T}$-bordism of $X_2$ with the identification of the boundaries $X_2\times\{0\}\sim X_2\times\{1\}$ like in the previous cases. We can see such manifold in Figure \ref{fig:MTT22}, where we depicted also $\Sigma^T_{22}$, found with the same reasoning of before. In this case the surface is just a torus represented by the string of $1$-cycles 
\begin{equation}
    acb^{-1}a^{-1}c^{-1}b\sim e^{-1}de d^{-1},
\end{equation}
where $d=a^{-1}c^{-1}$ and $e=bc^{-1}$ are a symplectic basis of $H^1(\Sigma^T_{(0/1,1)},\Z_2)$.
The next step is defining the transition function which changes the canonical basis of the tangent space given by the coordinates used in the top slice $X_2 \times \{1\}$ to the ones of the bottom slice $X_2 \times \{0\}$. Here the transition function will be lifted to elements of $\widetilde{\SL}(3,\mathbb{R})$, the double cover of $\SL(3,\mathbb{R})$. Under a $T$ transformation
\begin{equation}
\begin{pmatrix}
    \partial'_\chi & \partial'_\theta
    \end{pmatrix} = \begin{pmatrix}
    \partial_\chi & \partial_\theta
    \end{pmatrix} \begin{pmatrix}
    1 & 1 \\ 0&1 \\
    \end{pmatrix}^T,\quad
    \begin{pmatrix}
    \theta'\\ \chi'
    \end{pmatrix} = \begin{pmatrix}
    1 & -1 \\ 0&1 \\
    \end{pmatrix} \begin{pmatrix}
    \theta\\ \chi
    \end{pmatrix}.
    \label{ttransf}
\end{equation}
If we start from a vector $v \in T_p MT((T_{(0/1,1)})^2_2)$ with $p \in X_2 \times\{0\} \simeq X_2 \times \{1\}$ and write it in the  canonical basis of $X_2\times\{1\}$, then by applying $\mathsf{T}_T = H(1)$, where
\begin{equation}
 H(t):= \begin{pmatrix}
    1&t&0\\ 0&1&0\\ 0&0&1
    \end{pmatrix}
\end{equation}
we have:
\begin{equation}
\begin{pmatrix}
v_\theta\\ v_\chi\\ v_z
\end{pmatrix}= \begin{pmatrix}
1&1&0\\0&1&0\\0&0&1
\end{pmatrix}\begin{pmatrix}
v'_\theta\\ v'_\chi\\ v'_z
\end{pmatrix}.
\end{equation}
For the lift $\mathsf{T}_T\rightarrow\tilde{\mathsf{T}}_T\in\widetilde{\SL}(3,\mathbb{R})$ one notes that $\SL(3,\mathbb{R})$ can be continuously retracted to $SO(3)$ via the Gram-Schmidt procedure. If  $\gamma:\SL(3,\mathbb{R})\rightarrow SO(3)$ is such retraction, then its lifting $\tilde\gamma$ has to make the following diagram commute:
\begin{equation}
\begin{tikzcd}
\widetilde{\SL}(3,\mathbb{R}) \arrow{r}{\tilde{\gamma}} \arrow[swap]{d}{\pi} & SU(2) \arrow{d}{\pi} \\%
\SL(3,\mathbb{R}) \arrow{r}{\gamma}& SO(3)
\end{tikzcd}.
\end{equation}
Here $\pi$ denotes the projections. It follows that $\tilde{\gamma}$ maps the center $\{\pm \mathrm{id} \}$ to itself\footnote{The commutativity of the diagram would still be satisfied if the center of $\widetilde{\SL}(3,\mathbb{R})$ were mapped to the identity in $SU(2)$. However, one needs to require also surjectivity, otherwise $\tilde\gamma(\widetilde{\SL}(3,\mathbb{R}))= SO(3)$.}.
From the fact that $\gamma(H(t))=\mathrm{id}$ $\forall t\in \mathbb{R}$, then the lifting of the transition function can be  $\widetilde{\mathsf{T}}_{T}=\widetilde{H}_{\pm}(1)$, defined by the property $\tilde\gamma(\widetilde{H}_{\pm}(1))=\pm \mathrm{id} \in SU(2)$. For consistency with the choice made for $S$, we choose\footnote{The other choice is still legitimate, but would correspond to a redefinition $T\rightsquigarrow T (-1)^\CF$.} the lift $\widetilde{\mathsf{T}}_{T}=\widetilde{H}_+(1)$.\\
We can now turn to computing the value of $q$ for the cycles $d$ and $e$.
\begin{itemize}
\let\labelitemi\labelitemii
    \item For the loop $e$ we start from $x$, which has the coordinates $(\theta,\chi)=(0,1/2)$. We then go to a point $p$ in the middle of the segment $c$ connecting $x$ to $y=(\theta'=0,\chi'=1/2)$ while doing a $2 \pi n$ rotation along the direction $\partial_z$. From here we reach $y$ by doing a continuous transformation $H(-t)$ with $t\in [0,1]$. At this point the cycle can be smoothed so that by following it we do a $\pi$ rotation along the $\partial'_\chi$. Then $x'=(\theta'=1/2,\chi'=1/2)$ is reached without doing any rotation to the reference frame.
    Applying the transition function we are back to $x$ with the original basis of the tangent space. So taking everything into account we see that a vector $v$ goes to itself by
\begin{equation}
    v' = \underbrace{H(1)}_{\mathsf{T}_T}R_2(-\pi)R_2(\pi)H(-1)R_3(2 \pi n)v = v.
    \label{t11so3}
\end{equation}
The lift of this transformation is\footnote{In $\widetilde{\SL}(3,\mathbb{R})$ we formally represent elements of $SU(2)\subset \widetilde{\SL}(3,\mathbb{R})$ by their matrix representation in the fundamental of $SU(2)$.}
\begin{equation}
      \underbrace{(-\mathrm{id})}_{(-1)^\CF}\underbrace{\widetilde{H}_+(1)}_{\widetilde{\mathsf{T}}_T} e^{i \frac{\pi}{2}\sigma_2}e^{-i \frac{\pi}{2}\sigma_2}\widetilde{H}(-1)e^{- i n \pi} \xrightarrow{\tilde\gamma}\underbrace{\tilde\gamma (\widetilde{H}_-(1))}_{\tilde\gamma((-1)^\CF\widetilde{\mathsf{T}}_T)}\tilde\gamma(\widetilde{H}(-1)(-\mathrm{id})^n)= (-\mathrm{id})^{n+1}.
     \label{t11su2}
\end{equation} 
For the deformation contraction one has to group together all the components that define the transition function between the two bases of the tangent space of $X_2$, so in this case $(-1)^\CF$ and $\tilde{\mathsf{T}}_T$, and the transformations done by moving along the loop. The equality on the right side follows by remembering that $\widetilde{H}(-1)$ is the ending point of a continuous lift $\widetilde{H}(-t)$ with $t\in[0,1]$, $\widetilde{H}(0)=\mathrm{id}$ and using also the fact that for any element $A\in\widetilde{\SL}(3,\R)$ it holds $\tilde{\gamma}(AB)=\tilde{\gamma}(A)\tilde{\gamma}(B)$ if $B\in Z(\widetilde{\SL}(3,\R))$. At this point we can simply set $n=0$ to have an even framing, which means the normal bundle does not twist with respect to it and $q(e)=0\mod4$.
\item With the same approach along the cycle $c$ any vector $v$ rotates by
\begin{equation}
    \underbrace{H(1)}_{\mathsf{T}_T}R_2(\pi)R_2(-\pi)H(-1)R_3(2\pi n) = \mathrm{id}.
\end{equation}
The deformation contraction of its lift gives again the same element of \eqref{t11su2}, so we can conclude that $q(d)=q(e)$.
\end{itemize}
This means that for both classes $(0/1,1)$ we have
\begin{equation}
    (T_{(0,1)})^2_2=(T_{(1,1)})^2_2=1.
\end{equation}

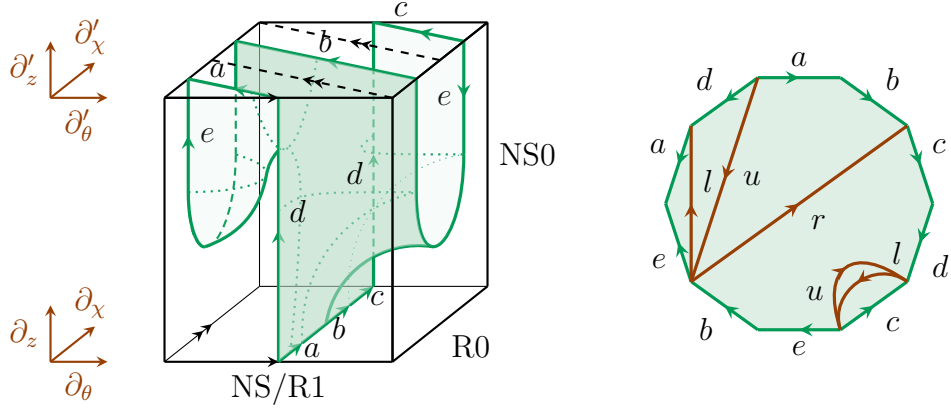
\begin{figure}
    \centering
    \begin{tikzpicture}[scale=0.5,>=stealth,baseline=(current  bounding  box.center)]
    \coordinate (P1) at (-3,-3);
    \coordinate (P2) at (3,-3);
    \coordinate (P3) at (3,4);
    \coordinate (P4) at (-3,4);
    \coordinate (P5) at (-0.5,-1);
    \coordinate (P6) at (5.5,-1);
    \coordinate (P7) at (5.5,6);
    \coordinate (P8) at (-0.5,6);
    \coordinate (R1) at (-3,1);
    \coordinate (R2) at (3,1);
    \coordinate (R3) at (-0.5,3);
    \coordinate (R4) at (5.5,3);
    \coordinate (R5) at (-2.375,1.5);
    \coordinate (R6) at (-1.125,2.5);
    \coordinate (R7) at (3.625,1.5);
    \coordinate (R8) at (4.875,2.5);
    \coordinate (R9) at (0,1);
    \coordinate (R10) at (2.5,3);
    \coordinate (A1) at (4.25,0);
    \coordinate (A2) at (-1.75,0);
    \coordinate (C1) at (0.625,1.5);
    \coordinate (C2) at (1.875,2.5);
    \coordinate (C3) at (-0.5,1.5);
    \coordinate (C4) at (0.5,2);
    \coordinate (C5) at (-2.375,-1);
    \coordinate (C6) at (-1.125,0);
    \coordinate (C7) at (3.625,-1);
    \coordinate (C8) at (4.875,0);
    \coordinate (C9) at (-0.1,3);
    \coordinate (C10) at (0.6,3);
    \coordinate (A3) at (-0.3,1.75);
    \coordinate (U1) at (-2.375,0.5);
    \coordinate (U2) at (-2.1,0);
    \coordinate (U3) at (3.625,0.5);
    \coordinate (U4) at (3.85,0);
    \coordinate (U5) at (4.55,0);
    \coordinate (U6) at (4.875,1.5);
    
    \coordinate (AxO) at (-6,-3);
    \coordinate (Ax1) at (-4.5,-3);
    \coordinate (Ax2) at ($(AxO)+(1.171,0.937)$);
    \coordinate (Ax3) at (-6,-1.5);
    
    \coordinate (AxO') at ($(AxO)+(0,7)$);
    \coordinate (Ax1') at ($(Ax1)+(0,7)$);
    \coordinate (Ax2') at ($(Ax2)+(0,7)$);
    \coordinate (Ax3') at ($(Ax3)+(0,7)$);    
    
    \draw[RawSienna,thick,->] (AxO) -- (Ax1) node[pos=0.5,anchor=north]{$\partial_\theta$};
    \draw[RawSienna,thick,->] (AxO) -- (Ax2) node[pos=0.9,anchor=south]{$\partial_\chi$};
    \draw[RawSienna,thick,->] (AxO) -- (Ax3) node[pos=0.5,anchor=east]{$\partial_z$};    
    
    \draw[RawSienna,thick,->] (AxO') -- (Ax1') node[pos=0.5,anchor=north]{$\partial'_\theta$};
    \draw[RawSienna,thick,->] (AxO') -- (Ax2') node[pos=0.9,anchor=south]{$\partial'_\chi$};
    \draw[RawSienna,thick,->] (AxO') -- (Ax3') node[pos=0.5,anchor=east]{$\partial'_z$};               
    
    \draw[thick] (P1) -- (P2) coordinate[pos=0.5](Q1) node[pos=0.5,sloped,rotate=-90]{\tikz\draw[->,thick](0,0);} node[pos=0.5,anchor=north]{$\ns/\r 1$}; 
    \draw[thick] (P2) -- (P3);
    \draw[thick] (P3) -- (P4) coordinate[pos=0.5](Q2);
    \draw[thick] (P4) -- (P1);
    \draw[thick] (P4) -- (P8) coordinate[pos=0.25](Q5) coordinate[pos=0.75](Q6);
    \draw[thick] (P7) -- (P8) coordinate[pos=0.5](Q4);
    \draw[thick] (P7) -- (P6) node[pos=0.5,anchor=west]{$\ns 0$};
    \draw[thick] (P6) -- (P2) node[pos=0.5,anchor=north west]{$\r 0$};
    \draw[thick] (P3) -- (P7) coordinate[pos=0.25](Q7) coordinate[pos=0.75](Q8);
    \draw[thin] (P8) -- (P5); 
    \draw[thin] (P5) -- (P6) coordinate[pos=0.5](Q3);
    \draw[thin] (P1) -- (P5) node[pos=0.5,sloped,rotate=-90]{\tikz\draw[->>,thick](0,0);};;
    \draw[ForestGreen,very thick,->-,name path=Q1Q2] (Q1) -- (Q2);
    \draw[ForestGreen,very thick,->-, name path=Q2Q5] (Q2) -- (Q5) coordinate [pos=0.3](W1) node[black,pos=0.45,anchor=south east]{$a$}; 
    \path[ForestGreen] (Q1) -- (Q3) coordinate[pos=0.25](W3) coordinate[pos=0.5](W4) coordinate[pos=0.75](W5) coordinate[pos=0.125](W6)
    node[black,pos=0.15,anchor=west]{$a$} node[black,pos=0.45,anchor=west]{$b$} node[black,pos=0.85,anchor=west]{$c$};
    \draw[ForestGreen,very thick,->] (Q1) -- (W3);
    \draw[ForestGreen,very thick,->] (W3) -- (W5);
    \draw[ForestGreen,very thick,->] (W5) -- (Q3);
    \draw[ForestGreen,thick,densely dashed,->-, name path=Q3Q4] (Q3) -- (Q4); 
    \draw[ForestGreen,very thick,->-] (Q8) -- (Q4) node[black,pos=0.7,anchor=south]{$c$}; 
    \draw[ForestGreen,very thick,->-] (Q8) -- (R8) node[black,pos=0.5,anchor=east]{$e$}; 
    \draw[ForestGreen,very thick] (R7) -- (Q7);  
    \draw[ForestGreen,very thick,->-, name path=Q6Q7] (Q7) -- (Q6) coordinate[pos=0.55](W2) node[black,pos=0.4,anchor=south east]{$b$}; 
    \draw[ForestGreen,thick,densely dashed, name path=Q6R6] (Q6) -- (R6); 
    \draw[ForestGreen,very thick,->-] (R5) -- (Q5) node[black,pos=0.5,anchor=west]{$e$};
    \draw[ForestGreen,thick,dotted] (R9) .. controls (C1) .. (R7) coordinate[pos=0.35](W7);
    \draw[ForestGreen,thick,dotted] (R10) .. controls (C2) .. (R8);
    \draw[ForestGreen,very thick] (R8) .. controls (C8) and (C7) .. (R7) coordinate[pos=0.6](A1); 
    \draw[ForestGreen,thick, densely dashed] (R6) .. controls (C6) and (C5) .. (R5) coordinate[pos=0.6](A2);
    \draw[ForestGreen,thick, dotted] (R5) .. controls (C3) and (C4) .. (R6) coordinate[pos=0.6](A3);
    \draw[ForestGreen,very thick] (A2) to [bend right=30] (A3);
    \draw[ForestGreen,thick, dotted, name path=A3W7] (A3) .. controls (C9) and (C10) .. (W7) coordinate[pos=0.5](A4);
    \draw[ForestGreen,thick, dotted] (W6) to [bend right=2] (W7);
    \draw[ForestGreen,very thick, name path=A1W4] (A1) to [bend right=45] (W4);
    \draw[ForestGreen,thick, dotted] (W3) to [bend left=40] (R7);
    \draw[ForestGreen,thin, dotted] (W5) to [bend left=45] (R8);
    \draw[ForestGreen,thick, dotted] plot [smooth, tension=1] coordinates {(W1) (A4) (W2)};
    \path[name intersections={of=Q1Q2 and A3W7,by=A5}];
    \path[name intersections={of=Q3Q4 and Q6Q7,by=A6}];
    \path[name intersections={of=Q3Q4 and A1W4,by=A7}];
    \path[name intersections={of=Q2Q5 and Q6R6,by=A8}];
    \draw[ForestGreen, very thick] (A3) to [bend left=15] (A5);
    \draw[ForestGreen, very thin,fill=ForestGreen!30,fill opacity=0.1] (R5) .. controls (U1) and (U2) .. (A2) to [bend right=30] (A3) to [bend left=15] (A5) --(Q2) -- (Q5) -- (R5);
    \draw[ForestGreen, very thin,fill=ForestGreen!30,fill opacity=0.6] (Q1) -- (Q3) -- (A7) to [bend left=15] (A1) .. controls (U4) and (U3) .. (R7) -- (Q7) -- (Q6) -- (A8) -- (Q2) -- (Q1);
    \draw[ForestGreen, very thin,fill=ForestGreen!30,fill opacity=0.1] (A1) .. controls (U4) and (U3) .. (R7) -- (Q7) -- (A6) -- (Q4) -- (Q8) -- (R8) .. controls (U6) and (U5) .. (A1);
    \draw[ForestGreen,very thick] (Q3) -- (A7);
    \draw[ForestGreen,very thick] (Q4) -- (A6);
    \draw[ForestGreen,very thick] (R5) .. controls (U1) and (U2) .. (A2);
    \draw[ForestGreen,very thick] (Q6) -- (A8);
    
    \draw[thick,dashed] ($(P4)!0.5!(P8)$) -- (P3) node[pos=0.5,sloped,rotate=90]{\tikz\draw[->>,thick](0,0);};
    \draw[thick,dashed] ($(P3)!0.5!(P7)$) -- (P8) node[pos=0.5,sloped,rotate=90]{\tikz\draw[->>,thick](0,0);};
    
    \node[right] at (0,1) {$d$};
    \node at (2,2) {$d$};    
    \draw[thick,->-] (P4) -- (P3);
    \draw[thick] (P3) -- (P7);
    \draw[thick] (P3) -- (P2);

\end{tikzpicture} \qquad
\begin{tikzpicture}[scale=0.5,>=stealth,baseline=(current  bounding  box.center)]

\node[draw=none,minimum size=3.5cm,regular polygon,regular polygon sides=10,fill=ForestGreen!30,fill opacity=0.4] (SIGMA) {};

\draw[ForestGreen,very thick,->-] (SIGMA.corner 2) -- (SIGMA.corner 1) node[black,pos=0.5,anchor=south]{$a$};
\draw[ForestGreen,very thick,->-] (SIGMA.corner 2) -- (SIGMA.corner 3) node[black,pos=0.5,anchor=south east]{$d$};
\draw[ForestGreen,very thick,->-] (SIGMA.corner 3) -- (SIGMA.corner 4) node[black,pos=0.5,anchor=south east]{$a$};
\draw[ForestGreen,very thick,->-] (SIGMA.corner 5) -- (SIGMA.corner 4) node[black,pos=0.5,anchor=north east]{$e$};
\draw[ForestGreen,very thick,->-] (SIGMA.corner 6) -- (SIGMA.corner 5) node[black,pos=0.5,anchor=north east]{$b$};
\draw[ForestGreen,very thick,->-] (SIGMA.corner 7) -- (SIGMA.corner 6) node[black,pos=0.5,anchor=north]{$e$};
\draw[ForestGreen,very thick,->-] (SIGMA.corner 7) -- (SIGMA.corner 8) node[black,pos=0.5,anchor=north west]{$c$};
\draw[ForestGreen,very thick,->-] (SIGMA.corner 9) -- (SIGMA.corner 8) node[black,pos=0.5,anchor=north west]{$d$};
\draw[ForestGreen,very thick,->-] (SIGMA.corner 10) -- (SIGMA.corner 9) node[black,pos=0.5,anchor=south west]{$c$};
\draw[ForestGreen,very thick,->-] (SIGMA.corner 1) -- (SIGMA.corner 10) node[black,pos=0.5,anchor=south west]{$b$};

\draw[RawSienna,very thick,->-] (SIGMA.corner 5) -- (SIGMA.corner 10) node[black,pos=0.5,anchor=north west]{$r$};
\draw[RawSienna,very thick,->-] (SIGMA.corner 2) -- (SIGMA.corner 5) node[black,pos=0.4,anchor=north west]{$u$};
\draw[RawSienna,very thick,->-] (SIGMA.corner 5) -- (SIGMA.corner 3) node[black,pos=0.5,anchor=south west]{$l$};

\coordinate (U1) at ($(0,0)!0.5!(SIGMA.corner 7)$);
\coordinate (U2) at ($(0,0)!0.5!(SIGMA.corner 8)$);
\coordinate (L1) at ($(0,0)!0.7!(SIGMA.corner 7)$);
\coordinate (L2) at ($(0,0)!0.7!(SIGMA.corner 8)$);

\draw[RawSienna,very thick,->-] (SIGMA.corner 7) .. controls (U1) and (U2) .. (SIGMA.corner 8) node[black,pos=0.1,anchor=south east]{$u$};
\draw[RawSienna,very thick,->-] (SIGMA.corner 8) .. controls (L2) and (L1) .. (SIGMA.corner 7) node[black,pos=0.1,anchor=south]{$l$};

\end{tikzpicture}
\caption{Left: the mapping torus $MT((T_{(0/1,1)}^T)^3_1)$ with the embedded surface $\Sigma_{31}^T$.\\ Right: the polygon representation of $\Sigma_{31}^T$ and its generators.}
\label{fig:MTT31}
\end{figure}

Finally, we are left with the computation of $(T_{(0/1,1)})^3_1$. In this case the $3$-manifold is given by identifying the two boundaries $X_1$ of the $T^2$-bordism $\mathcal{T}_{\{\r0,\ns/\r1\}}^{\{\r0,\ns/\r1\}}:\{\r0,\ns/\r1\}\rightarrow\{\r0,\ns/\r1\}$. The surface $\Sigma_{13}^T$ is found by looking for a smooth interpolation between its projection on $X_1\times\{0\}$ and $X_1\times\{1\}$. A schematic representation is given by 
\begin{equation}
    \begin{tikzpicture}[scale=0.4,baseline=(current  bounding  box.center)]
    \draw (0,0) rectangle (4,4);
    \draw[ForestGreen,thick] (2,0) -- (2,4);
    \node[below] at (2,0) {$X_1\times\{0\}$};    
    \end{tikzpicture}    \quad \rightarrow \quad
    \begin{tikzpicture}[scale=0.4,baseline=(current  bounding  box.center)]
    \draw (0,0) rectangle (4,4);
    \draw[ForestGreen,thick] (2,0) to[out=90,in=180] (4,1);
    \draw[ForestGreen,thick] (2,4) to[out=-90,in=180] (4,3);
    \draw[ForestGreen,thick] (0,3) to[out=0,in=90] (2,2); 
    \draw[ForestGreen,thick] (0,1) to[out=0,in=-90] (2,2);
    \node[below] at (2,0) {$X_1\times\{t\}$};        
    \end{tikzpicture} \quad \rightarrow \quad
    \begin{tikzpicture}[scale=0.4,baseline=(current  bounding  box.center)]
    \draw (0,0) rectangle (4,4);
    \draw[ForestGreen,thick] (2,0) to[out=90,in=0] (0,1);
    \draw[ForestGreen,thick] (2,4) to[out=-90,in=180] (4,3);
    \draw[ForestGreen,thick] (0,3) to[out=0,in=90] (2,2); 
    \draw[ForestGreen,thick] (4,1) to[out=180,in=-90] (2,2);   
    \node[below] at (2,0) {$X_1\times\{t'\}$};        
    \end{tikzpicture}\quad \rightarrow \quad
    \begin{tikzpicture}[scale=0.4,baseline=(current  bounding  box.center)]
    \draw (0,0) rectangle (4,4);
    \draw[ForestGreen,thick] (2,0) -- (0,1);
    \draw[ForestGreen,thick] (2,4) -- (4,3);
    \draw[ForestGreen,thick] (0,3) -- (2,2); 
    \draw[ForestGreen,thick] (4,1) -- (2,2);     
    \node[below] at (2,0) {$X_1\times\{1\}$};       
    \end{tikzpicture}\;
\end{equation}
with $0<t<t'<1$, while in Figure \ref{fig:MTT31} we can see its embedding in the $3$-manifold. As usual we can use some chains to write the string presentation of the surface, which in this case is 
\begin{equation}
    abcdc^{-1}ebea^{-1}d^{-1} \sim rr uu ll,
\end{equation}
with $r=abea^{-1}d^{-1},\,u=dae^{-1}c$ and $l=c^{-1}ea^{-1}$. 
Therefore $\Sigma^T_{13}=\mathbb{RP}^2_{r}\#\mathbb{RP}^2_{u}\#\mathbb{RP}^2_{l}$.\\ To find the value of its $\ABK$ invariant we compute the enhancement $q$ for $l,\,l+u$ and $l+u+r$. For our purpose it comes at hand to consider the portions of these cycles on $X_1\times \{1\}$ as if they were instead on the slice $X_1\times \{t'\}$. This does not pose any problem, because the value of $q$ is robust under any continuous deformation of the surface.
\begin{itemize}
\let\labelitemi\labelitemii
    \item We begin with $l\sim a e^{-1}c$, which we divide as usual in various pieces. From the starting point $p$ on $X_1\times\{1\}$ with coordinates $(\theta',\chi')=(1/2,0)$ we go to $q=(\theta'=0,\chi'=1/4)$ while doing a $\pi/2$ rotation around $\partial_z$. At $q$ we smooth the cycle and do a $-\pi/2$ rotation around $\partial'_\chi$. From $q$ to $r=(\theta'=0,\chi'=3/4)$ along $e^{-1}$ we do first a $\pi$ rotation around the $1^{\mathrm{st}}$ axis describing a u-turn, followed by a $(2n+1)\pi$ rotation around $\partial_z$. Like before, at $r$ we have to do a second $-\pi/2$ rotation around $\partial'_\chi$ and, while going back to $p$, one last $-\pi/2$ rotation around $\partial'_z$. This means that $v\in T_p MT((T_{0/1,1})^3_1)$ transforms as 
\begin{equation}
    v'= R_3(-\pi/2)R_2(-\pi/2)R_3((2n+1)\pi)R_1(\pi)R_2(-\pi/2)R_3(\pi/2)v=v.
    \label{ltr2}
\end{equation}
The lift is given by
\begin{equation}
    [-\mathrm{id}]e^{+i\frac{\pi}{4}\sigma_3}e^{i\frac{\pi}{4}\sigma_2}e^{-i\frac{(2n+1)\pi}{2}\sigma_3}e^{-i\frac{\pi}{2}\sigma_1}e^{i\frac{\pi}{4}\sigma_2}e^{-i\frac{\pi}{4}\sigma_3}=(-\mathrm{id})^{n[+1]},
\end{equation}
where the first $[-\mathrm{id}]$ is present only for the bordism class $(0,1)$. This means that $n=1$ and $q(l)=-3\mod4=1\mod4$ for $(1,1)$ and $n=0,\,q(l)=3\mod4$ for $(0,1)$.
\item For the cycle $l+u+r\sim e^{-1}b^{-1}$ we start from $q=(\theta'=0,\chi'=1/4)$ on $X_1\times\{1\}$. Here running along $e^{-1}$ we do the same rotations as before. The difference in the computation is that once we arrive at the point $r$ we have instead to do a $\pi/2$ rotation around $\partial'_\chi$, then go back to $q$  and conclude repeating doing a second $\pi/2$ rotation. The final result is that from $v \in T_q MT((T_{0/1,1})^3_1)$ we arrive to 
\begin{equation}
    v'=R_2(\pi/2)R_2(\pi/2)R_3((2n+1)\pi)R_1(\pi)v=v.
\end{equation}
This is essentially the exact expression of \eqref{ltr2} up to changing the orientation of the rotations around the $2^{\mathrm{nd}}$ axis, so it is no surprise that the lift in $SU(2)$ gives $(-\mathrm{id})^{n+1[+1]}$. Therefore  $q(l+u+r)=3\mod4$ for $(1,1)$ and $q(l+u+r)=1\mod4$ for $(0,1)$.
\item Finally it is the turn of $l+u\sim d$. In this case going from the $X_1\times\{0\}$ to $X_1\times\{1\}$ we do a $2\pi n$ rotation around $\partial_z$, followed by a continuous transformation $H(-2t),\, t\in[0,1]$. Thus a vector transforms as 
\begin{equation}
v'= \underbrace{H(1)H(1)}_{\mathsf{T}_{T^2}}H(-2)R_3(2\pi n)v = v. 
\end{equation}
The lift of the transformation in $\widetilde{\SL}(3,\mathbb{R})$ is then
\begin{equation}
    \underbrace{(-\mathrm{id})}_{(-1)^\CF}\widetilde{H}_+(1)\widetilde{H}_+(1)\widetilde{H}(-2)e^{-i\pi n \sigma_3}=(-\mathrm{id})^{n+1}.
\end{equation}
We conclude that $n=0$ and $q(l+u)=0\mod4$.
\end{itemize}
With these explicit computations one can arrive to the last results that determine the $T$ matrix, i.e.
\begin{equation}
    (T_{(0,1)})^3_1 =e^{i\frac{\pi}{4}\nu},\qquad (T_{(1,1)})^3_1 =e^{-i\frac{\pi}{4}\nu}.
\end{equation}

\section{Details on modular bootstrap}
\label{app:bootstrap}

Here we review in detail the linear functional method used to find the bounds presented in section \ref{subsec:results}, based on the original papers \cite{Friedan_2013,Collier:2016cls}. We also present the free fermion CFTs that almost saturate the bounds found for the kinks of Figures \ref{fig:J1}, \ref{fig:J3}.

In order to apply the linear functional method to \eqref{eq:crossingequation} one needs to expand the partition functions in terms of Virasoro characters:
\begin{equation}
    \sum_{h,\bar{h},j} n_{h,\bar{h},j}\left(\delta_i^j \chi_h(-1/\tau)\bar{\chi}_{\bar{h}}(-1/\bar{\tau})-S^j_i\chi_h(\tau)\bar{\chi}_{\bar{h}}(\bar{\tau})\right)=0,\qquad \forall\,i.
\end{equation}
Recall that in the proper basis the degeneracies $n_{h,\bar{h},j}$ are positive. Then, by defining $\vec{M}^j$ the vectors with entries
\begin{equation}
    M^j_i = \delta_i^j \chi_h(-1/\tau)\bar{\chi}_{\bar{h}}(-1/\bar{\tau})-S^j_i\chi_h(\tau)\bar{\chi}_{\bar{h}}(\bar{\tau}),
\end{equation}
we simply look for a functional $\alpha$ which returns a real function of $\Delta$ and $s$ that satisfies
\begin{equation}
\begin{cases}
    \alpha[\vec{M}^1](0,0)>0,&\\
    \alpha[\vec{M}^j](|s|,s)\ge 0,&\qquad \forall j,\,s\ne 0,\\
    \alpha[\vec{M}^j](\Delta,s)\ge 0,&\qquad \forall j,\,s,\, \Delta>\Delta^*_j.\\
\end{cases}
\label{eq:functionalconditions}
\end{equation}
Here $s$ is supposed to be any admitted value of the spins on $\mathcal{H}_j$, with the exception of $s=0$ for the second condition if it is a priori allowed in the corresponding $\mathcal{H}_j$. Instead $\Delta^*_j$ will vary depending on the bound we want to find.

The first condition asks for the functional to be strictly positive when evaluated on the vacuum, so that by applying it on the modular crossing equation always guarantees to rule out the corresponding spectra.

In our analysis we will consider a derivative basis for it around the point $\tau=i$. By introducing the variable $z$ such that $\tau=i \exp z$ we can then expand it as follows:
\begin{equation}
    \alpha[\vec{M}^j](\Delta,s)= \sum_{n,m,i} \gamma_{n,m,i} \partial_z^n \partial_{\bar{z}}^m M^j_i\rvert_{z=\bar{z}=0}.
\end{equation}
We now turn into explaining what are the values $\Delta^*_j$ for which we try to find a functional $\alpha$ that satisfies \eqref{eq:functionalconditions}:
\begin{enumerate}
    \item In order to find the lightest non-degenerate primary on the spectrum of $\mathcal{H}_{j_0}$, the definition is
    \begin{equation}
        \Delta^*_j=
        \begin{cases} 
        \mathrm{max}\left(\Delta^{j_0}_\gap,|s|\right),&\mathrm{if}\, j = j_0,\\
        |s|,&\mathrm{otherwise}.
        \end{cases} 
    \end{equation}
    \item In order to find the same bound, but for scalar primaries, we have
    \begin{equation}
        \Delta^*_j=
        \begin{cases} 
        \Delta^{j_0}_\scal,&\mathrm{if}\,j=j_0,\,s=0,\\
        |s|,&\mathrm{otherwise}.
        \end{cases}
    \end{equation}    
\end{enumerate}
We mention here that numerically it is convenient to allow for the possibility of having the limit $\Delta^j_{\mathrm{gap/scal}}\rightarrow |s|$ for non-degenerate Virasoro characters. In this case the contribution to the partition function (assuming for example $s>0$) is 
\begin{equation}
    \lim_{\Delta\rightarrow s} \chi_{\frac{\Delta+s}{2}}(\tau) \bar{\chi}_{\frac{\Delta-s}{2}}(\bar{\tau})= \chi_s(\tau)(\bar{\chi}_0(\bar{\tau})+\bar{\chi}_1(\bar{\tau})),
\end{equation}
i.e. it is equivalent to the contribution of a primary with $\Delta=s+1$ and a conserved current. We call this particular limit case a non-degenerate Virasoro character of generic type.

With these premises, the algorithm that determines the bound on the primaries is simple: one starts with some fixed value of $\Delta^{j_0}_{\gap/\scal}$ and then searches for a functional that has the properties discussed. Whether one exists or not then determines if the bound considered can be lowered or raised. The procedure in then repeated until one reaches the wanted precision.

To numerically implement this search, we must truncate the basis of the linear functional up to some derivative order $\Lambda$, i.e. $n+m\le \Lambda$. For us the choice will be set at $\Lambda=10$.

Moreover, we note that the usual procedure to utilize parity invariant partition functions is of no use here, with the exception of when $\nu=0,4\mod8$. Indeed, these are the only values for which the spin selection rules for the sectors $\mathcal{H}_i$ are symmetric under the reflection $s\mapsto -s$. Thus, if for other values of $\nu$ we were to restrict to parity invariant partition functions, the bounds we would find would not be as strict as possible, since we would mix $\tilde{Z}_i$ and $\tilde{\bar{Z}}_i$, that have different spin selection rules.
This means that computationally we are able to reduce the basis of functionals to a symmetric one only for the aforementioned cases $\nu=0,4\mod8$. 

Finally, we recall that by performing the computations with the SDPB solver, while we are actually able to consider for each $i$ a continuum spectrum of operators with $\Delta > \Delta^*_i$, we can instead enforce the conditions \eqref{eq:functionalconditions} only up to some value of spin $s\le s_{\mathrm{max}}$. Albeit a priori this might be a significant problem, one usually finds that at fixed $\Lambda$ for a reasonable value of $s_{\mathrm{max}}$ numerical stability is reached and the bounds converge within some error $\Delta_\delta$. Of course $s_{\mathrm{max}}$ will vary depending on the precision we want to meet, but generally is found to be of the same order of magnitude of $\Lambda$. In our case we decided to set $\Delta_\delta=0.01$, for which  numerical stability is reached by considering values of the spins up to $s_{\mathrm{max}}=40$.\\
We report also the other relevant parameters with which the numerical analysis has been performed, i.e.
\begin{table}[h]
\centering
\begin{tabular}{l|r}
\texttt{precision}     &  $700$\\
\texttt{primalErrorThreshold}     & $10^{-30}$ \\
\texttt{dualErrorThreshold}     & $10^{-30}$ \\
\texttt{maxComplementarity} & $10^{100}$\\
\texttt{feasibleCenteringParameter} & $0.1$\\
\texttt{infeasibleCenteringParameter} & $0.3$\\
\texttt{stepLengthReduction} & $0.7$
\end{tabular}
\end{table}

\subsection{Free fermions kinks}
Here we recall some basic facts about stacks of free fermion CFTs, in order to show that these theories saturate the bounds at the kinks found from the numerical analysis of section \ref{subsec:results}.

A generic stack of Majorana fermions 
\begin{equation}
    \mathcal{L} = \sum_{a=1}^n \psi_a \bar{\partial} \psi_a +  \bar{\psi}_a \partial \bar{\psi}_a 
\end{equation}
describes a spin-CFT with value of central charge $c=n/2$.
For us it is sufficient to consider fermions on a torus with periodicity conditions
\begin{equation}
    \psi(\theta +1,\chi)=e^{2\pi i \alpha}\psi(\theta,\chi),\qquad\psi(\theta ,\chi+1)=e^{2\pi i \beta}\psi(\theta,\chi),
\end{equation}
where $\alpha,\,\beta =0,1/2$ are the sum$\mod 1$ of the periodicity conditions defined by the spin structure and the $\Z_2$ global symmetry. We can treat these together as an element in $H^1(\mathbb{T}^2,\Z_2)$. Then the holomorphic partition function contribution of each Majorana fermion is of the form\footnote{Note that the square root of Jacobi functions presents branch-cuts, so after modular transformation one has to pay attention on how to move between different sheets. This behaviour takes place only when we have unpaired Majorana fermions, which, as we will see, usually contribute to the total anomaly by a $1\mod 8$ factor. Indeed, this precisely reflects the fact that $\{(-1)^F,(-1)^Q\}=0$ on $\mathcal{H}_{\r0}$ when $\nu=1\mod 2$.}
\begin{equation}
\begin{aligned}
   &\begin{tikzpicture}[scale=0.3,baseline= {($(current bounding box.base)-(-5pt,-5pt)$)}]
    \draw (0,0) rectangle (2,2);
    \node[left] at (0,1) {$0$};
    \node[below] at (1,0) {$0$};
   \end{tikzpicture}&= \sqrt{\frac{\vartheta_1(\tau)}{\eta(\tau)}},
   &\qquad
  \begin{tikzpicture}[scale=0.3,baseline= {($(current bounding box.base)-(-5pt,-5pt)$)}]
    \draw (0,0) rectangle (2,2);
    \node[left] at (0,1) {$0$};
    \node[below] at (1,0) {$1$};
   \end{tikzpicture}&=\sqrt{\frac{\vartheta_4(\tau)}{\eta(\tau)}},
   \\
   &\begin{tikzpicture}[scale=0.3,baseline= {($(current bounding box.base)-(-5pt,-5pt)$)}]
    \draw (0,0) rectangle (2,2);
    \node[left] at (0,1) {$1$};
    \node[below] at (1,0) {$0$};
   \end{tikzpicture}&= \sqrt{\frac{\vartheta_2(\tau)}{\eta(\tau)}},
   &\qquad
  \begin{tikzpicture}[scale=0.3,baseline= {($(current bounding box.base)-(-5pt,-5pt)$)}]
    \draw (0,0) rectangle (2,2);
    \node[left] at (0,1) {$1$};
    \node[below] at (1,0) {$1$};
   \end{tikzpicture}&=\sqrt{\frac{\vartheta_3(\tau)}{\eta(\tau)}},
  \end{aligned}
\end{equation}
where $\vartheta_i(\tau)$ are the Jacobi Theta functions. We remember they satisfy the following identities:
\begin{align}
    \begin{aligned}
    &\vartheta_1(\tau)=0,\\
    &\frac{\vartheta_2(-1/\tau)}{\eta(-1/\tau)}= \frac{\vartheta_4(\tau)}{\eta(\tau)}, &&\quad\frac{\vartheta_2(\tau+1)}{\eta(\tau+1)}= e^{i\pi/6}\frac{\vartheta_2(\tau)}{\eta(\tau)},\\
    &\frac{\vartheta_3(-1/\tau)}{\eta(-1/\tau)}= \frac{\vartheta_3(\tau)}{\eta(\tau)}, &&\quad\frac{\vartheta_3(\tau+1)}{\eta(\tau+1)}= e^{-i\pi/12}\frac{\vartheta_4(\tau)}{\eta(\tau)},\\
    &\frac{\vartheta_4(-1/\tau)}{\eta(-1/\tau)}= \frac{\vartheta_2(\tau)}{\eta(\tau)}, &&\quad\frac{\vartheta_4(\tau+1)}{\eta(\tau+1)}= e^{-i\pi/12}\frac{\vartheta_3(\tau)}{\eta(\tau)}.
    \end{aligned}
\end{align}
If one considers a single Majorana fermion charged under both $(-1)^F$ and $(-1)^Q=(-1)^{F_L}$, its total contribution to the partition function in $\mathcal{H}_{\ns 1}$ will be
\begin{equation}
    \begin{tikzpicture}[scale=0.3,baseline= {($(current bounding box.base)-(-5pt,-5pt)$)}]
    \draw (0,0) rectangle (2,2);
    \node[left] at (0,1) {$\ns 0$};
    \node[below] at (1,0) {$\ns 1$};
   \end{tikzpicture} =
    \begin{tikzpicture}[scale=0.3,baseline= {($(current bounding box.base)-(-5pt,-5pt)$)}]
    \draw (0,0) rectangle (2,2);
    \node[left] at (0,1) {$1$};
    \node[below] at (1,0) {$0$};
   \end{tikzpicture}\cdot
   \overline{
    \begin{tikzpicture}[scale=0.3,baseline= {($(current bounding box.base)-(-5pt,-5pt)$)}]
    \draw (0,0) rectangle (2,2);
    \node[left] at (0,1) {$1$};
    \node[below] at (1,0) {$1$};
   \end{tikzpicture}}\quad\overset{T^2}{\longrightarrow}\quad e^{i \pi/4}\begin{tikzpicture}[scale=0.3,baseline= {($(current bounding box.base)-(-5pt,-5pt)$)}]
    \draw (0,0) rectangle (2,2);
    \node[left] at (0,1) {$1$};
    \node[below] at (1,0) {$0$};
   \end{tikzpicture}\cdot
   \overline{
    \begin{tikzpicture}[scale=0.3,baseline= {($(current bounding box.base)-(-5pt,-5pt)$)}]
    \draw (0,0) rectangle (2,2);
    \node[left] at (0,1) {$1$};
    \node[below] at (1,0) {$1$};
   \end{tikzpicture}},
\end{equation}
signaling that under stacking it increases the anomaly $\nu\mapsto \nu +1 \mod 8$. Instead, by considering it to be charged under $(-1)^Q=(-1)^{F_R}$ rather than $(-1)^{F_L}$, one gets a decrease in the anomaly $\nu \mapsto \nu-1 \mod 8$.
Therefore, for a free fermion theory with $c=n/2$ we can have arbitrary anomaly $\nu=\tilde{\nu}_L-\tilde{\nu}_R \mod8$, where $\tilde{\nu}_{L,R}\le n$ are the number of left/right-moving fermions charged under $(-1)^{Q}$.

We start to look at the case $c=1$. This means the possible anomalies are $\nu=0,\pm 1,\pm 2\mod 8$. The partition functions \eqref{eq:Z+F+-Q} for $\nu=0,1,2$ are the following:
\begin{align}
    \nu=0\,:&\qquad\left\{\begin{aligned}
    &Z_{\ns0}^{+F+Q} =\frac{1}{2}\frac{|\vartheta_3(\tau)|^2+|\vartheta_4(\tau)|^2}{|\eta(\tau)|^2},\\
    &Z_{\ns0}^{+F-Q}=0,
    \end{aligned}\right.\\
    \nu=1\,:&\qquad\left\{\begin{aligned}
    &Z_{\ns0}^{+F+Q} =\frac{1}{4}\frac{(\sqrt{\vartheta_3(\tau)}+\sqrt{\vartheta_4(\tau)})(\overline{\vartheta_3(\tau)}\sqrt{\vartheta_3(\tau)}+\overline{\vartheta_4(\tau)}\sqrt{\vartheta_4(\tau)})}{|\eta(\tau)|^2},\\
    &Z_{\ns0}^{+F-Q} =\frac{1}{4}\frac{(\sqrt{\vartheta_3(\tau)}-\sqrt{\vartheta_4(\tau)})(\overline{\vartheta_3(\tau)}\sqrt{\vartheta_3(\tau)}-\overline{\vartheta_4(\tau)}\sqrt{\vartheta_4(\tau)})}{|\eta(\tau)|^2},
    \end{aligned}\right.\\    
    \nu=2\,:&\qquad\left\{\begin{aligned}
    &Z_{\ns0}^{+F+Q} =\frac{1}{4}\left\lvert\frac{\vartheta_3(\tau)+\vartheta_4(\tau)}{\eta(\tau)}\right\rvert^2,\\
    &Z_{\ns0}^{+F-Q} =\frac{1}{4}\left\lvert\frac{\vartheta_3(\tau)-\vartheta_4(\tau)}{\eta(\tau)}\right\rvert^2.
    \end{aligned}\right.
\end{align}
The partition functions for $\nu=-1,-2$ cases are complex conjugated partition functions for $\nu=1,2$ respectively.

Expanding them in $q$ we learn that the lightest primary states are:
\begin{table}[h]
    \centering
    \begin{tabular}{lcc}
         $\nu$& $\mathcal{H}_{\ns0}^{+F+Q}$ & $\mathcal{H}_{\ns0}^{+F-Q}$  \\
         \hline
         \hline
         0&$(1/2,1/2)$&$\times$\\
         $\pm 1$ &$(1/2,1/2)$&$(1/2,1/2)$\\
         $\pm 2$&$(1,0),(0,1)$&$(1/2,1/2)$\\
         \hline
    \end{tabular}
    \label{tab:boundsc1}
\end{table}\\
As anticipated, these values almost saturate the bounds found from the numerical analysis presented in Figures \ref{fig:J1} and \ref{fig:J3}. Note that for $\nu=2$ we have a conserved current state. However, this is not unexpected, as its mix with the vacuum $(h,\bar{h})=(0,0)$ defines a non-degenerate Virasoro character of generic type.

Next we focus on the kinks we found at $c=n/2=4\pm \nu/2$ for $\mathcal{H}_{\ns0}^{+F+Q}$. These are simply a set of $n=8\pm\nu$ Majorana fermions charged under $(-1)^Q=(-1)^{F_L}$ for $n=8+\nu$ and $(-1)^Q=(-1)^{F_R}$ for $n=8-\nu$. For example, in light of what we said, the partition function for $c=4+\nu/2$ is just 
\begin{equation}
    Z_{\ns0}^{+F+Q}=\frac{1}{4}\left\lvert\frac{\vartheta_3^c(\tau)+\vartheta_4^c(\tau)}{\eta(\tau)}\right\rvert^2
\end{equation}
and its expansion confirms that the lightest scalar is indeed a marginal operator with $\Delta=2$.

Finally, we note that the set of $n=4$ real fermions charged under $(-1)^Q=(-1)^{F_L}$ describes also the kink at $c=2$ for $v=4 \mod 8$. In fact in this case the partition function is \begin{equation}
    Z_{\ns0}^{+F-Q}=\frac{1}{4}\left\lvert\frac{\vartheta_3^2(\tau)-\vartheta_4^2(\tau)}{\eta(\tau)}\right\rvert^2,
\end{equation}
so the lightest non-degenerate primary has again $(h,\bar{h})=(1/2,1/2)$.

\bibliographystyle{JHEP}
\bibliography{cob-mod}

\end{document}